%% file: thesis.tex
\documentclass[a4paper,oneside,12pt]{ociamthesis}

\usepackage{etoolbox}

\AtBeginEnvironment{subappendices}{%
  \addtocontents{toc}{\protect\addvspace{10pt}Chapter Appendices}
}
\makeatletter
\patchcmd{\@chapter}{\protect\numberline{\thechapter}#1}
{\@chapapp~\thechapter: #1}{}{}
\makeatother

\usepackage[
	style=authoryear, 
	backend=bibtex, 
	doi=false, 
	isbn=false, 
	url=false, 
	maxcitenames=2, 
	natbib=true
]{biblatex}

\DeclareFieldFormat{citehyperref}{%
  \DeclareFieldAlias{bibhyperref}{noformat}
  \bibhyperref{#1}}

\DeclareFieldFormat{textcitehyperref}{%
  \DeclareFieldAlias{bibhyperref}{noformat}
  \bibhyperref{%
    #1%
    \ifbool{cbx:parens}
      {\bibcloseparen\global\boolfalse{cbx:parens}}
      {}}}

\savebibmacro{cite}
\savebibmacro{textcite}

\renewbibmacro*{cite}{%
  \printtext[citehyperref]{%
    \restorebibmacro{cite}%
    \usebibmacro{cite}}}

\renewbibmacro*{textcite}{%
  \ifboolexpr{
    ( not test {\iffieldundef{prenote}} and
      test {\ifnumequal{\value{citecount}}{1}} )
    or
    ( not test {\iffieldundef{postnote}} and
      test {\ifnumequal{\value{citecount}}{\value{citetotal}}} )
  }
    {\DeclareFieldAlias{textcitehyperref}{noformat}}
    {}%
  \printtext[textcitehyperref]{%
    \restorebibmacro{textcite}%
    \usebibmacro{textcite}}}

\renewcommand*{\bibfont}{\raggedright\small}
\title{\LARGE \sc Cosmic structure formation in \\ the non-linear regime \\ \large  \sc Beyond Gaussian Statistics and Standard Cosmologies}
\author{Alexander Gough}

\degree{Doctor of Philosophy}
\degreedate{April 2024}

\include{mymacros}
\usepackage{titlesec}
\titlespacing*{\section}
{0pt}{1.5ex}{1ex}

\definecolor{blue}{HTML}{284FAF}
\definecolor{mygrey}{HTML}{AAAAAA}

\usepackage{hyperref}
\hypersetup{
colorlinks=true,
linkcolor=blue,
filecolor=red,
citecolor=blue,
urlcolor=blue
}

\begin{document}


\setlength{\textbaselineskip}{15pt plus1pt minus1pt}

\setlength{\frontmatterbaselineskip}{15pt plus1pt minus1pt}

\setlength{\baselineskip}{\textbaselineskip}

\setlength\abovedisplayskip{5pt}
\setlength\belowdisplayskip{5pt}


\setcounter{secnumdepth}{2}
\setcounter{tocdepth}{1}


\begin{romanpages}

\maketitle


\begin{abstract}
	\input{text/abstract}
\end{abstract}
%
%
\input{text/statement_of_pubs}

\input{text/acknowledgements}

\dominitoc 

\flushbottom

\tableofcontents

\listoffigures
	\mtcaddchapter

\listoftables
	\mtcaddchapter


\end{romanpages}


\flushbottom

\renewcommand{\chapterheadstartvskip}{\vspace*{-60pt}}
\renewcommand{\chapterheadendvskip}{\vspace{10pt}}

\part{Background}\label{part:background}
\include{text/chapter1-intro}

\include{text/chapter-2-part1-homog-cosmo}
\include{text/chapter2-structure-formation}

\part{Clustering Statistics}\label{part:statistics}
\include{text/chapter3-LDT-for-cosmology}

\include{text/chapter4-LDT-for-MG}
\include{text/chapter5-covariances-of-pdf}

\part{Clustering Dynamics}\label{part:dynamics}
\include{text/chapter7-wave-dynamics-in-1D}
\include{text/chapter8-how-classical-is-FDM}

\include{text/chapter9-conclusions}


\startappendices
\appendix
\begin{appendices}
\include{text/appendix-master}

\end{appendices}

\renewcommand{\chapterheadstartvskip}{\vspace*{40pt}}
\renewcommand{\chapterheadendvskip}{\vspace{20pt}}
\setlength{\baselineskip}{0pt} 

\renewcommand*{\bibfont}{\footnotesize}

{\renewcommand*\MakeUppercase[1]{#1}%
\printbibliography[heading=bibintoc,title={References}]}


%

\end{document}

%% file: mymacros.tex
\usepackage{amsfonts}
\usepackage{amsmath}
\usepackage{amssymb}
\usepackage{amsthm}
\usepackage{mathrsfs} 
\usepackage{appendix} 
\usepackage{booktabs}
\usepackage{bm}
\usepackage{caption}
\usepackage{color}
\usepackage{enumitem}
\usepackage{graphicx}
\graphicspath{ {figures/} } 
\usepackage{float}
\usepackage{subcaption}
\usepackage{physics}
\usepackage{tikz}
\usepackage{tkz-euclide}

\usepackage{cancel}

\definecolor{darkcyan}{rgb}{0,0.6,0.6}

\newcommand{\refsec}[1]{Section~\ref{#1}}
\newcommand{\refeqn}[1]{Equation~\ref{#1}}

	\renewcommand{\th}{\textsuperscript{th}} 

   \newcommand{\LCDM}{$\Lambda$CDM}

\newcommand{\hypgeo}[1]{{}_0F_1\left( #1 \right)} 

\newcommand{\seteq}{\overset{!}{=}} 
\makeatletter
\newcommand*{\defeq}{\mathrel{\rlap{%
                     \raisebox{0.3ex}{$\m@th\cdot$}}%
                     \raisebox{-0.3ex}{$\m@th\cdot$}}%
                     =}
\makeatother 
\renewcommand{\leq}{\leqslant}

\newcommand{\paren}[1]{\left(#1 \right)}
\newcommand{\brac}[1]{\left[ #1 \right]}
\newcommand{\curly}[1]{\left\{ #1 \right\}}

\newcommand{\del}{\partial}
\newcommand{\Del}{\nabla}

\newcommand{\Lap}{\nabla^2}

\newcommand{\dnx}{\dd[n]{\bm{x}}}

\newcommand{\dnkk}{\frac{\dd[n]{\bm{k}}}{(2\pi)^n}}
\newcommand{\dkk}{\frac{\dd{k}}{2\pi}}

\newcommand{\FFT}[1]{\mathcal{F}\left[#1\right]}
\newcommand{\IFFT}[1]{\mathcal{F}^{-1}\left[#1\right]}

\newcommand{\kk}{\bm{k}}
\newcommand{\xx}{\bm{x}}
\newcommand{\pp}{\bm{p}}
\newcommand{\uu}{\bm{u}}
\newcommand{\vv}{\bm{v}}
\newcommand{\xxi}{\bm{\xi}}
\newcommand{\mP}{\mathcal{P}} 

\newcommand{\pyLDT}{\texttt{pyLDT}}
\newcommand{\Mpch}{$\mathrm{Mpc}/h$}

\newcommand{\fR}[1]{|f_{R#1}|}

\newcommand{\Om}{\Omega_{\mathrm{m}}}

\newcommand{\Orc}{\Omega_{\mathrm{rc}}}

\newcommand{\newtau}{\delta_{\rm L}}

\makeatletter
\DeclareRobustCommand{\cev}[1]{%
  \mathpalette\do@cev{#1}%
}
\newcommand{\do@cev}[2]{%
  \fix@cev{#1}{+}%
  \reflectbox{$\m@th#1\vec{\reflectbox{$\fix@cev{#1}{-}\m@th#1#2\fix@cev{#1}{+}$}}$}%
  \fix@cev{#1}{-}%
}
\newcommand{\fix@cev}[2]{%
  \ifx#1\displaystyle
    \mkern#23mu
  \else
    \ifx#1\textstyle
      \mkern#23mu
    \else
      \ifx#1\scriptstyle
        \mkern#22mu
      \else
        \mkern#22mu
      \fi
    \fi
  \fi
}

\makeatother

\newcommand{\R}{\mathbb{R}}

\newcommand{\Z}{\mathbb{Z}}

\newcommand{\cov}{\operatorname{cov}}

%% file: text/abstract.tex
The cosmic large scale structure encodes the formation and evolution of a weblike network of dark matter and galaxies within the Universe. The cosmological information is wrapped up in non-Gaussian statistics requiring characterisation beyond two-point correlations. Accurate modelling of these non-Gaussian statistics and the underlying non-linear dynamics of gravitational collapse are key to extracting maximal information from ongoing and upcoming cosmological surveys.

This thesis centres on questions relating to clustering statistics, dynamics, and fundamental physics:
\begin{enumerate}[label=\Alph*.]
\item How can we efficiently characterise the statistics of the late time matter field?
\item How can we capture the non-linear phase-space dynamics of gravitational collapse?
\item How do changes to fundamental physics impact those clustering statistics and dynamics?
\end{enumerate}
Specifically we present four aspects addressing these questions:
\begin{enumerate}
\item We demonstrate the probability distribution function (PDF) of the matter density can be accurately predicted in modified gravity and dynamical dark energy models, and that it provides good complementarity to standard two-point analyses for detecting these features.

\item We demonstrate the joint PDF of densities in two cells can be used to predict the covariance of the one-point PDF in simple clustering models, providing estimates of the density dependent clustering and super-sample covariance missed in cosmological simulations.

\item We use a wave-based forward model of dark matter to demonstrate its capability to encode the full phase-space dynamics beyond a perfect fluid and determine certain universal scaling features in such models.

\item Using the wave dark matter forward model we analyse one-point statistics to complement existing analytic and numerical approaches in studying fundamentally wavelike dark matter.
\end{enumerate}

%% file: text/statement_of_pubs.tex
\section*{Statement of publications}

This thesis is based in part on the following papers. The first four are published, and the fifth is submitted at the time of writing.
\begin{enumerate}
\item \href{https://doi.org/10.1093/mnras/stac904}{\textit{The matter density PDF for modified gravity and dark energy with large deviations theory.}}   M. Cataneo, C. Uhlemann, C. Arnold, \textbf{A. Gough}, B. Li, C. Heymans, 2022 Monthly Notices of the Royal Astronomical Society, 513, 1623, arXiv: \href{https://arxiv.org/abs/2109.02636}{2109.02636 }
\item \href{https://doi.org/10.3390/universe8010055}{\textit{One-point statistics matter in extended cosmologies}.} \textbf{A. Gough} and C. Uhlemann, 2022 Universe, 8, 55, arXiv: \href{https://arxiv.org/abs/2112.04428}{2112.04428}
\item \href{https://doi.org/10.21105/astro.2206.11918}{\textit{Making (dark matter) waves: Untangling wave interference for multi-streaming dark matter}.} \textbf{A. Gough} and C. Uhlemann, 2022 The Open Journal of Astrophysics, 5, 14, arXiv: \href{https://arxiv.org/abs/2206.11918}{2206.11918}
\item \href{https://doi.org/10.21105/astro.2210.07819}{\textit{It takes two to know one: Computing accurate one-point PDF covariances from effective two-point PDF models}.} C. Uhlemann, O. Friedrich, A. Boyle, \textbf{A. Gough}, A. Barthelemy, F. Bernardeau, S. Codis, 2023 The Open Journal of Astrophysics, 6, 1, arXiv: \href{https://arxiv.org/abs/2210.07819}{2210.07819}
\item \href{https://arxiv.org/abs/2405.15852}{\textit{When to interfere with dark matter? The impact of wave dynamics on statistics}.} \textbf{A. Gough} and C. Uhlemann, arXiv: \href{https://arxiv.org/abs/2405.15852}{2405.15852}
\end{enumerate}
Figures in this thesis which are published elsewhere are cited in the corresponding captions. The work presented in this thesis is my own, with the following clarifications:

\begin{itemize}
\item Papers 1 and 2 are based on the same work, with the latter being a conference proceeding I wrote about paper 1. In paper 1 I performed the Fisher analysis and wrote the relevant sections of the paper, but was not involved with the simulation data (developed by the Durham team) or the code development of \texttt{pyLDT} (by M. Cataneo). I wrote the text of paper 2, incorporating input from C. Uhlemann and the referee. Chapter~\ref{chap:MG-PDFs} is based on these works (with Chapter~\ref{chap:LDT-intro} introducing the formalism used).

\item Paper 3 was led by me: I performed all the calculations, generated all the figures, and wrote all the text of the paper. C. Uhlemann provided guidance and feedback at various stages which was incorporated into the final paper. This work is presented in Chapter~\ref{chap:making-dm-waves}.

\item I contributed work to paper 4 and which uses a method established in \textcite{Bernardeau_2022_CovariancesDensity}. I contributed to calculations in Sections 3 and 5 of paper 4, involving the bias in the large separation limit and the 3D matter clustering. I also derived the closed form of the bias functions in the minimal tree model, presented in Appendix A of paper 4, and wrote that section of the paper. Chapter~\ref{chap:covariance} presents this work, though in the context of 3D matter clustering rather than weak lensing convergence which makes up the majority of paper 4. 

\item Paper 5 is not published, but has been submitted for review at the time of writing, with the pre-print linked above. This paper was lead by me: I performed all the calculations, ran the forward models, wrote all the code for extracting statistics, generated all the figures, and wrote the text of the paper. C. Uhlemann provided guidance and feedback at various stages which was incorporated into the final paper. This work is presented in Chapter~\ref{chap:how-classical}.

\end{itemize}

%% file: text/acknowledgements.tex
%


\begin{alwayssingle}
\begin{savequote}[8cm]
\textit{This is what a forest is, after all. Don't believe the lie of individual trees, each a monument to its own self-made success. A forest is an interdependent community. Resources are shared, and life in isolation is a death sentence.}
  \qauthor{--- Becky Chambers \textit{To be Taught, if Fortunate}}
\end{savequote}

\chapter*{Acknowledgements}

This is the first section of my thesis that I started writing, and the last and most difficult to finish. My PhD started about 6 months after Covid-19 was declared a pandemic, and as such I did not move to Newcastle for the entire first year of my PhD. I knew then, even more than usual, that getting to the end of this process would rely heavily on support and contributions large and small from other people. That said, like any acknowledgements section, it is not possible to adequately thank all of the people whose paths have intersected in a positive way with mine, so I hope those not acknowledged here know how much they mean to me.

The funding for my PhD was supported by the Engineering and Physical Sciences Research Council from UK Research \& Innovation, and I am grateful for their support in conducting this research.

First and foremost I'd like to thank my supervisor Cora Uhlemann for all of the support, advice, knowledge, and skills you have shared with me over the last four years. I am extremely grateful to have had your guidance through this journey. I could not have asked for a better supervisor and I'm honoured to have been your first full PhD student, I'm sure you'll go on to have many more. You have continued to build a wonderful little group and facilitate an inclusive atmosphere. Thank you to everyone in our group: Beth, Lina, and Carolyn, who have all been good friends to me, and I'm sure will go on to do great things. I hope parts of this thesis will be helpful to some of you when it comes time to write yours.

I have loved working with the entire Cosmology and Astro-Obs groups at Newcastle. It has been exciting to be part of such rapidly growing research groups. Thank you to all of the faculty who have provided me with advice over the years. To all the other current and past students from these groups, thank you for building and maintaining such a collaborative atmosphere. I enjoy talking with you all at astro coffee and wish I had had more of a chance to get to know many of you than I did. Thanks to Ashley, Kate, Patrick, Charlie, Carola, and Houda in particular for always providing a good break when I needed it.

Thank you to the various people I have collaborated with over the years for helping me work on my exciting science or letting me work on your exciting science. In particular thank you to Matteo Cataneo for letting me come onto your project in my first year, to really get to grips with proper research, to Aoife Boyle, Alexandre Barthelemy, and Oliver Friedrich for  discussions on statistics things and inviting me to talk about my dynamics work, and to Oliver Hahn and Cornelius Rampf for interesting discussions about theory related things.

Thank you to my friends who have supported me even when I hit long spans of being bad at staying in touch, your love and friendship means the world to me. Thank you to The Jams, Matt, Aoibhinn, and Charis in particular. Thank you to Sayyada for maintaining our D\&D campaign, which has been one of the most stable fixtures in my life over the last several years. I cannot overstate how instrumental this has been for my mental health to know that once a week I will talk, laugh, and cry with my friends, while making questionable magical decisions with no real life consequences. Thank you Rosie (Jo), Kit (Thari), and Robin (Fig) for playing alongside me and Wren. I promise we'll set you free soon Sayyada, one way or another.

I would like to thank Catherine, Jeff, Margo, and Abi for housing me in the early days of the pandemic and for welcoming me into your family. Thank you to also to Margo for always taking an interest in my research, and providing me ample practice in explaining it, especially in the early days of my PhD.

Thank you to my parents for your unending support and for raising me to value curiosity and collaboration, both of which have been essential in completing this work. Thank you to my sister Lucy for always providing a laugh when I needed one, and for the continual Simon updates. Thank you to my grandparents Sophie, Colin, Ruth, Geoff, Jim (Boppi), and Laura, for your role in shaping my values and outlook on life, and to the new grandparents I've acquired Margo, Carol, and Doug, for being welcoming as I joined your family.

Finally, thank you to my wonderful partner Charlie. You have always supported me in everything I do, and I am delighted to share my life with you. This work is dedicated to you as it would not have been possible without your company.

\end{alwayssingle}

%% file: text/chapter1-intro.tex
\chapter{Introduction}\label{chap:intro}
\minitoc

\section{Historical context}
Cosmology is a subject which seeks to describe the nature of the Universe on the largest length scales and time scales. Despite many of its central questions: \textit{What is the Universe made of? How old is the Universe (if it has an age)? Why does the night sky look the way it does?} being pondered deeply by people for millennia, it is only relatively recently that some of them can be answered based on scientific observation.

While the movements of the planets in the solar system have been observed for thousands of years, it wasn't until the $17^{\rm th}$ century that a consistent theory of gravitation could explain their motion around the Sun. This theory of Newtonian gravity was built upon in 1915 by Einstein in his development of General Relativity, which remains the most successful description of gravity to date. Data about the larger structure of the Universe, beyond planetary orbits and the dynamics of individual stars, didn't begin until the 1920s, when it was discovered that our galaxy, the Milky Way, is just one galaxy among many in the Universe, each with billions of their own stars. Measurements of the velocities of these other galaxies revealed that the Universe is expanding \parencite{1929PNAS...15..168H}. This established the beginnings of observational cosmology, with the large scale Universe being a dynamical object, rather than a static and eternal. 

In the century since this discovery, our understanding of the history, geometry, and content of the Universe has developed dramatically. In 1964 Penzias and Wilson detected a nearly uniform signal of microwaves from all directions in the sky \parencite{Penzias1965ApJ}. This Cosmic Microwave Background (CMB) is now known to be a snapshot of the baby Universe, light which has been travelling for nearly 14 billion years back from when the Universe was much smaller and much hotter than it is today.  Along with our understanding of fundamental particle physics, the observations of cosmic expansion and this CMB radiation signal form the basis of the Hot Big Bang model of the Universe. 

Measurements in the 1960s and 70s of the rotations of galaxies hinted that there must be additional matter in the Universe beyond that which we can see \parencite{Rubin1970ApJ, Freeman1970ApJ}. The evidence for this ``dark matter'' component of the Universe has continued to build since then. Measurements across a large range of scales and during different epochs of the Universe's history all point  towards the same amount of additional non-luminous matter. In 1998, measurements from extremely distant supernovae provided evidence that not only is the Universe expanding, but that it is expanding faster now than it was in the past \parencite{1998ApJ...507...46S, 1998AJ....116.1009R, 1999ApJ...517..565P}. The source of this accelerated expansion is called ``dark energy'' and its fundamental nature remains unknown.

The current standard model of cosmology, called the $\Lambda$CDM model, has been enormously successful at explaining and predicting cosmological observations over the last few decades. It is fundamentally built upon General Relativity extended to the largest scales in the Universe, together with a new form of gravitating matter---cold dark matter (CDM)---and a component which can accelerate the cosmic expansion---dark energy in the form of a cosmological constant, $\Lambda$.

We currently sit are on the verge of a new era of cosmology, with unprecedented large surveys of the Universe currently ongoing or scheduled in the near future, including \textit{Euclid} \parencite{Euclid_mission}, Rubin Observatory LSST \parencite{LSST_mission}, and DESI \parencite{DESI_mission}. These surveys promise an enormous amount of data  which will allow precision measurement of cosmological parameters such as the energy content of the Universe. In addition to tightening in the parameters of the current standard model of cosmology, this era of precision cosmology also has the potential to shine a light on new fundamental physics, deepening our understanding of the Universe as a whole.

\section{Motivation for this work}

\subsection{Information from fluctuations}

\begin{figure}[h!t]
\centering
\includegraphics[width=\textwidth]{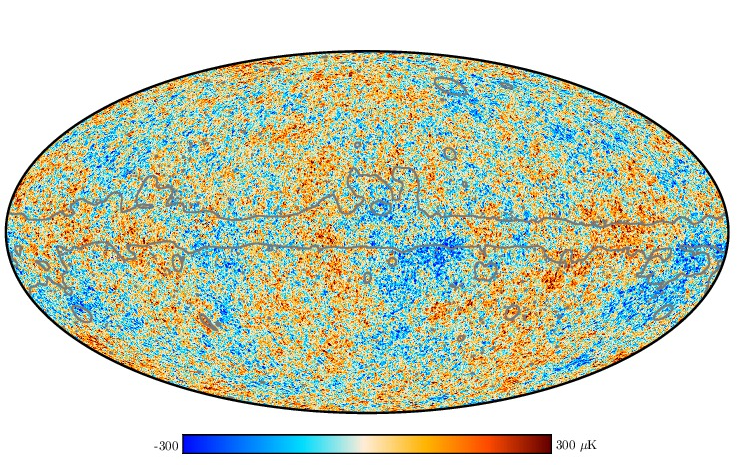}
\caption[Map of the Cosmic Microwave Background. Source: \parencite{PlanckCollaborationOverview}]{Map of the temperature fluctuations in the CMB measured by the \textit{Planck} satellite \parencite{PlanckCollaborationOverview}. The relative size of these fluctuations is $10^{-5}$ and their distribution is measured to be very close to Gaussian. These tiny fluctuations eventually grow to become the structures in the Universe we see in maps of galaxies such as in Figure~\ref{fig:2mass-galaxies}. }
\label{fig:cmb-temp-map}
\end{figure}

The original measurement of  the CMB observed it to be a nearly uniform temperature in all directions across the sky, with later measurements showing it to be a nearly perfect thermal blackbody with a temperature of 2.725 K. However, precision measurements of the CMB by space telescopes reveal small fluctuations about this average temperature, on the order of $10^{-5}$. The most recent map of these anisotropies, measured by the \textit{Planck} telescope \parencite{PlanckCollaborationOverview} are shown in Figure \ref{fig:cmb-temp-map}. The statistical correlations in these small fluctuations encode a huge amount of cosmological information, such that many of the strongest  constraints on cosmological parameters come from these precision measurements of the CMB. 

\begin{figure}[h!t]
\centering
\includegraphics[width=\textwidth]{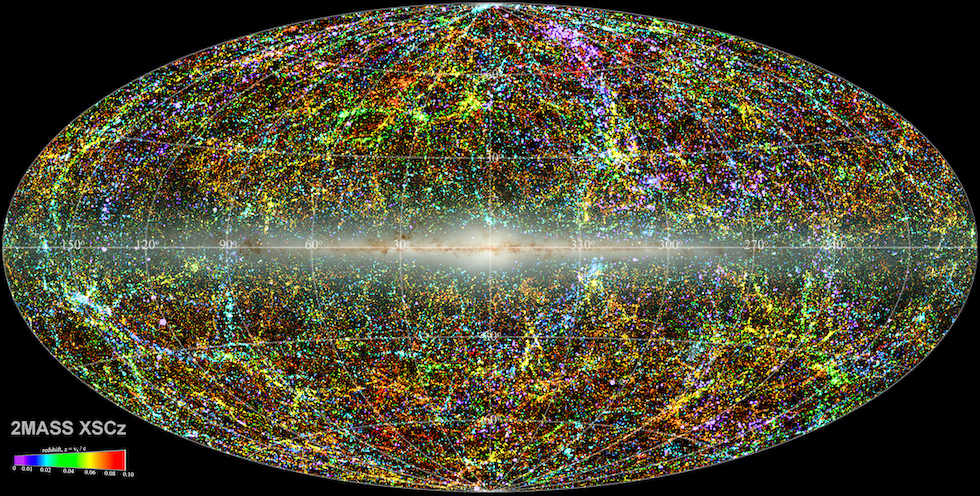}
\caption[Map of the 2MASS galaxy catalogue. Source: \parencite{Jarrett2004PASA}.]{Map of galaxy positions from the 2MASS Extended Source Catalogue with the Point Source Catalogue view of the Milky Way in the centre \parencite{Jarrett2004PASA}. }
\label{fig:2mass-galaxies}
\end{figure}

These tiny fluctuations in the temperature of the CMB correspond to tiny fluctuations in the density of the Universe, and they set the seeds for structure formation in the late Universe. Over time these small fluctuations in density of the early Universe collapse under gravity, drawing in more matter and becoming larger. The dark matter component of the Universe is an essential part of this process, both because it clusters earlier than ordinary matter (as it is not affected by the radiation pressure of the hot early Universe), and because it outweighs ordinary matter roughly five to one. This means that when galaxies and galaxy clusters eventually form, they trace out the skeleton of dark matter. Figure \ref{fig:2mass-galaxies} shows the positions of galaxies in the late Universe, distributed in a rich and complex structure called the cosmic web.

While data from the CMB has been extremely successful in constraining cosmological parameters, observations of the cosmic large-scale structure (LSS) by ongoing and future surveys will contain orders of magnitude more constraining power. This is because  the large-scale structure essentially acts as a video of structure formation over time, recording snapshots of galaxies at different times and tracking the evolution these fluctuations. In contrast, the CMB records only a single moment in the Universe's history, much like a single photograph.

Accurately extracting this additional information from observations of the LSS  is challenging however. Even by eye it is possible to see a qualitative difference in the statistical fluctuations in the CMB and the large-scale structure of the late Universe by comparing Figures \ref{fig:cmb-temp-map} \& \ref{fig:2mass-galaxies}. The fluctuations of the CMB are in some sense extremely simple, being extremely close to those of a Gaussian random field. Just as a zero-mean Gaussian distribution is fully characterised by its variance, a zero-mean Gaussian random field is fully characterised by its two-point statistics. Precise measurements of the CMB power spectrum (the Fourier space version of a two-point correlation function) therefore capture all of the statistical information available. In contrast, the distribution of matter in the late-time Universe is not well approximated by a Gaussian field. This is because the dynamics which govern gravitational collapse are non-linear, so even a perfectly Gaussian field will evolve into a non-Gaussian one.

\subsection{Research themes}

\begin{figure}[h!t]
\centering
\includegraphics[width=0.6\textwidth]{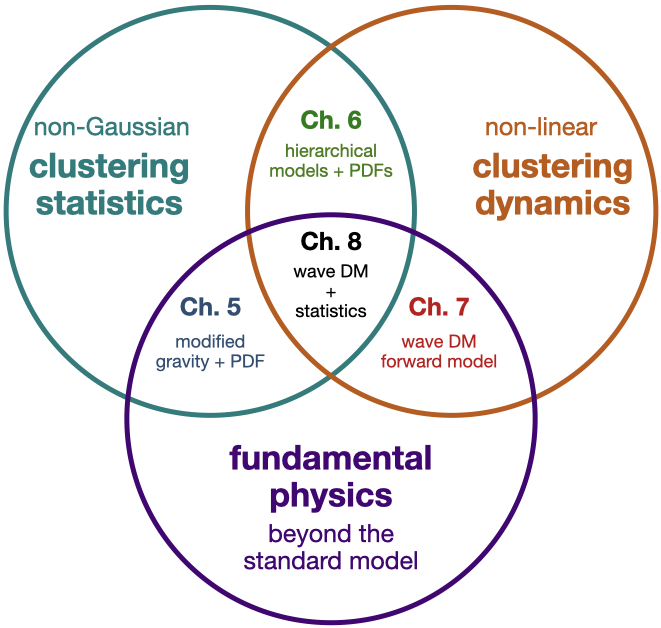}
\caption[Diagram of the main research themes of this thesis.]{Main themes of research in this thesis.}
\label{fig:venn_intro}
\end{figure}

Some of the main observables from cosmological surveys of large-scale structure are the positions and shapes of galaxies. These can be used to construct various statistics which are sensitive to cosmological parameters or new fundamental physics. Understanding these observables relies on accurate modelling of the total matter (both ordinary and dark matter)  field. Galaxies act as biased tracers of the underlying dark matter field, while gravitational lensing, which distorts and magnifies the light from distant galaxies, is sensitive to the total matter distribution. This thesis centres on three related problems in the late-time matter density field:

\begin{enumerate}
\item How can we efficiently characterise the statistics of the late-time non-Gaussian distribution of matter beyond its two-point statistics?
\vspace{-8pt}
\item How can we better model and solve the non-linear dynamical equations related to the gravitational collapse of matter?
\vspace{-8pt}
\item How do changes to fundamental physics change the late-time Universe? What statistics are sensitive to various extensions of $\Lambda$CDM, and how do different fundamental models of dark matter impact statistics already in use by cosmologists?
\end{enumerate}

The research Chapters in this thesis fall into different intersections between these research themes, summarised in Figure~\ref{fig:venn_intro}.

The particular statistic we make use of throughout is the probability distribution function (PDF) of the matter in spheres of radius $R$. This matter PDF is essentially a one-point function on the smoothed density field, which makes it relatively simple to measure and construct. We will show it can be accurately predicted in a wide variety of cosmologies in the quasi-linear regime. The matter PDF can be further compressed into its cumulants (the variance, skewness, kurtosis etc) which are simple averages of powers of the smoothed density field. We are interested in how new fundamental physics interacts with these statistics in two ways. First, given we have accurate predictions for these non-Gaussian statistics in the $\Lambda$CDM model, can we extend those to predict the same statistics beyond $\Lambda$CDM. For example, can they be predicted in the presence of modified gravity or different types of dark matter? Secondly, if these statistics can be applied in such cases, are they modified from their $\Lambda$CDM behaviour in such a way that we could detect the presence of this non-standard physics? If so, how do they compare to standard statistics in their constraining power?

In addition to predicting summary statistics of the cosmic density field, it is desirable to push our modelling of cosmic structure towards ``field-level'' analysis by better modelling the non-linear dynamics involved in structure formation. As we'll see in Chapter~\ref{chap:structure-formation}, the phase-space dynamics of a cold dark matter fluid involve an infinite hierarchy of (momentum) cumulants. Typically we truncate this hierarchy and model only the lowest order cumulants: the density, velocity, velocity dispersion etc. The research presented in Part \ref{part:dynamics} involves a forward modelling technique inspired by a classical-quantum correspondence, where the dark matter fluid is replaced by a wavefunction constructed to reproduce the CDM fluid properties in the semiclassical limit. This acts as an alternative closure scheme to the cumulant hierarchy of CDM dynamics, allowing us to push the analytic modelling of dark matter further into the non-linear regime. While this wavefunction picture of dark matter can be used simply as a modelling trick to go beyond linear behaviour, it is closely related to a certain classes of dark matter candidates such as ultralight axions. In such models, the dark matter field is fundamentally described by a coherent wavefunction which exhibits wave behaviour on astrophysical scales. Such ``fuzzy'' dark matter models are motivated by both particle physics models and small scale challenges to the $\Lambda$CDM model.

\section{Organisation of this thesis}


\textbf{Chapters \ref{chap:homog-cosmo} \& \ref{chap:structure-formation}} present standard cosmological theory, with  Chapter \ref{chap:homog-cosmo} focusing on the homogeneous ``background'' level cosmology while Chapter \ref{chap:structure-formation} presents the theory relevant for describing structure formation, including various perturbation theory schemes, the non-linear evolution of spherical collapse, and some essential statistical tools for characterising the cosmic density field. \\

\noindent \textbf{Chapter~\ref{chap:LDT-intro}} presents the model used for predicting the matter PDF, based on the mathematics of large deviations theory (LDT). We present an overview of the key definitions and theorems from LDT before presenting how the formalism is applied to the cosmic matter density. \\

\noindent \textbf{Chapter~\ref{chap:MG-PDFs}} is based on \textbf{\textcite{Cataneo.etal_2022_MatterDensity}} and \textbf{\textcite{Gough.Uhlemann_2022_OnePointStatistics}} which describe the same research work. We demonstrate that the LDT model of the matter PDF can be accurately applied in extended cosmological models (those with evolving dark energy and modified gravity). We then forecast how well we could detect the presence of these cosmologies beyond standard $\Lambda$CDM with access to a \textit{Euclid}-like survey volume. \\

\noindent \textbf{Chapter~\ref{chap:covariance}} is based on  \textbf{\textcite{Uhlemann.etal_2023_ItTakes}}, which presents a model for analytically computing covariance matrices for one-point statistics based on effective modelling of the joint PDF of two-cells. Accurate covariances for statistics are important for forecasting and parameter inference such as the forecasts presented in Chapter~\ref{chap:MG-PDFs}. We show that this model can be used to capture the ``super-sample covariance'', which simulation informed covariances struggle to capture, as well as deriving analytic results for density dependent ``bias'' functions in a hierarchical model of clustering. \\

\noindent \textbf{Chapter~\ref{chap:making-dm-waves}} is based on \textbf{\textcite{Gough.Uhlemann_2022_MakingDark}}. This presents a detailed analysis of a wave model of dark matter. We show that the interference patterns of the wave model can be ``unwoven'' into classical trajectories corresponding to regions where shell crossing has occurred. We investigate other features related to interference and present universal scaling features near regions of high density (caustics), which are expected to be present in other wave models of dark matter. \\

\noindent \textbf{Chapter \ref{chap:how-classical}} presents further work with the same wave-based perturbation theory as Chapter~\ref{chap:making-dm-waves}, now in three dimensions. At the time of writing this work is unpublished but has been submitted \textbf{\parencite{Gough2024arXiv}}. This Chapter synthesises all of the themes of my research, investigating non-Gaussian statistics in the cosmic web of both cold and wave dark matter. This allows us to separate the impact of dynamics and initial conditions on statistics of the cosmic web. This provides insights into numerical and theoretical approximations in the study of wavelike dark matter. \\

\noindent \textbf{Chapter \ref{chap:conclusions}} provides a summary of the contents of each of the research chapters (\ref{chap:MG-PDFs}--\ref{chap:how-classical}), and general conclusions about this work as a whole. \\

There are several appendices containing either notational conventions or technical derivations presented at the end of the thesis. They are not strictly necessary to follow the main body, however throughout my PhD I found myself thankful for other students who had written down explicit steps for how to do certain ``standard'' calculations, so I have tried to do the same here.

%% file: text/chapter-2-part1-homog-cosmo.tex
\chapter{Homogeneous cosmology}\label{chap:homog-cosmo}
\minitoc

In this Chapter we present the standard theory of homogeneous and isotropic cosmology, as well as a brief timeline of the universe and the parameters which enter into the $\Lambda$CDM model of cosmology. More detailed accounts of this theory can be found in standard introductions to cosmology, e.g. \textcite{Peebles_1980_LargescaleStructure,Mukhanov_2005, Dodelson_moderncosmology, Baumann_2022, Liddle2000cils, Schutz_2022, Choquet2015}, and this Chapter follows many of them in presentation. This homogeneous and isotropic part of the cosmological equations is referred to as the ``background'' cosmology (as opposed to the perturbations which are presented in Chapter \ref{chap:structure-formation}).

\section{General relativity}

Modern cosmology is built upon Einstein's theory of general relativity for gravity. This is a geometric theory which relates the curvature of spacetime to the mass and energy content of the universe. The central geometric object in general relativity is the spacetime metric, $\mathsf{g}$,  a rank 2 symmetric tensor that describes the lengths and angles between vectors in spacetime. In components\footnote{We use the convention of Greek indices running over all spacetime coordinates, while Latin indices run over just spatial components. We also use Einstein's summation convention on spacetime indices.} the infinitesimal spatial separation between two events in spacetime is
\begin{equation}
ds^2 = g_{\mu\nu} \dd{x}^\mu \dd{x}^\nu.
\end{equation}
For example, in Cartesian coordinates and $c=1$ units, flat Minkowski spacetime is described by,
\begin{equation}
g_{\mu\nu}^{\rm Minkowski}=\eta_{\mu\nu} = \mathrm{diag}(-1, 1, 1, 1).
\end{equation}
If we label spacetime events with the position 4-vector $x^{\mu}=(x^0, \xx)$ where $x^0=t$ and $\xx=(x^1, x^2, x^3)$, the spacetime interval between infinitesimally close events is
\begin{equation}
ds^2=-\dd{t}^2 + \delta_{ij}\dd{x}^i\dd{x}^j.
\end{equation}

In general relativity, freely falling test particles move along \emph{geodesics} of the spacetime, curves which generalise the notion of straight lines to curved manifolds. These curves describe the worldlines $X^\mu(\lambda)$ of both massless and massive particles according to the \emph{geodesic equation}
\begin{equation}\label{eq:geodesic-eq}
\dv[2]{X^\mu}{\lambda} + \Gamma^{\mu}_{\alpha\beta}\dv{X^\alpha}{\lambda}\dv{X^\beta}{\lambda} = 0.
\end{equation}
Here $\lambda$ is called an \emph{affine parameter}, which for massive particles can simply be taken as the proper time of the particle, but for massless particles such as photons, generalises the parametrisation such that the geodesic equation takes the same form. The symbols $\Gamma^\mu_{\alpha\beta}$ are known as Christoffel symbols  which encode additional geometric structure on the spacetime manifold (known as a \emph{connection}). In GR, the connection is taken to be the Levi-Civita connection, which can be constructed out of derivatives of the metric
\begin{equation}
\Gamma^\mu_{\alpha\beta} = \frac12 g^{\mu\lambda}\left[\del_\alpha g_{\beta\lambda} + \del_\beta g_{\alpha\lambda} - \del_\lambda g_{\alpha\beta}\right], \quad \del_\alpha \equiv \pdv{}{x^\alpha} \,,
\end{equation}
where $g^{\mu\lambda}$ represents the components of the inverse metric. The geodesic equation \eqref{eq:geodesic-eq} can be alternatively written in terms of the 4-momentum $P^\mu = (E, \bm{P})$,
\begin{equation}\label{eq:geodesic-eq-momentumform}
P^\alpha \dv{P^\mu}{x^\alpha} = - \Gamma^\mu_{\alpha\beta} P^{\alpha}P^{\beta}.
\end{equation}

The curvature of spacetime is encoded in the \textit{Riemann tensor} 
\begin{equation}
R^\alpha_{\phantom{\alpha}\beta\gamma\delta} = \del_\gamma \Gamma^{\alpha}_{\beta\delta} - \del_\beta \Gamma^{\alpha}_{\gamma\delta} + \Gamma^\alpha_{\gamma\epsilon} \Gamma^\epsilon_{\beta\delta} - \Gamma^\alpha_{\delta\epsilon} \Gamma^\epsilon_{\beta\gamma}\,,
\end{equation}
and the related \emph{Ricci tensor} and \emph{Ricci scalar}
\begin{equation}
R_{\mu\nu} = R^\lambda_{\phantom{\lambda}\mu\lambda\nu}\,, \quad R =  g^{\mu\nu}R_{\mu\nu}\,.
\end{equation}

The Einstein field equations relate this spacetime curvature to the distribution of matter and energy through the spacetime, encoded in the stress-energy tensor $T_{\mu\nu}$,
\begin{equation}
G_{\mu\nu} + \Lambda g_{\mu\nu} \defeq R_{\mu\nu} -\frac{1}{2}g_{\mu\nu}R + \Lambda g_{\mu\nu} = 8\pi G T_{\mu\nu}\,.
\end{equation}
The tensor $G_{\mu\nu}$ is called the Einstein tensor, which describes the spacetime geometry. The $\Lambda$ term here is referred to as the ``cosmological constant'' which is able to accelerate the expansion of the universe as we shall shortly see. The Einstein field equations can be derived  via a weak field equivalence to Newtonian gravity, or via an action principle with the Einstein-Hilbert action
\begin{equation}
S = \frac{1}{8 \pi G}\int \dd[4]{x} \sqrt{-g} (R - 2\Lambda) \,,
\end{equation}
where $g$ is the determinant of the metric.

\section{The Friedmann equations}

The Einstein field equations are generally difficult to solve, as they are coupled non-linear partial differential equations which must obey certain constraints to ensure general covariance. As such, spacetimes with a high degree of symmetry are desirable for simplifying the process of finding exact solutions. Fortunately, the universe appears to have some degree of symmetry on large scales, namely that it appears isotropic (the same in all directions). This isotropy is observed in the cosmic microwave background, as well as in the distribution of distant objects such as galaxies and quasars. If we couple this observation with the Cosmological Principle, which assumes that we are not somehow ``special'' observers in the universe, then the spacetime we live in should be isotropic about every point. This results in a space which is homogeneous and isotropic on large scales.

Requiring homogeneity and isotropy selects a preferred foliation of spacetime, one where every spacelike hypersurface of constant time is described by a maximally symmetric 3-dimensional manifold. This most general metric of this form is called the Friedmann-Lema\^{i}tre-Robertson-Walker (FLRW) metric, 
\begin{align}\label{eq:FLRW-metric}
g_{\mu\nu} \dd{x}^\mu \dd{x}^{\nu} &= - \dd{t}^2 + a^2(t) \gamma_{ij}\dd{x}^i \dd{x}^j \\
&= - \dd{t}^2 + a^2(t) \left[\frac{\dd{r}}{1-kr^2} + r^2 \dd{\theta} + r^2 \sin^2\theta\dd{\phi}\right] .
\end{align}
The function $a(t)$ is called the \emph{scale factor}, which encodes how the universe expands over time. We choose to set the value of $a(t_0)=1$ today. The tensor $\gamma_{ij}$ describes the geometry of maximally symmetric spaces corresponding to hypersurfaces of constant $t$. These can be globally flat, positively curved, or negatively curved. In the second line we have written the metric in polar coordinates, and the global curvature $k$ (positive provides a closed, spherical surface, negative provides an open, hyperbolic surface, and zero provides a flat, Euclidean three dimensional space) appears explicitly. The coordinates $\xx$ are called \textit{comoving coordinates}  and the time variable $t$ is referred to as \textit{coordinate time} or \textit{cosmic time}. Current observations indicate that the global curvature of the universe is very close to zero so for the remainder of this work we will specialise to a \emph{flat} universe with $k=0$. This is consistent with the approach taken in the $\Lambda$CDM model, which treats spatial curvature as an extension to the base model.

Another common choice of time variable in cosmology is \textit{conformal time} $\tau$ defined via $\dd{t} = a\dd{\tau}$, in which the form of the FLRW metric becomes\footnote{Noting that $\gamma_{ij}=\delta_{ij}$ for spatially flat universes.}
\begin{equation}
g_{\mu\nu} \dd{x}^\mu \dd{x}^{\nu} = a(\tau)^2 \left[- \dd{\tau}^2 + \delta_{ij}\dd{x}^i\dd{x}^j\right],
\end{equation}
where the scale factor simply multiplies the Minkowski flat spacetime metric. We will use the fairly standard notation
\begin{equation}
\dot{f} = \dv{f}{t}, \quad f' = \dv{f}{\tau},
\end{equation}
throughout this work unless otherwise mentioned. However, the cosmology literature has many different time variables which are useful in different contexts beyond simply these two, including the scale factor $a$ itself (or $\ln(a)$) and the linear growth factor $D_+$ which we will introduce later. Similarly, unlabelled spatial derivatives should be interpreted as with respect to comoving spatial coordinates.

It is worth noting that the comoving coordinates used in the FLRW metric are not physical observables. The physical coordinates are given by $\bm{r}_{\rm phys} =a(t)\xx$. Considering how this physical coordinate of an object changes in time, we can obtain the physical velocity 
\begin{align}
\bm{v}_{\rm phys} &= \dv{\bm{r}_{\rm phys}}{t} = \frac{\dot{a}}{a}\bm{r}_{\rm phys} + a\dot{\xx} \nonumber  \\
&= H\bm{r}_{\rm phys} + \bm{U} \\
&= \mathcal{H}\xx + \bm{U}\,,
\end{align}
where $\bm{U} = a \dot{\xx} = \xx'$ is called the \emph{peculiar velocity} of the object. The term $H=\dot{a}/a$ is called the Hubble parameter, which describes how the size of the universe changes over time (the term $\mathcal{H}=a'/a = aH$ is called the conformal Hubble parameter\footnote{We also note the useful relationships $\dot{f} =a^{-1} f'$ and $ \ddot{f} = a^{-2} [f'' - \mathcal{H} f']$, or $f' = a\dot{f}$ and $f'' = a^2[\ddot{f} + H \dot{f}]$.}). The first term in the physical velocity above is proportional to the distance between the object and the origin, and describes how the object is dragged along with the expansion of the universe in the so called \textit{Hubble flow}.

The FLRW metric \eqref{eq:FLRW-metric} determines the geometry on the left hand side of the Einstein field equations. To determine the form and evolution of the scale factor $a$, we need to supply the matter and energy component of the universe into the right hand side. We consider the stress-energy tensor associated with a perfect fluid with 4-velocity $U^\mu = \dv*{x^\mu}{\lambda}$, energy density $\rho$, and pressure $P$, is
\begin{equation}
T_{\mu\nu} = \left(\rho + P \right) U_\mu U_\nu  + P g_{\mu\nu}\,.
\end{equation}
In the frame where the universe appears isotropic, this fluid should have no 3-velocity, leaving the stress-energy tensor 
\begin{equation}
T^\mu_{\phantom{\mu}\nu} = \mathrm{diag}(-\rho, P, P, P).
\end{equation}
Most simple energy sources can be parametrised by the equation of state $w=\rho / P$. Non-relativistic matter\footnote{Also called ``dust'' to a cosmologist.} corresponds to $w=0$, while radiation corresponds to an equation of state $w=1/3$.

For a universe filled with a single fluid species according to this perfect-fluid stress energy tensor, the Einstein field equations reduce to the Friedmann equations which describe the evolution of the scale factor
\begin{subequations}
\begin{align}
H^2 \defeq \left(\frac{\dot{a}}{a}\right)^2 &= \frac{8\pi G}{3}\rho + \frac{\Lambda }{3} \,, \\
\frac{\ddot{a}}{a} &= -\frac{4\pi G}{3}\rho \left(1+3w\right) + \frac{\Lambda }{3}.
\end{align}
\end{subequations}
The first Friedmann equation determines the dynamics of the Hubble parameter $H$, while the second determines the acceleration of the universe through $\ddot{a}$. For a polytropic fluid, $\rho$ and $w$ should be interpreted as summed over each species (e.g. matter, radiation, dark matter). By defining the density $\rho_{\Lambda} = \Lambda/8\pi G$ and $p_{\Lambda} = -\rho_\Lambda$ (corresponding to $w_\Lambda = -1$), the contribution of the cosmological constant can be brought into the stress-energy term. 

These Friedmann equations can be combined to yield the continuity equation for fluid species $i$
\begin{equation}
\dot{\rho}_i + 3H\rho_i (1+w_i) = 0\,,
\end{equation}
which yields the evolution $\rho_i\propto a^{-3(1+w_i)}$ for the homogeneous background density of different fluids. This continuity equation could alternatively have been derived from the conservation of the stress-energy tensor $\nabla_\mu T^{\mu\nu}=0$.\footnote{The $\nabla_\mu$ here is the \emph{covariant derivative}, here expressed as $\nabla_\mu T^{\alpha\beta} = \del_\mu T^{\alpha\beta} + \Gamma^\alpha_{\mu\gamma}T^{\gamma\beta}+ \Gamma^{\beta}_{\mu\gamma}T^{\alpha\gamma}$.} Non-relativistic (cold) matter ($w=0$) simply dilutes with the volume expansion of the universe $\rho_m\propto a^{-3}$. Radiation and relativistic (hot) matter  ($w=1/3$) by contrast dilute more quickly with cosmic expansion $\rho_r\propto a^{-4}$. The cosmological constant $\Lambda$ is the particular case when $w=-1$ and thus does not dilute its energy density $\rho_{\Lambda}=\rm constant$. Any component with $w<-1/3$ will cause $\ddot{a}$ to be positive, causing accelerated expansion. 

Rather than referring to the absolute density, it is useful to refer to the fractional energy densities of different species by introducing the critical density $\rho_{\rm crit}(a) = 3 H^2(a) / 8 \pi G$, which is the density of a flat universe with no cosmological constant. The value of the critical density today is equal to 
\begin{equation}
\rho_{\rm cr,0}=1.9  \times 10^{-29}  \, h^2\, \mathrm{g/cm^3} = 2.8\times 10^{11} \, h^2 \, M_\odot/{\rm Mpc}^3.
\end{equation}
The fractional density of species $i$ is defined as $\Omega_i(a) = \rho_i(a) / \rho_{\rm crit}(a)$. We often refer to quantities evaluated at the present time ($a=1$) which we denote with a 0 label, e.g. $\Omega_m^0$ is the matter fraction today. Together the first Friedmann equation can be written as 
\begin{align}
H &= H_0\sqrt{\Omega_r^0 a^{-4} + \Omega_m^0 a^{-3} + \Omega_\Lambda^0} \\
&= H_0\sqrt{\Omega_r^0 (1+z)^4 + \Omega_m^0 (1+z)^3 + \Omega_\Lambda^0}\,,
\end{align}
for a universe with radiation, matter, and cosmological constant. In the second line we have written the Hubble parameter in terms of the cosmological redshift $z$, which is related to the scale factor by $a=1/(1+z)$. The present day value of the Hubble parameter is usually parametrised as $H_0 = 100 h \ \rm km/s/Mpc$ where $1 \rm \ Mpc \approx 3.086 \times 10^{22} \ m$.

\section{A timeline of the universe}

With an understanding of the basic dynamics of cosmic expansion, we present a brief timeline of the key events cosmic history according to the current standard model of cosmology: $\Lambda$CDM supplemented with a period of cosmic inflation in the early universe.  

Solving the Friedmann equations with currently measured cosmological parameters predicts a time $t_0\approx 13.7$ billion years ago where the scale factor $a$ vanishes. This would predict a universe of zero size, a spacetime singularity. We will take our measurement of time to begin here, setting $t=0$ at this singularity and expressing the history of the universe forward in time. \\

\noindent \textbf{Quantum gravity?}
It is currently unknown what happened during the period of $t\lesssim 10^{-34}$ seconds. At these very early times, the universe is so small and so hot that a description of quantum gravity is probably necessary to describe this period. The true nature of the spacetime singularity is not known. \\

\noindent  \textbf{Cosmic inflation} 
At $t \gtrsim 10^{-34}$ seconds, it is believed the universe entered a period of exponential expansion known as inflation. This was introduced in \textcite{Guth1981PhRvD} to address two issues in cosmology. 
\begin{itemize}
\item \textit{The flatness problem.} The Friedmann equation including curvature can be written
\begin{equation}
\abs{\Omega(a)-1} = \abs{\Omega_k(a)} =  \frac{\abs{k}c^2}{a^2 H^2}\,,
\end{equation} 
where $\Omega = \sum_{i}\Omega_i$ is the sum of all other energy sources. Given that the present value curvature is close to flat ($\Omega_k^0 = 0.001 \pm 0.002$ from \cite{Planck:2018}), it must have been even smaller in the past. The flatness problem questions the origin of this fine tuning.
\item \textit{The horizon problem.} Measurements of the CMB indicate its temperature is extremely homogeneous. This is surprising as the time the CMB was emitted, it should have consisted of over 40,000 causally disconnected patches which shouldn't have had time to communicate and thermalise with one another. The horizon problem questions how this homogeneity and the ``super horizon'' correlations in the early universe are explained.
\end{itemize}
Inflation proposes a period of time which is dominated by some new scalar field, called the inflaton, which drives exponential expansion of the scale factor, shrinking the Hubble sphere $\mathcal{H}^{-1} = (aH)^{-1}$ for a period of time. This rapid expansion acts to stretch out any initial curvature, providing a nearly spatially flat universe, and can solve the horizon problem by so rapidly inflating causally connected regions that they take up the entire observable universe today.

The physics and details of inflation are still quite speculative, but is the current  framework in setting the initial conditions of the universe. At the end of inflation is a period known as \textbf{reheating} which transfers energy from the inflaton field to the fields of the standard model particles, starting the \textbf{hot big bang}. This period is also thought to provide the origin of fluctuations which will eventually grow to become cosmic structure, by amplifying quantum fluctuations in the early universe to macroscopic scales. \\

\noindent  \textbf{Radiation domination.}
The early universe is dominated by the radiation component of the energy density, such that the scale factor evolves as $a_{\rm RD}(t) \propto t^{1/2}$.  \\

\noindent  \textbf{Phase transitions.} 
After inflation ends, the universe consists of a hot, dense plasma of standard model particles, which cool while the universe expands.  As the universe cools, various phase transitions occur in the fundamental forces. At $t\approx 10^{-11}$ seconds, when the thermal energy of the plasma is $\sim\! 100$ GeV, the electromagnetic and weak nuclear forces separate. At $t\approx 10^{-6}$ seconds ($\sim\! 150$ MeV) the quark-gluon plasma cools enough that it can condense into nucleons, known as the QCD phase transition. \\

\noindent  \textbf{Neutrino decoupling.} 
When the universe has cooled to $\sim\! 1$ MeV,  ($t\approx 1$ second,  $z\approx 6\times 10^9$), neutrinos fall out of equilibrium with the rest of the thermal bath of particles, a process called decoupling. They are the first to do so as they only interact via the weak force, so other particles maintain equilibrium via other interactions. These neutrinos free stream across the universe, creating a cosmic neutrino background (C$\nu$B) analogous to the cosmic microwave background, though it is so low energy it has not yet been detected. \\

\noindent  \textbf{Big bang nucleosynthesis.} At $t\approx 3$ minutes (100 keV, $z=4\times 10^8$) the universe has cooled enough to form light atomic nuclei, mostly in the form of hydrogen, deuterium, helium, and lithium. The abundance ratios of these light nuclei can be accurately predicted and are one of the key successes of the current cosmological model \parencite{Cyburt2016RvMP, Fields2020JCAP, Burns2024EPJC}. From this stage onward things become much better understood and well constrained by current models of physics.  \\

\noindent  \textbf{Matter-radiation equality.} At $t\approx 60,000$ years  ($z\approx 3400$), the universe has expanded sufficiently that the contribution of radiation and matter to the over all energy density is now equal. Going forward, we move from a period of radiation domination into matter domination, were the scale factor grows as $a_{\rm MD}(t) \propto t^{2/3}$. \\

\noindent  \textbf{Recombination and photon decoupling.} After 380,000 years ($z\approx 1100$) the universe has cooled sufficiently that the free electrons can combine with the nuclei to form electrically neutral atoms. This allows the photons to scatter one final time off the charged ions before decoupling and freely streaming throughout the universe. These photons are what we now see as the cosmic microwave background, and after this point the universe is finally transparent to light. 

Before photon decoupling, the dynamics of photons and baryons are tightly-coupled together. In the presence of the small inhomogeneities in the early universe, gravity causes the baryonic matter to infall towards overdensities. It then heats up and is supported by photon pressure, oscillating in sound waves through the primordial plasma. These acoustic oscillations are observed in the shape of the CMB power spectrum, with correlations occurring on certain characteristic scales. After photon decoupling, this oscillatory feature gets imprinted into the matter density at a characteristic length scale, which will act as a sort of ``standard ruler'' and imprint on the statistics of galaxy clustering which happens billions of years later. \\

\noindent  \textbf{Reionisation.} After the CMB is released, the universe mostly consists of neutral gas, which does not emit light.  Unsupported by photon pressure now however, baryonic matter can begin collapsing under gravity. At $t\approx 100$ million years ($z\approx 50$), enough time has passed that the initial perturbations can collapse into the first stars. These stars inject high energy photons back into the universe, ionising the neutral gas throughout the universe. \\

\noindent  \textbf{Structure formation.}  As dark matter does not couple to electromagnetism, it is not supported by the photon pressure before photon decoupling and so falls into the gravitational wells of the early universe, deepening them. Once baryons are free to collapse rather than oscillate, they are drawn to these seeds of structure, collapsing to form galaxies, then galaxy clusters and larger structures. This continues over billions of years, developing the cosmic web of galaxies, which trace the underlying skeleton of dark matter. 

The $\Lambda$CDM model of cosmology has allowed us to accurately predict several features of this large-scale structure. The characteristic scale from baryon acoustic oscillations (BAO) in the early universe can be extracted from the two-point correlation statistics of galaxy positions \parencite{Beutler2017MNRAS, Alam2017MNRAS}. We can also predict and observe the statistical distribution of galaxy positions and shapes due to gravitational lensing, and from those extract the cosmological parameters \parencite[e.g.][]{DES-y3kp:2021, DESICollaboration2024arXiv_cosmoconstraints}. \\

\noindent  \textbf{Matter-$\Lambda$ equality.} At $t\approx 9$ billion years 
($z\approx 0.3$) the energy density of the matter component and the dark energy ($\Lambda$) component of the universe are equal. Going forward we enter the $\Lambda$-dominated era, where the scale factor grows exponentially $a_{\Lambda \rm D}(t) \propto e^{t\sqrt{\Lambda/3}}$. \\

\noindent \textbf{Present day.} We find ourselves at $t\approx 13.8$ billion years 
($z=0$).

\section{Cosmological parameters in $\Lambda$CDM}

The $\Lambda$CDM model of cosmology consists of three different energy components. 
\begin{itemize}
\item \textbf{Matter.} This refers to any component whose energy density scales as $\Omega_m(a)\propto a^{-3}$. This includes \textbf{baryonic}\footnote{To a cosmologist, all standard model interactions are referred to as ``baryonic'' in contrast to the particle physics classification of baryons as composite particles made of quarks. The phrase ``baryonic physics'' is to be contrasted with the ``dark'' physics of dark matter and dark energy. } \textbf{matter}, which describes matter which couples to electromagnetism, and \textbf{cold dark matter} (CDM), some currently unknown matter component which interacts only via gravity.
\vspace{-8pt}
\item \textbf{Radiation.} This refers to any component whose energy density scales as $\Omega_r(a)\propto a^{-4}$. This includes photons and a contribution from light neutrinos. For most of this thesis we ignore the role of radiation, as even by the time of the CMB the universe was dominated by matter, so for structure formation the relevant components are primarily matter and dark energy.
\vspace{-8pt}
\item \textbf{Cosmological constant/Dark energy ($\Lambda$).} This refers to the component whose energy density remains constant over time. Dark energy more generally refers to any component with equation of state $w<1/3$ which provides accelerated expansion.
\end{itemize}
A particular model\footnote{Both when considering $\Lambda$CDM and extensions to it, we will sometimes use the shorthand ``a chosen cosmology'' to mean ``a chosen cosmological model and a chosen set of parameters for that model''.} of $\Lambda$CDM requires 6 parameters: the baryon and total matter densities $\Omega_b, \Omega_m$, two parameters describing the primordial fluctuations $A_s$ (an amplitude) and $n_s$ (the scalar spectral index), the optical depth to reionisation $\tau$, and then one parameter to encode the effect of dark energy, commonly $\Omega_\Lambda$, the Hubble parameter $h$, or something observationally motivated such as the angular acoustic scale $\theta_*$ in CMB observations. The values of these parameters as measured from the CMB temperature anisotropies are shown in Table \ref{tab:planck-cosmo-params}.

\begin{center}
\begin{table}[h!]
\centering
\begin{tabular}{ccccccc}
\hline
$\Omega_c^0$ & $\Omega_b^0$ & $\Omega_r^0$ & $\Omega_\Lambda^0$ & $n_s$ & $A_s$ & $h$ \\
\hline
0.262 & 0.04897 & $3.65\times 10^{-5}$ & 0.6889 & 0.9665 & $3.047\times 10^{10}$ & 0.6766 \\
\hline
\end{tabular}
\caption[Cosmological best fit parameters for the  $\Lambda$CDM model from \textit{Planck}.]{Cosmological parameters from \textit{Planck} temperature and polarisation data + lensing + BAO \parencite{Planck:2018}.}
\label{tab:planck-cosmo-params}
\end{table}
\end{center}

To these standard parameters, some additional useful parameters are sometimes added. The parameter $\sigma_8 \approx 0.81$  \parencite{Planck:2018} is the variance of linear density fluctuations in spheres of radius $8 \ h^{-1} \ \mathrm{Mpc}$  which can be used instead of the primordial amplitude $A_s$ to normalise the size of density fluctuations. This is a particularly useful parameter in studying the late time universe through gravitational clustering and lensing, and the parameter $S_8 = \sigma_8 \sqrt{\Omega_m^0/0.3}$ is in weak tension from measurements of the early (CMB) and later (weak lensing) universe \parencite{Douspis:2019, DiValentino:2021, Perivolaropoulos:2021}. The sum of neutrino masses $\sum m_\nu \approx 0.06 \ \mathrm{eV}/c^2$ is also used as a parameter of a slightly extended $\Lambda$CDM model. Technically neutrinos are not part of the $\Lambda$CDM model, but given they are known to exist their effect on cosmology is important as they imprint a distinctive signature since they begin as relativistic (acting as radiation) then slow and cool to behave as non-relativistic matter.

%% file: text/chapter2-structure-formation.tex

\chapter{Structure formation in cosmology}\label{chap:structure-formation}
\minitoc

The actual universe is of course not perfectly homogeneous and isotropic, and how these inhomogeneities grow into the cosmic large-scale structure we see is the central interest of this work. In this Section we briefly present the relativistic treatment of these perturbations about the homogeneous background cosmology, focusing on a pressureless fluid component, appropriate for describing the matter (dark and baryonic) component of the universe on large scales. For the remainder of this Chapter, and this thesis except where otherwise noted, we work within the Newtonian approximation for both the dynamical equations (which can be approached with Eulerian or Lagrangian perturbation theory, and exactly solved in the case of spherical symmetry) and the statistical description of the matter density field. 

This Chapter largely follows the notation and presentation of \textcite{Ma1995ApJ, Dodelson_moderncosmology, Baumann_2022} for the relativistic treatment, and follows  \textcite{Peebles_1980_LargescaleStructure, Bertschinger_1995_CosmologicalDynamics, Bernardeau:2002} for the Newtonian treatment.

\section{Relativistic perturbation theory}

To describe the inhomogeneous universe we write both the metric and the stress-energy tensor as the sum of a homogeneous and isotropic part plus a perturbation
\begin{subequations}
\begin{align}
g_{\mu\nu}(t,\xx) &= \bar{g}_{\mu\nu} + \delta g_{\mu\nu}(t,\xx) , \\
T_{\mu\nu}(t,\xx) &= \bar{T}_{\mu\nu} + \delta T_{\mu\nu}(t,\xx) ,
\end{align}
\end{subequations}
where $\bar{g}_{\mu\nu}$ and $\bar{T}_{\mu\nu} $ describing the homogeneous and isotropic solutions from Chapter \ref{chap:homog-cosmo}. The procedure is then to expand the equations $\nabla_\mu T^{\mu\nu}=0$ and $G_{\mu\nu}=8\pi G T_{\mu\nu}$ in these perturbations to derive equations for their components order by order in the small fluctuations. In our case we will only consider leading order (linear) contributions from these perturbations. However, since GR requires general covariance, this subject can get more nuanced and calculations complicated quickly even at leading order. There are also a wide variety of (often conflicting) conventions in both signs and letters used for various perturbations, so when comparing results check carefully how they have been defined. Here I mostly follow the notation used in \textcite{Baumann_2022}.

\subsection{Metric perturbations}

The most general perturbation to the FLRW metric (written in conformal time) can be written
\begin{equation}
g_{\mu\nu}\dd{x}^{\mu}\dd{x}^{\nu} = a^2(\tau) \left[-(1+2A)c^2\dd{\tau}^2 + 2 B_i c \dd\tau\dd{x}^i + (\delta_{ij}+2E_{ij})\dd{x}^i \dd{x}^j\right],
\end{equation}
where $A$, $B_i$, and $E_{ij}$ are all functions of space and time. Spatial indices are raised and lowered with $\delta_{ij}$, such that e.g. $B_i = \delta_{ij} B^j$. It is useful to further decompose these perturbations into purely scalar, vector, and tensor parts as follows
\begin{subequations}
\begin{equation}
B_i = \del_i B + \hat{B}_i\,, 
\end{equation}
\begin{equation}
E_{ij} = C \delta_{ij} + \left(\del_i \del_j - \frac{1}{3}\delta_{ij}\nabla^2\right)E + \frac{1}{2}\left(\del_i \hat{E}_j + \del_j \hat{E}_i\right) + \hat{E}_{ij}\,,
\end{equation}
\end{subequations}
where $B, C,E$ are scalar degrees of freedom, $\hat{B}_i, \hat{E}_i$ are vector degrees of freedom, and $\hat{E}_{ij}$ is a tensor degree of freedom. The hatted vectors are divergenceless $\del^i \hat{B}_i =\del^i \hat{E}_i  = 0$, and the tensor degree of freedom is transverse $\del^i\hat{E}_{ij} = 0$ and traceless $\delta^{ij}\hat{E}_{ij}=0$. These cosmological perturbations are thus broken into
\begin{itemize}
\item 4 scalars degrees of freedom: $A, B, C, E$,
\vspace{-8pt}
\item 4 vector degrees of freedom: $\hat{B}_i, \hat{E}_i$ (reduced from 6 by the divergenceless restriction),
\vspace{-8pt}
\item 2 tensor degrees of freedom: $\hat{E}_{ij}$ (reduced from 9 by the transverse and traceless restriction).
\end{itemize}
The equations for scalars, vectors, and tensors from the Einstein field equations do not mix at linear order in these perturbations, allowing them to be treated separately. We will mostly focus on scalar perturbations, though vector perturbations can be important for correctly deriving the Newtonian limit \parencite{Kopp2014JCAP}, and tensor perturbations describe the propagation of gravitational waves.

\subsection{Perturbations to stress-energy}
The perturbations the the stress energy tensor are written
\begin{subequations}
\begin{align}
T^0_{\phantom{0}0} &= -\left(\bar{\rho} + \delta\rho\right) = -\bar{\rho}(1+\delta)\,, \\
T^{i}_{\phantom{i}0} &= -\left(\bar{\rho} + \bar{P} \right) U^i \equiv q^i \,, \\
T^i_{\phantom{i}j} &= \left(\bar{P} + \delta P \right)\delta^i_j + \Pi^i_{\phantom{i}j}, \quad \Pi^i_{\phantom{i}i} = 0 \,,
\end{align}
\end{subequations}
where $U^i$ is the bulk (peculiar) velocity of the fluid and $\Pi^i_{\phantom{i}j}$ is the anisotropic stress. The combination appearing in $T^{i}_{\phantom{i}0}$ defines the momentum density $q^i$. For a polytropic fluid, the stress-energy tensors of each fluid species simply add. The vector components can be decomposed into scalar and divergence free vector parts as
\begin{equation}
U_i = \del_i U + \hat{U}_i, \quad q_i = \del_i q + \hat{q}_i\,.
\end{equation}
We also introduce the \emph{overdensity} $\delta=\delta\rho/\bar{\rho}$ such that the total density is given by $\rho = \bar{\rho}(1+\delta)$. 

\subsection{Choice of gauge}

The perturbations as defined above are not unique, as choosing different coordinates can change perturbations of one type into another, or introduce perturbations which don't actually exist due to a poor coordinate choice. Thus one must be careful not to assign too much direct meaning to the perturbation variables as they have been defined, when working in the full relativistic context.

Consider for example homogeneous FLRW spacetime and change coordinates to a new time coordinate $\tau \to \tilde\tau = \tau + \xi^0(\tau,\xx)$. The homogeneous value of the density then gets perturbed
\begin{align}
\rho(\tau) &= \rho(\tilde\tau - \xi^0(\tilde\tau,\xx)) \nonumber \\
&= \bar{\rho}(\tilde{\tau})-\bar{\rho}'(\tilde\tau) \xi^0(\tilde\tau)\,.
\end{align}
Thus, just by changing our time coordinate, we have introduces something which looks like a density perturbation $\delta\rho = -\bar{\rho}'\xi^0$. This can also be done in reverse, removing a real perturbation in the energy density by choosing our spacelike hypersurface accordingly. 

People approach this gauge problem in two different ways. One approach is to only work with specific combinations of these perturbations which do not depend on the choice of gauge, which allows changing coordinates freely. Some particularly useful gauge invariant quantities are the Bardeen variables \parencite{Bardeen1980PhRvD}
\begin{subequations}
\begin{align}
\Psi_B &= A + \mathcal{H}(B-E') + (B-E')', \quad \hat{\Phi}_i=\hat{B}_i - \hat{E}_i, \quad \hat{E}_{ij}\,, \\
\Phi_B &= -C + \frac{1}{3}\nabla^2E -\mathcal{H}(B-E')\,,
\end{align}
\end{subequations}
which are related to the Newtonian potential, and the comoving density contrast $\Delta$,
\begin{equation}
\bar{\rho}\Delta = \delta \rho + \bar{\rho}'(v + B)\,,
\end{equation}
which is the gauge invariant density perturbation.

The other approach is to choose a gauge and stick to it, tracking perturbations in both the metric and the stress-energy tensor. Here we list a few common gauges, and refer to \textcite{Clifton2020PhRvD} for gauge choices applied to non-linear cosmology.
\begin{itemize}
\item The \textbf{(conformal) Newtonian gauge}, defined by $B=E=0$, such that the perturbed metric takes the form
\begin{equation}
g_{\mu\nu}\dd{x}^\mu\dd{x}^\nu = -(1+2\Psi_B)\dd{t}^2 + a(t)^2 (1-2\Phi_B)\delta_{ij}\dd{x}^i\dd{x}^j.
\end{equation}
Note that the signs of the perturbations in the Newtonian gauge (and which potential is called which) is quite inconsistent across the literature. This gauge is the one we take when discussing relativistic effects, as it clearly relates to the Newtonian limit, and the geometry is straightforward to interpret. We note also the \textbf{Poisson gauge} here, which for scalar perturbations is parametrised
\begin{equation}
g_{\mu\nu}\dd{x}^\mu\dd{x}^\nu = a(\tau)^2[-e^{2\Psi} \dd{\tau}^2 + e^{-2\Phi}\delta_{ij}\dd{x}^i\dd{x}^j]\,,
\end{equation}
which to linear order in metric perturbations is the same as the Newtonian gauge.

\item The \textbf{spatially flat gauge} takes $C=E=0$, removing the spatial perturbation.

\item The \textbf{synchronous gauge} removes the perturbation in the time variable, setting $A=B=0$. 

\item The \textbf{uniform density gauge} sets the total density perturbation to zero, $\delta\rho =0$.

\item The \textbf{comoving gauge} sets the velocity perturbation in the stress-energy tensor to zero, $U+B=0$. 
\end{itemize}

\subsection{Linearised Einstein equations for collisionless matter}

We now consider only scalar perturbations and work in the Newtonian gauge. As in this thesis we are primarily interested in the dynamics of the matter component of the universe, we take $\bar{P}=\delta P=0$ (corresponding to an equation of state $w=0$) and $\Pi_{ij}=0$. The space-space part of the Einstein equations implies that in the absence of anisotropic stress, the scalar potentials $\Phi_B$ and $\Psi_B$ are equal to one another, which we will now assume. Inserting the perturbed metric into the Einstein equations and working to linear order in these potentials results in a set of fluid equations 
\begin{subequations}\label{eq:lin-einstein}
\begin{align}
\delta' + \grad \cdot \bm{U} - 3 \Phi_B' &= 0 \,, \\
\bm{U}' + \mathcal{H} \bm{U} - \grad\Phi_B &= 0 \,, \\
\nabla^2 \Phi_B - 3\mathcal{H}(\Phi'_B + \mathcal{H} \Phi_B) &= 4 \pi G a^2 \bar{\rho} \delta\,. \label{eq:lin-ein-poisson}
\end{align}
\end{subequations}
We will see fluid equations like these in the Newtonian treatment in the following Section, but these are valid on all scales. Before moving forward, it is worth reiterating what assumptions have gone into writing down the fluid equations \eqref{eq:lin-einstein}. 
\begin{itemize}
\item We have assumed GR is the correct description of gravity on all scales.
\vspace{-8pt}
\item We have written them only for a perfect fluid energy source described with equation of state $w=0$.
\vspace{-8pt}
\item We are working in the Newtonian gauge.
\vspace{-8pt}
\item We have assumed no anisotropic stress, and thus the perturbation potentials $\Phi_B$ and $\Psi_B$ are equal.
\vspace{-8pt}
\item We have written the fluid equations to \emph{linear order} in both the potentials \emph{and} the fluid perturbations $\delta$ and $\bm{U}$.
\end{itemize}

\section{Newtonian cosmology}

The treatment in the previous Section determined the dynamics of a collisionless perfect fluid in expanding space to linear order in the gravitational potential $\Phi_B$ and linear order in the perturbed fluid variables on all cosmological scales. While this is a good description of matter in the early universe, before primordial fluctuations have had time to grow under gravitational collapse, to understand structure growth it would be preferable to extend the fluid equations to non-linear order in the fluid variables. We would also like to consider models flexible enough to study energy sources beyond  perfect fluids.

On the scales relevant for cosmological structure formation in the late universe, it is possible to achieve this non-linear description of the distribution of matter through simply using Newtonian gravity, which we will briefly argue below. This is extremely useful, as Newtonian gravity is structurally much simpler than GR, removing issues like choice of gauge and fictitious perturbations, which allows the treatment of the non-linear equations more easily. On a pragmatic note, many results about the non-linear nature of the late-time universe come from $N$-body cosmological simulations \parencite[see e.g.][for a review]{AnguloHahn2022} which evolve a collection of particles according to Newtonian gravity. The papers \textcite{Fidler2015PhRvD, Fidler2016JCAP, Fidler2017JCAP_b, Fidler2017JCAP} establish the ``$N$-body gauge'' to aid in interpreting the results from $N$-body simulations as results in the context of GR.


We still use the Friedmann equations to determine the background evolution of the cosmic expansion. In some cases we make use of the Einstein-de Sitter (EdS) universe, which is a flat universe with only matter, as it allows for some expressions to have closed analytic forms. This is often a good approximation for cosmic structure formation as most of structure formation occurs during the matter-dominated era (with dark energy only becoming dominant from $z\approx 0.3$).

\subsection{The Newtonian limit}

The motivation for the Newtonian approach to structure formation can be seen from the Poisson-like linearised Einstein equation \eqref{eq:lin-ein-poisson}. Transforming to Fourier space, the dynamics of the potential are 
\begin{equation}
-k^2 \Phi_B - 3\mathcal{H} (\Phi_B' + \mathcal{H} \Phi_B) = 4 \pi G a^2 (\rho-\bar{\rho}).
\end{equation}
On small scales, where $k\gg \mathcal{H} = (aH)$, the first term of this equation dominates, reducing to a Poisson equation sourced by the difference in density to the homogeneous background
\begin{equation}\label{eq:poisson-lin-ein-fourier-subhor}
-k^2 \Phi_B = 4\pi G a^2  (\rho-\bar{\rho}).
\end{equation}

The distance $(aH)^{-1}$ defines the \textit{Hubble sphere}, and is often referred to as the cosmological horizon.\footnote{This name is a slight misnomer. For FLRW spacetimes, the Hubble sphere roughly coincides with the comoving particle horizon, defined as the maximum (comoving) distance light could travel since some early initial time. This particle horizon determines the history of causal connection. The Hubble sphere instead determines whether a separation could be in causal contact \emph{now}. See \textcite{Baumann_2022} for more discussion.} Perturbations with large wavelengths $\lambda \sim 1/k \gg (aH)^{-1}$ are known as \emph{super-horizon modes}, while small scale perturbations with wavelengths $\lambda \sim 1/k \ll (aH)^{-1}$ are called \emph{sub-horizon modes}. The present day (comoving) size of the Hubble sphere is
\begin{equation}
\left(\frac{c}{aH}\right)_0 = 3000 \ h^{-1} \ \mathrm{Mpc} \approx 4.2\times 10^{3} \ \mathrm{Mpc}\,.
\end{equation}
On these sub-horizon scales, the density perturbation can be shown to be gauge independent \parencite{Hahn2016PhRvD_GRscreen}. To correctly derive the Newtonian fluid equations from the linearised Einstein equations we also require that velocities are non-relativistic. We can interpret the Poisson equation \eqref{eq:poisson-lin-ein-fourier-subhor} as valid on \emph{all scales} if the density perturbation is interpreted in the synchronous gauge. A more complete discussion of the Newtonian limit in this context can be found in \textcite{Kopp2014JCAP, Milillo2015PhRvD}.

\subsection{Particle motion in Newtonian framework}

We now derive the motion of particles in expanding space in the Newtonian framework. These trajectories could alternately be obtained by taking the limit of the geodesic equations \eqref{eq:geodesic-eq} or \eqref{eq:geodesic-eq-momentumform}. We work in comoving coordinates $\xx$ which are related to physical distances, $\bm{r}$ by the scale factor $\bm{r}=a(t)\xx$. We've seen the peculiar velocity of a particle is $\bm{U}=a\dot{\xx}=\xx'$. 

To account for the effect of Hubble expansion we write derivatives with respect to constant comoving position:
\begin{subequations}
\begin{align}
\nabla_{\bm{r}} &= \frac{1}{a}\nabla_{\xx} \,, \\
\left(\pdv{}{t}\right)_{\bm{r}} = \left(\pdv{}{t}\right)_{\xx} + \left(\pdv{\xx}{t}\right)_{\bm{r}} \cdot \nabla_{\xx} &= \left(\pdv{}{t}\right)_{\xx} + H \xx \cdot \nabla_{\xx}.
\end{align}
\end{subequations}
We omit the subscript $\xx$ from spatial derivatives from now on except where it would cause confusion. The physical acceleration of a particle comes from the Newtonian gravitational potential $\phi_{\rm tot}$,
\begin{equation}
\ddot{\bm{r}} = - \nabla_{\bm{r}} \phi_{\rm tot}.
\end{equation}
The equation for peculiar acceleration is most conveniently written in conformal time,
\begin{equation}
\ddot{\bm{r}} = \frac{1}{a}(\mathcal{H}' \xx + \mathcal{H} \xx' + \xx'') = -\frac{1}{a}\grad \phi_{\rm tot}\,,
\end{equation}
the $\mathcal{H}'\xx$ represents a spatial dependence of this acceleration induced by our choice of origin. Introducing the background potential $\phi_{bg}$ such that
\begin{equation}
-\frac{1}{a}\grad \phi_{bg} = \frac{1}{a}\mathcal{H}'\xx \,,
\end{equation}
we can remove this term by splitting the gravitational potential into $\phi_{\rm tot} = \phi_{bg} + \Phi_N$. The peculiar potential $\Phi_N$ is the potential of interest for us, being sourced by fluctuations away from the homogeneous value of the density
\begin{equation}\label{eq:cosmological-poisson}
\nabla^2 \Phi_N = 4\pi G a^2 \bar{\rho} \delta = \frac{3}{2a}\Omega_m^0 H_0^2 \delta = \frac{3}{2}\mathcal{H}^2(a)\Omega_m(a) \delta \,,
\end{equation}
while the background potential is sourced by the mean matter density\footnote{There are some technical details swept under the rug here, referred to as the \emph{Jeans swindle}, as an infinite homogeneous matter density cannot be in equilibrium \parencite[see e.g. Section 5.5.2 of ][for a discussion]{BinneyTremaine2008}. However it can be safely applied to  Newtonian gravity in a cosmological context \parencite{Kiessling1999arXiv}. }
\begin{equation}
\nabla^2 \phi_{bg} = 4\pi G a^2 \bar{\rho}.
\end{equation}
Notice the similarity between \eqref{eq:cosmological-poisson} and the sub-horizon linearised Einstein equation \eqref{eq:poisson-lin-ein-fourier-subhor}. In the linearised Einstein equation \eqref{eq:poisson-lin-ein-fourier-subhor} both $\Phi_B$ and $\delta$ are restricted to being small, while in the Poisson equation \eqref{eq:cosmological-poisson} in the Newtonian framework, we make  no such assumption, allowing the density perturbation to be large. 

The equations of motion for a particle moving in expanding spacetime are then (written in both conformal and coordinate time)
\begin{subequations}\label{eq:particle-equations-of-motion}
\begin{align}
\xx'' + \mathcal{H} \xx' &= - \grad \Phi_N, \\
\ddot{\xx} + 2 H \dot{\xx} &= -\frac{1}{a^2}\grad \Phi_N ,
\end{align}
\end{subequations}
together with the cosmological Poisson equation \eqref{eq:cosmological-poisson}.

\subsection{Vlasov-Poisson equations}\label{sec:vlasov-poisson}

To describe the distribution of matter in the universe we need to go beyond single particles. Instead of starting from the stress-energy tensor for a perfect fluid as we did in the relativistic case, we consider a collection of particles. The state of a collection of $N$ particles can be encoded in their distribution in phase-space. The 1-particle phase-space distribution function $f(\xx,\pp,t)$ encodes the state of the system by providing the number of particles in an infinitesimal phase-space volume at some time $t$ by
\begin{equation}
\dd{N} = f(\xx,\bm{p},t) \dd[3]{\xx}\dd[3]{\bm{p}}.
\end{equation}
Here the canonical momentum\footnote{Another justification for this form of the canonical momentum comes from the spatial part of the 4-momentum. Parametrised by coordinate time, the 4-momentum of a massive particle is $P^\mu=m\dot{x}^\mu = (P^0, m\dot{\xx})$. The canonical momentum is properly thought of as a dual vector/1-form in Hamiltonian mechanics, with a lowered index $P_\mu \defeq \pdv*{\mathscr{L}}{X^\mu}$ with Lagrangian $\mathscr{L}$. Lowering the index with the unperturbed FLRW metric, the spatial part of the canonical 4-momentum is $P_i=a^2 m \dot{x}^i$, so we define the 3-vector $(\bm{p})_i=P_i$.} variable $\bm{p}$ is defined $\bm{p}=a m \bm{U} = a^2 m \dot{\xx}$, which makes the phase-space measure $\dd[3]{\xx}\dd[3]{\bm{p}} = \dd[3]{\bm{r}}\dd[3]{\bm{U}}$ a proper quantity. The equations of motion for such particles are then
\begin{align}\label{eq:newton-eom}
	\dot{\xx} = \frac{\bm{p}}{a^2 m}, \quad \dot{\bm{p}} = -m \nabla \Phi_N \,,
\end{align}
where $\Phi_N$ is the peculiar gravitational potential defined in equation~\eqref{eq:cosmological-poisson}. In the absence of collisions and two-body correlations,
Louiville's theorem requires that the 1-particle distribution function is conserved in time
\begin{equation}
	\dv{f}{t} = \pdv{f}{t} + \dot{\xx} \cdot\nabla_{\! \xx}f + \dot{\pp} \cdot \nabla_{\! \pp} f = 0\,,
\end{equation}
which can be expressed via the equations of motion \eqref{eq:newton-eom} as 
\begin{subequations}\label{eq:vlasov}
\begin{align}
	\pdv{f}{t} &= -\frac{\pp}{ma^2}\cdot \nabla_{\! \xx} f + m \nabla_{\! \xx} \Phi_N \cdot \nabla_{\! \pp} f \\
	&= \left[ \frac{p^2}{2ma^2} + m\Phi_N(\xx)\right] \left( \cev{\del}_x\vec{\del}_p + \cev{\del}_p \vec{\del}_x \right) f \\
	&= \{\mathscr{H},f\} \,,
\end{align}
\end{subequations}
where $\cev{\del},\vec{\del}$ indicates whether the derivatives act on functions to their left or right. The final line expresses the evolution of the phase space distribution as a Poisson bracket with a Hamiltonian $\mathscr{H}$. This equation is referred to as the \emph{Vlasov} equation and when coupled to the Poisson equation~\eqref{eq:cosmological-poisson}, the \emph{Vlasov-Poisson} equations. The Vlasov equation is a special case of the Boltzmann equation from statistical mechanics, where the effects of collisions (which would enter on the right side as a non-conservation of phase-space volume) have been neglected. As such, the Vlasov equation is also called\footnote{This equation could also be known as the Jean's equation, as noted in  \textcite{Henon_1982_VlasovEquation}, since Jeans applied its use to stellar dynamics 23 years before Vlasov applied it in the context of collisionless plasmas.} the \textit{collisionless Boltzmann} equation.

It is possible to derive a Vlasov/Boltzmann equation in the relativistic context as well. The stress-energy tensor is related to the phase-space distribution by
\begin{equation}
T_{\mu\nu} \defeq \frac{1}{\sqrt{-g}}\int \dd[3]{\pp} \frac{P_\mu P_\nu}{P^0}f(\xx, \pp, t)\,,
\end{equation}
where $P^\mu$ is the four-momentum. One can then write down the evolution of the phase-space distributions of the different fluid species (baryons, CDM, photons, neutrinos, etc) including the interaction terms between them, resulting a set of coupled equations called the (Einstein)-Boltzmann hierarchy. Codes which solve this coupled set of equations, commonly referred to as Boltzmann codes, such as \textsc{CAMB} \parencite{CAMB} or \textsc{CLASS} \parencite{Lesgourgues2011arXiv} are essential for predicting the statics of the CMB for example. Note that the Boltzmann equation derived in this way requires that the distribution function is perturbatively close to its homogeneous value. The Vlasov equation \eqref{eq:vlasov} in the Newtonian context does not require $f$ to be close to its homogeneous value, extending our description of the matter content into the non-linear regime.

If we don't neglect 2-body interactions in the Vlasov equation, the evolution of the 1-particle distribution function is related to 2-particle distribution function, and in general the evolution of the $n$-particle distribution function depends on the $(n+1)$-particle distribution function. This infinite cascade is called the BBGKY (Bogoliubov-Born-Green-Kirkwood-Yvon) hierarchy, and is \emph{separate} from the Vlasov cumulant hierarchy we'll see shortly. See Section 3.2 of \textcite{Bertschinger_1995_CosmologicalDynamics} for a discussion of constructing the BBGKY from the exact ``spiky'' Klimontovich distribution function of a collection of particles. 

Solving the Vlasov-Poisson system analytically or even numerically is generally very difficult as it is a non-linear partial integro-differential equation (since the potential is sourced by the density, an integral over $f$) in 6+1 dimensions.

\subsection{Cosmological fluid equations}\label{sec:cosmo-fluid-equations}

Generally we do not need to know the full phase space distribution of our system, instead focusing on the spatial distribution. This allows us to solve for moments of the phase space distribution, where we integrate out momentum information, rather than solving for $f$ fully.

The first few moments of $f$ with respect to momentum\footnote{Note that there are differing conventions for normalising these moments, whether the scale factor is included or not for example determines whether the background level equations are made to be comoving or not.} are
\begin{subequations}\label{eq:vlasov-moments}
\begin{align}
    M^{(0)}(\xx,t) &=  \frac{1}{a^3}\int \dd[3]{p} f(\xx,\pp,t)  = \rho(\xx,t)\,,\\
    M^{(1)}_i(\xx,t) &= \frac{1}{a^3} \int \dd[3]{p} \frac{p_i}{a m} f(\xx,\pp,t)  = \rho(\xx,t) U_i(\xx,t)\,, \\
    M^{(2)}_{ij}(\xx,t) &=\frac{1}{a^3} \int \dd[3]{p} \frac{p_ip_j}{(a m)^2} f(\xx,\pp,t)  = \rho(\xx,t) U_i(\xx,t) U_j(\xx,t) + \rho(\xx,t) \sigma_{ij}(\xx,t) \,,
\end{align}
\end{subequations}
where $\rho$ is the fluid density, $\bm{U}=a\dot{\xx}$ is the  peculiar velocity, and $\sigma_{ij}$ is a velocity dispersion. The parts of these moments which are not simply products of lower moments are called the \emph{cumulants} of the phase space distribution. For example, the second moment consists of the product of the first moments plus the term involving the velocity dispersion. We will discuss moments and cumulants in the context of the density field later, for discussion of them in the context of general probability distributions see Appendix~\ref{app:prob-distributions}. Taking appropriate moments of the Vlasov equation yields evolution equations in these variables, referred to as the cosmological fluid equations, 
\begin{subequations}\label{eq:cosmological-fluids-newtonian}
\begin{align}
	\dot{\rho} + 3H\rho + \frac{1}{a}\div(\rho \bm{U}) &= 0\,, \\
	\dot{\bm{U}} + H \bm{U} + ( \bm{U} \cdot \grad )\bm{U} &= -\frac{1}{a}\grad \Phi_N - \frac{1}{a\rho}\grad \cdot (\rho\bm{\sigma}) \,, \\
	\nabla^2 \Phi_N = 4 \pi G a^2 \bar{\rho} \delta &= \frac{3}{2a}\Omega_m^0 H_0^2 \delta = \frac{3}{2}\Omega_m(a)\mathcal{H}^2(a)\delta \,,
\end{align}
\end{subequations}
where $(\grad\cdot \bm{A} )_i = \nabla_j A_{ji}$. These are in order the continuity equation (conservation of mass), the Euler equation (conservation of momentum), and the Poisson equation for gravity. These are very similar to the linearised Einstein equations \eqref{eq:lin-einstein}, but are not restricted to the fluid variables being small, allowing for non-linear evolution.

In general it is possible to determine an evolution equation for these moments/cumulants. The Vlasov equation can be rewritten as an equation for the moment/cumulant generating function  \parencite[following e.g.][]{Uhlemann2014, Uhlemann2018finitelygenerated} resulting in the moment evolution equation\footnote{Noting that the $n^{\rm th}$ moment defined in those papers is related to ours by a factor of $a^{3+n}$.} 
\begin{equation}
[\del_t + (n+3)H(t) ] M^{(n)}_{i_1,\dots i_n} =  -\frac{1}{a^2}\nabla_j M^{(n+1)}_{i_1,\dots i_n, j} - M^{(n-1)}_{(i_1,\dots i_{n-1}} \cdot \nabla_{i_n)}\Phi_N \,,
\end{equation}
where indices enclosed in parentheses indicate symmetrisation $a_{(i}b_{j)}=a_ib_j + a_jb_i$. Thus, the evolution of the $n^{\rm th}$ moment depends on the value of the $(n+1)^{\rm th}$ moment, cascading into an infinite hierarchy. A self-consistent solution to the Vlasov-Poisson equations therefore requires a strategy to close this hierarchy. 

The most obvious closure strategy is that of \emph{truncation}, where we simply only calculate the cumulants up to some given order $n$, declaring all higher order ones to be zero. This strategy is only self-consistent if the cumulants are truncated at second order, which in the case of the Vlasov equation is equivalent to modelling a perfect fluid, described by a density and velocity field only. As soon as velocity dispersion $\sigma_{ij}$ is present, all higher cumulants are also sourced dynamically, and a truncation scheme is no longer self-consistent. The presence of velocity dispersion can also source vorticity in the velocity field, even if the flow began as a purely potential flow. We return to the closure of the Vlasov cumulant hierarchy later in Chapter~\ref{chap:making-dm-waves}.

\begin{figure}[p]
\centering
\includegraphics[width=\textwidth]{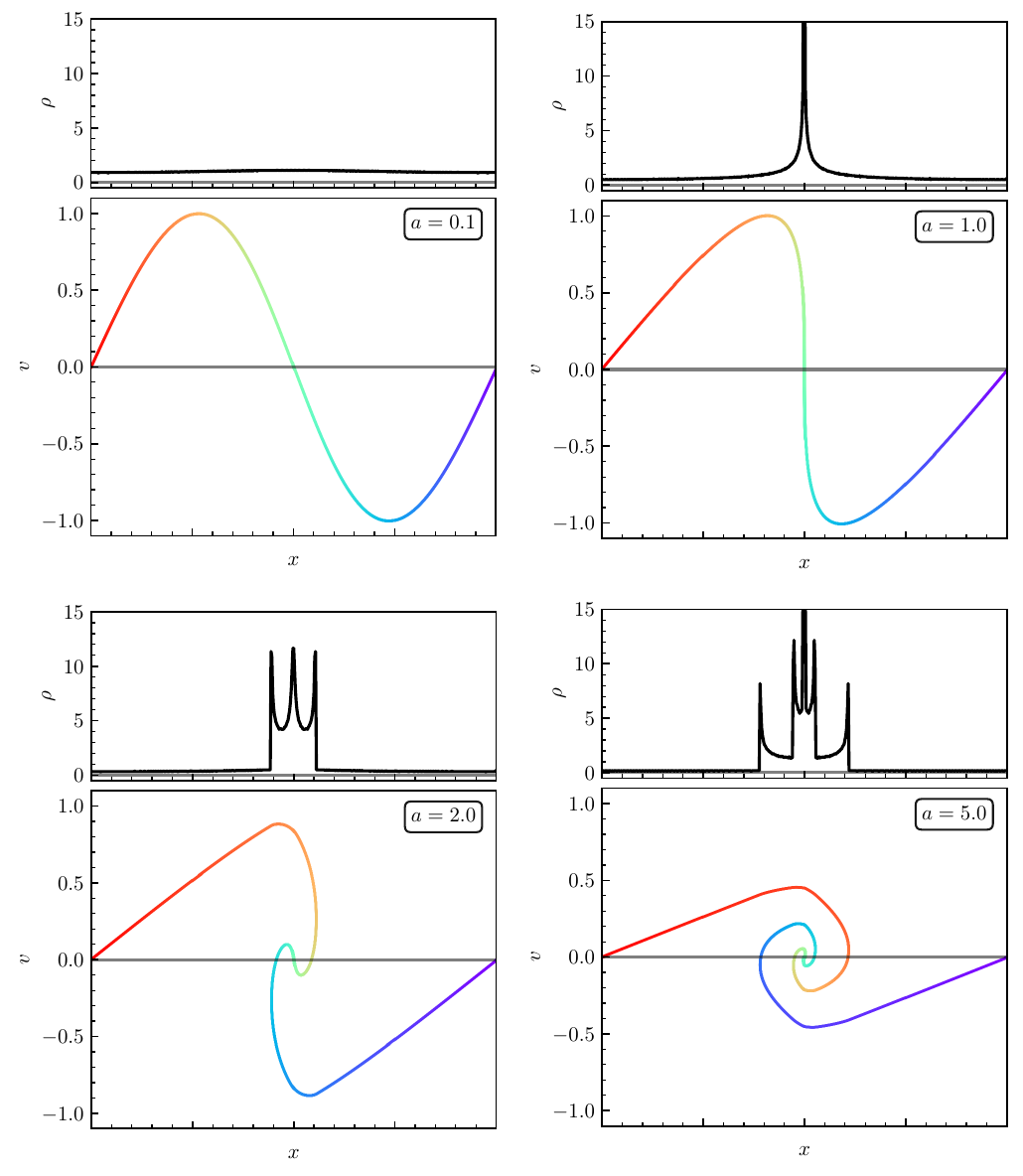}
\caption[One dimensional cosmological $N$-body simulation.]{The phase space evolution and density of a one-dimensional $N$-body simulation of 8192 particles in an Einstein-de Sitter background. The phase-space velocity variable here is $v = a^{-3/2}p/m$ where $p$ is the standard canonical momentum. Particles are coloured according to their initial position. At $a=1$ the phase space sheet becomes vertical and the density diverges in an event called shell crossing, leading to a region of 3 streams. As the system continues to evolve further shell crossing occurs, leading to  5 streams and eventually a virialised structure. }
\label{fig:nbody-shellcrossing}
\end{figure}

Figure \ref{fig:nbody-shellcrossing} shows the phase-space distributions and fluid densities in a one dimensional $N$-body simulation.\footnote{Produced with \url{https://bitbucket.org/ohahn/cosmo_sim_1d/src/master/}.} Initially we consider particles which are uniformly spaced and given a sinusoidal velocity profile. At $a=1$, the phase-space sheet becomes vertical at the origin, as many particles pile up, in an event called shell crossing. The density at this point formally becomes infinite. Since we consider a  collisionless fluid, past this shell crossing event, the different streams of particles flow through one another and a three-stream region develops. Further evolution leads to more shell crossing and tighter fluid streams until a virialised object forms. This multistreaming allows for flows which begin without velocity dispersion or vorticity to develop them naturally, as a product of stream averaging.

\section{Eulerian perturbation theory}\label{sec:intro-SPT}

With the cosmological fluid equations \eqref{eq:cosmological-fluids-newtonian}, we can now split the fluid variables into background and fluctuation parts, and proceed with perturbation theory which goes beyond linear order in the fluid variables. Assuming the fluid fluctuation variables are small (in a statistical sense) then we can solve the cosmological fluid equations order by order in these fluid variables in what is called Eulerian Perturbation Theory or simply Standard Perturbation Theory (SPT). The background-fluctuation split for the density is
\begin{equation}
	\rho(\xx,t) = \bar{\rho}(t) [1+\delta(\xx,t)]\,,
\end{equation}
where $\bar{\rho}$ is the homogeneous density and $\delta$ is the \emph{overdensity}. The velocity is already a perturbation about the Hubble expansion $\bm{v}_{\rm phys} = \dot{a}\xx + \bm{U}$. The gravitational potential is also treated as a perturbative quantity, as it was in the relativistic case. The continued treatment of the gravitational potential as small is well justified as typical fluctuations in $\Phi$ are  $\lesssim \order{10^{-4}}$ on all scales.

At background level, the cosmological fluid equations read
\begin{equation}
	\dot{\bar{\rho}} + 3 H \bar{\rho} = 0 \,,
\end{equation}
which recovers the standard dilution of matter, $\bar{\rho}\propto a^{-3}$. The full equations for the fluctuations about the background are
\begin{subequations}\label{eq:fluid-equations-coordinate-time}
\begin{align}
	\dot{\delta} + \frac{1}{a} \div((1+\delta)\bm{U}) &= 0 \,, \\
	\dot{\bm{U}} + H(t) \bm{U} + (\bm{U} \cdot \grad)\bm{U} &= -\frac{1}{a}\grad\Phi_N \, , \\
	\nabla^2 \Phi_N &= \frac{3}{2a}\Omega_m^0 H_0^2 \delta.
\end{align}
\end{subequations}

\subsection{Linear SPT}

To examine the linear order behaviour of the fluid perturbations we decompose the velocity into its divergence $\theta = \grad\cdot \bm{U}$ and curl $\bm{w} = \grad\times\bm{U}$ modes. The linearised fluid equations are then 
\begin{subequations}
\begin{align}
	\dot{\delta} + \frac{1}{a} \theta &\overset{\rm lin}{=} 0 \, , \label{eq:linear-continuity} \\
	\dot{\theta} + H(t) \theta &\overset{\rm lin}{=} -\frac{3}{2a^2} \Omega_m^0 H_0^2 \, , \\
	\dot{\bm{w}} + H \bm{w} &\overset{\rm lin}{=} 0 \, ,
\end{align}
\end{subequations}
where we've folded the Poisson equation into the equation for the velocity divergence. From these equations we see that the vorticity (curl mode of the velocity) decays in the linear regime $\bm{w}\propto a^{-1}$. Since initial vorticity decays with cosmic expansion, we will take it to be 0 at initial time. If we also set the velocity dispersion to zero, this becomes self-consistent, as vorticity modes cannot be generated without velocity dispersion, which keeps the vorticity zero until shell crossing occurs. In this multistreaming region, velocity dispersion of the different streams will then source vorticity which then cascades the entire Vlasov hierarchy.

The linearised equations for the fluid variables can be further combined into a second order equation for the density contrast
\begin{equation}
\ddot{\delta} + 2 H(t) \dot{\delta} - \frac{3}{2a^3}\Omega_m^0 H_0^2 \delta \overset{\rm lin}{=} 0.
\end{equation}
This linearised equation for overdensities has two solutions (as it is second order in time) which we call $D_\pm(t)$ such that the linear density contrast obeys
\begin{equation}
\delta(\xx,t) \overset{\rm lin}{=} D_+(t) A(\xx) + D_-(t) B(\xx)\,,
\end{equation}
where $A(\xx), B(\xx)$ are arbitrary spatial functions set by the initial conditions. The function $D_+$ is called the growing mode of the linear growth factor, while $D_-$ is the called the decaying mode. The linear velocity divergence is given by
\begin{equation}
\theta(\xx,t) \overset{\rm lin}{=} - a H(t) [f_+ A(\xx) + f_- B(\xx)]\,,
\end{equation}
where $f_\pm = \dv{\ln D_\pm}{\ln a}$ are the growth \textit{rates}.

In an Einstein-de Sitter universe these linear growth factors are
\begin{equation}
D_+^{\rm EdS} = a \propto t^{2/3} , \quad D_-^{\rm EdS} = a^{-3/2}\propto t^{-1}\,,
\end{equation}
so to first order, perturbations grow linearly with the scale factor. In a universe with only $\Lambda$ and CDM, the linear growth factors can be solved exactly in terms of hypergeometric functions, see Appendix~\ref{app:growth_hypgeometric}.

The linear density field $\delta_{\rm L}$ is simply a rescaling of the density field given at some early reference time if we allow enough time for the decaying mode of the initial conditions to decay away. This perturbative ansatz allows us to write $\delta_{\rm L}(\kk,t) = D_+(t) \delta_0(\kk)/D_+(t_{\rm ref})$ where $t_{\rm ref}$ is some early reference time. We typically take the reference time to be sufficiently early that the rescaled field $\delta_0(\kk)$ is Gaussianly distributed, such as the time of the CMB.

The matter power spectrum describes the correlations between two Fourier modes of the density, and fully characterises the statistics of a Gaussian random field. The linear power spectrum is defined
\begin{equation}
\ev{\delta_{\rm L}(\kk)\delta_{\rm L}(\kk')} = (2\pi)^3 \delta_D(\kk+\kk') P_{\rm L}(k)\,,
\end{equation}
which is related to its value at the reference time by $P_{\rm L}(k,t) = D_+(t) P_0(k) / D_+(t_{\rm ref})$. We discuss the power spectrum and other statistics in more detail in Section \ref{sec:corr-functions}

\begin{figure}[h!t]
\centering
\includegraphics[width=0.7\textwidth]{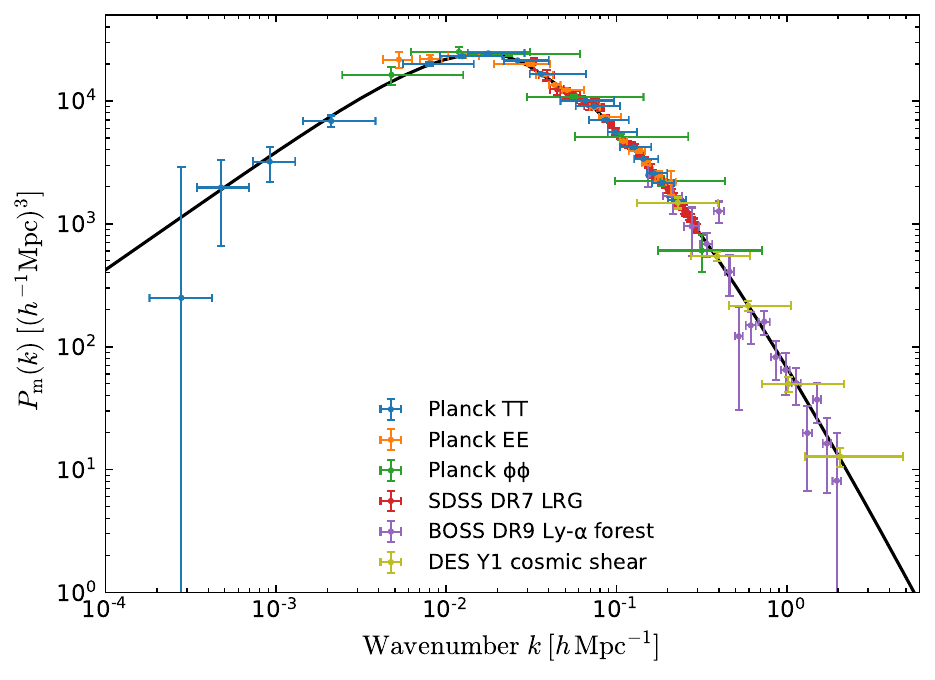}
\caption[The linear matter power spectrum. Source: \parencite{PlanckCollaborationOverview}]{The linear matter power spectrum probed by a variety of different cosmic probes, by ESA and the Planck Collaboration \parencite{PlanckCollaborationOverview}. The turn over in the spectrum occurs at roughly $k_{\rm eq}$, with low $k$ (large scales) scaling as $k^{n_s}\sim k$ and high $k$ (small scales) scaling as $k^{n_s-4}\sim k^{-3}$. }
\label{fig:linear-pk}
\end{figure}

The linear power spectrum is usually parametrised in the following way
\begin{equation}
P(k,t) = A(t) P_{\rm primord}(k) T^2(k,t) = A_s D_+^2(t) \left(\frac{k}{k_0}\right)^{n_s} T^2(k,t)\,,
\end{equation}
where the transfer function $T(k,t)$ parametrises growth on different scales. The amplitude $A_s$ and index $n_s$ parametrise the size and scale dependence of the fluctuations in the early universe, while the scale $k_0$ is some ``pivot scale'' chosen for observational convenience. Inflation predicts $n_s\approx 1$ --- a so called \emph{scale invariant spectrum} --- which is indeed what is measured by \textit{Planck}. We do not present a detailed analysis of the form of the transfer function here, however we do note the rough scalings for convenience. The shape of the matter transfer function depends on the growth of fluctuations both inside and outside the horizon during the radiation and matter dominated eras, with a turnover at the wavenumber corresponding to the size of the horizon at matter-radiation equality ($z_{\rm eq}\approx 3400$)
\begin{equation}
T^2(k) \propto \begin{cases} 
1 & k<k_{\rm eq}  \\
\frac{k_{\rm eq}^4}{k^4}\left(1 + \log(\frac{k_{\rm eq}}{k})\right)^2 & k >k_{\rm eq}\,,
\end{cases}
\end{equation}
which means that the asymptotic behaviour of the linear matter power spectrum is $P_{\rm L}(k)\propto k^{n_s}\sim k$ on large scales, while on small scales $P_{\rm L}(k)\propto k^{n_s-4}\sim k^{-3}$. The precise form of the transfer function can be calculated by solving the relativistic Boltzmann hierarchy with Boltzmann solvers.

\subsection{Non-linear SPT}

We will now make the self-consistency approximation that vorticity, $\bm{w}=\grad \times \bm{U}$, and velocity dispersion, $\bm{\sigma}$, initially vanish. We return to a discussion of multistreaming and velocity dispersion in Chapter~\ref{chap:making-dm-waves}.

In the discussion of non-linear SPT we will proceed in an EdS spacetime where $D_+=a$ as this simplifies the discussion, and we work in the perturbative regime where we neglect the role of the decaying mode. We make the \emph{ansatz} that the density and velocity fields can be expanded as a power series in $a$
\begin{subequations}
\begin{align}\label{eq:PT-expansions}
\delta(\xx,t) &= \sum_{n=1}^\infty \delta^{(n)}(\xx,t) = \sum_{n=1}^\infty a^n(t)\delta^{(n)}(\xx)\,, \\
 \theta(\xx,t) &= \sum_{n=1}^\infty \theta^{(n)}(\xx,t) = -\mathcal{H}(t) \sum_{n=1}^\infty a^n(t) \theta^{(n)}(\xx) \,,
\end{align}
\end{subequations}
where this is also an expansion in the linear fields, e.g. $\delta^{(p)}=\order{(\delta^{(1)})^p}$. In the full $\Lambda$CDM case, the fields do not factorise in space and time in this way, but a generalisation of these series --- $a\mapsto D_+$ and $\mathcal{H}\mapsto \mathcal{H}f$ --- produces typical errors at the sub-percent level so it remains a useful formalism \parencite{Bernardeau:2002}.

For Gaussian initial conditions, the linear density field is also Gaussian, and therefore this expansion a series in powers of Gaussian fields, which will allow us to make use of Wick's theorem to compute correlation functions. Inserting this perturbative ansatz into the fluid equations allows the $n^{\rm th}$ order non-linear density and velocity divergence to be written as integrals over $n$ copies of the linear density contrast with some ``perturbation theory kernels'' $F_n$ and $G_n$
\begin{subequations}\label{eq:F_n_kernels}
\begin{align}
\delta^{(n)}(\kk,t) &=  \prod_{i=1}^n\left[\int \frac{\dd[3]{\kk_i}}{(2\pi)^3} \delta^{(1)}(\kk_i,t)\right] F_n(\kk_1, \dots, \kk_n) \  (2\pi)^3 \delta_{\rm D}\left(\kk-\sum_{j=1}^n \kk_j\right), \\
\theta^{(n)}(\kk,t) &=  \prod_{i=1}^n\left[\int \frac{\dd[3]{\kk_i}}{(2\pi)^3} \delta^{(1)}(\kk_i,t)\right] G_n(\kk_1, \dots, \kk_n) \  (2\pi)^3 \delta_{\rm D}\left(\kk-\sum_{j=1}^n \kk_j\right) ,
\end{align}
\end{subequations}
where explicit forms of $F_n, G_n$ and recursion formulae for them can be found in e.g. \textcite{Jain.Bertschinger_1994_SecondOrderPower, Bernardeau:2002}.

Another quantity which will be useful for us in a few different contexts is the angular average of these kernel functions, called ``perturbation vertices''
\begin{subequations}
\begin{align}
\nu_n &=n! \int \frac{\dd{\Omega_1}\dots \dd{\Omega_n}}{(4\pi)^n} F_n(\kk_1, \dots, \kk_n)\,, \label{eq:SPT_nu_n_def}\\
\mu_n &=n! \int \frac{\dd{\Omega_1}\dots \dd{\Omega_n}}{(4\pi)^n} G_n(\kk_1, \dots, \kk_n)\,,
\end{align}
\end{subequations}
which are closely related to spherical collapse dynamics and are only weakly dependent on cosmology. We will see how these can be used to  compute statistics of the density field in Section \ref{sec:tree-order-skewness-SPT}.

\section{Lagrangian perturbation theory}\label{sec:intro-LPT}

The perturbation theory thus far has treated the fluid variables themselves as the small quantities, using a set of fixed, Eulerian, coordinates which do not move with the fluid. An alternative approach to describing fluids is to track the displacements of parcels of the fluid (or displacement of particles) from their initial positions, writing their finial (Eulerian) position as a function of their initial (Lagrangian) position. 

The map $\bm{q} \mapsto \xx_t(\bm{q}) = \bm{q} + \xxi(\bm{q},t)$ which takes initial positions $\bm{q}$ to final positions $\xx$ is called the Lagrangian map. The function $\xxi(\bm{q},t)$ is called the displacement field\footnote{In other texts the displacement field is often called $\psi$ or $\Psi$. We avoid that here as we will later be discussing LPT-like dynamics in the context of wavefunctions, so we reserve $\psi/\Psi$ for wavefunctions.} as it encodes how much the parcel of fluid is displaced from its initial position in time $t$. In Lagrangian Perturbation Theory (LPT) this displacement field is the object which is treated perturbatively. Solutions in LPT are particularly important in setting initial conditions for cosmological $N$-body simulations. Particles are displaced according to the transfer functions output by Boltzmann solvers to some intermediate redshift, before significant shell crossing and non-linear collapse occurs (typically between $z\sim 25$--$100$ depending on the particular simulation), before switching to numerically solving the positions with the $N$-body code.

Note that the Eulerian and Lagrangian formulations of fluid flow have different time derivatives, related by
\begin{equation}\label{eq:convective-deriv}
\partial_t^L = \partial_t^E + \frac{1}{a}\bm{U} \cdot \bm{\nabla}_{\xx}\,,
\end{equation}
where $\del_t^L$ is the Lagrangian time derivative and $\partial_t^E$ is the Eulerian space time derivative. This is the standard form of the ``convective'' derivative from fluid dynamics. However, note that the Eulerian space formulation we've already presented is ``Lagrangian'' in the sense that we're comoving with the Hubble expansion, while the Lagrangian formulation comoves with the full peculiar velocity.

The density is recovered from this formulation by the conservation of mass. Assuming that this mapping is one-to-one (which is true before shell crossing) the mass in a final region of Eulerian space is simply the mass enclosed in the initial region in Lagrangian space times the volume ratio between the regions, encoded in the Jacobian of this mapping
\begin{align}
\rho(\xx,t) \dd[3]{\xx} &= \rho(\bm{q}) \det[\pdv{\xx(\bm{q},t)}{\bm{q}}] \dd[3]{\bm{q}}  ,\\
&=  \rho(\bm{q}) \det[1 + \pdv{\xxi(\bm{q},t)}{\bm{q}}] \dd[3]{\bm{q}} ,
\end{align}
resulting in the density 
\begin{equation}
1+\delta(\xx,t) = \frac{1}{\det[1 + \bm{\nabla}_{\! \bm{q}}\cdot\xxi(\bm{q},t)] } = \frac{1}{\mathcal{J}}\,,
\end{equation}
where $\mathcal{J}$ is the Jacobian of the Lagrangian mapping. In the multistreaming case we would have to sum over the different streams, including contributions from all $\bm{q}$ which map to the same $\xx$. 

\begin{figure}[h!t]
\centering
\includegraphics[width=\textwidth]{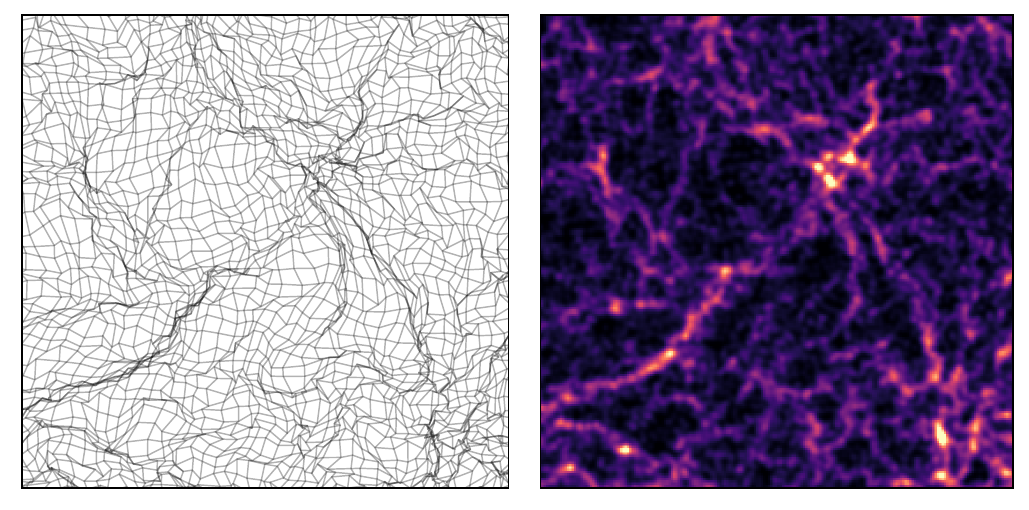}
\caption[2D deformed grid and density field from a Gaussian displacement field. ]{An initially uniform grid displaced by a Gaussian displacement field with $P(k)\propto k^{-3}$. The left panel shows how each initial (Lagrangian) grid square deforms, while the right figure shows the density of particles, assigned to a regular grid by  a cloud-in-cell mass assignment scheme and smoothed with a Gaussian filter. The relationship between  the volume scaling factor (Jacobian) of the displacement and  the density of particles in Eulerian space can clearly be seen.}
\label{fig:zeldo-displacement}
\end{figure}

This inverse relationship between the Jacobian and the density can be seen in Figure~\ref{fig:zeldo-displacement}, where regions which expand more end up less dense in Eulerian space.

LPT then proceeds by expanding  $\xxi$ in a perturbative series $\xxi = \sum \xxi^{(n)}$ in the same way as in SPT, where $\xxi^{(n)}(\bm{q},t)=a^n(t)\xxi^{(n)}(\bm{q})$ in EdS spacetime. The Jacobian $\mathcal{J}$ then gains an expansion as perturbative series \parencite[see e.g.][]{Bouchet1995A&A} $\mathcal{J}=1+\sum \mathcal{J}^{(n)}$ where
\begin{equation}
\mathcal{J}^{(1)} = \sum_i  \xi^{(1)}_{i,i}, \quad \mathcal{J}^{(2)} = \sum_i  \xi^{(2)}_{i,i} + \frac12 \sum_{i\neq j} \left[\xi^{(1)}_{i,i}\xi^{(1)}_{j,j} - \xi^{(1)}_{i,j}\xi^{(1)}_{i,j}\right], \quad \dots 
\end{equation}
where $\xi_{i,j}=\del_{q_j}\xi_i$. These expansions can then be entered into the particle equation of motion \eqref{eq:particle-equations-of-motion}
\begin{equation}
\mathcal{J} \grad_{\! \xx} \cdot \left[\dv[2]{\xxi}{t} + 2H\dv{\xxi}{t}\right] = \frac{3}{2a^2}\Omega_m^0 H_0^2 (\mathcal{J}-1)\,,
\end{equation}
and solved order by order. To write the equations fully in Lagrangian space, the spatial derivative can be replaced $\nabla_i= (\delta_{ij}+\xi_{ij})^{-1}\nabla^{\rm L}_j$ where $\nabla^{\rm L}_{j} = \partial_{q_j}$ is the derivative with respect to Lagrangian coordinates. Note also that the time derivative here is still the Eulerian time derivative (keeping $x$ fixed) rather than the Lagrangian/convective time derivative \eqref{eq:convective-deriv}.

Lagrangian Perturbation Theory has the advantage over SPT that to fixed perturbative order the dynamics are more accurate to the full non-linear dynamics. Essentially LPT resums information encoded at higher SPT order as it captures information with is non-local in Eulerian position. This has been formally demonstrated in 1D \parencite{McQuinn.White_2016_CosmologicalPerturbation, Pajer.vanderWoude_2018_DivergencePerturbation}.

\subsection{Zel'dovich approximation}\label{sec:intro-zeldovich-approximation}

Expanding the relation between the Eulerian density and Jacobian of the Lagrangian mapping for small displacements
\begin{align}
1 + \delta(\xx,t) &= \frac{1}{\det[1 + \bm{\nabla}_{\! \bm{q}}\cdot\xxi] }  \nonumber \\
&= \frac{1}{1 + \bm{\nabla}_{\! \bm{q}}\cdot\xxi + \order{(\bm{\nabla}_{\! \bm{q}}\cdot\xxi )^2}}  \nonumber \\
&\simeq 1 - \bm{\nabla}_{\! \bm{q}}\cdot\xxi + \order{(\bm{\nabla}_{\! \bm{q}}\cdot\xxi )^2}\,,
\end{align}
such that the lowest order displacement from LPT is determined by
\begin{equation}
\delta(\xx,t) = - \bm{\nabla}_{\! \bm{q}(\xx)} \cdot \xxi^{(1)}(\bm{q}(\xx),t).
\end{equation}
The displacement can be written as
\begin{equation}
\xx(\bm{q},t) = \bm{q} + D_+ \nabla_{\! \bm{q}} \chi \,,
\end{equation}
for some velocity potential $\chi$ which is related to the initial gravitational potential. We can find the relationship between $\chi$ and the gravitational potential through the velocity field. The peculiar velocity $\bm{U}$ is obtained from this displacement mapping as
\begin{equation}
\bm{U} = a\dot{\xx} = a\dot{D}_+ \nabla_{\! \bm{q}} \chi = a H D_+ f \nabla_{\! \bm{q}} \chi \,,
\end{equation}
where $f=\dv*{\log D}{\log a}$ is the growth rate. In the linear regime, which applies at early times, the velocity field is related to the gravitational potential by
\begin{equation}\label{eq:linear-velocity-relation}
\bm{U} \overset{\rm lin}{=} - \frac{2 f}{3 \Omega_m(a) H(a)} \frac{\nabla \Phi_N}{a}.
\end{equation}
By equating these two expressions at early time, the velocity potential is given by
\begin{equation}
\chi = - \frac{2}{3\Omega_m(a)H^2(a) a^2 D_+}\Phi_N^{\rm (ini)}.
\end{equation}

The Zel'dovich approximation  (ZA) \parencite{Zeldovich1970} asserts that this displacement holds even into the region of non-linear densities. It has the nice property that in 1D the Zel'dovich approximation is exact before shell crossing. On a mathematical level this happens because
\begin{equation}
\det[1+\dv{\xi}{q}] = 1 + \dv{\xi}{q}\,,
\end{equation}
making the mapping
\begin{equation}
1 + \delta(\xx,t) = \frac{1}{\det[1 + \bm{\nabla}_{\! \bm{q}}\cdot \xxi]} \simeq \frac{1}{1 +\bm{\nabla}_{\! \bm{q}}\cdot \xxi }\,,
\end{equation}
and exact relationship in one dimension. More physically, this is because gravity in one dimension is independent of distance (consider Gauss' law on an infinite sheet of matter). 

If we change our velocity variable away from the standard peculiar velocity $\bm{U}$ and instead define $\vv = \dv*{\xx}{D_+}$ the displacement equation for 1LPT/Zel'dovich approximation becomes particularly simple
\begin{equation}
\xx^{\rm ZA}(\bm{q},D_+) = \bm{q} + D_+ \dv{\xx}{D_+} = \bm{q} + D_+ \vv \,,
\end{equation}
such that we can interpret the Zel'dovich approximation as simply ballistic, constant velocity trajectories in these coordinates, treating $D_+$ as our time variable and $\vv$ as our velocity variable. In an EdS universe this simplifies as $D_+ = a$. Another interpretation of the Zel'dovich approximation is a system satisfying the cosmological fluid equations \eqref{eq:fluid-equations-coordinate-time} but with the Poisson equation replaced with the relationship between velocity and gravitational potential in the linear regime, equation~\eqref{eq:linear-velocity-relation}.

\section{Correlation functions and density cumulants}\label{sec:corr-functions}

Having presented the standard dynamical descriptions of the dark matter fluid, we now present the standard statistical tools which are used to quantify the fluctuations in the density field.

When measuring cosmic fields, we do not have access to the full ``microscopic'' density distribution which obeys the fluid equations we have laid out thus far, we will always measure the density field smoothed over some physical scale $R$. If the density field is smoothed with a function $W_R(\xx)$ (called a window or filter function), this corresponds to a convolution in real space, or multiplication in Fourier space
\begin{subequations}
\begin{align}
\delta_R(\xx) &= \int \dd[3]{\xx'} W_R(\xx-\xx')\delta(\xx') = (W_R * \delta)(\xx), \\
\delta_R(\kk) &= \tilde{W}_R(\kk)\delta(\kk).
\end{align}
\end{subequations}
In this work we smooth fields with spherical top-hats, such that the density $\delta_R(\xx)$ is simply the average value of $\delta(\xx)$ in a sphere of radius $R$ about point $\xx$. This corresponds to the Fourier space filter
\begin{equation}
\tilde{W}_R(\kk) \equiv W(kR) = \frac{3}{(kR)^3}\left[\sin(kR)-kR\cos(kR)\right].
\end{equation}
Another common window function is to smooth with a Gaussian, see Appendix~\ref{app:window-functions}.   

Motivated by the fact that the early universe is very close to a Gaussian random field, we define the two-point correlation function, which entirely determines the other statistics of a Gaussian field. This two-point correlation function, and its Fourier counterpart the power spectrum, form the basic toolkit of statistical cosmology. The two-point correlation function\footnote{Strictly, the two-point correlation function is the connected part of this ensemble average $\ev{\delta_1\delta_2}_c = \ev{\delta_1\delta_2} - \ev{\delta_1}\ev{\delta_2}$. For fields with 0 mean, such as the overdensity, the connected part of the second moment is equal to the second moment.} of the matter overdensity is
\begin{equation}
\xi(\abs{\xx_1-\xx_2}) = \ev{\delta(\xx_1)\delta(\xx_2)}.
\end{equation}
This depends only on the magnitude of the separation of the points by homogeneity and isotropy. In principle the angle brackets should represent an ensemble average over different random initial conditions. However, since we only have access to one realisation of the universe, we make use of the Ergodic hypothesis, replacing ensemble averages with spatial averages. This assumes that the correlation features in our field decay sufficiently rapidly that the sampled volume consists of several statistically independent regions. 

The power spectrum is defined\footnote{There are varying conventions for the factors of $2\pi$ in the power spectrum, see Appendix \ref{app:sec:power-spectra}. Some people prefer to work with the dimensionless power spectrum $\Delta^2(k)=k^3 P(k)/2\pi^2$, which is the power in a shell around wavenumber $k$. }  analogously in Fourier space
\begin{equation}
\ev{\delta(\kk)\delta(\kk')} =  (2\pi)^3 \delta_{\rm D}(\kk+\kk') P(k) \,,
\end{equation}
where the Dirac delta appearing here comes from statistical isotropy. These functions are related to one another by  Fourier transform 
\begin{equation}
\xi(r) = \int \frac{\dd[3]{\kk}}{(2\pi)^3} e^{i\kk\cdot \bm{r}} P(k).
\end{equation}

The variance of the smoothed density field is given as an integral over the power spectrum
\begin{equation}\label{eq:var_from_Pk}
\sigma^2(R) = \langle \delta_R^2 \rangle = \int_0^\infty \frac{\dd{k}}{2\pi^2} k^2 W(kR)^2 P(k),
\end{equation}
which quantifies the typical amplitude of fluctuation with spatial size $R$. The root-mean-square fluctuation in the density in spheres of radius $R$ is then given by $\sqrt{\sigma^2(R)}$, if the smoothing window is taken to be a spherical top-hat.

These sorts of statistics naturally generalise to the $n$-point correlation functions, which we now discuss. Higher order correlation functions are defined as the connected part of correlations between more copies of the density field at different points, denoted with a subscript $c$ on the average brackets
\begin{equation}
\xi_n(\xx_1, \dots, \xx_n) = \ev{\delta(\xx_1) \dots \delta(\xx_n)}_c.
\end{equation}
The connected part of an ensemble average is defined by subtracting off products of lower order correlations, summed over all permutations of points. This is analogous to the construction of cumulants from moments (indeed, the cumulants of the density field are simply these correlation functions evaluated at a single point). These connected parts encode the new, independent information at each order, and are generally simpler to work with mathematically. For example, the 3-point correlation function can be expressed  (using the shorthand $\delta_i \equiv \delta(\xx_i)$):
\begin{equation}
\xi_3(\xx_1, \xx_2, \xx_3) = \ev{\delta_1\delta_2\delta_3} -  \ev{\delta_1\delta_2}\ev{\delta_3} - \ev{\delta_1\delta_3}\ev{\delta_2} - \ev{\delta_2\delta_3}\ev{\delta_1}.
\end{equation}
Since $\delta$ is a field with 0 average, the singlet terms $\ev{\delta_i}$ vanish, and the connected part simply equal to the full ensemble average. However, for the 4 point correlation function
\begin{equation}
\xi_4(\xx_1, \xx_2, \xx_3,\xx_4) = \ev{\delta_1\delta_2\delta_3\delta_4} - \left[\ev{\delta_1\delta_2}\ev{\delta_3\delta_4} + \rm perm\right],
\end{equation}
produces products over 2 point correlation functions, which are non-zero.

Related to these $n$-point correlation functions are the cumulants of the density field $\langle\delta^n\rangle_c$ (see Appendix~\ref{app:prob-distributions} for a general discussion of moments and cumulants). The cumulants of the density field are simply the $n$-point correlation function evaluated at a single point $\xx_1=\dots=\xx_n$. Note that with a real space spherical smoothing window, the $n^{\rm th}$ cumulant of the smoothed density field $\langle\delta_R^n\rangle_c$ is simply the average value of the $n$-point correlation function within that sphere. By statistical homogeneity, the cumulants are independent of position. These cumulants characterise the one-point probability distribution (PDF) of the matter field, which will be one of the principle statistics of interest in this work for information beyond the power spectrum.

\subsection{Diagrams for correlation functions}
Here we briefly sketch how one would predict the non-linear two-point correlation function  (or power spectrum) from perturbation theory to illustrate the machinery developed thus far. 

The calculation of correlation functions in perturbation theory can be represented in a diagrammatic way, analogously to Feynman diagrams in quantum field theory. This expansion can be performed in real space for the $n$-point correlation functions, or in Fourier space for the poly-spectra. The procedure is as follows. Given a perturbative expansion for the non-linear density $\delta= \sum_p \delta^{(p)}$, where $\delta^{p}$ is $p^{\rm th}$ order in $\delta^{(1)}\propto \delta_{\rm L}$, insert the expansion into the relevant correlation function average. 
\begin{figure}[h!t]
\centering
\includegraphics[width=\textwidth]{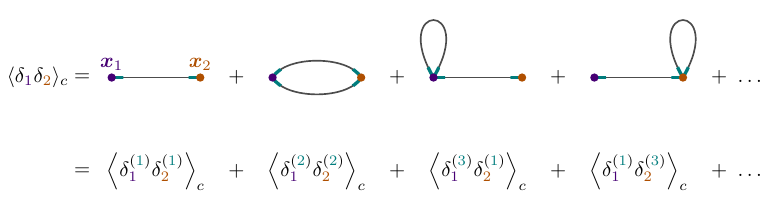}
\caption[Diagrammatic expansion of the two-point correlation function to one loop.]{The 2 point correlation function or power spectrum expanded to 1 loop order in perturbation theory. We use the shorthand $\delta_i=\delta(\xx_i)$, so subscripts indicate the spatial point, while superscripts represent the perturbative order. The three 1-loop diagrams here are of order $(\delta^{(1)})^4$, and correspond to integrals over products of 2 linear power spectra, coupled by the kernel functions $F_2, F_3$.}
\label{fig:pk_to_1loop}
\end{figure}

As an example we predict the non-linear two-point function by
\begin{equation}
\ev{\delta(\xx_1)\delta(\xx_2)}_c = \ev{(\delta^{(1)}(\xx_1) + \delta^{(2)}(\xx_1)+\dots)(\delta^{(1)}(\xx_2) + \delta^{(2)}(\xx_2)+\dots)}_c.
\end{equation}
For Gaussian initial conditions, we can apply Wick's theorem, which states that the average of an odd number of zero-mean Gaussian fields is 0, and of an even number of fields is the product of all pairs of 2-point averages, summed over permutations. Since $\delta^{(p)}\sim \delta_{\rm L}^p$, the only terms which are non-zero are those where the sum of perturbative order of the fields is even
\begin{align}
\ev{\delta(\xx_1)\delta(\xx_2)}_c = \phantom{+}&\langle \delta^{(1)}(\xx_1)\delta^{(1)}(\xx_2) \rangle  \nonumber \\
+ &\langle \delta^{(2)}(\xx_1)\delta^{(2)}(\xx_2) \rangle  \nonumber \\
+ &\langle \delta^{(3)}(\xx_1)\delta^{(1)}(\xx_2) \rangle + \langle \delta^{(1)}(\xx_1)\delta^{(3)}(\xx_2) \rangle + \dots\,.
\end{align}
The first of these is the ``tree-order'' result, which is simply the related to the linear power spectrum. The next two lines are the next-to-leading-order (NLO) contributions, which contain 4 copies of $\delta_{\rm L}$ each. These correspond to diagrams with 1-loop in them, as shown in Figure~\ref{fig:pk_to_1loop}. To be consistent in this perturbative calculation, note that we must work to loop order, rather than fixed order in the expansion of the density (i.e. the 1-loop correction to the power spectrum contains both $\delta^{(2)}$ and $\delta^{(3)}$ terms). These higher order densities $\delta^{(n)}$ can then be written as integrals over the kernels $F_n$ \eqref{eq:F_n_kernels}, which will result in integrals involving these kernels at the vertices of the diagrams, and lines in the diagrams corresponding to linear power spectra arising from $\langle\delta^{(1)}(\kk)\delta^{(1)}(\kk')\rangle$.

\subsection{Tree order skewness}\label{sec:tree-order-skewness-SPT}

As another simple example, we calculate the skewness of the unsmoothed density field $\langle \delta^3 \rangle_c$ in EdS spacetime to tree order. To tree order, there is only one diagram connecting 3 points, consisting of one vertex with two edges and two vertices with one edge connected to them. As we're calculating a cumulant, these points will end up being labelled the same, so the 3 versions of this diagram contribute the same amount to the skewness (a symmetry factor of 3). Thus, the tree order skewness is
\begin{equation}
\langle\delta^3\rangle_c = 3\ev{\delta^{(1)}\delta^{(1)}\delta^{(2)}}_c + \rm loops\,.
\end{equation}
This could also be obtained by simply completing the algebraic expansion (remember that the diagrams act simply as a mnemonic to organise these calculations)
\begin{align}
\langle\delta^3\rangle_c &= \ev{(\delta^{(1)}+\delta^{(2)}+\dots)^3}_c \nonumber\\
&= \ev{(\delta^{(1)}+\delta^{(2)}+\dots)(\delta^{(1)}+\delta^{(2)}+\dots)(\delta^{(1)}+\delta^{(2)}+\dots)}_c \nonumber \\
&= \ev{\delta^{(1)}\delta^{(1)}\delta^{(2)} + \delta^{(1)}\delta^{(2)}\delta^{(1)} + \delta^{(2)}\delta^{(1)}\delta^{(1)} + \dots}_c  \nonumber \\
&= 3\ev{\delta^{(1)}\delta^{(1)}\delta^{(2)}}_c + \rm loops\,.
\end{align}
 We then express the $2^{\rm nd}$ order density $\delta^{(2)}(\xx)$ in terms of the perturbation theory kernel (evaluating at $\xx=0$ by homogeneity)
\begin{align}
\delta^{(2)}(\xx) &= \int \dd[3]{\kk}\frac{\dd[3]{\kk}_1\dd[3]{\kk}_2}{(2\pi)^3 (2\pi)^3} \delta_{\rm D}(\kk-\kk_{12}) \delta_{\rm L}(\kk_1) \delta_{\rm L}(\kk_2) F_2(\kk_1,\kk_2) \,,
\end{align}
where $\kk_{12}=\kk_1+\kk_2$. Inserting this into our correlation gives
\begin{align}
\langle\delta^3\rangle_c^{\rm tree} = 3\int  \frac{\dd[3]{\kk}\dd[3]{\kk}_{1,2,3,4}}{(2\pi)^{12}}   &(2\pi)^3 \delta_{\rm D}(\kk+\kk_{12}) \times \nonumber \\
& \ev{ \delta_{\rm L}(\kk_1)\delta_{\rm L}(\kk_2)\delta_{\rm L}(\kk_3)\delta_{\rm L}(\kk_4) }_c \, F_2(\kk_1,\kk_2) \,,
\end{align}
Applying Wick's theorem, if $\kk_1$ and $\kk_2$ are paired, then the integral vanishes (see the structure of $F_2$ in Appendix~\ref{app:sec:f2-nu2}) while the other two pairings contribute equally to produce linear power spectra. After integrating out the Dirac deltas this leaves,
\begin{align}
\langle\delta^3\rangle_c^{\rm tree} = 6 \int \frac{\dd[3]{\kk_1}\dd[3]{\kk_2}}{(2\pi)^6} P_{\rm L}(\abs{\kk_1+\kk_1}) P_{\rm L}(k_1) F_2(\kk_1, \kk_1 + \kk_2) \,.
\end{align}
Changing to polar coordinates, we can integrate out the angular part of this integral, as the power spectra are independent of angle
\begin{align}
\langle\delta^3\rangle_c^{\rm tree} &= 6 \int \frac{\dd{k_1}\dd{k_2}}{(2\pi)^6} k_1^2 P_{\rm L}(k_1) \, k_2^2 P_{\rm L}(k_2) \int \dd{\Omega_1}\dd{\Omega_2} F_2(\kk_1, \kk_2)  \nonumber \\
&= 6 \int \frac{\dd{k_1}\dd{k_2}}{(2\pi)^6} k_1^2 P_{\rm L}(k_1) \, k_2^2 P_{\rm L}(k_2)  \times \frac{1}{2!}(4\pi)^2 \nu_2 \nonumber \\
&= 3 \nu_2 \left[ \int \frac{\dd{k}}{2\pi^2} k^2 P_{\rm L}(k) \right]^2 = 3 \nu_2 \sigma_{\rm L}^2\,,
\end{align}
where in the final line we have identified the linear variance $\sigma_{\rm L}^2=\langle \delta_{\rm L}^2\rangle_c$ and the angular average of the perturbation theory kernel $\nu_2$ defined in equation \eqref{eq:SPT_nu_n_def}. For the SPT perturbation theory kernels, these $\nu_n$ are independent of the wavevector magnitude, and in EdS spacetime they are pure numbers. The scaling of the tree order cumulants as powers of the variance is a characteristic behaviour, which motivates defining the \emph{reduced cumulants}
\begin{equation}\label{eq:define-Sn}
S_n = \frac{\langle \delta^n \rangle_c}{\langle \delta^2 \rangle_c^{n-1}} = \frac{\langle \delta^n \rangle_c}{\sigma^{2(n-1)}}.
\end{equation}
These reduced cumulants are then simply pure numbers to tree order in EdS spacetime, and in general $\Lambda$CDM cosmologies they only have a weak cosmology dependence. In SPT the  $\nu_2 = 34/21$ (see Appendix \ref{app:sec:f2-nu2}), leading to the tree order reduced skewness
\begin{equation}
S_3^{\rm tree, \, SPT} = 3 \nu_2 = \frac{34}{7}.
\end{equation}

In general we calculate the cumulants of the smoothed density field (if using spherical top-hat filters, this is equivalent to calculating the average of the cumulant in spheres). The method is the same as above, we simply have to account for the window functions in the integration. For spherical top-hat smoothing, the tree order skewness retains its ``bare'' structure, simply acquiring a smoothing term related to the linear variance (see Appendix~\ref{app:sec:S3-smoothing-SPT} for the SPT case)
\begin{equation}
S_3^{\rm tree}(R) = 3 \nu_2 + \dv{\log \sigma_{\rm L}^2(R)}{\log R}.
\end{equation}
Higher order cumulants behave similarly, though may have additional types of smoothing terms.

\section{Spherical collapse}

One of the few cases where the non-linear dynamics can be analytically treated  is the collapse of a spherically symmetric region. In the case of an EdS universe spherical collapse has an exact analytic solution, which is useful even in $\Lambda$CDM and extended cosmologies. We'll see in Chapter~\ref{chap:LDT-intro} that predicting the matter PDF with large deviations theory requires knowledge of spherical collapse dynamics.

\subsection{Derivation: overdense case}

We sketch the derivation of the solution for the density in an initially spherically overdense region. The approach in the underdense case is similar. 
Before any shells cross, the mass in a shell is conserved and Birkhoff's theorem guarantees that the mechanical energy of the shell is conserved. This total mechanical energy, $\epsilon$, is negative in the case of an overdensity, as the system is gravitationally bound. For a shell of (physical) radius $r$ enclosing a total mass $M$, 
\begin{equation}
\frac{1}{2}\left(\dv{r}{t}\right)^2 - \frac{GM}{r} = \epsilon.
\end{equation}
By changing coordinates 
\begin{equation}
\dd{t} = \left(-\frac{1}{2\epsilon}\right)^{1/2} r_*\dd\theta, \quad r_* = -\frac{GM}{2\epsilon}, \quad r = r_*y\,,
\end{equation}
the collapse equation reduces to
\begin{equation}
\left(\dv{y}{\theta}\right)^2 = 2y - y^2\,,
\end{equation}
which is solved by $y(\theta) = 1-\cos(\theta)$. With this, the radius and time can be expressed parametrically in terms of $\theta$
\begin{subequations}
\begin{align}
r(\theta) &= -\frac{GM}{2\epsilon}(1-\cos\theta)\,, \\
t(\theta) &= \frac{GM}{(-2\epsilon)^{3/2}}(\theta - \sin\theta).
\end{align}
\end{subequations}
We would like to write the mapping from the linearly extrapolated density to the true non-linear density. The mass at some initial scale factor $a_i$ is 
\begin{equation}
M = \frac{4\pi}{3}\rho_i a_i^3 R_i^3\,,
\end{equation}
where $\rho_i, R_i$ are the initial density and comoving radius of the  overdensity, which is conserved.  In an Einstein-de Sitter background, the scale factor evolves as
\begin{align}\label{eq:SC-scalefactor-theta}
a &= \left(\frac{8\pi G}{3}\rho_i a_i^3\right)^{1/3} \left(\frac32 t\right)^{3/2} \nonumber \\
&= \left(\frac{9}{2}\right)^{1/3} \frac{4\pi G}{(-6\epsilon)}\rho_ia_i^3 R_i^2 (\theta-\sin\theta)^{2/3}\,,
\end{align}
where in the second line we have substituted the parametric expression for $t$ in terms of $\theta$. The comoving radius $R=r/a$ of the region is given by
\begin{equation}
R(\theta) = R_i \left(\frac29\right)^{1/3}\frac{1-\cos\theta}{(\theta-\sin\theta)^{2/3}}.
\end{equation}
From the scale factor and comoving radius, we can obtain the density contrast as
\begin{equation}
1+\delta(R) = \frac{3 M }{4\pi r^3 \bar{\rho}} = \left(\frac{R_i}{R}\right)^3\,,
\end{equation}
where $\bar{\rho}$ is the background mean density of the universe, given by $\bar{\rho}=\rho_i a_i^3 / a^3$. Substituting for $R(\theta)$ above we obtain the non-linear density in the region
\begin{equation}\label{eq:SC-deltaNL-theta}
\delta(\theta) = \frac{9}{2}\frac{(\theta - \sin\theta)^2}{(1-\cos\theta)^3} - 1.
\end{equation}
The linear density is obtained by relating the mechanical energy $\epsilon$ to the initial conditions. The scale factor and density expressions \eqref{eq:SC-scalefactor-theta}, \eqref{eq:SC-deltaNL-theta} can be expanded in small $\theta$ leading to
\begin{equation}
\epsilon = -\frac{5}{3}\frac{\delta_i(R_i)}{a_i}\frac{4\pi G}{3}\rho_i a_i^3 R_i^2\,,
\end{equation}
where $\delta_i(R_i)$ is the density contrast at the initial time. In an EdS universe, the linear growth factor $D_+=a$, so the linear density is simply $\delta_{\rm L} = a \delta_i / a_i$, given by
\begin{equation}
\delta_{\rm L}(\theta) = \frac{3}{20}[6(\theta-\sin\theta)]^{2/3}.
\end{equation}

\begin{figure}[h!t]
\centering
\includegraphics[scale=1]{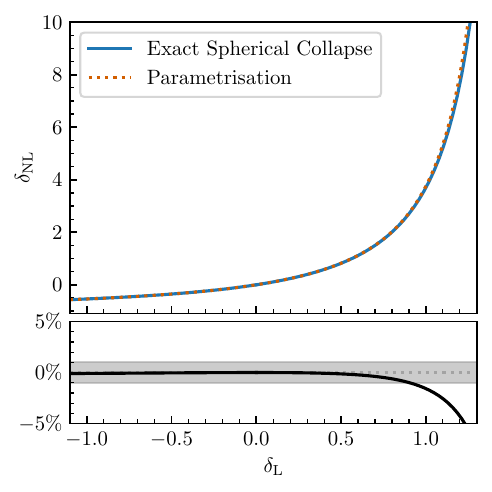}
\caption[Spherical collapse mapping between linear and non-linear densities.]{Comparison of the analytic parametrisation and the exact spherical collapse mapping between linear and non-linear densities before shell crossing. We use the spherical collapse index $\nu_{\rm SC}=21/13$ to match the skewness from SPT. The lower panel shows the relative error of this parametrisation to the exact solution, the shaded region at 1\%. }
\label{fig:spherical-collapse-exact}
\end{figure}

In the case of an initial underdensity, the result is 
\begin{subequations}
\begin{align}
\delta_{\rm NL} &= \frac{9}{2}\frac{(\sinh\theta-\theta)^2}{(\cosh\theta-1)^3} -1\,, \\
\delta_{\rm L} &= -\frac{3}{20} [6( \sinh\theta - \theta)^{2/3}].
\end{align}
\end{subequations}
This exact mapping between linear and non-linear densities is shown in Figure \ref{fig:spherical-collapse-exact}.

An alternative derivation for the evolution of spherical collapse is to consider the fluid equations in the case of spherical symmetry, assuming no shear or vorticity in the velocity field. This leads to the differential equation
\begin{equation}\label{eq:LCDM-spherical-collapse-diffeq}
\ddot{\delta} + 2 H(a) \dot{\delta} - \frac{4}{3} \frac{\dot{\delta}^2}{1+\delta} = 4 \pi G \bar{\rho}(a) (1+\delta) \delta = \frac{3}{2}\Omega_m(a) H^2(a) (1+\delta)\delta\,,
\end{equation}
for the overdensity in a spherically collapsing sphere.

\subsection{Empirical formula for spherical collapse}

In practice, there is a very simple and accurate parametrisation of this spherical collapse 
\begin{equation}\label{eq:SC-fitting-formula}
\mathcal{F}(\delta_{\rm L}) = \rho_{\rm NL}^{\rm SC}(\delta_{\rm L}) = \left( 1 - \frac{\delta_{\rm L}}{\nu_{\rm SC}} \right)^{-\nu_{\rm SC}},
\end{equation}
where $\nu_{\rm SC}$ is a free parameter which can be tuned depending on the use case. The use of of the letter $\nu$ here is suggestive, as the vertex factors of a perturbation theory are related to spherical collapse in the following way. As we've seen, knowledge of all the angular averaged perturbation kernels $\nu_n$ gives knowledge of the all the cumulants (to tree order in SPT). \textcite{Bernardeau_1992_GravityinducedQuasiGaussian} showed that the generating function\footnote{They refer to the generating function of the vertex factors as $\zeta(\tau)$. We refer to it as $\mathcal{F}(\delta_{\rm L})$ for consistency with later Chapters.} for these $\nu_n$, 
\begin{equation}
\mathcal{F}(\delta_{\rm L}) = \sum_{n}\frac{\nu_n}{n!}\delta_{\rm L}^n\,,
\end{equation} 
obeys the dynamical equations of spherical collapse \eqref{eq:LCDM-spherical-collapse-diffeq}. In this sense, the leading order dynamics of Eulerian perturbation theory is spherical collapse.

In general if a perturbation scheme has vertex factors $\nu_n$, then to match the tree-order skewness from that perturbation scheme with the parametrised spherical collapse require (see Appendix~\ref{app:spherical_collapse_to_skew})
\begin{equation}
\nu_{\rm SC} = \frac{1}{\nu_2 - 1}.
\end{equation}
For SPT in 3D,  $\nu_2=34/21$ which gives a spherical collapse index of $\nu_{\rm SC}=21/13\approx 1.6$.

\section{Hierarchical clustering models}

A particular class of non-Gaussian fields are the \emph{hierarchical models}, which provide a set of consistent relations between the $n$-point correlation functions. When represented diagrammatically, the correlation functions in hierarchical models can be expressed as 
\begin{equation}
\xi_n(\xx_1, \dots, \xx_N) = \sum_{\mathsf{T}\in  \rm trees} Q_N(\mathsf{T}) \prod_{\mathrm{lines}\in \mathsf{T}} \xi(\xx_i, \xx_j)\,,
\end{equation}
such that each contribution to the $n$-point function can be constructed only by considering connected diagrams on $n$-points with no loops (trees). 
\begin{figure}[h!t]
\centering
\includegraphics[width=0.8\textwidth]{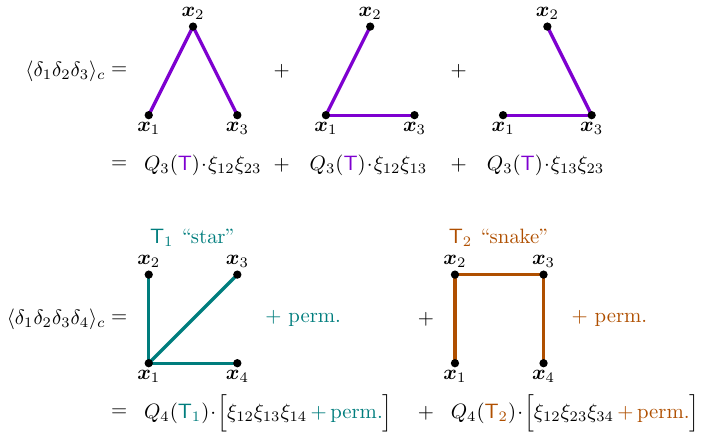}
\caption[Tree diagrams for 3- and 4- point correlation functions in a hierarchical model.]{The tree diagrams for the 3- and 4-point correlation functions in a hierarchical model. The there are two reduced correlation functions $Q_4$ corresponding to the two topologies of the diagrams. We use the shorthand $\delta_i=\delta(\xx_i), \xi_{ij}=\xi(\xx_i,\xx_j)$ and perm. refers to permuting the labels of the points in all possible combinations.}
\label{fig:heirarchical_4point}
\end{figure}
Each diagram has the edges weighted by the two-point correlation between those points $\xi(\xx_i,\xx_j)$, and each diagram is then given a weight depending only on its overall topology $Q_n(\mathsf{T})$. We show this in Figure~\ref{fig:heirarchical_4point}, as the 3 point function only has one topology of tree diagram (corresponding to a single value for $Q_3$), while the 4-point correlation function has two distinct topologies, $Q_4(\mathsf{T}_1), Q_4(\mathsf{T}_2)$.

For power law two-point correlations functions
\begin{equation}\label{eq:power-law-xi}
\xi(r) = \left(\frac{r}{r_0}\right)^{-\gamma},
\end{equation}
these hierarchical models exhibit a scaling relation
\begin{equation}\label{eq:hierarchical-scaling}
\xi_n(\lambda \xx_1, \dots, \lambda \xx_n) = \lambda^{-\gamma (n-1)} \xi_n(\xx_1, \dots, \xx_n)\,.
\end{equation}
This form of the two-point correlation function \eqref{eq:power-law-xi} is expected for initial power spectra $P(k)\propto k^n$ which gives $\xi(r)\propto r^{-(n+3)}$, and the specific form \eqref{eq:power-law-xi} is observed to fit galaxy data over $r$ between $0.1$--$10$ Mpc/$h$, with $r_0\simeq 5$ Mpc/$h$ and $\gamma\simeq 1.8$ \parencite{Groth1977ApJ, Davis1977ApJS, Peebles_1980_LargescaleStructure}. This sort of scaling is anticipated from the scale free nature of Newtonian gravity, making hierarchical models an attractive class of non-Gaussian fields to model structure formation, even in the non-linear regime.

In particular, the scaling relation \eqref{eq:hierarchical-scaling} guarantees that the $n^{\rm th}$ cumulant in a hierarchical model scales as $\bar{\xi}_n = \langle\delta^n\rangle_c \propto \ev{\delta^2}_c^{n-2}=\sigma^{2(n-1)}$. This is the same scaling we saw in the tree-order skewness from SPT in Section~\ref{sec:tree-order-skewness-SPT}. In hierarchical models the reduced cumulants $S_n$ are entirely determined by the functions $Q_n(\mathsf{T})$ and the product of two-point correlations. If we assume the product of the averages is well approximated by the average of the products then the reduced cumulants are given by
\begin{equation}
S_n \approx \sum_{\mathsf{T}\in  \rm trees} Q_n(\mathsf{T})\,,
\end{equation}
in such models.

We can see that tree-order SPT is hierarchical in precisely this way, as the tree-order contribution to $\langle \delta^p\rangle_c$ features kernels functions $F_n$ of at most degree $p-1$, which couple $p-1$ copies of the variance $\sigma^2 = \langle\delta^2\rangle_c$. 

A further subclass of hierarchical models, called \emph{tree models}, are those where the contribution from each tree diagram $Q(\mathsf{T})$ is computed locally, depending only on the number of lines into each vertex
\begin{equation}
Q_n(\mathsf{T}) = \prod_{\mathrm{vertices}\in \mathsf{T}} \nu_p\,,
\end{equation}
where the $\nu_p$ is a weight associated to a vertex with $p$ incoming lines and we take $\nu_0 = \nu_1 = 1$ for completion. In the case of the 4 point correlation function as shown in Figure~\ref{fig:heirarchical_4point}, this would produce
\begin{equation}
S_4 = Q_4(\mathsf{T}_1) + Q_4(\mathsf{T}_2) = 4 \nu_3 + 12 \nu_2^2\,,
\end{equation}
as there are 4 ``star'' diagrams with one vertex connected to the remaining 3 points, and 12 ``snake'' diagrams where no point is connected to more than 2 points. In this way, $S_p$ in a hierarchical model is entirely fixed by the vertex factors of at most order $p-1$.

As we saw in Section~\ref{sec:tree-order-skewness-SPT}, the vertex factors defined in this way for SPT are precisely the angle averaged perturbation theory kernels defined in equation~\eqref{eq:SPT_nu_n_def}. Thus the leading order dynamics of Eulerian perturbation theory produce a hierarchical tree model for the correlation functions of the density field, where the vertex weights are angular averages of the perturbation theory kernels. We will meet another specific hierarchical model in Chapter \ref{chap:covariance} known as the ``minimal tree model''.

%% file: text/chapter3-LDT-for-cosmology.tex

\chapter{Large deviations theory for cosmology} \label{chap:LDT-intro}

\minitoc

This Chapter introduces the mathematical framework of Large Deviations Theory which we used to predict the one-point matter PDF in following Chapters. 

\section{Introduction}

The study of the frequency of rare events is a mathematically rich field of probability theory. Large Deviations Theory (LDT) describes the asymptotic behaviour of the tails of probability distributions, far away from regions where the familiar central limit theorem applies. For our purposes it is useful to think of LDT as a formalism for obtaining the exponential behaviour of probability distributions, essentially as an extension to the law of large numbers and central limit theorem. In a physics context, modelling rare events arises in any sort of non-equilibrium system, making LDT and related concepts applicable in statistical mechanics, disordered systems, and even some quantum systems, in addition to the cosmological context presented here.

The rigorous mathematics of LDT can be difficult treading, and is not necessary to understand at the level of application to the cosmological matter PDF. We present but do not prove the main definitions and theorems needed to construct the matter PDF, neglecting technical details of topology and measure theory. \textcite{Touchette_2012, Touchette_2009_LargeDeviation}  present a physicist styled introduction to LDT as applied to statistical mechanics, with Appendix B of \textcite{Touchette_2009_LargeDeviation} providing details on some of the rigorous formulation of the theory.

The end result of this Chapter is a formalism for predicting the one-point PDF of matter densities averaged over spheres of radius $R$, based on LDT. The rareness in the LDT sense is here controlled by the inverse variance of matter densities thus limiting the reach of the theory to radii above $\sim \! 10$ Mpc/$h$ where $\sigma\lesssim 1$. It is worth stressing here that while LDT as a formalism is concerned with the behaviour of the tails of probability distributions, this does not mean that the LDT formalism for the matter PDF is concerned with predicting the PDF deep in the tails. Instead, we apply the LDT theorems to ``rare fluctuations'' in the path integral mapping statistics from initial/linear to final/non-linear densities. This enables us to predict the matter one-point PDF from Gaussian initial conditions. Indeed, when the matter PDF is used for statistical forecasting or inference (as we do in Chapter~\ref{chap:MG-PDFs}) the deep tails of the matter PDF are often cut, with the constraining power of the PDF coming from the bulk near the peak.

\subsection{Development of LDT for cosmology}

The ground work for the application of LDT to the PDF of the cosmic matter field was developed in  several works \parencite{Bernardeau_1992_GravityinducedQuasiGaussian, Bernardeau94, Bernardeau_1994_EffectsSmoothinga, Bernardeau2000A&A, Valageas02}. In these works certain effects which become better understood within the LDT formalism, such as the exponential decay of the PDF past the critical density (which we discuss in Section \ref{sec:LDT-technical-stuff}), were noted but not fully formalised. The formalism as it is used today was initially laid out in \textcite{LDPinLSS} and since then there have been several refinements and applications to different cosmological problems.

The formalism was further refined in \textcite{Uhlemann16} in ways which extend the range of validity we discuss later. The LDT formalism has been adapted to go beyond the statistics of matter in a single sphere, capturing both two-point statistics through density dependent bias functions  (which we touch on in Chapter~\ref{chap:covariance})  \parencite{Uhlemann17Kaiser, Uhlemann2017MNRAS, Bernardeau1996A&A, Codis2016MNRAS}. It has also been adapted to biased tracers such as halos and galaxies \parencite{Uhlemann:2018b, Friedrich_2021}, the weak lensing statistics including the convergence PDF and aperture mass functions \parencite{Reimberg2018PhRvD, Uhlemann:2018a, Barthelemy:2020,Barthelemy2021MNRAS, Barthelemy2022PhRvD, Barthelemy2023arXiv, Boyle.etal_2023_CumulantGenerating}. 

It has also been shown to be flexible enough to work in non-standard cosmologies, and to provide useful information about new physics, including primordial non-Gaussianity \parencite{Uhlemann18pNG, Friedrich20pNG}, massive neutrinos \parencite{Uhlemann:2020, Boyle_2020}, evolving dark energy and modified gravity \parencite{Boyle_2020, Cataneo.etal_2022_MatterDensity, Gough.Uhlemann_2022_OnePointStatistics}. Chapter \ref{chap:MG-PDFs} presents the PDF predicted by LDT in modified gravity and dark energy cosmologies. Large deviations theory has also been applied to early universe physics including inflation and primordial black holes \parencite{Cohen2023PhRvD}.

\section{Large deviations theory}

\subsection{A motivating example}

To motivate the definitions and ideas of large deviations theory, we start with the analysis of a simple random variable.

\begin{figure}[h!t]
\centering
\includegraphics[scale=1]{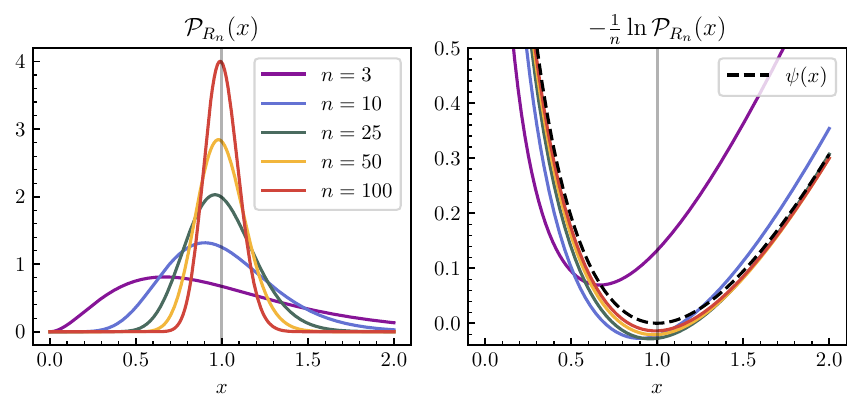}
\caption[Sample mean for an exponentially distributed random variable and the associated rate function. ]{(Left) Distribution of the sample mean of $n$ independent random variables drawn from the same exponential distribution with mean $\mu=1$. As the number of events grows the sample mean concentrates around the distribution mean. In the vicinity of the mean the distribution becomes closer and closer to a Gaussian distribution. However far away from the mean, in the tails of the distribution, the PDF is not well approximated by a Gaussian. (Right) The convergence of $-\frac{1}{n}\ln\mP_{R_n}(x)$ to the rate function $\psi(x)$ as the driving parameter $n$ grows. The rate function is minimised at the mean of the distribution. }
\label{fig:LDT_exponential_example}
\end{figure}

Consider a random variable $X$ which is exponentially distributed. Such a random variable describes, for example, the distribution of times/distances between occurrences of a Poisson process. The probability density function of the random variable $X$ is then
\begin{equation}
\mP_X(x) = \begin{cases}
\frac{1}{\mu}\exp(-\frac{x}{\mu}) & x>0 \\
0 & x\leq 0\,,
\end{cases}
\end{equation}
which has mean $\ev{X} = \mu$. Familiar to all scientists is the sample mean, where we take the average of $n$ samples drawn from the same distribution 
\begin{equation}
R_n = \frac{1}{n}\sum_{i=1}^n X_i\,,
\end{equation}
where $X_i$ are independent random variables which are all exponentially distributed. The distribution of observed sample means can be obtained by successive convolutions of the PDF of $X$, or by using the cumulant generating function associated to this PDF. The resulting distribution for the sample mean $R_n$ is given by the \emph{Erlang distribution},
\begin{equation}\label{eq:Erlang_dist}
\mP_{R_n}(x) = \left(\frac{n}{\mu}\right)^n x^{n-1} \frac{\exp[-nx/\mu ]}{(n-1)!}.
\end{equation}
Due to the central limit theorem, as $n\to \infty$ the distribution of $R_n$ centralises around the true mean $\mu$ and in fact approaches a Gaussian around the true mean, as can be seen in Figure~\ref{fig:LDT_exponential_example}. However, far away from the mean, in the tails of the distribution of $R_n$, the distribution is no longer well approximated by a Gaussian. Large deviations theory is concerned with the leading order, exponential behaviour of these tails in the $n\to \infty$ limit.

To examine the tails of the PDF in equation \eqref{eq:Erlang_dist} we consider its asymptotic behaviour in the limit of large $n$
\begin{subequations}
\begin{align}
\mP_{R_n}(x) &\sim \exp[-n\psi(x)] \,, \\
\psi(x) &\defeq \lim_{n\to \infty} - \frac{1}{n}\ln\mP_{R_n}(x) = \frac{x}{\mu}-1-\ln(\frac{x}{\mu}),
\end{align}
\end{subequations}
where the function $\psi(x)$ is called the \emph{rate function}. Expanding this about $x=\mu$ the rate function can be written
\begin{equation}
\psi(x) = \frac{1}{2}\left(\frac{x-\mu}{\mu}\right)^2 + \sum_{n=3}^\infty \frac{(-1)^n }{n} \left(\frac{x-\mu}{\mu}\right)^n,
\end{equation}
which shows the leading order quadratic behaviour anticipated from the central limit theorem, along with higher order corrections as we move away from the peak.

In Figure~\ref{fig:LDT_exponential_example} we show that convergence to this rate function happens rapidly in $n$. This rate function has a minimum at the true mean $\mu$ and in the case of a Gaussian distribution would be simply quadratic. The asymmetry of the rate function $\psi(x)$ about its minimum is precisely what determines the behaviour in the tails of the distribution. This rate function is the dominant contribution to the asymptotic behaviour of $\mP_{R_n}$ with other corrections being sub-exponential.

\subsection{Key definitions and theorems of LDT}

For our purposes we restrict ourselves to the case of continuous random variables, though the mathematics of LDT is capable of handling both discrete and continuous random variables. We say that a family of random variables $\{X_n\}$, each with associated probability density function $\mP_{X_n}$, which depends on some \emph{driving parameter} $n$ obeys a large deviation principle (LDP) if the following limit exists and is not zero everywhere
\begin{equation}
	\lim_{n\to \infty} -\frac{1}{n}\ln\mP_{X_n}(x) = \psi(x).
\end{equation}
If $X_n$ obeys an LDP, then the function $\psi(x)$ is called its \emph{rate function}, and the PDF for the random variable is asymptotically 
\begin{equation}
	\mP_{X_n}(x) \sim \exp[-n \psi(x)],
\end{equation}
where here $\sim$ denotes an asymptotic equality in the $n\to \infty$ limit.

The key theorems in large deviations theory for our purposes relate the rate function $\psi(x)$ to the \emph{scaled cumulant generating function} (SCGF) of the random variable, and the rate functions of one random variable $X_n$ to a related random variable $Y_n$.

The scaled cumulant generating function of a random variable in this context is
\begin{equation}
\varphi_{X_n}(\lambda) = \lim_{n\to \infty} \frac{1}{n}\log \ev{e^{n\lambda X_n}},
\end{equation}
see Appendix~\ref{app:prob-distributions} for further discussion of how this is related to the standard cumulant generating function.

We now state the 3 theorems we need to apply large deviations theory to the cosmic matter field. 

\begin{enumerate}
\item \textbf{The G\"artner-Ellis Theorem}~\\
With knowledge of the scaled cumulant generating function of a random variable, $\varphi_{X_n}(\lambda)$, the G\"artner-Ellis Theorem allows calculation of the rate function $\psi(x)$.

If a random variable $X_n$ has a SCGF $\varphi_{X_n}(\lambda)$ which is differentiable, then $X_n$ satisfies a large deviation principle with rate function given by the \emph{Legendre-Fenchel transform}  of the SCGF
\begin{equation}
\psi(x) = \sup_{\lambda} [\lambda x - \varphi_{X_n}(\lambda)].
\end{equation}
When the SCGF is additionally strictly convex, then the \emph{Legendre-Fenchel transform}  reduces to the \emph{Legendre transform}\footnote{This might be more familiar in the context of Lagrangians and Hamiltonians, under the replacements where $\varphi \to L$, $\lambda \to \dot{x}$, $x\to p$ and $\psi\to H$. \textcite{Zia.etal_2009_MakingSense} presents an instructive and pedagogical approach to Legendre transforms.}
\begin{equation}\label{eq:GE-theorem}
\psi(x) = \lambda x - \varphi_{X_n}(\lambda), \quad x = \dv{\varphi_{X_n}}{\lambda}\,,
\end{equation}

Note that there is a specific version of this theorem, \emph{Cram\'{e}r's theorem}  (the original 1938 paper was translated into English in \textcite{Cramer2018arXiv_english}) which applies to the sample mean of independent and identically distributed (i.i.d.) random variables. In this case, the scaled cumulant generating function is simply the cumulant generating function as, for $X$ which are i.i.d.
\begin{equation}
\varphi_{X_n}(\lambda) = \lim_{n\to \infty} \frac{1}{n}\log\ev{e^{\lambda\sum_i^n X_i}} = \lim_{n\to \infty} \frac{1}{n}\log \prod_{i=1}^n\ev{e^{\lambda X_i}} = \log \ev{e^{\lambda X}} = \phi_X(\lambda). 
\end{equation}

\item \textbf{Varadhan's Theorem}~\\
Varadhan's theorem goes in the opposite direction to the G\"artner-Ellis theorem. Continuous functions $f$ in the same form as the SCGF (including the SCGF itself) can be constructed from the rate function $\psi(x)$,
\begin{equation}
\varphi[f] \defeq \lim_{n\to \infty} \frac{1}{n}\ln\ev{e^{nf(X_n)}} = \sup_{x}[f(x) - \psi(x)]\,.
\end{equation}
Taking the specific choice of a linear function $f(x) = \lambda x$ allows construction of the SCGF from the rate function via
\begin{equation}\label{eq:Varadhan-theorem}
\varphi(\lambda) = \sup_{x}[\lambda x - \psi(x)].
\end{equation}

\item \textbf{The Contraction Principle}~\\
The contraction principle allows the computation of rate functions for random variables in terms of the rate variable of a different random variable. If  a random variable $X_n$  satisfies an LDP with rate function $\psi_{X_n}$, then  the related $Y_n=\mathcal{F}(X_n)$ also satisfies an LDP if the mapping $\mathcal{F}$ is continuous (it need not be single valued). The rate function of $Y_n$ is related to the rate function of $X_n$ in the following way
\begin{equation}\label{eq:contraction-principle-general}
\psi_{Y_n}(y) = \inf_{\{x:\mathcal{F}(x)=y\}} \psi_{X_n}(x).
\end{equation}
In the case when $\mathcal{F}$ is many-to-one, we are contracting information about the rate function of one random variable down to another. This has a nice physical interpretation: the improbable fluctuations in $Y_n$ are sourced by the most probably of the improbable fluctuations in $X_n$. In the case where $\mathcal{F}$ is bijective, then the rate functions are related by simply $\psi_{Y_n}(y) = \psi_{X_n}(\mathcal{F}^{-1}(x))$.

\end{enumerate}


\section{Predicting the matter PDF with LDT}
\subsection{The matter PDF}

\begin{figure}[h!t]
\centering
\includegraphics[scale=1]{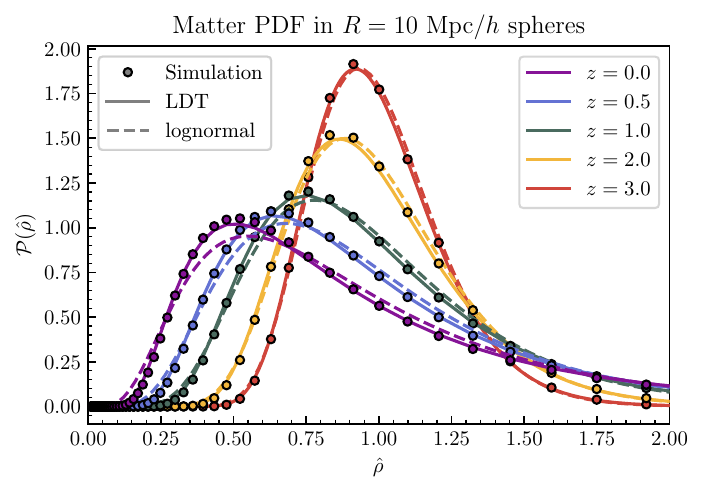}
\caption[The matter PDF in 10 Mpc/$h$ spheres.]{The PDF of the normalised cosmic density, measured in $10 \ h^{-1} \ \rm Mpc$ spheres at redshifts $z=0$--$3$. The data points are measured from the \textsc{Quijote} simulations \parencite{Villaescusa-Navarro:2020}. Solid lines show the LDT model for the matter PDF. Dashed lines show a lognormal model for the PDF with the same variance. }
\label{fig:quijote_pdfs}
\end{figure}

The goal of this modelling is to apply large deviations theory to the PDF of the late time cosmic density field when filtered on some scale $R$, shown in Figure~\ref{fig:quijote_pdfs} along with the predicted PDF based on large deviations theory (solid lines) and a lognormal model (dashed lines). The matter PDF is a particularly simple statistic, and the non-Gaussianity of the late time matter field can clearly be seen in the asymmetry of the PDF at low redshift. This allows the matter PDF to capture essential information beyond traditional two-point statistics, which we will see explicitly in Chapter \ref{chap:MG-PDFs}.

This kind of one-point statistic has a long history in cosmology, with Hubble noting that the two-dimensional count-in-cells of galaxies is approximated by a lognormal distribution \parencite{Hubble1934ApJ}. This idea was extended in \textcite{Coles1987MNRAS} which showed that a wide variety of non-Gaussian fields can be obtained by non-linear transformations of Gaussian fields. In particular, the lognormal model of the density field takes the matter density to be 
\begin{equation}
\rho(\xx) = \exp[\mu(\xx)]\,,
\end{equation}
 with $\mu(\xx)$ being Gaussian \parencite{Coles1991MNRAS}. Mathematically, this lognormal model is one of the simplest self-consistent random fields which always has $\rho>0$ and for which analytic properties can be derived. Kinematically, this lognormal model corresponds to extrapolating the continuity equation to non-linear densities, while assuming the velocity divergence remains Gaussian. This is unable to reproduce shell crossing (as in the Zel'dovich approximation) but does reduce to linear order perturbation theory for small densities. In addition to providing a better fit to the PDF at low redshift, the LDT model of the matter PDF is advantageous over the lognormal and similar models as it is directly sensitive to the cosmological parameters.  The lognormal model is known to be insufficient in predicting the response of the PDF to cosmological parameters needed for inference \parencite{Uhlemann:2020}. However, simple models for non-Gaussian fields such as these are still valuable in a different way, providing fast numerical recipes to generate many realisations of a density field which can be used to approximate covariance matrices, as we discuss in Chapter \ref{chap:covariance}.

\subsection{Applying the LDT theorems}

\begin{figure}[h!t]
\centering
\includegraphics[scale=1]{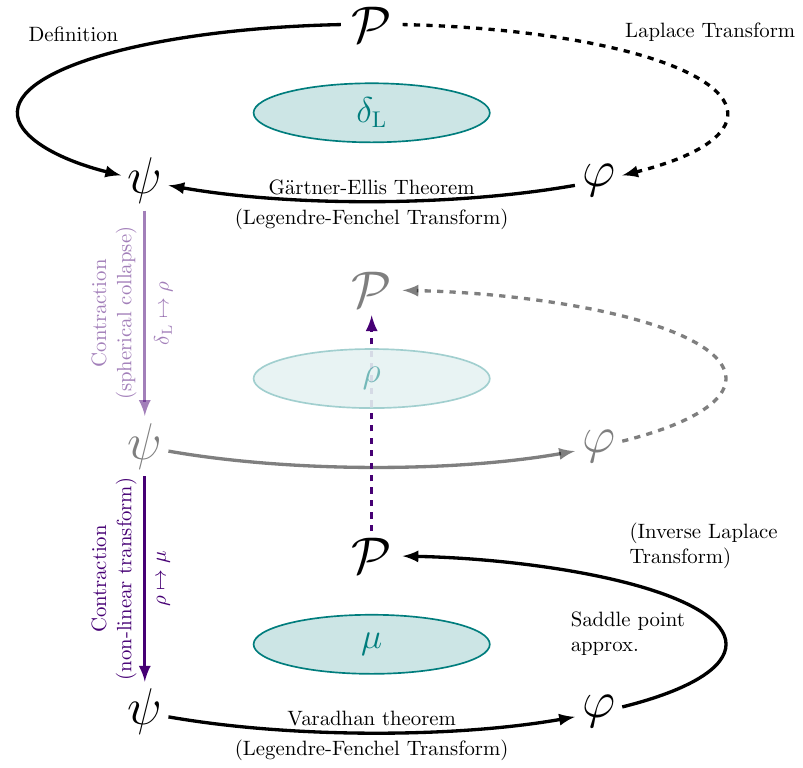}
\caption[Schematic diagram for the main theorems of Large Deviations Theory as applied to cosmology.]{Schematic diagram for the main theorems of Large Deviations Theory as applied to cosmology. The G\"artner-Ellis Theorem is equation~\eqref{eq:GE-theorem}, Varadhan's theorem equation \eqref{eq:Varadhan-theorem}, contraction principle \eqref{eq:contraction-principle-general}. Beginning from knowledge of the PDF of the linear density PDF, we can follow the relevant theorems to the PDF of the non-linear matter density $\rho$. By following an additional non-linear transformation $\rho \mapsto \mu$ we can increase the range of densities for which this model is accurate, as discussed in  Section \ref{sec:LDT-technical-stuff}.}
\label{fig:LDT_flowchart}
\end{figure}

The role of the driving parameter $n$ in LDT is played by the inverse non-linear variance $1/\sigma^2(z,R)$, such that the ``large deviations limit'' is the limit $\sigma^2\to 0$, either large scales or early times. We also will sometimes write down the ``decay rate functions'' $\Psi$ which are related to the rate functions $\psi$ by absorbing the variance 
\begin{equation}
\Psi_{R,\rho}(\rho) = \frac{\psi_{R, \rho}(\rho)}{\sigma^2(R,z) }\,,
\end{equation}
such that the exponential behaviour of the PDFs is given by
\begin{equation}
\mP_{R,\rho}(\rho) \propto \exp[-\Psi_{R, \rho}(\rho)] = \exp[- \frac{\psi_{R,\rho}(\rho)}{\sigma^2(R)}],
\end{equation}

For Gaussian initial conditions, the PDF, $\mP^{\rm lin}(\delta_{\rm L})$, of the linear matter density contrast, $\delta_{\rm L}$, in a sphere of radius $r$ is a Gaussian distribution with width given by the linear variance at scale $r$ and redshift $z$
\begin{equation}\label{eqn:initial_pdf}
    \mP^{\rm lin}_{r}(\delta_L) = \sqrt{\frac{\Psi^{\rm lin \prime \prime}_{r}(\delta_{\rm L})}{2\pi}}\exp\left[-\Psi^{\rm lin}_{r}(\delta_{\rm L})\right], \quad \Psi^{\rm lin}_{r}(\delta_L) = \frac{\delta_{\rm L}^2}{2\sigma_{\rm L}^2(r,z)}.
\end{equation}

We then consider all mappings\footnote{Note that in some of the other papers using the LDT model of the matter PDF \parencite[e.g.][]{Bernardeau94, LDPinLSS}, the mapping from linear density contrasts to non-linear densities via spherical collapse is called $\zeta(\tau)$ where $\tau$ is the linear density contrast. This is due to the relationship between the generating function for the vertex factors $\nu_n$ and spherical collapse dynamics, as noted at the end of Chapter \ref{chap:structure-formation}. We will generally use $\delta_{\rm L}$ for the linear density contrast and $\mathcal{F}(\delta_{\rm L})$ as the spherical collapse mapping function.} $\rho = \mathcal{F}(\delta_{\rm L})$ from the linear density to the final density. Due to the spherical symmetry of the matter PDF the dynamics which dominate this mapping are those which respect this spherical symmetry \parencite{Valageas02, Ivanov2019JCAP} and thus the mapping $\mathcal{F}$ is that of spherical collapse. This can be understood in the context of path integrals as stating that the most dominant evolution from linear densities to non-linear densities which affects the matter PDF is that of spherical collapse. It is important to note that this does not require that an initial sphere evolves via spherical collapse, simply that the statistics map via the dynamics of spherical collapse. 
We often will make use of the parametrisation $\mathcal{F}(\delta_{\rm L})$ in equation~\eqref{eq:SC-fitting-formula}, though of course one could also numerically solve for the spherical collapse mapping for any cosmology of interest.

Mass conservation allows us to link the initial sphere size $r$ to the final sphere size
$R$ by $r = R\rho^{1/}\footnote{This footnote should also be interpreted as part of the exponent. The 1/3 exponent applies for 3D applications only.}$, where $\rho=1+\delta$ is the \textit{normalised} non-linear density field given by the spherical collapse mapping $\rho(\delta_{\rm L})=\mathcal{F}(\delta_{\rm L})$.

To obtain the PDF of the late time density we apply the contraction principle to find 
\begin{equation}\label{eq:LDT_rate_function_rho}
\Psi_\rho(\rho) = \frac{\sigma_{\rm L}^2(R,z)}{\sigma_{\rm NL}^2(R,z)}\frac{\delta_{\rm L}^{\rm SC}(\rho)^2}{2 \sigma_{\rm L}(R\rho^{1/3},z)}\,,
\end{equation}
where $\delta_{\rm L}^{\rm SC}(\rho) = \mathcal{F}^{-1}(\rho)$ is the linear density contrast associated with non-linear density $\rho$ via spherical collapse. From this rate function, the exponential behaviour of the matter PDF is given by
\begin{equation}
\mP_\rho(\rho) \sim \exp[-\frac{\psi_\rho(\rho)}{\sigma_{\rm NL}^2(z,R)}] = \exp[-\Psi_\rho(\rho)].
\end{equation}
To obtain the prefactor to this expression, we would compute the scaled cumulant generating function $\varphi_\rho(\lambda)$ associated with the final density via Legendre-Fenchel transform then perform an inverse Laplace transform on the SCGF to obtain the PDF. This can be done analytically, as in \parencite{Bernardeau14}, but an excellent approximation can be achieved by approximating the inverse Laplace transform with a saddle point approximation in the $\sigma^2\to 0$ limit. This takes the form
\begin{equation}\label{eq:rhoPDF-saddle-point}
\mP_\rho(\rho) \approx \sqrt{\frac{\Psi_\rho''(\rho)}{2\pi}}\exp[-\Psi_\rho(\rho )].
\end{equation}
The saddle point approximation was used in a similar way in \textcite{Fry_1985_CosmologicalDensity}, though they use a hierarchical model for the cumulant generating function rather than spherical collapse mapping.

\subsection{Technical considerations in LDT for the matter PDF}\label{sec:LDT-technical-stuff}

Two technical points are in order regarding applying LDT to the calculation of the matter PDF. The first is the issue of applying the LDT theorems in the case of non-zero variance, as Varadhan's theorem constructs the scaled cumulant generating function $\varphi_\rho(\lambda)$ from the rate function $\psi_\rho(\rho)$ in the limit $\sigma_{\rm NL}^2\to 0$
\begin{equation}\label{eq:SCGF-rho-LDT}
\varphi_{\rho}(\lambda) = \sup_{\delta_{\rm L}} \left[ \lambda \rho(\delta_{\rm L}) -  \frac{\sigma_{\rm L}^2(R)}{2 \sigma_{\rm L}^2(R\rho(\delta_{\rm L})^{1/3})}\delta_{\rm L}^2\right],
\end{equation}
where $\rho(\delta_{\rm L})$ is the spherical collapse mapping. The PDF is however defined as the inverse Laplace transform of the cumulant generating function $\phi_\rho(\lambda)$,
\begin{equation}
\mP_\rho(\rho) = \int_{-i\infty}^{i\infty} \frac{\dd{\lambda}}{2\pi i} e^{-\lambda \rho + \phi_\rho(\lambda)}.
\end{equation}
The CGF for $\rho$ at finite variance $\sigma^2$ is chosen to be related to the SCGF \eqref{eq:SCGF-rho-LDT} by matching them the asymptotic limit
\begin{equation}
\phi_\rho(\lambda) = \frac{1}{\sigma_{\mathrm{NL}, \delta}^2}\varphi_\rho(\sigma_{\mathrm{NL}, \delta}^2 \lambda).
\end{equation}
The SCGF $\varphi_\rho$ is the generating function for the reduced cumulants $S_n$ (defined in equation \eqref{eq:define-Sn}) in the limit of 0 variance.

It is worth noting the power already present in this method over calculating the PDF from Eulerian perturbation theory. It is not known how to compute the matter PDF in SPT directly, instead relying on calculating the cumulants one by one and then approximating the PDF from those cumulants such as with an Edgeworth expansion. In this LDT prescription, from the non-linear variance, the reduced cumulants $S_n$ (via the SCGF), and spherical collapse dynamics we are able to construct something equivalent to an infinite Edgeworth expansion.

With this prescription for applying the results at non-zero variance, we can justify the form of the saddle point approximation \eqref{eq:rhoPDF-saddle-point} more fully. The full inverse Laplace transform expression for the matter PDF is
\begin{align}
\mP_\rho(\rho) &= \int_{-i\infty}^{i\infty} \frac{\dd{\lambda}}{2 \pi i} e^{-\lambda \rho + \phi(\lambda)}  = \int_{-i\infty}^{i\infty} \frac{\dd{\lambda}}{2 \pi i} e^{-\lambda \rho + \frac{1}{\sigma^2}\varphi(\sigma^2\lambda)} ,
\end{align}
we now define $\tilde\lambda = \sigma^2\lambda$ and rewrite
\begin{equation}
\mP_\rho(\rho) = \int_{-i\infty}^{i\infty} \frac{\dd{\tilde\lambda}}{2 \pi i \sigma^2} \exp[-\frac{1}{\sigma^2}(\tilde\lambda \rho - \varphi(\tilde\lambda))].
\end{equation}
The stationary point to apply in the saddle point approximation (Appendix \ref{app:SPA}) is given by $\dv*{\varphi}{\lambda} = \rho(\lambda)$
which is precisely the stationary condition in the G\"artner-Ellis theorem when $\varphi$ is convex. The direct application of the saddle point approximation on this stationary point is
\begin{equation}
\mP_\rho(\rho)\approx \sqrt{\frac{1}{2\pi \sigma^2 \varphi''(\lambda_*)}} \exp[-\frac{\lambda_*\rho - \varphi(\lambda_*)}{\sigma^2}],
\end{equation}
where $\lambda_* = \lambda_*(\rho)$ as defined by the stationary condition. We can then recognise $\lambda_*\rho - \varphi(\lambda_*)=\psi(\rho)$ by Varadhan's theorem. The second derivative of $\varphi$ can be expressed in terms of $\psi$ by the fact that they are related by Legendre transform (in the region where $\varphi$ is convex). Convex functions which are related by Legendre transform have first derivatives which are inverses \parencite{Zia.etal_2009_MakingSense} thus the second derivatives are related $\varphi'' = 1/\psi''$, leaving the final form \eqref{eq:rhoPDF-saddle-point}.

The second technical point relates to the convexity of the decay rate function $\Psi$. Generally there will exist some critical density $\rho_c$ such that $\Psi_{\rho}''(\rho_c)=0$ above which the Legendre transform of $\Psi_\rho$ is not defined \parencite{Bernardeau_1994_EffectsSmoothinga}. However, we are free to apply a further transformation to the final density field, and can simply apply the contraction principle \eqref{eq:contraction-principle-general} again (corresponding to the lower layer in Figure~\ref{fig:LDT_flowchart}). If $\mu(\rho)$ is a monotonic transformation of the density field, their rate functions are equal $\psi_\mu(\mu)=\psi_\rho(\rho(\mu))$. A non-linear transformation on the density field can increase the region where the associated rate function is convex, extending the applicability of the saddle point approximation. \textcite{Uhlemann16} advocates for using the log-density $\mu=\log(\rho)$ in the context of the matter PDF. The associated SCGF is then given by
\begin{equation}
\varphi_\mu(\lambda) = \sup_{\delta_{\rm L}} \left[ \lambda \mu(\delta_{\rm L}) -  \frac{\sigma_{\rm L}^2(R)}{2 \sigma_{\rm L}^2(R\rho(\mu)^{1/3})}\delta_{\rm L}^2\right],
\end{equation}
where $\rho(\mu) = e^{\mu}$, $\mu(\delta_{\rm L}) = \log(\rho(\delta_{\rm L}))$, and $\rho(\delta_{\rm L})$ is the spherical collapse mapping. In extending the LDT results to the case of finite variance, there is now a choice of whether to keep the driving parameter as the non-linear variance of the density field $\sigma_{\mathrm{NL},\rho}^2$ or to instead use the non-linear variance of the transformed density $\sigma_{\mathrm{NL}, \mu}^2$. In an EdS background in the absence of smoothing, this amounts to whether the reduced cumulants of the density $S_n^\rho$ or the log-density $S_n^\mu$ ought to be constant to tree order. From a theoretical perspective there is no perturbation theory motivation for these choices, and in the limit of vanishing variance this distinction vanishes as guaranteed by LDT. When using the log-density transformation we follow \textcite{Uhlemann16} and take the variance of the log-density field $\sigma_{\mathrm{NL}, \mu}^2$ as the driving parameter.

Applying the saddle point approximation to the PDF for $\mu$ (simply replacing every instance of $\rho$ with $\mu$ in equation~\eqref{eq:rhoPDF-saddle-point}), and noting that the $\rho$ PDF is related to the $\mu$ PDF by $\mP_\rho (\rho) = \mP_\mu (\mu)\dv*{\mu}{\rho} = \mP_\mu (\log \rho)/\rho$, the raw matter PDF from the saddle point approximation is 
\begin{equation}\label{eq:raw-mu-PDF}
\tilde{\mathcal{P}}_\rho(\rho) \approx \sqrt{\frac{\psi_\rho''(\rho) + \psi_\rho'(\rho)/\rho}{2\pi \sigma_{\mathrm{NL},\mu}^2}}  \exp(-\frac{\psi_\rho(\rho )}{\sigma_{\mathrm{NL},\mu}^2} ).
\end{equation}
The additional terms in the prefactor extend the range of applicability of this approximation, owing to the additional convexity granted from the log-transform. 

The final requirement for the obtaining the matter PDF in this manner is to ensure correct normalisation and mean. If the PDF were calculated from the full inverse Laplace transform this would be ensured, however, since we approximate it using the saddle point approximation the function in equation~\eqref{eq:raw-mu-PDF} is not guaranteed to have $\ev{\rho}=1$. By specifying the mean of the log-density $\ev{\mu}=\ev{\log\rho}$, the ``raw'' PDF in equation~\eqref{eq:raw-mu-PDF} can be rescaled to the final PDF by
\begin{equation}\label{eq:PDF-renorm}
\mP_\rho(\rho) = \tilde{\mathcal{P}}_\rho\left( \rho \cdot \frac{\ev{\rho}}{\ev{1}} \right) \cdot \frac{\ev{\rho}}{\ev{1}^2}\,,
\end{equation}
where here $\ev{f(\rho)}$ denotes integration against the ``raw PDF'' $\ev{f(\rho)} = \int \dd{\rho} \tilde{\mathcal{P}}(\rho) f(\rho)$.

\subsection{Summary of LDT for matter PDFs}

The LDT model of the matter PDF relies on only three ingredients
\begin{itemize}
\item the time- and scale- dependence of the linear variance $\sigma_{\rm L}^2(R,z)$.
\vspace{-6pt}
\item the mapping between linear density contrast and non-linear densities from spherical collapse, $\mathcal{F}:\delta_{\rm L}\mapsto \rho$. This can either be solved numerically, or can be well approximated by the fitting function for the Einstein-de Sitter exact solution
\begin{equation*}
\rho_{\rm SC}(\delta_{\rm L}) = \mathcal{F}(\delta_{\rm L}) = \left( 1-\frac{\delta_{\rm L}}{\nu_{\rm SC}}\right)^{-\nu_{\rm SC}}.
\end{equation*}
\vspace{-6pt}
\item the non-linear variance of the log-density field $\sigma_{\mathrm{NL},\mu}$.
\end{itemize}

The linear variance is straightforwardly calculated in a given theory of cosmology, requiring only linear dynamics. In $\Lambda$CDM cosmologies, the mapping for spherical collapse can be obtained by numerical integration of equation~\eqref{eq:LCDM-spherical-collapse-diffeq}, or by use of the parametrisation~\eqref{eq:SC-fitting-formula}. In cosmologies beyond $\Lambda$CDM, the parametrised EdS collapse is still remarkably accurate in the context of predicting the matter PDF, as we will see in Chapter~\ref{chap:MG-PDFs}. The non-linear variance can either be viewed as a free parameter of this model, or can be measured from simulations.

With these ingredients, the rate function for the density PDF can be constructed
\begin{equation*}
\psi_\rho(\rho;R,z) = \frac{[\delta_{\rm L}(\rho)]^2}{2}\frac{\sigma_{\rm L}^2(R,z)}{\sigma_{\rm L}^2(R\rho^{1/3},z)}\,,
\end{equation*}
where $\delta_{\rm L}(\rho) = \mathcal{F}^{-1}(\rho)$ is the linear density contrast associated to the non-linear density $\rho$ by spherical collapse. Approximating the inverse Laplace transform with the saddle point approximation, and using the log-density as an intermediate step to extend the range of validity, the (raw, unnormalised) matter PDF is given by
\begin{equation*}
\tilde{\mathcal{P}}_\rho(\rho) \approx \sqrt{\frac{\psi_\rho''(\rho) + \psi_\rho'(\rho)/\rho}{2\pi \sigma_{\mathrm{NL},\mu}^2}}  \exp(-\frac{\psi_\rho(\rho )}{\sigma_{\mathrm{NL},\mu}^2} )\,,
\end{equation*}
which can then be properly normalised by equation \eqref{eq:PDF-renorm}.

%% file: text/chapter4-LDT-for-MG.tex

\chapter{The matter PDF in extended cosmologies with LDT}\label{chap:MG-PDFs}

\minitoc

This Chapter is based on work presented in \textcite{Cataneo.etal_2022_MatterDensity} and \textcite{Gough.Uhlemann_2022_OnePointStatistics}. We show that the LDT formalism for the matter PDF introduced in Chapter~\ref{chap:LDT-intro} can be successfully adapted to predict the matter PDF in ``extended'' cosmologies with modified gravity and evolving dark energy. We consider two modifications to gravity --- Hu-Sawiki $f(R)$ gravity and the normal branch of the Dvali-Gabadaze-Porrati (DGP) braneworld --- and the parametrised $w_0 w_a$CDM dark energy model. A Fisher forecast is presented to quantify the information content in the matter PDF compared to the 3D matter power spectrum for a \textit{Euclid}-like survey volume. The PDF is found to halve the uncertainty on parameters for an evolving dark energy model relative to the power spectrum on its own. The PDF contains enough non-linear information to substantially increase the detection significance of departures from general relativity, with improvements up to six times the power spectrum alone.

\section{Introduction}

As presented in Part \ref{part:background}, $\Lambda$CDM has solidified as the standard model of cosmology, successfully predicting a wide range of cosmological observations, with the most accurate measurements of cosmological parameters currently coming from measurements of the CMB.
However, while CMB data is very valuable in extracting cosmological information, in the push to sub-percent measurements of standard cosmological parameters, and in testing non-standard cosmologies, the large-scale structure (LSS) of the universe is the most promising complementary tool. In particular, the large-scale structure provides a window to the expansion history of the universe, which makes it an exciting probe of dynamical dark energy and modifications to gravity.
 
Modifications to the standard cosmological model are motivated through several routes. Observationally, in recent years mild to severe tensions between early- and late-time probes of the growth of structure and background expansion have arisen \parencite[recently reviewed in e.g.][]{Douspis:2019, DiValentino:2021, Perivolaropoulos:2021} which strain the $\Lambda$CDM model. On a more fundamental level, thorough testing of GR---on which $\Lambda$CDM is based---has only been conducted on small astrophysical scales and in the strong field regime \parencite{Will:2014,Abbott:2017,Abbott:2019}, which leaves significant theory space open for modifications to the field equations on cosmological scales \parencite{DES-y3kp:2021,Ishak:2019,Ferreira:2019,Troester:2021,Raveri:2021,Pogosian:2021}. The nature of the two key ingredients in $\Lambda$CDM---dark matter and dark energy---is also unknown at a fundamental level. In this Chapter we consider modifications to GR and the (late-time) dynamics of dark energy, which we collectively refer to as ``extended cosmologies''. We return to the fundamental nature of dark matter in Chapters \ref{chap:making-dm-waves} and \ref{chap:how-classical}.

Current analysis of large-scale structure data searching for deviations from $\Lambda$CDM are largely built upon two-point statistics \parencite{Simpson:2013,Song:2015,Amon_2018_Eg,DES-y1MG:2019,Troester:2021,Lee:2021,Muir:2021,Chudaykin:2021,Vazsonyi:2021}.  A large amount of theoretical work has gone into modelling these two-point statistics beyond their linear behaviour in modified gravity and dark energy cosmologies, as a large amount of cosmological information is locked away in the small, non-linear scales  \parencite[e.g.,][]{Koyama:2009,Takahashi:2012,Brax:2012,Heitmann:2014,Zhao:2014,Mead:2016,Casarini:2016,Cusin:2018,Cataneo:2019,Winther:2019,Ramachandra:2021,EE2:2021}. However, we have seen that the matter distribution in the late universe is non-Gaussian, due to the non-linear collapse of structures under gravity, so combining these two-point analyses with non-Gaussian statistics is crucial for unlocking the full information content of the late-time universe. Non-Gaussian statistics respond strongly to modified gravity and dark energy cosmologies through the induced changes in the higher order moments of the cosmic density field, which leads to a remarkable complementarity with traditional two-point constraints \parencite{Shirasaki:2017,Peel:2018,Sahlen:2019,Liu:2021}. In particular, modifications to gravity induce an additional density dependence on the dynamics of matter, so statistics such as the marked power spectrum and correlation function (which weight two-point statistics based on density environments) can be sensitive to modifications to gravity and other fundamental physics such as massive neutrinos \parencite{Armijo2018MNRAS, Philcox2020PhRvD, Massara2021PhRvL, Aviles2021arXiv, Massara2023ApJ}. We note that some other non-Gaussian statistics which have been investigated in extended cosmologies are the bispectrum \parencite{Brax:2012,Munshi:2017,Yamauchi:2017,Crisostomi:2020,Bose:2020b}, higher-order weak lensing spectra \parencite{Munshi:2020}, the halo mass function \parencite{Lam_2012,Cataneo:2016,Hagstotz:2019,McClintock:2019,Bocquet:2020}, the void size function \parencite{Perico:2019,Verza:2019,Contarini:2021} and Minkowski functionals \parencite{Kratochvil:2012,Fang:2017}.

The probability density function of the three-dimensional matter density field smoothed on a given scale is a particularly simple non-Gaussian statistic that has observable counterparts in galaxy counts-in-cells and the weak lensing PDF. As presented in Chapter \ref{chap:LDT-intro}, the matter PDF can be accurately predicted in the mildly non-linear regime to extract cosmological information using large deviations theory. In this Chapter we build upon that formalism, demonstrating that the LDT formalism can be extended to apply in modified gravity (MG) and dark energy (DE) cosmologies using Einstein-de Sitter spherical collapse dynamics together with knowledge of the linear theory for the extended cosmologies. Modified gravity and evolving dark energy are known to imprint characteristic changes on the skewness, kurtosis, and higher cumulants of the PDF from $N$-body simulations \parencite{Li_2012halosvoidsfR,Hellwing:2013,Hellwing:2017,Shin:2017}. These are naturally recovered by predicting the full shape of the matter PDF. Through Fisher forecasts we quantify for the first time the ability of the PDF to detect distinct departures from GR and to constrain the dark energy equation of state, especially when combined with the matter power spectrum.

\section{Selected extended models of cosmology}
In this Section we introduce the specific extensions to $\Lambda$CDM which are considered for the remainder of the Chapter.

\subsection{Evolving dark energy}

One of the simplest extensions to the standard $\Lambda$CDM is replace the cosmological constant $\Lambda$ with some evolving dark energy component. We consider the simplest parametrisation for an evolving dark energy
\parencite{Chevallier:2001,Linder:2003}
\begin{align}\label{eq:DEeos}
    w_{\rm eff}(a) = w_0 + w_a(1-a) \, ,
\end{align}
where $\{w_0, w_a\}$ are phenomenological parameters. This parametrisation simply expands the dark energy equation of state about its present day value in a generic way, without specifying an underlying mechanism for why the dark energy is dynamical. In a universe with only matter and dark energy obeying this equation of state, the Hubble rate is given by
\begin{equation}
H(a) = H_0 \left[ \Omega_m^0 a^{-3} +  \Omega_{\rm DE}^0 a^{-3(1+w_0+w_a)} e^{-3w_a(1-a)} \right]^{1/2}.
\end{equation}

We refer to models with constant but $w_0\neq -1$ equations of state as $w$CDM cosmologies, while referring to those with $w_a$ non-zero as $w_0w_a$CDM cosmologies. 

We note that during the writing of this thesis the first results analysing the Baryon Acoustic Oscillations in galaxies, quasars, and Lyman-$\alpha$ tracers from DESI  \parencite{DESICollaboration2024arXiv_cosmoconstraints} were released, including an analysis of evolving dark energy parametrised in this way. Their analysis finds a slight preference for evolving dark energy over a constant $w_0$, between 2.9$\sigma$ and $3.5\sigma$ significance depending on which supernovae data are also included in the analysis.

\subsection{Modified gravity}

The field of modified gravity is enormous, and we do not present a comprehensive overview of the landscape of possible modifications to GR. We direct the reader to reviews such as \textcite{Clifton.etal_2012_ModifiedGravity, Joyce.etal_2015_CosmologicalStandard, Ferreira_2019_CosmologicalTests} for more complete overviews of cosmological modifications to gravity. We consider two models in this work, Hu-Sawiki $f(R)$ \parencite{Hu:2007, DeFelice:2010} and the normal branch of Dvali-Gabadaze-Porrati (DGP) braneworld gravity \parencite{Dvali:2000}.

Generally when modifying the theory of gravity, the relationship between the two gravitational potentials in the perturbed FLRW metric, as well as the Poisson equation relating the gravitational potential to the matter density are modified. New terms which arise in the Poisson equation \eqref{eq:cosmological-poisson} which are not GR-like are referred to as a ``fifth force'' as these dynamics represent some new fundamental physics not described by the standard four fundamental forces.

Modified gravity theories typically require ``screening mechanisms'' which allow the effects of the modification to vanish in high density environments. This removes the effect of the fifth force in solar system like environments to be consistent with constraints on beyond GR parameters on those scales. This screening can occur via the gravitational potential \parencite[as in chameleon screening][]{Khoury2004PhRvD}, its gradient \parencite[as in k-mouflage][]{Babichev2009IJMPD}, or the density field directly (second derivatives of the potential) as in Vainshtein screening \parencite{Vainshtein1972PhLB}. Such mechanisms typically arise from the presence of non-linear terms in the equations of motion, and naturally lead to additional density dependence of gravity in modified gravity cosmologies.

\subsubsection{DGP gravity}

DGP gravity arises as a higher dimensional theory of gravity  \parencite{Dvali:2000}. The action of DGP gravity takes the same form as the standard Einstein-Hilbert action of GR, integrated over both the four-dimensional (brane) spacetime and the full five-dimensional (bulk) spacetime the brane spacetime is embedded in
\begin{equation}
S=\int_{\rm brane} \dd[4]{x}\sqrt{  -g^{(4)}  } \left[\frac{ R^{(4)} }{  2\kappa^2  } +\mathcal{L}_m (g^{ (4) }_{\mu\nu},\psi)\right] + \int_{\rm bulk} \dd[5]{x} \sqrt{  -g^{(5)}  } \frac{  R^{(5)}  }{  2(\kappa^{(5)})^2  }.
\end{equation}
Here $\mathcal{L}_m$ represents the Lagrangian of matter, which only appears in the brane, and $\psi$ denotes the matter fields. The metric $g^{(5)}$ is the five-dimensional spacetime metric and $g^{(4)}$ is the restriction of the bulk metric to the four-dimensional brane. The five-dimensional Ricci scalar $R^{(5)}$ is divided by the gravitational strength $(\kappa^{ (5) })^2=8\pi  G^{ (5) } /c^4$ (with $\kappa^2$ being the standard four-dimensional gravitational strength), which allows us to define the \emph{crossover scale} as the ratio of the two different gravitational strengths
\begin{equation}
r_{\rm c} = \frac{1}{2}\frac{  G^{(5)}   }{G}.
\end{equation}
Below this crossover scale gravity looks four dimensional, while at larger lengths the higher dimensional nature of this theory begins to enter. In order to be viable this crossover scale must be of order the Hubble radius. 

The Hubble parameter in DGP gravity (in the absence of a cosmological constant) has two branches
\begin{equation}\label{eq:Om_rc}
H(a) = H_0 \sqrt{\Omega_m^0 a^{-3} + \Omega_{\rm rc}} \pm \sqrt{\Omega_{\rm rc}}, \quad \Omega_{\rm rc} \equiv \frac{1}{4(r_{\rm c}H_0)^2}\,,
\end{equation}
where we have introduced an effective energy density contribution $\Omega_{\rm rc}$ \parencite[see, e.g.,][]{Lombriser:2009}. The standard growth of GR is recovered in the limit $\Omega_{\rm rc}\to 0$. The branch corresponding to the positive sign above is called the ``self accelerating branch'' (sDGP), and was originally promising as an alternative to dark energy in explaining cosmic acceleration. However this self accelerating branch has theoretical instabilities, due to degrees of freedom without minimum energy states (so called ``ghosts''), and is generally not considered viable for this reason.
In this work we consider the so called ``normal branch'' of DGP (nDGP), corresponding to the minus sign in equation~\eqref{eq:Om_rc}, together with a smooth dark energy component which is chosen such that the background expansion is identical to $\Lambda$CDM \parencite{Schmidt:2009},
\begin{equation}
H^2 = H_0^2\left[\Omega_m^0 a^{-3} + \Omega_\Lambda\right].
\end{equation}

The modified Poisson equation for nDGP gravity takes the form
\begin{subequations}
\begin{align}
\nabla^2\Phi = \nabla^2 \Phi_{\rm GR} + \frac{1}{2}\nabla^2 \phi_{\rm DGP} ,
\end{align}
where $\phi_{\rm DGP}$ can be considered a scalar degree of freedom called the \emph{brane bending mode}, which produces an attractive fifth force. On scales smaller than the Hubble horizon and crossover scale \parencite{Schmidt2009:DGPgrav, Brito2014PhRvD, Winther2015PhRvD}, and in the quasi-static regime, the brane bending mode obeys \parencite{Koyama:2009}
\begin{equation}
\nabla^2\phi_{\rm DGP} + \frac{r_{\rm c}^2}{3\beta a^2}\big[(\nabla^2\phi_{\rm DGP})^2 - (\nabla_i\nabla_j\phi_{\rm DGP})^2\big] = \frac{8\pi G a^2}{3\beta}\bar{\rho}\delta\,.
\end{equation}
\end{subequations}
where 
\begin{equation}
\beta(a) \defeq 1 + 2r_{\rm c} H\left(1 + \frac{\dot H}{3H^2 }\right).
\end{equation}

\subsubsection{$f(R)$ gravity}

Another natural way to generalise GR phenomenologically is to add an arbitrary function of the Ricci scalar, $R$, to the standard Einstein-Hilbert action. The resulting action is
\begin{equation}\label{eq:fR-action}
S =  \int \dd[4]{x}\sqrt{-g} \frac{R+f(R)}{2\kappa^2} + \mathcal{L}_m(g_{\mu\nu}, \psi_m)\,.
\end{equation}
In this form the functional form $f(R) = 0$ reduces to standard GR, and $f(R) = - 2\Lambda$ with constant $\Lambda$ produces GR with a cosmological constant $\Lambda$. We take the Hu-Sawiki parametrised form\footnote{The specific parametrisation presented here is as in e.g. \textcite{Cataneo:2016}.} \parencite{Hu:2007}
\begin{equation}
f(R) = -2 \Lambda \frac{R^n}{R^n + \mu^{2n}}\,,
\end{equation}
with free parameters $\Lambda>0$, $\mu^2$, and $n>0$. In the high curvature regime $R \gg \mu^2$, this can be expanded
\begin{align}
f(R) &= -2 \Lambda + 2\Lambda\frac{\mu^{2n}}{R^n} \\
&= -2\Lambda  - \frac{f_{R0}}{n}\frac{\bar{R}^{n+1}_0}{R_n}\,,
\end{align}
where $f_{R0} = -2n\Lambda \mu^{2n} / \bar{R}^{n+1}_0$ is the present day value of $\dv*{f}{R}$ and $\bar{R}$ is the background value of the Ricci scalar.

This derivative $f_R  =\dv*{f}{R}$ can be considered a scalar degree of freecom often referred to as the \emph{scalaron} and the $f(R)$ action can be brought into the standard scalar-tensor form via suitable transformations \parencite[see e.g.][for details]{Sotiriou2010RvMP, Carroll.etal_2006_ModifiedsourceGravity, deFelice2011PhLB}. Because of this we prefer to discuss the parametrisation in terms of the present day value of the field $f_{R0}$. The scalar field formulation of $f(R)$ gravity allows it to be absorbed into the class of Horndeski models \parencite{Horndeski:1974} which are the most general scalar-tensor theories which result in second order equations of motion for the field.

For $\abs{f_{R0}}\ll 1$ the high curvature approximation is appropriate for all redshifts $z\geq 0$, as there are two curvature scales set by $\Lambda \sim \order{\bar{R}_0}$ and $\mu^2$. \textcite{Hu:2007} demonstrate that this model produces $\order{\abs{f_{R0}}}$ deviations from a cosmological constant. Given current upper bounds from galaxy clusters limit $\abs{f_{R0}}\lesssim 10^{-5}$ \parencite{Cataneo2015PhRvD} in this work we consider $\abs{f_{R0}}=10^{-5}, 10^{-6}$ for which the background evolution is very close to $\Lambda$CDM and we adopt an effective equation of state $w_{\rm eff}=-1$.

The modification to the Poisson equation for $f(R)$ gravity is
\begin{subequations}
\begin{align}
\nabla^2 \Phi &= \nabla^2 \Phi_{\rm GR} - \frac{1}{2}\nabla^2 \delta f_R\,, \\
\nabla^2 \delta f_R &= -\frac{a^2}{3}\Big[\delta R(f_R) + 8\pi G \bar{\rho}\delta \Big],
\end{align}
\end{subequations}
where $\delta R(f_R) = R - \bar{R}$ is the perturbation to the Ricci curvature and $\delta f_R = f_R(R) - f_R(\bar{R})$ is the fluctuation in the scalaron field.

\subsection{Parametrising perturbations in extended cosmologies}

For scalar-tensor theories within the Horndeski class \parencite{Horndeski:1974},  the evolution of linear density perturbations on sub-horizon scales can be parametrised by \parencite{Gleyzes:2013,Bellini:2014}
\begin{align}\label{eq:MG-linear-growth}
    \partial_a^2 D+ \frac{3}{2a}\left[ 1-w_{\rm eff}(a)\Omega_{\rm eff}(a) \right] \partial_a D - \frac{3\Om(a)}{2a^2} [ 1+ \epsilon(k,a) ] D = 0 \, ,
\end{align}
where $D$ is the linear growth function such that the final linear density fluctuation $\hat{\delta}_{\rm L}(k,z) = D(k,z)\delta^{\rm ini}$, $\Omega_{\rm eff}(a)$ and $w_{\rm eff}(a)$ are the energy density and equation of state, respectively, of the effective dark energy fluid driving the background acceleration, and $\epsilon(k,a)$ represents the scale- and time-dependent fractional deviation from the gravitational constant. The standard $\Lambda$CDM growth equations are recovered by setting $w_{\rm eff}\!=\!-1$, $\epsilon\!=\!0$.

This linear growth equation is appropriate for all the extensions considered here even though they are not all obviously of Horndeski type. Smooth dark energy such as $w_0w_a$CDM are accounted for in \eqref{eq:MG-linear-growth} simply by $\epsilon=0$. For the scales we consider, nDGP gravity enters mainly as a modification to the background dynamics, similarly to a dark energy model. This can be more formalised by writing DGP gravity as a four-dimensional scalar-tensor theory on scales between the Hubble radius and about 1000 km, as done in \textcite{Nicolis:2004,Park:2010}. The scalar degree of freedom here is called the  brane-bending mode. $f(R)$ gravity can be manifestly put into Horndeski form by considering $f_R$ as a scalar field.  See \textcite{Cataneo:2019} for the explicit functions to represent DGP and $f(R)$ gravity in Horndeski form.

The $k$-dependence of $\epsilon(k,a)$ means that the linearised density fluctuations no longer separate space and time as they do in $\Lambda$CDM, even in the linear regime modified gravities produce scale dependent growth. 

The evolution of a spherical top-hat density fluctuation, $\delta$, can be parametrised \parencite[see, e.g.,][]{Schmidt:2009b}
\begin{align}\label{eq:SCeqn-MG}
    \ddot\delta + 2 H \dot\delta - \frac{4}{3} \frac{\dot{\delta}^{2}}{(1+\delta)} = \frac{3}{2} H^2 \Om (1 + \mathscr{F}) (1+\delta)\delta \, ,
\end{align}
where dots denote derivatives with respect to cosmic time, $H$ is the Hubble parameter, and for simplicity we have omitted the time dependence from all quantities. Here, $\mathscr{F}$ is a function describing departures from GR which also incorporates a generic screening mechanism to restore standard gravity in high-density environments \parencite[see, e.g.,][]{Koyama:2018,Lombriser:2018}. Note that in the limit of small linear fluctuations $\mathscr{F} \rightarrow \epsilon$, and Equation~\eqref{eq:SCeqn-MG} reduces to Equation~\eqref{eq:MG-linear-growth}. Note that in the limit $\mathscr{F}=0$ we recover the GR spherical collapse equation~\eqref{eq:LCDM-spherical-collapse-diffeq}.

\section{LDT ingredients in extended cosmologies }

For simplicity we refer the non-$\Lambda$CDM cosmologies we consider in this work, both modified gravity and evolving dark energy as \emph{extended} cosmologies. We denote quantities in these extensions with ``ext''. Quantities evaluated in a standard $\Lambda$CDM cosmology are labelled simply with $\Lambda$ for brevity in some places. We now wish to examine the effect of each of these extensions on the ingredients for computing the matter PDF via LDT.

\subsection{Linear variance}

The linear variance is the most straightforward of the LDT ingredients to update in extended cosmologies, simply requiring the change to linear growth as described in equation~\eqref{eq:MG-linear-growth} for the specific theory. In this work the only extensions we consider are those which affect only the late universe, e.g. we require that $D_{\rm ext} \to D_{\Lambda}$ at sufficiently early times such that the initial conditions and primary CMB anisotropies are unchanged. Hence, the linear power spectrum (and thus linear variance) can be obtained by a simple rescaling of the $\Lambda$CDM linear power spectrum
\begin{equation}\label{eq:Pk_ext}
P_{\rm L}^{\rm ext}(k,z) = \left[\frac{D_{\rm ext}(k,z)}{D_{\Lambda}}\right]^2 P^{\Lambda}_{\rm L}(k,z_i)\,,
\end{equation}
where $z_i$ is some reference redshift taken deep in the matter-dominated era. The linear variance is then defined relative to this linear power spectrum in the usual way, c.f equation~\eqref{eq:var_from_Pk}.

For $w_0w_a$CDM cosmologies, the growth is only modified by the presence of an effective $w_{\rm eff}$ in equation~\eqref{eq:MG-linear-growth}, corresponding to setting $\epsilon(k,a)=0$. 

In DGP gravity, the linearised perturbations are characterised by 
\begin{equation}\label{eq:eps_DGP}
\epsilon_{\rm DGP}(a) = \frac{1}{3\beta(a)}, \quad \beta(a) = 1 + 2 r_{\rm c}H \left(1 + \frac{a\partial_a H}{3H}\right).
\end{equation}

In $f(R)$ gravity, the scalar field $f_R$ has a mass $m_{f_R}$ which defines an effective Compton wavelength. This wavelength determines the range of the fifth force interactions. For linear fluctuations this Compton wavelength is
\begin{align}\label{eq:compton}
    \lambda_{\rm C}(a) \equiv m_{f_R}^{-1} = \sqrt{3 c^2 (n+1) |f_{R0}| \frac{\bar R^{n+1}(a=1)}{\bar R^{n+2}(a)}} \, ,
\end{align}
where $n$ and $f_{R0}$ are free parameters of the theory, $c$ is the speed of light, and $\bar{R}$ is the background Ricci scalar. The dynamics of the linear growth modifications is controlled by 
\begin{align}\label{eq:eps_fr}
    \epsilon_{f(R)}(k,a) = \frac{(k \lambda_{\rm C}/a)^2}{3[ 1 +  (k \lambda_{\rm C}/a)^2]} \, ,
\end{align}
with $\epsilon_{f(R)} \approx 0$ for $k\lambda_{\rm C}/a \ll 1$, and reaches a maximum of $\epsilon_{f(R)} \approx 1/3$ for $k\lambda_{\rm C}/a  \gg 1$. GR is restored on all scales for $|f_{R0}| = 0$.

\subsection{Spherical collapse mapping}

The spherical collapse mapping which enters the LDT model of the matter PDF is \textit{a priori} difficult to account for in modified gravity, as violation of mass conservation and shell crossing, complicate whether exact solutions to equation~\eqref{eq:SCeqn-MG} exist, even when neglecting screening mechanism.

We find that by neglecting non-linear screening mechanisms, in the mildly non-linear regime ($R \gtrsim 10 \ h^{-1} \ \rm Mpc$) any modified gravity and dark energy effect on the spherical collapse/expansion can be accurately captured by the following approximation
\begin{align}\label{eq:SCapprox}
    \newtau^{\rm ext}(\rho,z) \approx \frac{\sigma_{\rm L}^{\Lambda}(R\rho^{1/3},z)}{\sigma_{\rm L}^{\rm ext}(R\rho^{1/3},z)} \delta_{\rm L}^{\rm EdS}(\rho)  ,
\end{align}
where $\delta_{\rm L}^{\rm EdS}(\rho)$ corresponds to the parametrised EdS spherical collapse mapping~\eqref{eq:SC-fitting-formula}. That is, for the purposes of constructing the rate function to enter the matter PDF, the non-linear spherical collapse density in an extended cosmology is well approximated by simply rescaling the Einstein-de Sitter spherical collapse mapping by the change in the \emph{linear} variance from the extended cosmology.

Note that the definition of $\delta_{\rm L}^{\rm ext}$ in equation~\eqref{eq:SCapprox} does not match the linear contrast solution to equation~\eqref{eq:MG-linear-growth}, $
\hat{\delta}_{\rm L}$, since we use the $\Lambda$CDM growth $D_{\Lambda}$ to extrapolate the initial density to the final redshift rather than the modified growth. For scale-independent late-time extensions (where $\sigma_{\rm L}^{\Lambda}/\sigma_{\rm L}^{\rm ext}=D_\Lambda/D_{\rm ext}$ and $\sigma_{\rm L}^{\rm ext, ini}\approx \sigma_{\rm L}^{\Lambda, \rm ini}$ we could instead use the modified growth for the extrapolation $\tilde{\delta}_{\rm L}^{\rm ext}\defeq D_{\rm ext}\delta_{\rm ext}^{\rm ini}$ leading to $\tilde{\delta}_{\rm L}^{\rm ext} = \delta_{\rm L}^{\rm EdS}(\rho)$ in place of equation~\eqref{eq:SCapprox}. We opt for equation~\eqref{eq:SCapprox} as it explicitly accounts for scale-dependent modifications, such as $f(R)$ gravity, for scale independent modifications these approximations produce the same rate function.  

This approximation was already argued in the dark energy case in \textcite{Codis:2016}, and is justified \textit{a posteriori} by comparison to simulations in \textcite{Cataneo.etal_2022_MatterDensity}.

\begin{figure}[h!t]
\centering
\includegraphics[width=0.7\textwidth]{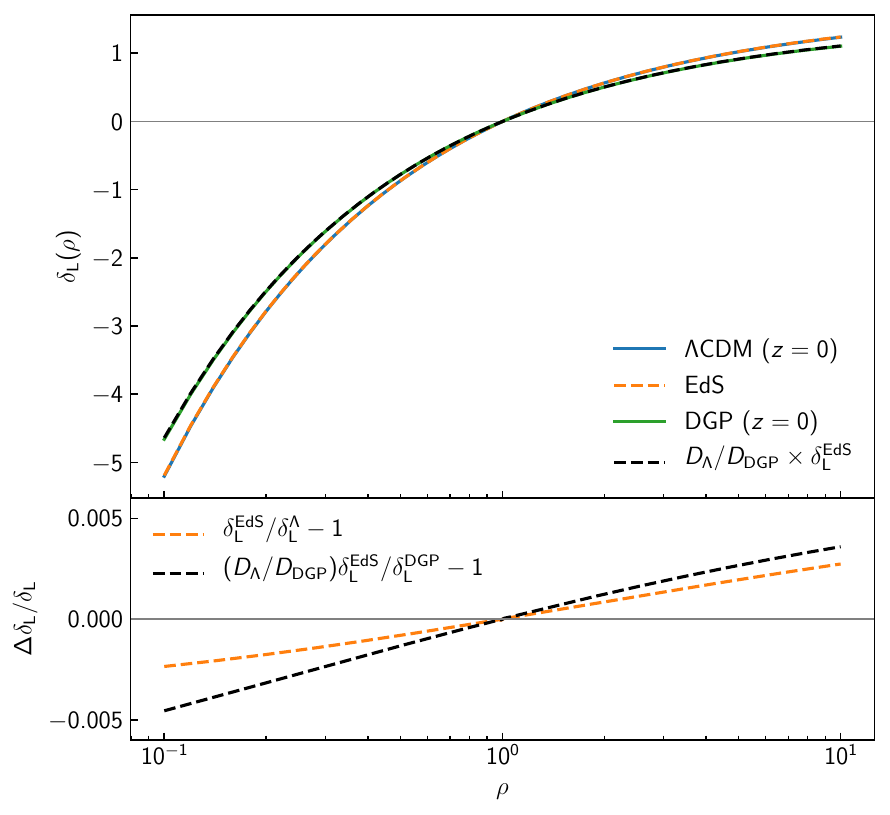}
\caption[Spherical collapse mapping for DGP gravity compared to $\Lambda$CDM and EdS spacetimes.]{Mapping between the final density $\rho$ (normalised by the mean density) and the initial linearly-scaled density fluctuation $\delta_{\rm L}$ for a spherical top-hat perturbation. (Upper panel) The curves show the density evolution for different background models. For $\Lambda$CDM (blue) and Einstein-de Sitter (EdS, dashed orange) gravity is described by GR, while for DGP (green) gravity is modified by equation~\eqref{eq:dgp_sc_eq} with $r_{\rm c}H_0 = 0.5$ (or $\Omega_{\rm rc}=0.25$). Here the $\Lambda$CDM and DGP are evaluated at $z=0$, while the dashed black line represents the EdS mapping rescaled by the ratio of the $\Lambda$CDM-to-DGP linear growth ratio at $z=0$. (Lower panel) The fractional difference of the EdS  mapping from the $\Lambda$CDM prediction (dashed orange) and the rescaled EdS from the DGP evolution (dashed black). The rescaled $\delta_{\rm L}^{\rm EdS}$ reproduces the DGP modified gravity phenomena by better than 0.5\%, and it is seen that most of the difference comes from the difference between the EdS and $\Lambda$CDM mappings. This Figure was published in \textcite{Cataneo.etal_2022_MatterDensity}. }
\label{fig:DGP_spherical_collapse}
\end{figure}

In the non-linear regime the evolution of spherical top-hat over-densities in DGP is correctly described by equation~\eqref{eq:SCeqn-MG}. For under-densities, instead, the same function $\mathscr{ F}$ incorporating the Vainshtein screening \parencite[see, e.g.,][]{Schmidt:2010} produces either unphysical solutions or a strength of the fifth force exceeding the expected linear limit for voids \parencite{Falck:2015}. Here, we neglect the Vainshtein screening by linearising the modification to gravity which accounts for most of the difference between EdS and DGP spherical evolution. This is validated against simulations in Section 4 of \parencite{Cataneo.etal_2022_MatterDensity}. An effective spherical collapse mapping for DGP can then be obtained by setting $\mathscr{F}=\epsilon_{\rm DGP}$ and extrapolating from the initial redshift $z_i$ via 
\begin{align}\label{eq:dgp_sc_eq}
\frac{\delta^{\rm (ini)}(\rho,z)}{\sigma_{\rm L}^{\Lambda }(R\rho^{1/3},z_i)} = \frac{ D_{\Lambda}(z) \delta^{\rm (ini)}(\rho,z)  }{  D_{\Lambda }(z) \sigma_{\rm L}^{\Lambda }(R \rho^{1/3},z_i) } = \frac{\delta_{\rm L}^{\rm DGP}(\rho,z) }{  \sigma_{\rm L}^{\Lambda }(R \rho^{1/3},z_i) } \,.
\end{align}
In practice, the $\delta_{\rm L}^{\rm DGP}(\rho)$ mapping obtained this way is approximated to within 0.5\% by the EdS spherical collapse mapping, as shown in Figure~\ref{fig:DGP_spherical_collapse}.

As $f(R)$ is a scale dependent modification to gravity, the non-linear evolution of the density of matter in spheres in $f(R)$ gravity is complicated by the violation of mass conservation and shell crossing \parencite{Brax:2012,LiEfstathiou_2012, Borisov:2012,Kopp:2013,Lombriser:2013}. Therefore, equation~\eqref{eq:SCeqn-MG} cannot be used to find the exact $\delta_{\rm L}^{f(R)}(\rho)$ mapping even when neglecting the chameleon screening. Instead, \textcite{Cataneo.etal_2022_MatterDensity} compares the reduced cumulants $S_3$, $S_4$,  of the matter density field in  $f(R)$ simulations against the NLO predictions for the reduced cumulants \parencite{Uhlemann16}
\begin{subequations}\label{eq:cumulants}
\begin{align}
    S_3^{\rm NLO} &= S_3^{\rm tree} + \sigma_{\mathrm{NL},\rho}^2 \left[ \frac32 S_4^{\rm tree} - 4 S_3^{\rm tree} - 2 (S_3^{\rm tree})^2 + 7 \right] , \\
    S_4^{\rm NLO} &= S_4^{\rm tree} + \sigma_{\mathrm{NL},\rho}^2 \left[ 2 S_5^{\rm tree} - \frac{17}{2}S_4^{\rm tree} + 66 S_3^{\rm tree} - 12 (S_3^{\rm tree})^2 - 3 S_4^{\rm tree}S_3^{\rm tree} - 45 \right],
\end{align}
\end{subequations}
where all quantities vary with smoothing scale and redshift, $\sigma_\rho^2$ is the non-linear variance of the density field.  The standard tree-level expressions can be calculated from standard perturbation theory as seen in Chapter \ref{chap:structure-formation}. These cumulants are in good agreement, suggesting that the Einstein-de Sitter prescription works well in the $f(R)$ case on mildly non-linear scales. See \textcite{Cataneo.etal_2022_MatterDensity} for further details and discussion on the validation of these results against simulations.

\subsection{Non-linear variance}

The final ingredient entering the LDT formalism for predicting the matter PDF is the non-linear density of the log-density field. For an arbitrary cosmology this could be measured from a suite of simulations, however this is numerically expensive and provides the non-linear variance only at one cosmology. An effective approximation approximates the log-density non-linear variance at some arbitrary cosmology in terms of the linear variance and the non-linear variance at some fiducial cosmology \parencite{Uhlemann:2020}
\begin{equation}
\label{eq:siglogcos}
    \sigma_{\ln \rho}^2(R,z) \simeq \frac{\ln\left[1 + \sigma^2_{\rm L}(R,z)\right]}{\ln\left[1 + \sigma^2_{\rm L, fid}(R,z)\right]} \sigma_{\ln \rho, \rm fid}^2(R,z) \,.
\end{equation}
This relation allows us to predict the non-linear variance for arbitrary cosmologies given the measured non-linear variance at one fiducial $\Lambda$CDM cosmology, $\sigma^2_{\rm L}$. \textcite{Cataneo.etal_2022_MatterDensity} finds that this approximation has a  typical accuracy of 0.2--1\% for the extensions studied in this work. In terms of the matter PDF, for densities $|\ln\rho - \langle \ln\rho \rangle| < 2\sigma_{\ln\rho}$ the log-normal approximation \eqref{eq:siglogcos} returns predictions that are within 2\% of those based on the non-linear variance measured from the simulations. 

Alternatively, the non-linear variance of the log-density could also be chosen to reproduce a predicted non-linear variance of the density, $\sigma_{\rho}^2(R,z)$, obtained from matter power spectrum fitting functions such as \texttt{halofit}  \parencite{HALOFIT}, \textsc{Hmcode} \parencite{Mead:2021} or \textsc{Respresso} \parencite{Nishimichi17}.

\subsection{Summary of modifications to LDT for extended cosmologies}

In summary, the matter PDF in modified gravity and dark energy cosmologies can be predicted using the LDT formalism discussed in Chapter \ref{chap:LDT-intro} under the following replacements to the rate function in equation~\eqref{eq:LDT_rate_function_rho}
\begin{itemize}
\item $\sigma_{\rm L}\to \sigma_{\rm L}^{\rm ext}$ via the modified linear power spectrum in equation~\eqref{eq:Pk_ext},
\vspace{-6pt}
\item $\sigma_{\rm NL} \to \sigma_{\rm NL, \ln\rho}^{\rm ext}$ either by measuring directly from simulation, or by the lognormal approximation \eqref{eq:siglogcos},
\vspace{-6pt}
\item $\delta_{\rm L}(\rho,z) \to \delta_{\rm L}^{\rm EdS}(\rho)$ where $\delta_{\rm L}^{\rm EdS}(\rho)$ is the spherical collapse mapping described by equation~\eqref{eq:SC-fitting-formula}.
\end{itemize}

\subsection{\texttt{pyLDT}}

The large deviation theory predictions under the prescription described above for moving to extended cosmologies has been implemented in \texttt{pyLDT}\footnote{\url{https://github.com/mcataneo/pyLDT-cosmo}}, released alongside \textcite{Cataneo.etal_2022_MatterDensity}. The linear growth for $f(R)$ gravity and DGP is obtained by solving equation~\eqref{eq:MG-linear-growth}, while the linear power spectrum for the standard cosmology, as well as for the evolving dark energy models, is computed with CAMB\footnote{Note that the common approximation for the linear growth $D(z)\propto H(a)\int_0^a {\rm d} a' (a' H(a'))^{-3}$ [quoted in equation (6) of \textcite{Codis:2016} and (A1) of \textcite{Uhlemann:2020}] is not accurate enough to estimate the response of the PDF to changing $w$ beyond a cosmological constant.} \parencite{CAMB}. Extensions to other modified gravity theories only require either to add a specific function describing changes to the gravitational constant, $\epsilon(k,a)$, or to couple the code to dedicated Einstein-Boltzmann solvers such as \texttt{hi\_class} \parencite{Zumalacarregui:2017,Bellini:2020} and EFTCAMB \parencite{Hu:2014}. By default, \texttt{pyLDT} applies the lognormal approximation~\eqref{eq:siglogcos} to calculate the non-linear variance at arbitrary cosmology.

Matter PDFs presented in this Chapter are all calculated with \texttt{pyLDT}.



\section{Extended cosmology simulations}\label{sec:mg_simulations}

In this Section we describe the modified gravity and dark energy simulations used in \textcite{Cataneo.etal_2022_MatterDensity} to validate the accuracy of predicting the matter PDF with \texttt{pyLDT} as described above.

\begin{table}[h!t]
\centering
\begin{tabular}{@{}lllllll@{}}
\toprule
    & $\Omega_m^0$ & $\Omega_b^0$ & $h$ & $n_s$ & $A_s \times 10^{9}$ & $\sigma_8^\Lambda$ \\ \midrule
DGP & 0.3072      & 0.0481      & 0.68     & 0.9645      &    2.085   &  0.821  \\
$f(R)$ & 0.31315      & 0.0492      & 0.6737     &  0.9652     &   2.097 &  0.822      \\
$w_0w_a$CDM    & 0.26      & 0.044      & 0.72     & 0.96      & 2.082   &  0.79    \\ \bottomrule
\end{tabular}
\caption[Baseline cosmological parameters used in cosmological simulations of modified gravity and dark energy.]{Baseline $\Lambda$CDM cosmological parameters for the three simulation suites used in \textcite{Cataneo.etal_2022_MatterDensity} and this Chapter. The first column refers to the extension investigated within that particular suite. $\Omega_m^0$ and $\Omega_b^0$ are, respectively, the present-day background total matter and baryon density in units of the critical density, $h = H_0/100$ is the dimensionless Hubble constant, $A_s$ and $n_s$ are the amplitude and slope of the primordial power spectrum, and $\sigma_8^\Lambda$ is the amplitude of mass fluctuations for the baseline $\Lambda$CDM cosmology.}
\label{tab:lcdm_cosmos}
\end{table}

\subsection{Evolving dark energy simulations}

For the evolving dark energy cosmologies we used the publicly available matter density PDFs\footnote{\url{https://astro.kias.re.kr/jhshin/}} measured from a suite of single-realisation $N$-body simulations with $N_{\rm p} = 2048^3$ and $L_{\rm box} = 1024$ \Mpch{} described in \textcite{Shin:2017}. The baseline flat $\Lambda$CDM cosmology has the parameters listed in Table~\ref{tab:lcdm_cosmos}, and for the $w_0w_a$CDM cosmologies we have the four pairs $\{w_0,w_a\} = \{-1.5,0\}$, $\{-0.5,0\}$, $\{-1,-1\}$, and $\{-1,+1\}$. The power spectrum normalisation at $z=0$ is fixed to its baseline value for all dark energy extensions except for $\{w_0,w_a\} = \{-1,+1\}$, which we found to have a somewhat smaller  $\sigma_8$\footnote{Because the linear theory normalisation cancels out the rate function, knowledge of $\sigma_8$ is irrelevant for the LDT predictions when measurements of the variance of the simulated density field are available. In fact, the non-linear variance carries information on $\sigma_8$ so that, ultimately, its impact on the theory PDF is properly accounted for.}.

\subsection{DGP simulations}

The DGP simulations used in this work were first presented in \textcite{Cataneo:2019}, and they were carried out using the \textsc{Ecosmog} code \parencite{Li:2013nua,Li:2011vk}, which is based on the publicly-available Newtonian cosmological $N$-body and hydrodynamical simulation code \textsc{Ramses} \parencite{Teyssier:2001cp}. This code solves the non-linear equation of motion of the scalar field in the DGP model using adaptively refined meshes, where a cell in the mesh splits into 8 son cells when the effective particle number of simulation particles in it exceeds 8. They ran one realisation with box size $L_{\rm box}=512$ Mpc$/h$ and particle number $N_{\rm p}=1024^3$ for each of the following: a baseline $\Lambda$CDM cosmology with cosmological parameters listed in Table~\ref{tab:lcdm_cosmos}, two DGP models with $\Orc = 0.25$ (DGPm) and $\Orc = 0.0625$ (DGPw), and the corresponding pseudo cosmologies with final output redshifts $z_{\rm f} = 0$ and $z_{\rm f} = 1$. These runs adopt a domain grid, i.e., a regular base grid with uniform resolution that covers the entire simulation domain, with $1024^3$ cells. Although it has been shown that, for many of the usual statistics of matter and dark matter halo fields, very fine simulation meshes are not necessary for the DGP model \parencite[][]{Barreira:2015xvp}, in these runs we have not set an upper limit of the highest refinement level, given that they were designed to be used to study novel statistics. At late times, the most refined regions in the simulation domain have a cell size that is $1/2^6$ times the domain grid cell size; this corresponds to an effective force resolution (twice the cell size) of $\simeq15.3$ kpc$/h$ in those regions. 

The ICs of these simulations are generated using \texttt{2LPTIC} \parencite{lpt}, with an initial redshift $z_{\rm ini}=49$. This is lower than the initial redshift used for the $f(R)$ runs described below ($z_{\rm ini}=127$), but the second-order Lagrangian perturbation theory is still a good approximation at $z=49$. Since the effect of modified gravity is negligible at $z>49$, it is neglected in the ICs.

\subsection{$f(R)$ gravity simulations}

The simulations in $f(R)$ gravity used for the analysis in this work were carried out with the \textsc{AREPO} cosmological simulation code \parencite{springel2010, weinberger2020} employing the MG extension introduced in \textcite{arnold2019}. The simulation suite consists of 8 independent realisations, each run for a baseline $\Lambda$CDM cosmology (see Table~\ref{tab:lcdm_cosmos} for the selected parameter values), and for  $f(R)$ Hu-Sawicki models with $n=1$ and $\fR0 = 10^{-5}$ (F5), $10^{-6}$ (F6). The suite is completed by two \emph{pseudo} cosmology runs per $f(R)$ model, one for the final output redshift $z_{\rm f} = 0$ and the other for $z_{\rm f} = 1$. In short, a pseudo cosmology is a $\Lambda$CDM cosmology with initial conditions adapted so that its linear matter power spectrum at a later epoch, $z_{\rm f}$, matches that of the \emph{real} beyond-$\Lambda$CDM cosmology of interest \parencite{Mead:2017,Cataneo:2019}, 
\begin{align}\label{eq:pseudo_cosmo}
    P_{\rm L}^{\rm pseudo}(k,z_{\rm f}) = P_{\rm L}^{\rm real}(k,z_{\rm f}) \, . 
\end{align} 
Each simulation uses $N_{\rm p} = 1024^3$ dark matter particles in a $L_{\rm box} = 500$ \Mpch{} side-length box.

The initial conditions (ICs) of the independent realisations were selected such that the large-scale sample, or cosmic, variance in the 3D matter power spectrum is minimal when averaged over the simulations. In order to implement this we created 100 independent initial conditions using \texttt{2LPTIC} \parencite{lpt} and measured their 3D matter power spectrum. We then considered all possible pairs of these ICs and selected the four `best' pairs according to the following criteria \parencite[this follows the procedure outlined in][to find ICs with approximately opposite modes on large scale]{slics}:
\begin{itemize}
    \item each individual power spectrum of a selected pair, as well as their average power spectrum, should deviate as little as possible from the desired linear theory power spectrum for $k < k_{\rm Ny}/2 = \pi N_{\rm p}/ 2 L_{\rm box}$;
    \vspace{-6pt}
    \item and the relative difference of each individual power spectrum to the theory spectrum should fluctuate around zero on large scales rather than being positive or negative over large $k$-ranges to avoid a leakage of power from large to small scales.
\end{itemize}

To simulate structure formation in $f(R)$ gravity the simulation code has to solve both the standard Newtonian forces and the fifth force. \textsc{AREPO} computes the standard gravity forces using a Tree Particle-Mesh algorithm in our simulations. The $f(R)$ gravity forces are computed employing an iterative solver on an adaptively refining mesh which ensures increased resolution in high density regions \parencite[see][for details]{arnold2019}. 

Due to the very non-linear behaviour of the scalar field in $f(R)$ gravity, tracking its evolution is computationally very expensive. To keep the computational cost of the simulations as small as possible, \textsc{AREPO} therefore employs an adaptive timestepping scheme which only updates the MG forces when necessary \parencite{arnold2019}. The standard gravity accelerations are largest (and change most frequently) within large halos, so that they have to be updated with a very small timestep. However, these very same regions in $f(R)$ gravity are largely screened for $\fR0 \lesssim 10^{-5}$. Therefore, the maximum MG acceleration will typically be much smaller than the maximum standard gravity acceleration, allowing a larger MG timestep without compromising the accuracy of the simulations.

\section{The matter PDF}

In this Section we present a comparison between the matter PDF in modified gravity and that predicted by \texttt{pyLDT}. 
\begin{figure}[h!t]
\centering
    \includegraphics[width=0.7\textwidth]{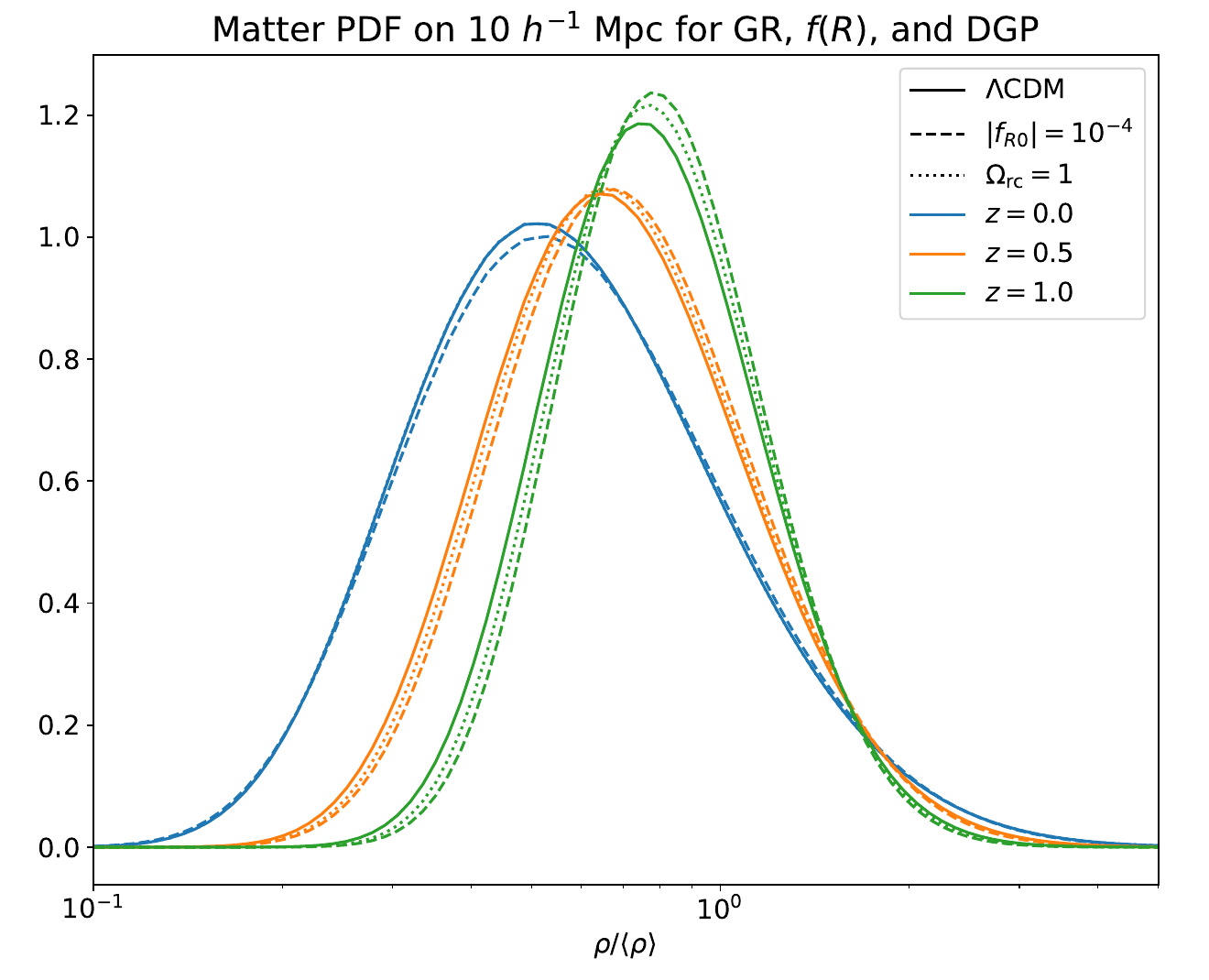}
    \caption[Comparison of the matter PDF in 10 Mpc/$h$ spheres between $\Lambda$CDM, $f(R)$, and DGP gravity with fixed variance.]{Comparison of the matter PDF in  $R=10 \ h^{-1}\rm \ Mpc$ spheres in $\Lambda$CDM, $f(R)$, and DGP cosmologies. The cosmological parameters are chosen such that the clustering amplitude $\sigma_8$ is the same in all cases at redshift 0, rather than matching the initial amplitude $A_s$. This normalises the overall width of the PDF. The resulting difference in tilt and redshift dependence is due to the change to gravity. This Figure was published in \textcite{Gough.Uhlemann_2022_OnePointStatistics}.}
    \label{fig:PDF_comparison}
\end{figure}

\begin{figure}[h!]
    \centering
     \includegraphics[width=\textwidth]{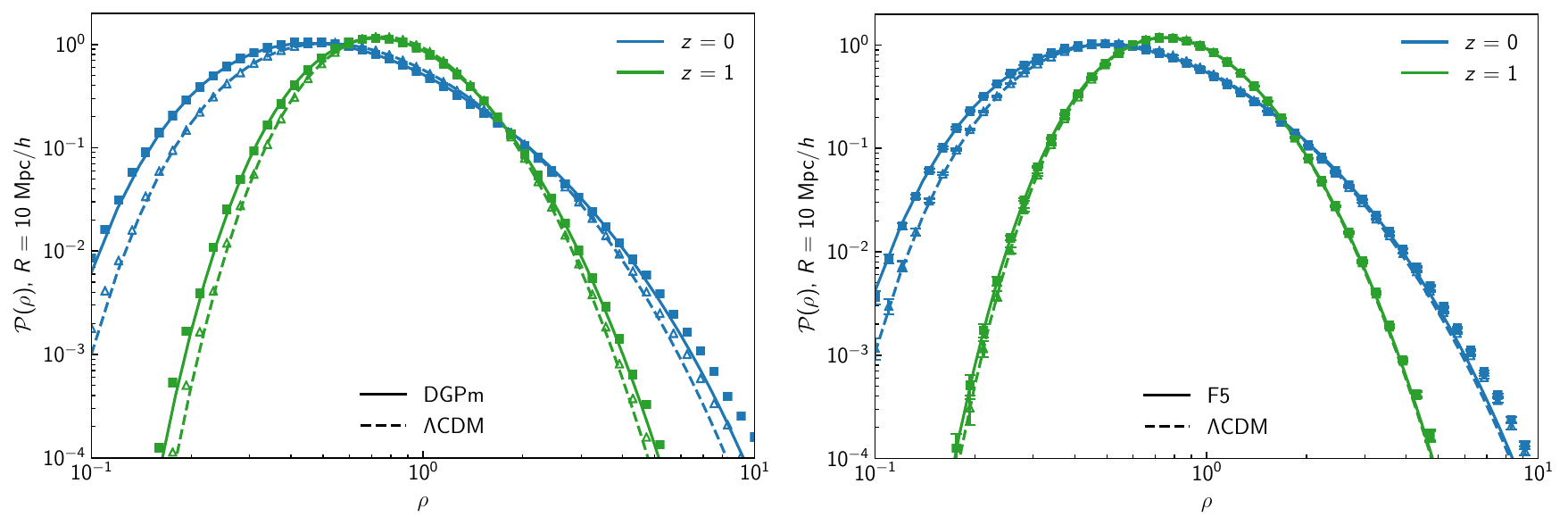}
    \caption[Comparison matter PDF in 10 Mpc/$h$ spheres in $\Lambda$CDM, $f(R)$, and DGP gravity, run on the same initial conditions.]{Matter PDF in spheres of radius $R = 10$ \Mpch{} at $z=0$ (blue) and $z=1$ (green) for $\Lambda$CDM (dashed) and modified gravity (solid). (Left panel) Data points are the simulation measurements from a single realisation (triangles for $\Lambda$CDM and squares for DGPm) and lines represent the theory predictions. For DGP the primary effect of the enhanced growth is that of increasing the variance of the distribution, which in turn results in heavier tails, i.e. more under/over-dense structures compared to the standard cosmology. (Right panel) Data points and corresponding uncertainties are the mean and error on the mean measured from 8 realisations (triangles for $\Lambda$CDM and squares for F5). Note that, contrary to the DGP cosmology, the modified growth in $f(R)$ gravity substantially affects the skewness of the distribution, thus leading to an asymmetric enhancement over $\Lambda$CDM. This Figure was published in \textcite{Cataneo.etal_2022_MatterDensity}.}
    \label{fig:mg_pdf}
\end{figure}

Figures~\ref{fig:PDF_comparison} and \ref{fig:mg_pdf} show the matter PDF in $R=10$ \Mpch{} spheres for $\Lambda$CDM, nDGP, and $f(R)$ gravity. Figure~\ref{fig:PDF_comparison} shows the theoretically predicted PDF calculated using \texttt{pyLDT} in these three theories of gravity, with cosmological parameters chosen to produce the same $\sigma_8$ at redshift $z=0$. Generically, introducing modified gravity will change both the width and the shape of the PDF (as we see in Figure \ref{fig:mg_pdf}). Since $\sigma_8$ sets the width of the PDF, normalising the cosmologies to have the same $\sigma_8$ at redshift 0 allows us to isolate the distinct features of modified gravity on the PDF, as done in Figure~\ref{fig:PDF_comparison}. Note that the value of $\Omega_{\rm rc}$ and $\abs{f_{R0}}$ used here are larger than those used in the simulations described in Section~\ref{sec:mg_simulations} to make this effect more visible.

We see the residual difference in the shape of the matter PDF, as well as a distinct redshift dependence in the modified gravity PDFs compared to the $\Lambda$CDM PDF. These differences in shape and redshift dependence (as well as a difference in scale dependence not shown in Figure~\ref{fig:PDF_comparison} \& \ref{fig:mg_pdf}) are what allows the PDF to break degeneracies between standard cosmological parameters and modified gravity, as we will examine in the following Section. The effect of an evolving dark energy on the matter PDF is similar to that of DGP gravity, entering mostly by modifying the expansion history, as a scale-independent modification to gravity. We see that at low redshift, the DGP and $\Lambda$CDM lines are very similar in Figure \ref{fig:PDF_comparison}, as DGP is acting as a scale independent modification to gravity, and we have required the variances to match at redshift $z=0$ for this Figure.

Figure \ref{fig:mg_pdf} shows the comparison between the PDFs as measured from the simulations described in the previous section (for details on the extraction of these PDFs of the simulations refer to \textcite{Cataneo.etal_2022_MatterDensity}) and the predicted PDFs from LDT generated with \texttt{pyLDT}. Because these simulations all have initial conditions with similar values of $A_s$, the changes in theory of gravity affect the overall variance of the PDF, which changes the width in addition to other features characterising its shape.

We see that in the both cases, the MG and $\Lambda$CDM PDFs become more similar at high redshift when evolved from the same initial conditions (the rate of this convergence dependant on the details of the model). However, there are clear differences that reflect the finite or infinite extent of the fifth force. In DGP, structures on all scales are subject to the same modification and changes to the higher moments of the distribution are driven mostly by increases in its variance. This follows immediately from equation~\eqref{eq:cumulants} as the tree-level terms are identical for DGP and $\Lambda$CDM. In $f(R)$ by contrast, density fluctuations evolve in different gravity conditions dependent on their size.  For example, the present-day Compton wavelength in our F5 cosmology is approximately 8 \Mpch{} (and smaller at earlier times). Therefore, spherical over-densities reaching a final radius $R=10$ \Mpch{} experience very little fifth force for most of their collapse history. In contrast, spherical under-densities have sizes comparable to or smaller than the Compton wavelength (at the same epoch) and thus experience the fifth force in the later stages of their expansion (i.e. $z \lesssim 2$), with the emptiest regions experiencing a full 33\% enhancement of the gravitational force. It is this asymmetric behaviour that contributes to the increased skewness of the PDF in $f(R)$ gravity compared to $\Lambda$CDM \parencite[see also][]{Hellwing:2013}. The LDT model is able to capture this behaviour thanks to the linear variance term $\sigma_{\rm L}^2(R)$ probing different scales $r=R\rho^{1/3}$ depending on the density of the sphere, as discussed in Chapter \ref{chap:LDT-intro}. We will see a similar density dependence effect inducing skewness in the context of wave dark matter in Chapter \ref{chap:how-classical}.

Equivalent versions of Figure \ref{fig:mg_pdf} for the DGPw and F6 simulations, as well as scale dependence and residuals between the simulations and theory can be found in \textcite{Cataneo.etal_2022_MatterDensity}.

\section{Forecasting constrating power with the Fisher formalism}\label{sec:Fisher}

This Section presents forecasts for DGP and $f(R)$ gravity and $w_0w_a$CDM combining the matter PDF and the matter power spectrum on mildly non-linear scales. For the MG models we determine the ability of future experiments to detect relatively small deviations from GR (i.e. F6 and DGPw) at a statistical significance $> 5\sigma$ (see Table~\ref{tab:MGdetection}), while for evolving DE we will be interested in the FoM using \LCDM\ as fiducial cosmology (see Table~\ref{tab:DEconstraints}).

\subsection{Fisher formalism}

To forecast the errors on a set of cosmological parameters, $\bm{\theta}$, we use the Fisher matrix formalism \parencite[see e.g.][for an introduction to this technique]{Verde2010LNP, Heavens2009arXiv, Euclid:2020}. The Fisher matrix given a (set of) summary statistics in the data vector $\bm{S}$ is 
\begin{equation}
F_{ij}= \sum_{\alpha,\beta}\frac{\partial S_\alpha}{\partial \theta_i}C^{-1}_{\alpha \beta}\frac{\partial S_\beta}{\partial \theta_j} \Big\vert_{\theta_{\rm ref}}  +  \frac12 \mathrm{tr}\left[\pdv{\mathsf{C}}{\theta_i} \mathsf{C}^{-1} \pdv{\mathsf{C}}{\theta_j} \mathsf{C}^{-1}\right]\Big\vert_{\theta_{\rm ref}} ,
\label{eq:Fisher}
\end{equation}
where $S_\alpha$ is the $\alpha^{\rm th}$ element of the data vector $\bm{S}$ and $\mathsf{C}^{-1}$ denotes the matrix-inverse of the (data)-covariance matrix $\mathsf{C}$, whose components are 
\begin{equation}
\label{eq:covariance}
C_{\alpha \beta} = \langle (S_\alpha-\bar{S}_\alpha)(S_\beta - \bar{S}_\beta) \rangle\,,\quad \bar{S}_\alpha = \langle S_\alpha \rangle~.
\end{equation}
If the data-covariance is independent of the cosmological parameters, the second term in equation \eqref{eq:Fisher} vanishes. The parameter covariance matrix $\mathsf{C}(\bm{\theta})$ is then obtained as inverse of the Fisher matrix.
In the Fisher formalism, marginalisation over a subset of parameters is achieved by simply selecting the appropriate sub-elements of the parameter covariance. The Fisher formalism becomes exact in the approximation that the data likelihood is Gaussian and the prior distributions are uninformative.

We consider three data vectors for our forecasts, corresponding to the three sets of constraints in Figures~\ref{fig:fisher_mg_both}, and \ref{fig:fisher_w}. These are the PDF alone, the matter power spectrum alone, and a stacked data vector which combines both the PDF and the matter power spectrum. For the PDF data vector, we only use the central region of the PDF around the peak (located in underdense regions), removing the lowest $3\%$ and highest 10\% of densities \parencite[as advocated in][]{Uhlemann:2020}. We choose this approach in order to limit the impact of small-scale effects (like baryonic feedback, non-linear galaxy bias, shot noise and redshift-space distortions) that are more severe for rare events and would otherwise degrade the constraining power when moving from the 3D matter PDF to an actual observable like the spectroscopic tracer PDF.
For the matter power spectrum data vector, we limit ourselves to mildly non-linear scales up to $k_{\rm max}=0.2\, h$/Mpc to ensure the accuracy of theoretical derivatives from fitting functions. We found the conservative scale cut for the power spectrum to be crucial to facilitate an agreement between parameter constraints and degeneracy directions from predicted and simulated derivatives, especially when considering the full set of cosmological parameters.
For all cosmological parameters, we compute partial derivatives from two-point finite differences
\begin{equation}
\frac{\partial \bm{S}}{\partial \theta_i}\simeq\frac{\bm{S}(\theta_i+\dd{\theta}_i)-\bm{S}(\theta_i-\dd{\theta}_i)}{2\dd\theta_i}\,.
\label{eq:derivatives}
\end{equation}
We rely on partial derivatives determined from theoretical predictions for the matter PDF from \pyLDT{} and the matter power spectrum from \texttt{ReACT} \parencite{Bose:2020} combined with \textsc{HMCODE} \parencite{Mead:2021}, which provides flexibility to compute constraints or the detection significance for extended models at the desired fiducial cosmology. The step sizes have been chosen to ensure convergence of the derivatives, and agree with the step sizes used in the {\sc Quijote} simulation suite for the set of $w_0$CDM parameters. The theory generated derivatives for $w_0$CDM parameters are validated with measurements from the {\sc Quijote} simulations as discussed in \textcite{Cataneo.etal_2022_MatterDensity}. We adopt Gaussian priors for  $\{\Omega_b^0,n_s\}$ to ensure compatibility of the matter power spectrum derivatives between simulations and theoretical predictions. The prior widths correspond to $\sigma[n_s] = 0.0041$ \parencite{Planck:2018} and $\sigma[100\Omega_b^0 h^2] = 0.052$ \parencite{Cooke:2016, Abbott:2018}. 

In this work, we use the covariance matrix obtained from a set of 15000 simulations of the {\sc Quijote} $N$-body simulation suite \parencite{Villaescusa-Navarro:2020} using the fiducial $\Lambda$CDM cosmology ($\Omega_m^0 = 0.3175$,  $\Omega_b^0 = 0.049$, $H_0 = 68$ km/s/Mpc, $n_s = 0.96$, $\sigma_8 = 0.834$). The joint covariance matrix of the mildly non-linear matter PDF and the matter power spectrum is described in \textcite{Uhlemann:2020}, see particularly their Figure~12. We make the approximation that the covariance matrix of the matter PDF and matter power spectrum in the mildly non-linear regime is independent of cosmology and theory of gravity and well-captured by the 15000 simulations of the {\sc Quijote} simulation suite. To mitigate potential effects of modified gravity on the covariance, we fix the standard cosmological parameters to the values of the fiducial {\sc Quijote} cosmology. In particular, we set $A_s=2.13\times 10^{-9}$ such that $\sigma_8$ increases only slightly for the modified gravity cosmologies, that is, by 1.6\% for F6 and 3.8\% for DGPw. As those are small perturbations from the fiducial $\Lambda$CDM cosmology, they will only induce a small error on the true covariances and hence only marginally affect parameter constraints. As this error will affect both the PDF and power spectrum covariance in a similar way, comparisons of their respective constraining power are expected to be robust. We thus neglect the second term in equation \eqref{eq:Fisher} to calculate the Fisher matrices. For future high precision cosmology, covariance estimation for PDF-based observables from galaxy clustering and weak lensing can rely on tuned lognormal mocks \parencite{Gruen:2018,Boyle_2020}, potentially complemented with predictions for effects induced by variations in the local mean density \parencite{Jamieson_2020}. We will investigate modelling the covariance of the matter PDF further in Chapter \ref{chap:covariance}, based on effective two-point PDF models, to capture the super-sample effect which is difficult to obtain from simulations.

To correct for a potential bias depending on  the size of the data vector $N_{S}$ compared to the number of simulations $N_{\rm sim}$, we multiply the inverse of the simulated covariance matrix by the Kaufman-Hartlap factor \parencite{Kaufman67,Hartlap06}, $f_{\rm KH}=(N_{\rm sim} - 2 - N_{S})/(N_{\rm sim} - 1)$.  Since in our case the number of simulations for covariance estimation is very large (15000) compared to the maximal length of the data vector (218 for our three-redshift analysis of the PDF at three scales and the mildly non-linear power spectrum), this factor will be close to one throughout, $f_{\rm KH}\geq 0.985$. 
We mimic a \textit{Euclid}-like effective comoving survey volume of $V \approx 20 \, ({\rm Gpc}/h)^3$ split across three redshift bins of equal width $\Delta z=0.2$ located at $z=0,0.5,1$ by
multiplying the covariance at each redshift with the ratio of the comoving shell volume to the simulation volume $V_{\rm sim}=1 \, ({\rm Gpc}/h)^3$.

\subsection{Modified gravity forecast}

We now compare the constraining power of the matter PDF to that of  the matter power spectrum  (with $k_{\rm max} = 0.2 \ h \ \rm Mpc^{-1}$) for DGP and $f(R)$ gravity. In all cases, the forecasts shown are marginalised over all remaining $\Lambda$CDM parameters.

\begin{figure}[h!]
    \centering
    \includegraphics[width=\textwidth]{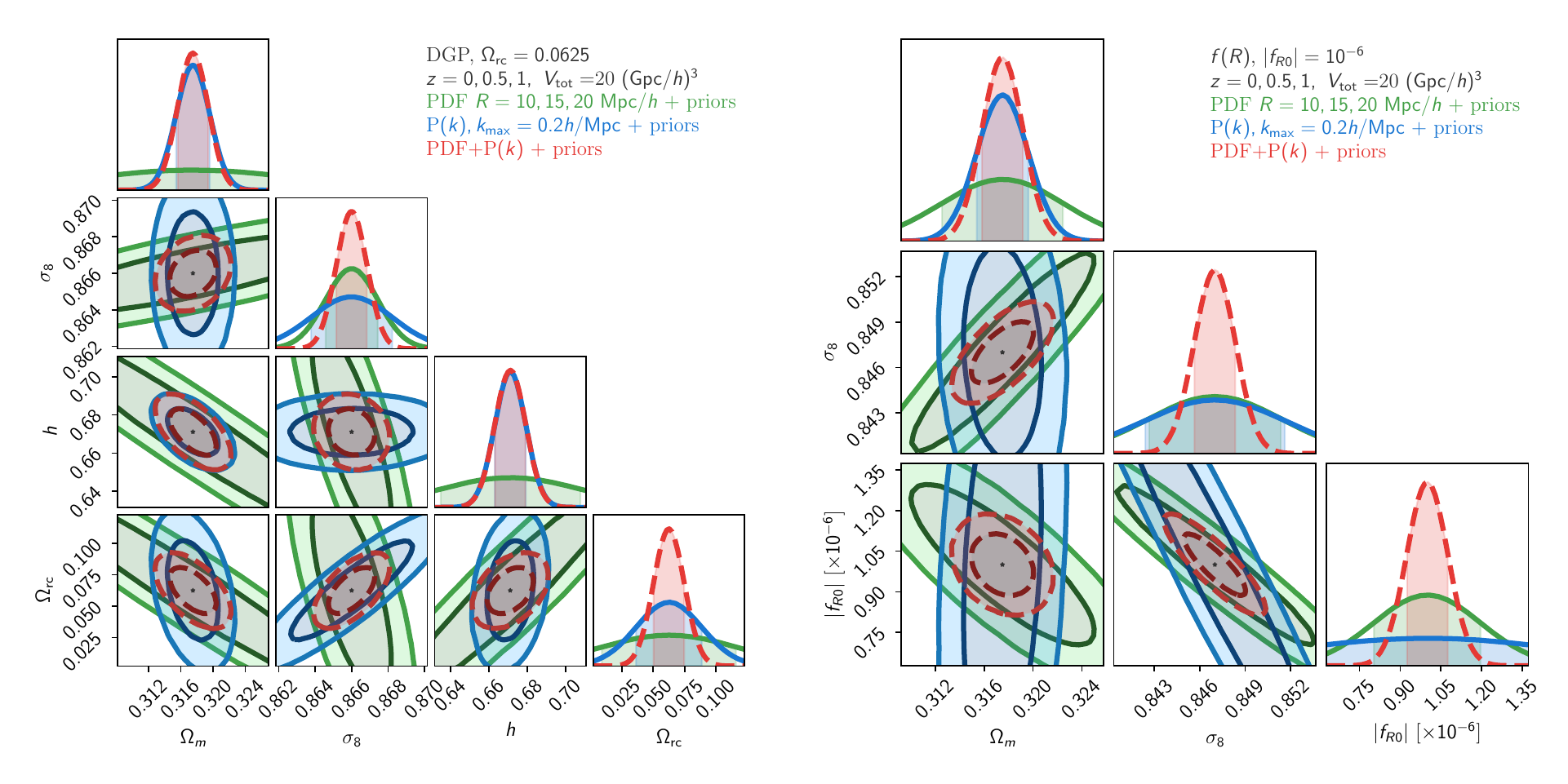}
    \caption[Parameter constraints from a Fisher forecast for DGP and $f(R)$ gravity.]{Marginalised Fisher forecast constraints on DGP (left) and $f(R)$ (right) gravity. We use an external prior on $n_s$ and $\Omega_b$ (as described in the text) for the DGPw and F6 fiducial cosmologies.  Contours correspond to the matter PDF at 3 scales and 3 redshifts (green), the matter power spectrum up to $k = 0.2 \ h \ \mathrm{Mpc}^{-1}$ (blue), and their combination, which includes the covariance between the PDF and power spectrum (red dashed). The panels of this Figure were published in \textcite{Cataneo.etal_2022_MatterDensity, Gough.Uhlemann_2022_OnePointStatistics}.}
    \label{fig:fisher_mg_both}
\end{figure}

\begin{figure}[h!]
    \centering
    \includegraphics[width=0.75\columnwidth]{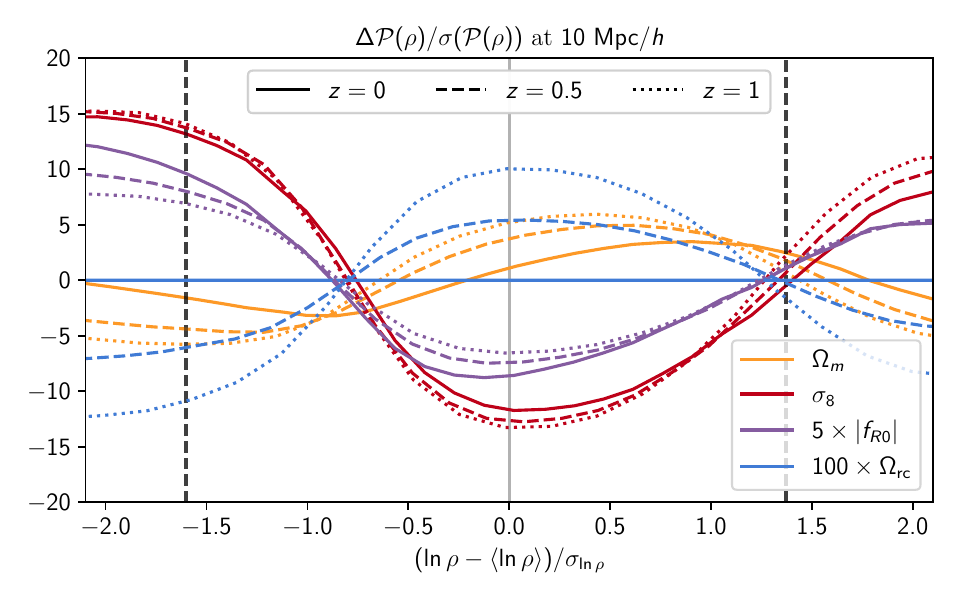}
    \caption[Derivatives of the matter PDF in modified gravity cosmologies.]{Comparison of PDF differences divided by the error on the PDF as estimated from the {\sc Quijote} simulations. Line style indicates the redshift, while colour indicates parameter being deviated. The vertical lines represent the region used at each redshift to construct the PDF data vector. While the shape of the $\sigma_8$ and $\Omega_{\rm rc}$ derivatives are similar, their different redshift dependence allows the degeneracy to be broken when combining redshifts. The $f_{R0}$ derivatives, in addition to having different redshift dependence, exhibit a skewness not present in the $\sigma_8$ derivatives, allowing significant information to be extracted even at a single redshift, including $z=0$. Note that the amplitudes between parameters should not be directly compared, as the $f_{R0}$ and $\Omega_{\rm rc}$ lines have been scaled up to be visible on the same scale as $\sigma_8$. This Figure was published in \textcite{Cataneo.etal_2022_MatterDensity}.}
    \label{fig:pdf_diff_comp_MG}
\end{figure}

Figure~\ref{fig:fisher_mg_both} shows the Fisher forecasts for DGP and $f(R)$ cosmologies. Table~\ref{tab:MGdetection} summarises the detection significance for particular flavours of these modified gravity models expressed in units of  standard deviation from GR. In a universe where the growth of structure is governed by DGP gravity with $\Omega_{\rm rc} = 0.0625$, a $5\sigma$ detection of modified gravity can still be reached by combining the matter PDF with the matter power spectrum. Combining the PDF and power spectrum as complementary probes is beneficial in both MG scenarios. In particular, for DGP the matter PDF is important for constraining $\sigma_8$, while the power spectrum is important for obtaining the correct value of $\Omega_m^0$. This is because while $\Omega_m^0$ has a distinctive signature in the power spectrum  \parencite[see Appendix C of][]{Cataneo.etal_2022_MatterDensity}, the matter PDF is sensitive to the total matter density only through its impact on the skewness and the linear growth factor, $D(z)$. The anti-correlation between the Hubble parameter, $h$, and $\Omega_m^0$ visible in Figure~\ref{fig:fisher_mg_both} can be explained by their similar impact on the skewness of the PDF \parencite[see Figure~9 in][]{Uhlemann:2020}. Evolving dark energy also presents this feature, although we do not show it in Figure~\ref{fig:fisher_w} as it does not create any unexpected degeneracy directions as in the DGP model.

The partial degeneracy in the PDF between $\sigma_8$ and the modified gravity parameters, $|f_{R0}|$ or $\Omega_{\rm rc}$, seen in Figures~\ref{fig:fisher_mg_both} is understood by noticing that the presence of modified gravity changes the width of the matter PDF, as can be seen in Figure~\ref{fig:pdf_diff_comp_MG}. However, the responses of the PDF to the presence of modified gravity or changes in $\sigma_8$ have different scale- and time-dependence, therefore by combining the information from different scales and redshifts we can break this degeneracy. 
Figure~\ref{fig:pdf_diff_comp_MG} shows that $\sigma_8$ and $\Omega_{\rm rc}$ have opposite effects on the PDF, which would lead to a positive correlation between these parameters. However, in Figure~\ref{fig:fisher_mg_both} the $\sigma_8$--$\Omega_{\rm rc}$ plane shows an anti-correlation for the PDF, which is indirectly induced by the strong positive correlation between $\Omega_{\rm rc}$ and $h$. We checked that when $h$ is fixed to its fiducial value, rather than marginalised over, the PDF contours do indeed display a positive correlation between $\sigma_8$ and $\Omega_{\rm rc}$, as indicated by the derivatives in Figure~\ref{fig:pdf_diff_comp_MG}.

In the case of $f(R)$ gravity, the matter PDF is particularly useful, reaching a $5\sigma$ detection before combining with the matter power spectrum. This is due to an additional skewness in the $|f_{R0}|$ derivatives sourced by the scale-dependent fifth force and the fact that the PDF holds information about deviations from $\Lambda$CDM even at redshift 0, unlike in DGP. However, as discussed in more detail in \textcite{Cataneo.etal_2022_MatterDensity},  using the non-linear variance predicted by equation~\eqref{eq:siglogcos} is a better approximation in DGP than in $f(R)$ gravity, and we thus expect the forecasted constraints to be more reliable for the DGP model than for $f(R)$ gravity .

\begin{table}[h!]
\centering
    \begin{tabular}{lccc}
    \hline
         ~ & F6 detection & DGPw detection\\\hline
         PDF, 3 scales + prior & $5.15\sigma$ & $1.17\sigma$ & \\
         $P(k)$, $k_{\rm max} = 0.2\ h /\rm Mpc$ + prior & $2.01\sigma$ & $2.42\sigma$ & \\
         PDF + $P(k)$ + prior & $13.40\sigma$ & $5.19\sigma$ & \\
    \hline
    \end{tabular}
    \caption[Parameter constraints from the matter PDF and matter power spectrum in $f(R)$ and DGP gravity. ]{Detection significance for a fiducial $f(R)$ gravity model with $|f_{R0}|=10^{-6}$, and a fiducial DGP model with $\Omega_{\rm rc}=0.0625$. The constraints on $f(R)$ gravity from the PDF are stronger than in DGP owing to the additional skewness produced by the scale-dependent fifth force, which is visible in the $|f_{R0}|$ derivatives shown in Figure~\ref{fig:pdf_diff_comp_MG}. Moreover, unlike DGP, where the approximate theory PDF matches the $\Lambda$CDM prediction at $z=0$, in $f(R)$ gravity the PDF differs from that of the standard cosmology at low redshifts, which allows even more non-linear information to be extracted.  }
    \label{tab:MGdetection}
\end{table}

\subsection{Dark energy forecast}

\begin{figure}[h!]
    \centering
     \includegraphics[width=0.5\columnwidth]{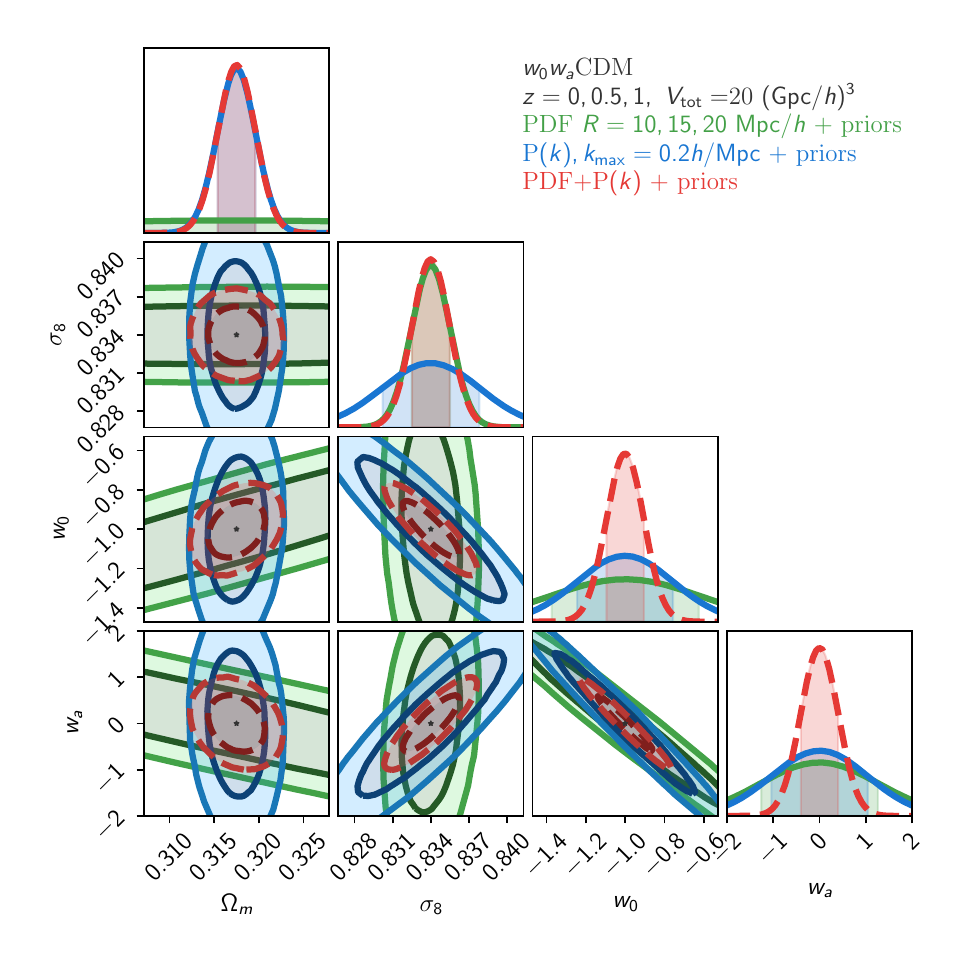}
    \caption[Parameter constraints from a Fisher forecast on $w_0w_a$CDM cosmology.]{Fisher forecast constraints on $\{\Omega_m, \sigma_8, w_0, w_a\}$ (marginalised over $\{\Omega_b,n_s\}$ using the external prior described in the text) for the $w_0$CDM model around the fiducial {\sc Quijote} $\Lambda$CDM cosmology. Contours correspond to the matter PDF at 3 scales and 3 redshifts (green), the matter power spectrum up to $k = 0.2 \ h \ \mathrm{Mpc}^{-1}$ (blue), and their combination which includes the covariance between the PDF and power spectrum (red dashed). This Figure was published in \textcite{Cataneo.etal_2022_MatterDensity} and \textcite{Gough.Uhlemann_2022_OnePointStatistics}. }
    \label{fig:fisher_w}
\end{figure}

In this Section we consider a dark energy fluid with an equation of state described by equation~\eqref{eq:DEeos}. 

Many of the features of the parameter constraints from the matter PDF and matter power spectrum are similar to the features seen for scale-independent modifications to GR. In particular, the matter PDF is much better at constraining $\sigma_8$ than the power spectrum, while the power spectrum more directly measures $\Omega_m^0$, as can be seen in Figure~\ref{fig:fisher_w}. A summary of constraints on $\sigma_8$, $w_0$, and $w_a$, along with the dark energy Figure of Merit (FoM) is shown in Table~\ref{tab:DEconstraints}. The FoM is calculated from the inverse of the error ellipse area in the $w_0$-$w_a$ plane as
\begin{equation}
    \mathrm{FoM} = \frac{1}{\sqrt{\det\mathsf{C}(w_0, w_a)}}\,,
\end{equation}
where $\mathsf{C}(w_0, w_a)$ is the parameter covariance matrix marginalised over all parameters except $w_0$ and $w_a$. The combined FoM for the matter PDF and matter power spectrum is a factor of 9 larger than the PDF alone, and  5 times better than the power spectrum when only information from the mildly non-linear regime is included. This combined FoM of 243 sits between the range of the pessimistic and optimistic predictions for combined galaxy clustering and weak lensing from \textit{Euclid} \parencite[see Table~13 from][]{Euclid:2020}. The PDF is sufficient to measure $\sigma_8$ to sub-percent accuracy, with the inclusion of the power spectrum improving this constraint only marginally. 

\begin{figure}[h!]
\centering
    \includegraphics[width=0.75\textwidth]{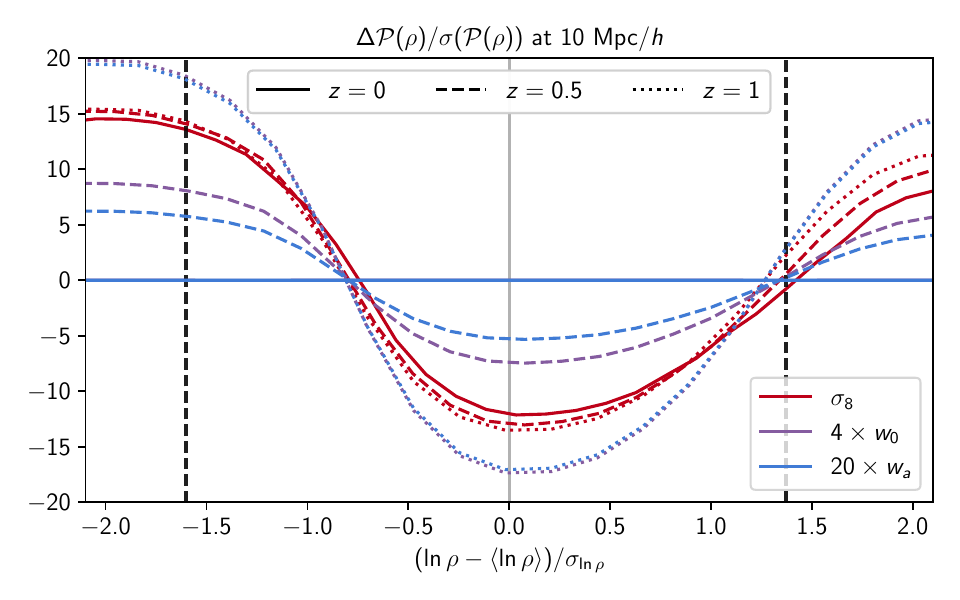}
    \caption[Derivatives of the matter PDF in an evolving dark energy universe.]{Derivatives of the matter PDF in an evolving dark energy universe. The dependence of the matter PDF on $\Omega_m$ is easily distinguished from the others by its distinct skewness (see Figure~\ref{fig:pdf_diff_comp_MG}) and hence not shown here. The $\sigma_8$, $w_0$, and $w_a$ derivatives are similar in shape, but have different redshift evolutions, which allows for degeneracy~breaking. This Figure was published in \textcite{Gough.Uhlemann_2022_OnePointStatistics}. }
    \label{fig:PDF_deriv_DE}
\end{figure}

Increasing either $w_0$ or $w_a$  increases the growth rate and hence the variances at $z>0$ (with marginal changes at $z=0$ due to fixed $\sigma_8$), which amounts to an anti-correlation between $w_0$ and $w_a$. Most of the other degeneracy directions in the $w_0w_a$CDM case can be understood by considerations of linear theory. Changing a single parameter at fixed $\sigma_8$ (or similarly when $\sigma_8$ is allowed to vary) induces a change in the growth rate. Suitable pairs of parameters can then produce growth rates close to the fiducial cosmology. For example, the positive correlation between $w_0$ and $\Omega_m$ arises from the suppression of the growth rate by increasing $\Om$ while keeping $\sigma_8$ fixed. While one would expect $w_0$ and $w_a$ to vary in the same way with $\Om$ and $\sigma_8$, they in fact vary in opposing directions as shown in Figure~\ref{fig:fisher_w}. However, when $w_0$ is fixed to its fiducial value, rather than marginalised over, the contours do indeed flip in sign to the direction expected, suggesting that the tight anti-correlation in the $w_0$-$w_a$ plane dominates the other degeneracies. Figure~\ref{fig:PDF_deriv_DE} shows that at fixed scale and redshift, the response of the PDF to $w_0$ or $w_a$ is very similar in shape to the response to $\sigma_8$. For this reason, for dynamical dark energy (and other scale independent modifications to gravity) the degeneracy is mainly broken by a difference in the redshift dependence.

\begin{table}
    \centering
    \begin{tabular}{lcccc}
        \hline
         ~ & $\sigma[\sigma_8]/\sigma_8^{\rm fid}$ & $\sigma[w_0]$ &  $\sigma[w_a]$ & FoM\\\hline
         PDF, 3 scales + prior & 0.18\% & 0.37 & 1.25 & 27\\
         $P(k), k_{\rm max}=0.2 h/$Mpc + prior & 0.45\% & 0.24 & 1.03 & 50\\
         PDF + $P(k)$ + prior & 0.17\% & 0.09 & 0.40 & 243\\ 
         \hline
    \end{tabular}
    \caption[Parameter constraints from the matter PDF and matter power spectrum in $w_0w_a$CDM. ]{Constraints from mildly non-linear scales on $\sigma_8$, $w_0$ and $w_a$ derived including a prior on $\{\Omega_b^0,n_s\}$, as well as dark energy Figure of Merit (FoM) for the matter PDF, power spectrum and their combination.}
    \label{tab:DEconstraints}
\end{table}

\section{Conclusions}

Standard two-point statistics are not sufficient to make full use of the information content in the cosmic large-scale structure, and would leave large amounts of data from current and upcoming galaxy surveys under utilised. The full shape of the matter density PDF in spheres has been shown to provide great complementarity to the standard two-point statistics, and allows extraction of information from the non-linear regime. The analytic framework described here has been successfully applied to $\Lambda$CDM universes along with extensions including primordial non-Gaussianity~\parencite{Friedrich20pNG} and massive neutrinos~\parencite{Uhlemann:2020}. This work demonstrates that the LDT formalism continues to work in modified gravity and dark energy scenarios, providing a powerful non-Gaussian probe of fundamental physics complementary to two-point statistics.

While the analysis presented here is idealised in that it relies on knowledge of the true matter distribution, it is encouraging for realistic scenarios. In the case of $\Lambda$CDM cosmologies, the LDT approach has been translated into several observable quantities, including weak lensing \parencite{Barthelemy:2020, Boyle_2020, Thiele_2020}, galaxy clustering \parencite{Repp_2020, Friedrich_2021}, and density-split statistics \parencite{Gruen:2018, Friedrich_2018}. In particular, the LDT approach developed for the $\Lambda$CDM lensing convergence PDF could be straightforwardly adapted to the entire class of scalar-tensor theories with lensing potential $\Phi_{\rm lens}^{\rm MG}\approx \Phi_{\rm lens}^{\rm GR}$ which includes the $f(R)$ and DGP gravity models presented in this Chapter. 

Given the theoretical information content in the matter PDF demonstrated here, extending the LDT framework to observables in the context of modified gravity would be a worthwhile endeavour for constraining both astrophysical (e.g., baryonic feedback, intrinsic alignment, and galaxy bias) and cosmological parameters to complement two-point statistics. Additionally, other the success of matter PDF and LDT in these extended cosmologies is encouraging for other tests of fundamental physics such as the nature of dark matter. Wavelike dark matter also introduces a scale dependent clustering and growth, which we have seen can be captured accurately by the LDT model of the matter PDF. We return to this in Chapter \ref{chap:how-classical}, where we examine one point statistics in a wavelike dark matter model.

%% file: text/chapter5-covariances-of-pdf.tex

\chapter{Covariances matrices of matter PDFs}\label{chap:covariance}
\minitoc

This Chapter is based on work presented in \textcite{Uhlemann.etal_2023_ItTakes}, using a method presented in \textcite{Bernardeau_2022_CovariancesDensity} for predicting the covariance of one-point statistics from models of the joint two-point PDF. \textcite{Uhlemann.etal_2023_ItTakes} mainly focuses on applying this method to the weak lensing convergence $\kappa$. In this Chapter we mainly focus on applying the same techniques to the three-dimensional matter (over)density field $\delta$ or $\rho$. We present the general theory laid out in \textcite{Bernardeau_2022_CovariancesDensity} which relies on an expansion in powers of the two-point correlation function between cells. We then apply this prescription to a few simple analytic models, presenting next-to-leading order results for general non-Gaussian models, a comparison between simulation and theoretically predicted results from large deviations theory, and results to all orders for a simple hierarchical clustering model. These results can supplement covariances from simulated boxes with fixed overall density to account for the ``super-sample covariance'' which standard $N$-body simulations are unable to capture.

\section{Introduction}

In the previous Chapters we have seen that the one-point PDF of the matter density field captures essential non-Gaussian information about the universe at late times and small scales. Extracting this cosmological information requires accurate modelling of the both PDF itself (including its response to changing cosmological parameters) as well as the covariance matrices associated with it and other statistics (as seen in the Fisher forecast in Chapter \ref{chap:MG-PDFs}). Such covariance matrices are often informed by cosmological simulations,  but getting accurate covariance matrices can prove difficult and require thousands of numerical simulations to ensure convergence \parencite{TaylorJoachimi2014CovariancesEstimation,Colavincenzo2019MockCovariancesBispectrum}. For statistics relying on higher-order correlation functions involving $N$ points, modelling the covariance generally requires assumptions for the shape of $2N$-point correlations. Even for Gaussian fields, this task has only been achieved recently at all orders $N$ \parencite{Hou2021}. 

To go beyond Gaussian fields, which are not well suited to describe the late-time or small-scale 3D distribution of matter, fast numerical recipes such as (shifted) lognormal models \parencite[e.g.][]{Xavier2016} have become a tool of choice in higher-order analyses such as studies of the bispectrum \parencite{Martin2012, Halder2021}, moments of cosmic random fields \parencite{Gatti19}, density split statistics \parencite{Friedrich_2018, Gruen:2018}, the full PDF of cosmic random fields \parencite{Uhlemann:2020, Friedrich20pNG, Boyle_2020}, and even map-based inference approaches \parencite{Sarma_Boruah2022}.  These lognormal models are popular as they are straightforward to implement numerically, and can be tuned to match both a desired power spectrum at all scales and a desired skewness at one given scale \parencite[c.f. Section 5.2 and Figure 4 of][]{Friedrich20pNG}. However, simulations based on these models are limited, as they potentially fail to realistically capture the hierarchy of moments beyond the skewness \parencite[for an extension see][]{Baratta2020} as well as the full shape of higher order $N$-point functions.

With the large data vectors that future multi-probe analyses of upcoming surveys like \textit{Euclid} \parencite{Euclid_mission, Euclid16} and the Rubin Observatory LSST \parencite{LSST_mission} are aiming for, estimated covariances --- whether from lognormal simulations or other mock data --- can lead to degradation of cosmological constraining power \parencite{DS2013, Friedrich2017precisionmatrixexp, Percival2022}. This makes it desirable to have an analytic understanding of measurement uncertainties. In this Chapter we present how the covariance matrix of one-point density statistics can be calculated from knowledge of the two-point PDF of densities in cells. This builds on the previously established large-separation expansion considered in \textcite{Codis.etal_2016_LargescaleCorrelations,Uhlemann17Kaiser,Repp2021indicator,Repp2021varcovar}, to develop the more realistic separation-dependent correlation. In particular we demonstrate that this technique can model the ``super-sample covariance'' (the effect of large scale modes on modes inside survey volume) which typical simulations struggle to capture. We also present results for the biasing functions, which encode the density dependence of two-point clustering, for all orders in the average cell correlation in a simple hierarchical model called the ``minimal tree model''. This method is applied in greater detail to the weak lensing convergence field in \textcite{Uhlemann.etal_2023_ItTakes}, where the cosmic fields are more Gaussian and the covariance matrices for one-point statistics are better approximated by with lognormal mock data.

\section{Covariance prediction from joint 2-point PDF}

Before calculating the covariance matrix of a PDF measurement, let us specify how such a measurement is conducted. We focus on three-dimensional quantities such as the matter density in this Chapter, which could be translated to galaxy counts in a spectroscopic galaxy survey \parencite{Friedrich_2021}.  \textcite{Bernardeau_2022_CovariancesDensity} establishes the method used here for generic observables, including three-dimensional quantities, but these results as we present them generalise easily to two-dimensional data such as the weak lensing convergence PDF, which is examined in detail in \textcite{Uhlemann.etal_2023_ItTakes}. Weak lensing observables are more amenable to analysis in this way, as the line-of-sight projection involved in the weak lensing convergence and photometric galaxy  density on mildly non-linear scales tends to Gaussianise the underlying field. However, we will consider some models of the joint two-point PDF which can capture key clustering properties.

\begin{figure}[h!t]
\centering
\includegraphics[width=0.75\columnwidth]{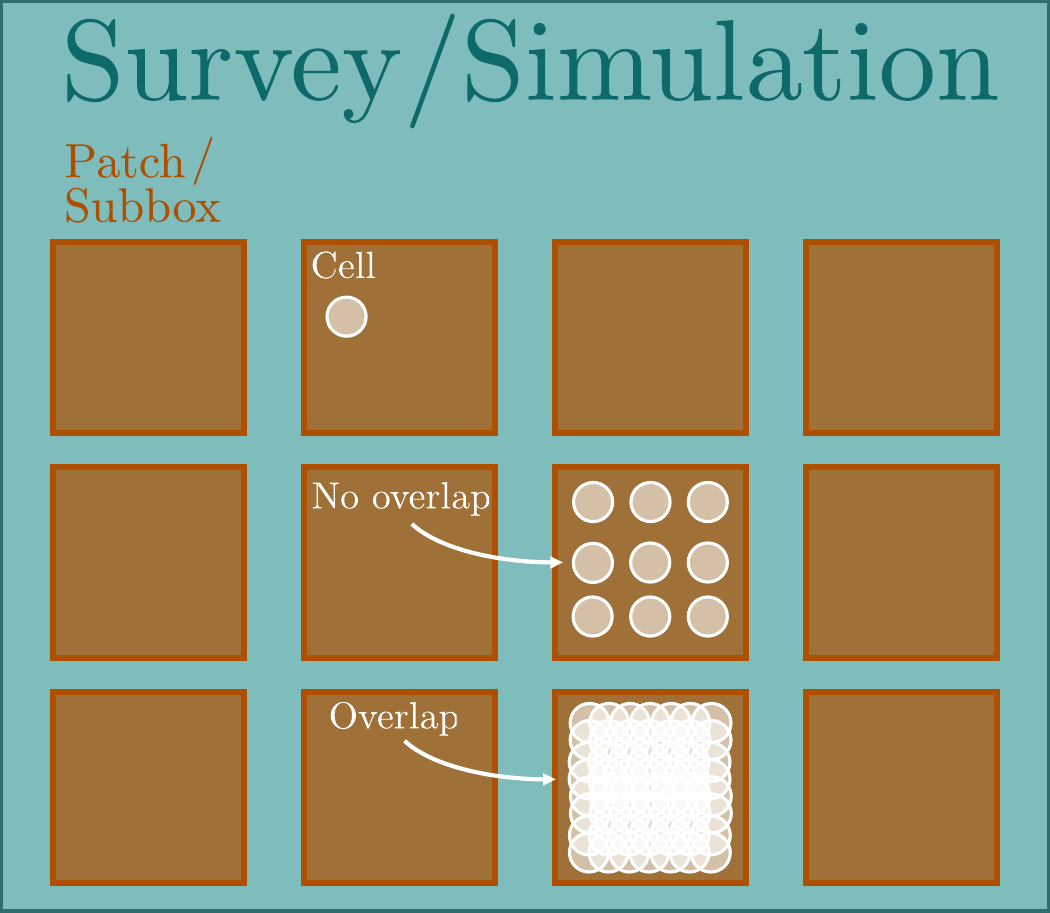}
\caption[Schematic diagram for different areas/volumes used to measure a cosmic field.]{Schematic depiction of the different areas/volumes which are discussed and their relation to one another. The largest is the entire survey or simulation box, which may be broken into smaller patches or subboxes/subvolumes. The covariance of the PDF of the densities in spherical cells within a (sub)volume is the key quantity of interest in this Chapter. }
\label{fig:survey_patches_cartoon}
\end{figure}

Assume we measure the field of interest within some survey volume $V_{\rm survey}$. We then smooth the field with a spherical aperture of radius $R$ (potentially cutting away parts of the edges of the survey such that the aperture falls entirely within the survey volume), and pixelise the resulting map. Each pixel of the final map contains the average value of the field in the sphere of radius $R$, which we refer to as cells, about the pixel's centre. For concreteness,  we wish to estimate the one-point probability density of a certain cosmological field, such as the matter density field $\rho$.  A schematic diagram of some key terms is shown in Figure \ref{fig:survey_patches_cartoon}. In a realistic survey, or if comparing to cosmological simulations, the survey or simulation might be broken into smaller pieces, which we refer to as patches (in the 2D case) or subboxes/subvolumes (in the 3D case). The spherical cells can be placed with differing degrees of overlap within the volume of the survey or simulation. The total number of cells in a (sub)volume, and their degree of overlap will have important effects on the covariance of the PDF of densities in those cells. 

In practice one will need to compute the matter PDF within a set of finite bins, which we assume are centred around a central value $\rho_i$ with constant bin width $\Delta$, which results in $\mathrm{bin}_i = [\rho_i-\Delta/2, \rho_i+\Delta/2]$. The PDF in each of these bins is then approximated as
\begin{equation}
\mP(\rho_i) \approx \frac{\mathbb{P}(\rho \in  [\rho_i-\Delta/2, \rho_i+\Delta/2])}{\Delta }\,,
\end{equation}
where $\mP(\rho)$ is the PDF (obtained in the limit of the bin widths going to 0), and $\mathbb{P}(X)$ is the probability of statement $X$. In the pixelised map, the right hand side of this expression can be approximated by the ratio of pixels in a certain density bin over the total number of pixels. We introduce the estimator $\hat\mP$,
\begin{equation}
\hat\mP(\rho_i) = \frac{\#\{\mathrm{pix \ with} \ \abs{\rho-\rho_i}\leq \Delta/2 \}}{\#\{\mathrm{pix}\}\Delta}.
\end{equation}
Following \textcite{Bernardeau_2022_CovariancesDensity}, we introduce the weight function
\begin{equation}
\chi_i(\rho) = \begin{cases} 1 & \mathrm{if} \ \rho \in  [\rho_i-\Delta/2, \rho_i+\Delta/2]) \\
0 & \rm else\,,
\end{cases}
\end{equation}
such that the estimator can be written
\begin{equation}
\hat\mP(\rho_i) = \frac{\sum_P \chi_i(\rho_P)}{N_{P}\Delta} \approx  \frac{1}{V_{\rm survey}} \int_{\rm survey}\dd[3]{\xx} \frac{\chi_i(\rho(\xx))}{\Delta}\,,
\end{equation}
where the sum index $P$ runs over all $N_P$ pixels, and $\rho(\xx)$ is the smoothed, continuous density field at position $\xx$. From this estimator, the second moment is given by
\begin{align}
\langle\hat\mP(\rho_i)\hat\mP(\rho_j)\rangle &\approx \frac{1}{V_{\rm survey}^2}\int_{\rm surv}\int_{\rm surv}\dd[3]{\xx_1}\dd[3]{\xx_2} \frac{\langle\chi_i(\rho(\xx_1)) \chi_j(\rho(\xx_2))\rangle}{\Delta^2} \nonumber \\
&\approx \frac{1}{V_{\rm survey}^2}\int_{\rm surv}\int_{\rm surv}\dd[3]{\xx_1}\dd[3]{\xx_2} \mP(\rho_1, \rho_2; \abs{\xx_1 - \xx_2})\,,
\end{align}
where $\mP(\rho_1, \rho_2; r)$ is the joint PDF of two points of the smoothed density field separated by distance $r$. This can be further simplified to integrating simply over the separation between the points
\begin{equation}\label{eq:covfromjointPDFmodel}
\langle\hat\mP(\rho_i)\hat\mP(\rho_j)\rangle = \int \dd{r} P_d(r) \mP(\rho_1, \rho_2; r)\,,
\end{equation}
where $P_d(r)$ indicates the distribution of distances between points in a given survey. The covariance of the PDF estimator can then be computed from this moment via
\begin{equation}
\cov(\mP(\rho_i),\mP(\rho_j)) = \ev{\hat{\mP}(\rho_i)\hat{\mP(\rho_j)}} - \overline{\mP}(\rho_i)\overline{\mP}(\rho_j)\,,
\end{equation}
where $\overline{\mP} = \ev{\mP} \approx \mP$. We see that by construction the covariance matrix subtracts off the trivial product of the two one-point expectation values, in the same way that cumulants and correlation functions are constructed to only include the non-trivial, connected parts. We see that covariance predictions rely on two key ingredients: 
\begin{itemize}
	\item the distance distribution between separations for a given survey,
	\vspace{-8pt}
	\item the two-point PDF of the smoothed field at a given separation.
\end{itemize} 
From this we see that effective models of the PDF in two cells allows us to predict the covariance of the one-point PDF for a given survey.

\subsection{Distance distributions}

While the distance distributions $P_d(r)$ depend on the details of the survey used, it is worth noting two simple cases here. For a square two-dimensional survey, the distribution of distances relative to the edge length of the square $r=\abs{\bm{x}_1-\bm{x}_2}/L_{\rm side}$ is given by
\begin{equation}\label{eq:distance_distrib_square}
P_{d}^{\rm square}(r) = \begin{cases} 
2r[\pi + (r-4)r] & r \in [0,1] \\
2r[\pi-2+4\sqrt{r^2-1}-r^2-4\sec^{-1}r] & r \in (1,\sqrt{2}].
\end{cases}
\end{equation}

\begin{figure}[h!t]
\centering
\includegraphics[scale=1]{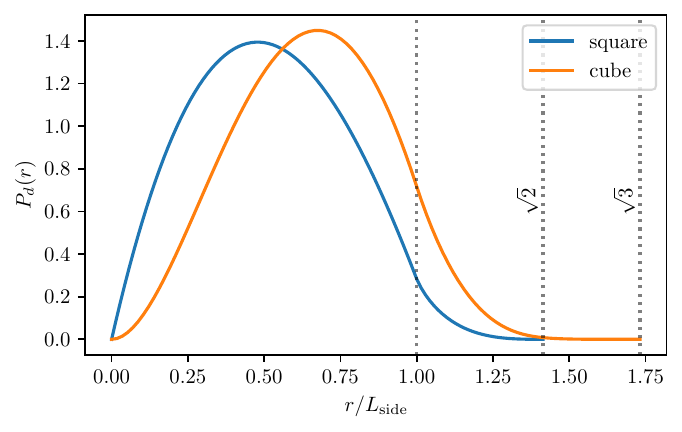}
\caption[Distance distributions for points within a square and cube survey area/volume.]{Distance distributions within a cube and a square survey volume/area, as given by equations \eqref{eq:distance_distrib_square} \& \eqref{eq:distance_distrib_cube}. Vertical lines are shown at $\sqrt{2}$ (the maximal separation in a square survey) and $\sqrt{3}$ (the maximal separation in a cubic survey).}
\label{fig:distance_dist}
\end{figure}

The case of 3D spatial separations is similar, but closed form expressions become unwieldy even for simple survey volumes. The distance distribution in a cubic box can be written in closed form for the unit cube, corresponding to the problem of  ``cube line picking'' for which a closed-form expression is available \parencite{Mathai1999cubelinepicking,WeissteinCubeLinePicking}. The distribution of distances (relative to the side length of the cube) is
\begin{equation}\label{eq:distance_distrib_cube}
P_{d}^{\rm cube}(r) = \begin{cases} 
-r^2 \left[ (r-8)r^2 + \pi (6 r - 4)\right] & r\in [0,1] \\[18pt]
\begin{aligned}[b]
2r \Big[\Big(&r^2 - 8\sqrt{r^2-1}+3\Big)r^2 - 4 \sqrt{r^2-1} \\ &+12 r^2 \sec^{-1}r + \pi (3-4r)- \frac12\Big] \end{aligned} & r\in(1,\sqrt{2}] \\[18pt]
\begin{aligned}[b]
r \Big[(1&+r^2)(6\pi + 8\sqrt{r^2-2}-5-r^2)  \\
&-16 r \csc^{-1}(r\sqrt{r^2-2})  \\ 
&- 24 (r^2+1)\tan^{-1}(\sqrt{r^2-2})\Big], \end{aligned} & r\in(\sqrt{2},\sqrt{3}].
\end{cases}
\end{equation}
The distance distributions for a square and cube survey patch are shown in Figure \ref{fig:distance_dist}. 

\subsection{Finite sampling (shot noise) effects}

For the remainder of this Chapter we will work in the continuum limit, rather than the more realistic pixelised case, however before we do so it is worth noting the effect of finite sampling on this prescription for applying the work to real data. The contribution to the covariance due to finite sampling of the underlying continuous field (also called the \emph{shot noise}) is \parencite{Codis2016MNRAS, Uhlemann.etal_2023_ItTakes}
\begin{equation}
\cov_{\rm FS}(\mP_i,\mP_j) = \frac{\overline{\mP}_i}{\Delta_i N_T}\delta_{i,j}\,,
\end{equation}
where $\Delta_i$ is the $i^{\rm th}$ bin width, $\mP_i\equiv \mP(\rho_i)$, and $N_T$ is the total number of cells. Notice that this term is proportional to the Kronecker delta, and thus only contributes to the diagonal of the covariance matrix. As this term scales inversely with the total number of cells, (or equivalently the survey volume) it can be rendered arbitrarily small by increasing the number of cells. For a fixed observation volume however this comes at the price of heavy cell overlaps at small separations, as seen in Figure \ref{fig:survey_patches_cartoon}. In this regime, a small-scale expansion with small $\Delta_\xi^{1/2}= \sqrt{\sigma^2 - \xi}\ll \xi \ll 1$ can be used to capture effect of heavy cell overlaps as discussed in \textcite{Bernardeau_2022_CovariancesDensity}. For this Chapter we consider closed form two-point PDF models which automatically include these overlap effects. 

\subsection{Super-sample covariance effects}

%

\begin{figure}[h!t]
\centering
\includegraphics[width=\columnwidth]{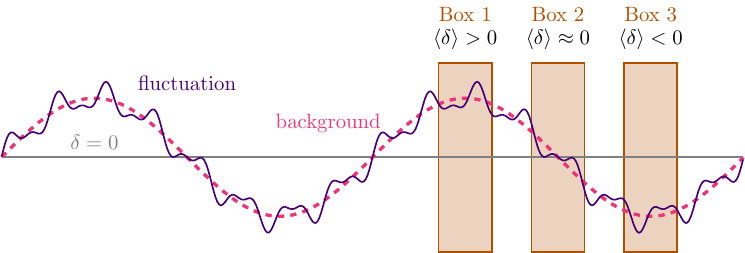}
\caption[The effect of super-sample covariance from box sizes smaller than background fluctuations.]{In the presence of fluctuations larger than the survey/simulation box size, the mean overdensity $\delta$ can vary from its global mean value of 0. }
\label{fig:peak-background-split}
\end{figure}

In cosmological $N$-body simulations \parencite[see e.g.][for a recent review]{AnguloHahn2022}, the mean overdensity is generally enforced to be zero over the volume of a simulation to be consistent with periodic boundary conditions. However, in the presence of large-scale background modes, with wavelengths larger than the simulation box size, the simulation box can acquire a mean density which differs from the global mean. This effect is illustrated in Figure \ref{fig:peak-background-split}. By setting this sample mean in each simulation volume to zero, the variance of large scale modes over an ensemble of simulations is artificially set to zero, despite them each having different initial random fields. This leads to an underestimate of the covariance which is due to large-scale modulations to the density and tidal fields, which is referred to as the super-sample covariance (SSC) \parencite{Akitsu2019PhRvD, Li2018JCAP, Klypin2018MNRAS, AnguloHahn2022, Takahashi2009ApJ}.  Since $N$-body simulations with comparable volumes ($\order{100} \ (\mathrm{Gpc}/h)^3$) and resolution (halo masses of $\order{10^{10} \ \rm M_\odot}$) to upcoming surveys are difficult to achieve, owing simply to the memory requirements of simulating a large number of particles, this super-sample covariance is a key effect which simulation informed covariances struggle to capture.

The SSC effect has been modelled previously in the context of the matter power spectrum \parencite{TakadaHu_SSC, Barreira_2017} and cosmic shear \parencite{Linke2024A&A}. In simulations, the presence of these large scale fluctuations can be accounted for in the ``separate universe simulations'' \parencite{Sirko2005ApJ, Baldauf_2011_galbias, Wagner_2014OUP, Li2014PhRvD, Baldauf_2016_longwavelength} where the effect of a non-zero overdensity is absorbed into the background evolution, such that it evolves as if it were a universe with modified cosmological parameters, similarly to how spherical overdensities can be treated by the Friedmann equations for a closed universe.

Since the super-sample covariance effect is difficult to capture in simulated covariances, lessons from analytic models are appealing. Moreover, covariance matrices  constructed from simulations require a large number of realisations to be reliable, as inversion of the covariance (which is needed for inference and forecasting, as seen in Chapter \ref{chap:MG-PDFs})  is easily biased from its true value if the number of realisations is not much larger than the length of the data vector \parencite{Hartlap06}.

In the context of PDFs, estimates of the large separation limit are possible even beyond specific (log)-normal models. In the case of well separated cells, where the correlation between their densities becomes small, the joint PDF for the densities in those cells is well approximated by \parencite{Bernardeau1996A&A, Codis.etal_2016_LargescaleCorrelations} 
\begin{equation}
\mP(\rho_i, \rho_j, d \gtrsim 2R) \simeq \mP(\rho_i)\mP(\rho_j)[1+\xi(d)b_1(\rho_i)b_1(\rho_j)]\,,
\end{equation}
with some bias function $b_1$ which describes how the two-point correlation, $\xi(d)$, is modulated by the local density.  A similar result holds for the weak lensing convergence and this functional form has been found to be very robust. When  the initial conditions are Gaussian, this modulation becomes independent of separation. We will see how these bias functions are derived for simple models in Section~\ref{sec:large-separation-cov}.

Under this factorisation, the super-sample covariance becomes 
\begin{equation}\label{eq:cov-SSC-factorisation-biasprod}
\cov_{\rm SSC}(\mP(\rho_i), \mP(\rho_j)) = \overline{\xi}\,  (b_1\mP)(\rho_i) \, (b_1\mP)(\rho_j)\,,
\end{equation}
where the mean correlation $\overline\xi$ is defined\footnote{Note that this differs from the definition of $\overline{\xi}$ used in \textcite{Bernardeau_2022_CovariancesDensity}, where they define it to be the average values of the correlation function within a \emph{cell}, which we denote $\sigma^2$, the variance of densities in cells.}
\begin{equation}\label{eq:mean-correlation}
\overline\xi = \int^{R_{\rm max}} d{r} P_d(r) \xi(r)\,,
\end{equation}
where $P_d(r)$ is the probability distribution of distances, and $R_{\rm max}$ is the maximum distance set by the size of the survey.  Note that if we subdivided our simulation volume, or considered a suite of cosmological simulations, the mean correlation, $\overline{\xi}$, defined in \eqref{eq:mean-correlation} would corresponds to the variance of the mean density $\rho$ measured across different simulations or subvolumes $\overline{\xi} = \sigma_{\bar{\rho}}^2$.

To leading order in the mean correlation function $\overline{\xi}$,  the covariance in the binned cases is then
\begin{equation}
    \label{eq:covPDFlargesep}
        \text{cov}(\mP(\rho_i),\mP(\rho_j)) = \overline\xi (\overline{ \mP b_1})(\rho_i)  (\overline{ \mP b_1})(\rho_j) + \delta_{ij} \frac{\overline\mP(\rho_i)}{\Delta_i N_T}\,,
    \end{equation}
    where $N_T$ is the total number of cells and the mean correlation $\bar\xi$ is given by equation~\eqref{eq:mean-correlation}.
Note that the shot-noise term only acts on the diagonal. The diagonal terms of the covariance give the variance of the PDF, which to leading order is,
\begin{equation}
\text{var}(\mP(\rho)) = \overline\xi (\overline{b_1 \mP})^2(\rho)+\frac{\bar\mP(\rho)}{\Delta N_T}\,,
\end{equation}
where $\Delta$ is the PDF bin width.

Let us also recall that for non-overlapping cells, typically the ``finite sampling'' effect dominates (and is well-described by Poisson noise). For overlapping cells, the number of cells is extremely large thus effectively removing the finite sampling term, but requiring the treatment of overlaps. The impact of overlapping cells can be captured by either closed-form expressions for the two-point PDF valid at small distances (like for the Gaussian and minimal tree model discussed in Sections \ref{sec:large-separation-cov} \& \ref{sec:minimal-tree-model},  for lognormal models discussed in \textcite{Uhlemann.etal_2023_ItTakes}), or a heavy overlap expansion complementary to the large separation one as described in \textcite{Bernardeau_2022_CovariancesDensity}. 

\subsection{Illustrative covariance measurements}

\begin{figure}[h!t]
\centering
\includegraphics[width=\columnwidth]{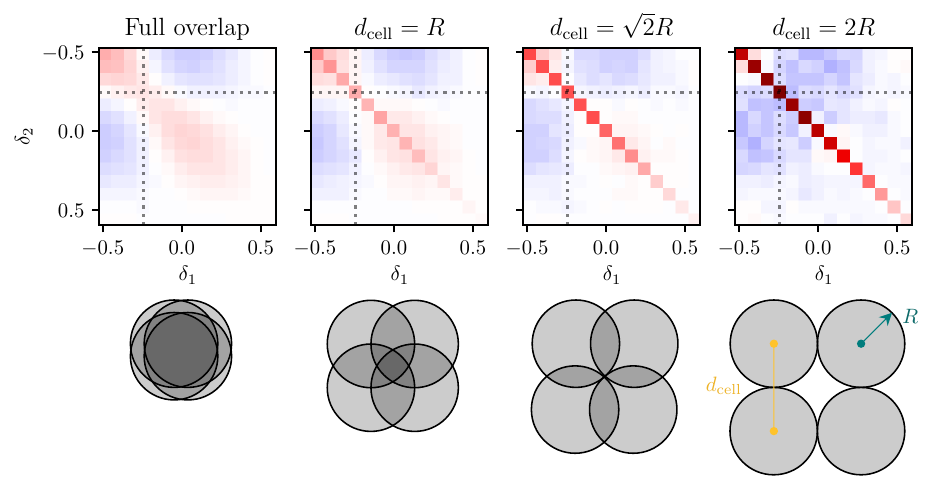}
\caption[Illustrative covariance measurements of the matter density field with different degrees of cell overlap.]{Covariance matrices between the bins of the matter PDF with differing degrees of overlap cells. These are measured from in $R=10\  \mathrm{Mpc}/h$ spheres at $z=1$ from 1000 realisations of the \textsc{Quijote} simulations. The dotted lines indicate the location of the peak of the mean PDF. The fully overlapping case takes spheres spaced 1.95 Mpc/$h$ apart, corresponding to a $(512)^3$ grid on the \textsc{Quijote} simulations. We also consider two partially overlapping sphere cases, one spaced such that the separation is one radius, another with the separation of $\sqrt{2}$ times the radius, such that diagonal neighbouring cells do not overlap. Finally we consider minimally non-overlapping cells, spaced by their diameter. We see that higher overlap decreases the overall magnitude of terms in the covariance matrix, but that this increases the relative contribution of the off diagonal terms. All panels are coloured on the same colour scale.}
\label{fig:rho_cov_overlap}
\end{figure}

We build a data vector $\bm{S}$ from the values of the PDF histogram of the matter density $\rho$ measured from the \textsc{Quijote} simulations ($L_{\rm box} = 1000 \ \mathrm{Mpc}/h$)  and smoothed with a spherical top-hat filter with radius $R= 10 \ \mathrm{Mpc}/h$. The covariance matrix of this binned $\rho$-PDF is then obtained as 
\begin{equation}
\label{eq:covariance}
C_{ij} = \langle (S_i-\bar{S}_i)(S_j - \bar{S}_j) \rangle\,,\quad \bar{S}_i = \langle S_i \rangle~\,,
\end{equation}
where $\langle\cdot\rangle$ indicates an ensemble average over 1000 different realisations. We extract 4 different data vectors corresponding to differing degrees of overlap, as illustrated in Figure \ref{fig:rho_cov_overlap}. The fully overlapping spheres corresponds to spheres placed on a $(512)^3$ grid (such that their centres are spaced $1.95$ Mpc/$h$ apart), while the other cases are spaced at factors of $1$, $\sqrt{2}$, and $2$ times the smoothing radius $R=10$ Mpc/$h$.

We show these correlation matrices in Figure~\ref{fig:rho_cov_overlap} in four cases with differing degrees of overlap between the cells, varying from the pixel level to twice the sphere radius.  The values along the diagonal represent the error on that bin of the PDF, while the off diagonal elements represent the correlation between different bins. We see directly that increasing the number of cells for the PDF measurement decreases the finite sampling (or shot-noise) term which contributes to the diagonal of the covariance matrix. This suppression comes at the cost of increasing the overlap between the cells which creates a positive correlation band around the diagonal.

\section{Leading order covariance expansions}\label{sec:large-separation-cov}

In this Section we preform an expansion of the joint two-point PDF in powers of the two-point cell correlation function $\xi_{12}(d)$,  with the help of the cumulant generation function (CGF) $\phi$. We will pay particular attention to the leading and next-to-leading order (NLO) contributions to the covariance matrix. We present these results for bivariate Gaussian PDFs to all orders, the leading order and NLO results for general non-Gaussian models, as well as results for fields which have been mean-subtracted or normalised by the mean field value within a subvolume, as these introduce changes to the biasing functions.

Throughout this Section we will focus on the density contrast $\delta$ as our zero-mean field of interest. However, these results also hold for the weak lensing convergence $\kappa$ with angular separation $\theta$ replacing the three-dimensional separation $d$. The results and further analysis on $\kappa$ can be found in \textcite{Uhlemann.etal_2023_ItTakes}.

\subsection{Bivariate Gaussian model}\label{sec:gaussian-model}

The study of multivariate Gaussian fields has a long history in cosmology \parencite[see e.g. ][]{Bardeen.etal_1986_StatisticsPeaks, 1987ApJ...323L.103B, 1991ApJ...380L...5H}. The results presented here are well known, but allow us to develop the notation and formalism to apply it to other models for the joint PDF.

A Gaussian joint PDF with zero mean is characterised by the one-point variance $\sigma^2$ and the two-point correlation between cells $\xi = \xi_{12}(d)$. The bivariate Gaussian PDF is given by
\begin{equation}\label{eq:joint_pdf_gaussian}
\mP_{\rm G}(\delta_1, \delta_2) \defeq \mP(\delta_1, \delta_2 \,\vert\, d) = \frac{1}{2\pi\sqrt{D}}\exp[-\frac{1}{2}\bm{\delta}^\top \cdot \mathsf{\Sigma}^{-1} \cdot \bm{\delta}]\,,
\end{equation}
where
\begin{equation}
\bm{\delta} = \begin{pmatrix}
\delta_1 \\ \delta_2
\end{pmatrix}, \quad \mathsf{\Sigma} = \begin{pmatrix}
\sigma^2 & \xi_{12}(d) \\
\xi_{12}(d) & \sigma^2
\end{pmatrix}\,,
\end{equation}
and $D=\det\mathsf{\Sigma} = \sigma^4 - \xi^2$, $\sigma^2$ is the variance of the smoothed density field, $\xi=\xi_{12}(d)$ is the two-point correlation function evaluated at this separation. 

A straightforward integration recovers that the marginal distributions $\mP_{\rm G}(\delta_1), \mP_{\rm G}(\delta_2)$ are themselves Gaussians with variance $\sigma^2$. We now wish to  calculate the CGF associated with this joint distribution. For a one-dimensional Gaussian with zero mean, the corresponding CGF is
\begin{equation}
\phi_0^{\rm G}(\lambda) = \sum_{n=1}^\infty \frac{\ev{\delta^n}_c}{n!}\lambda^n = \frac{1}{2}\sigma^2\lambda^2 \,,
\end{equation}
as all cumulants of third order and higher vanish for a Gaussian.  This is the CGF which generates the marginal distributions $\mP(\delta_i)$. For a joint Gaussian, the two-point CGF is defined via a series in the joint cumulants
\begin{subequations}
\begin{align}
\phi_{\rm 2D}^{\rm G}(\lambda_1, \lambda_2) &= \sum_{p,q}\frac{\ev{\delta_1^p\delta_2^q}}{p!q!} \lambda_1^p \lambda_2^q \\
&= \frac{1}{2}\ev{\delta_1^2}_c \lambda_1^2 + \frac{1}{2}\ev{\delta_2^2}_c \lambda_2^2 + \ev{\delta_1\delta_2}_c \lambda_1 \lambda_2 + \dots \\
&= \frac{1}{2}\sigma^2 (\lambda_1^2 + \lambda_2^2) + \xi \lambda_1 \lambda_2\,,
\end{align}
\end{subequations}
where all higher order terms vanish. Rewriting the joint CGF in terms of $\phi_0^{\rm G}$ and making it symmetric in our two variables,
\begin{align}\label{eq:2d-gaussian-cgf}
\phi_{\rm 2D}^{\rm G}(\lambda_1, \lambda_2) &= \frac{1}{2}\sigma^2\lambda_1^2 + \frac{1}{2}\sigma^2\lambda_2^2 + \xi \lambda_1 \lambda_2 \nonumber \\
&= \phi_0^{\rm G}(\lambda_1) + \phi_0^{\rm G}(\lambda_2) + \xi\phi_1^{\rm G}(\lambda_1)\phi_1^{\rm G}(\lambda_2)\,,
\end{align}
where $\phi_1^{\rm G}(\lambda)=\lambda$. If the joint distribution were non-Gaussian then there would also be terms higher order in $\xi$ generating $\phi_{n\geq 2}(\lambda)$.

Equipped with the CGFs, we can now relate the joint PDF to the individual PDFs via inverse Laplace transform (see Appendix \ref{app:prob-distributions}), expanding the final exponential in a series in $\xi$,
\begin{equation}
\mP_{\rm G}(\delta_1, \delta_2) = \int \frac{\dd{\lambda_1}\dd{\lambda_2}}{(2\pi i)^2} \exp[-\lambda_1\delta_1 - \lambda_2 \delta_2 + \phi_{\rm 2D}^{\rm G}(\lambda_1, \lambda_2)].
\end{equation}
Inserting the CGF from equation~\eqref{eq:2d-gaussian-cgf} and expanding in powers of $\xi_{12}$ 
\begin{subequations}
\begin{align}
\mP_{\rm G}(\delta_1, \delta_2) &=  \int \frac{\dd{\lambda_1}\dd{\lambda_2}}{(2\pi i)^2} \exp[-\lambda_1\delta_1 + \phi_0^{\rm G}(\lambda_1)] \exp[-\lambda_2\delta_2 + \phi_0^{\rm G}(\lambda_2)] \exp[\xi\phi_1^{\rm G}(\lambda_1)\phi_1^{\rm G}(\lambda_2)] \\
&= \int \frac{\dd{\lambda_1}\dd{\lambda_2}}{(2\pi i)^2} e^{-\lambda_1\delta_1 + \phi_0^{\rm G}(\lambda_1)} e^{-\lambda_2\delta_2 + \phi_0^{\rm G}(\lambda_2)} \left[1 + \xi\phi_1^{\rm G}(\lambda_1)\phi_1^{\rm G}(\lambda_2) + \order{\xi^2}\right]\,,
\end{align}
\end{subequations}
which defines a series of weighted integrals against the marginal distributions. This defines the bias functions 
\begin{equation}
\mP_{\rm G}(\delta_1, \delta_2) = \mP_{\rm G}(\delta_1)\mP_{\rm G}(\delta_2) \left[ 1 + \sum_{n=1}^\infty \frac{\xi^n}{n!} b_n^{\rm G}(\delta_1)b_n^{\rm G}(\delta_2) \right],
\end{equation}
where 
\begin{equation}
b_n^{\rm G}(\delta)\mP_{\rm G}(\delta) = \int_{-i\infty}^{i\infty} \frac{\dd{\lambda}}{2\pi i} [\phi_1^{\rm G}(\lambda)]^n \exp[-\lambda \delta + \phi_0^{\rm G}(\lambda)],
\end{equation}
define the bias functions which control the density dependent two-point clustering. In the Gaussian case since $\phi_1^{\rm G}(\lambda)=\lambda$ these integrals can either be performed directly via Wick rotation, or we can replace the $\lambda_i$ as derivatives of the exponential with respect to $\delta_i$, leaving
\begin{subequations}
\begin{align}
\frac{\mP_{\rm G}(\delta_1,\delta_2; d)}{\mP_{\rm G}(\delta_1)\mP_{\rm G}(\delta_2)} &= 1+ \sum_{n=1}^\infty \frac{\xi^n}{n!}b_{n,\rm G}(\delta_1) b_{n,\rm G}(\delta_2) \,, \\
b_{n, \rm G}(\delta) &= \frac{(-1)^n}{\mP_G(\delta)}\pdv[n]{\mP(\delta)}{\delta} = \frac{1}{\sigma^n}\operatorname{He}_n\left(\frac{\delta}{\sigma}\right),
\end{align}
\end{subequations}
where $\operatorname{He}_n(x)$ are the probabalist's Hermite polynomials defined by
\begin{equation}
\operatorname{He}_n(x) = (-1)^n \exp[-\frac{x^2}{2}]\dv[n]{}{x}\exp[-\frac{x^2}{2}].
\end{equation}
The leading order bias terms for this Gaussian model are then
\begin{equation}
b_{1,\rm G}(\delta) = \frac{\delta}{\sigma^2}, \quad b_{2,\rm G}(\delta) = \frac{\delta^2-\sigma^2}{\sigma^4}=b_{1,\rm G}^2(\delta) - \frac{1}{\sigma^2}\,,
\end{equation}
which we could alternatively have derived by simply expanding the joint PDF \eqref{eq:joint_pdf_gaussian} in $\xi$ directly. This leading order term  reproduces the linear ``Kaiser'' bias  function $b(\delta)=\delta/\sigma^2$, which is well known for a Gaussian field \parencite{Kaiser84,Codis2016MNRAS,Uhlemann17Kaiser}.

\begin{figure}
\centering
\includegraphics[scale=1]{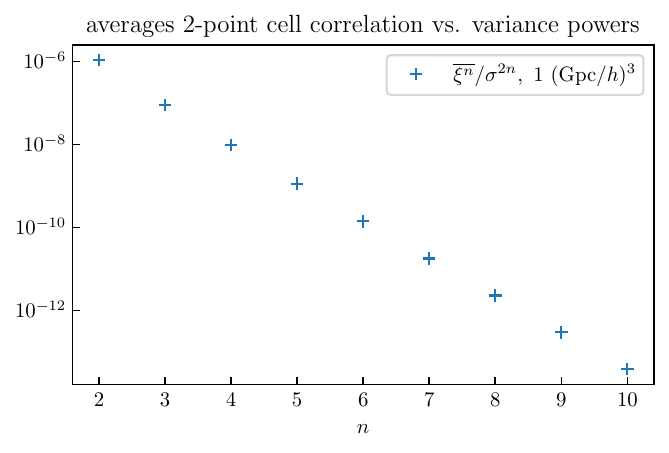}
\caption[The scaling of the average of powers of the two-point correlation function.]{Scaling of the average of powers of the two-point cell correlation function compared to the variance of $\delta$ across 1000 realisations in the \textsc{Quijote} simulations smoothed on $R=10$ Mpc/$h$ at $z=0$. }
\label{fig:average_xi_power_scaling}
\end{figure}

The large separation covariance is then given by integrating joint PDF against the distribution of separations as in equation \eqref{eq:covfromjointPDFmodel}
\begin{subequations}
\begin{align}
\cov(\mP_{\rm G}(\delta_1)\mP_{\rm G}(\delta_2)) &= \sum_{n=1}^\infty \frac{\overline{\xi^n}}{n!}\pdv[n]{\mP(\delta_1)}{\delta_1}\pdv[n]{\mP(\delta_2)}{\delta_2} \label{eq:gaussian_cov_sum}  \\
&= \sum_{n=1}^\infty \frac{\overline{\xi^n}}{n!\sigma^{2n}} \operatorname{He}_n\left(\frac{\delta_1}{\sigma}\right)\mP_{\rm G}(\delta_1)\operatorname{He}_n\left(\frac{\delta_2}{\sigma}\right)\mP_{\rm G}(\delta_2) \,,
\end{align}
where in the second line we have defined the average of powers of the two-point correlation function over the relevant survey/simulation volume
\begin{equation}
\overline{\xi^n} \defeq \int \dd{r} P_d(r) \xi_{12}^n(r)\,, \label{eq:meanxi}
\end{equation}
\end{subequations}
analogously to the mean in equation \eqref{eq:mean-correlation}.
The averages of powers of the correlation function $\overline{\xi^n}$ compared to the powers of the variance $\sigma^{2n}$ control the large separation expansion of the covariance and their hierarchy is shown for the density contrast in Figure \ref{fig:average_xi_power_scaling}. Similar hierarchical scaling can be seen in the $\kappa$ covariance described in  \textcite{Uhlemann.etal_2023_ItTakes}.

Due to the hierarchical ordering of averages of the correlation function we expect the two leading order terms in the regime $\delta\lesssim\mathcal O(\sigma)$ to be
\begin{subequations}
\begin{align}
\label{eq:covGauss_LO}
\frac{{\rm cov}(\mP_{\rm G}(\delta_1),\mP_{\rm G}(\delta_2))}{\mP_{\rm G}(\delta_1)\mP_{\rm G}(\delta_2)} &=\frac{\bar\xi}{\sigma^2} \frac{\delta_1}{\sigma}\frac{\delta_2}{\sigma} \\
\label{eq:covGauss_NLO}
&+\frac{\overline{\xi^2}}{2\sigma^4} \frac{\delta_1^2-\sigma^2}{\sigma^2} \frac{\delta_2^2-\sigma^2}{\sigma^2}\\
\notag &+\mathcal O\left(\frac{\overline{\xi^3}}{\sigma^6}\right),
\end{align}
\end{subequations}
which predicts a \emph{super-sample covariance} leading order term similar to the separate universe picture~\eqref{eq:cov_SSC_SU}, being  proportional to the variance of the background $\delta_s$ in different subvolumes or realisations, $\bar\xi=\sigma^2(\delta_s)$, and the derivative of the PDF. 

This large-separation expansion of the covariance can be connected to an eigendecomposition and is presented in Section 4 of \textcite{Uhlemann.etal_2023_ItTakes}, where the hierarchical ordering of $\overline{\xi^n}/\sigma^{2n}$ will reflect in the eigenvalues and the bias functions multiplied by the Gaussian PDF evaluated at a given $\delta$-value will yield the eigenvectors (after orthogonalisation and appropriate normalisation). In particular, the zero-crossings of the first bias function $b_{1,\rm G}(\delta)$ at $\delta=0$ and the second bias function $b_{2,\rm G}(\delta)$ at $\delta=\pm \sigma$ determine the zero crossings of the first two eigenvectors of the covariance matrix.

\subsection{General non-Gaussian case to next-to-leading order}\label{sec:non-Gauss-NLO-cov}

For a general non-Gaussian PDF described by its cumulant generating function we can proceed as follows. Expand the joint two-point CGF in powers of the two-point correlation $\xi$,
\begin{align}\label{eq:two-point-CGF-expansion}
\phi(\lambda_1,\lambda_2) = &\phi_0(\lambda_1) + \phi_0(\lambda_2) + \xi \phi_1(\lambda_1)\phi_1(\lambda_2) \nonumber \\
&+ \frac{\xi^2}{2}\left[\phi_1^2(\lambda_1)\phi_2(\lambda_2) + \phi_2(\lambda_1)\phi_1^2(\lambda_2)\right] + \order{\xi_{12}^3}.
\end{align}
The connected part of the joint PDF is then the inverse Laplace transform of this CGF, expanding the exponential with powers of $\xi$ down
\begin{align}
\mP(\delta_1,\delta_2)-\mP(\delta_1)\mP(\delta_2) = \int &\frac{\dd{\lambda_1}}{2\pi i} \frac{\dd{\lambda_2}}{2\pi i} \nonumber \\
&\exp[-\lambda_1\delta_1 -\lambda_2\delta_2 + \phi(\lambda_1,-\lambda_1) + \phi(\lambda_2, -\lambda_2)] \times  \nonumber \\
&\Big(\xi\phi_1(\lambda_1)\phi_1(\lambda_2) + \frac{\xi^2}{2}\phi_1^2(\lambda_1)\phi_1^2(\lambda_2) \nonumber \\
&+ \frac{\xi^2}{2} \left[\phi_1^2(\lambda_1)\phi_2(\lambda_2) + \phi_2(\lambda_1)\phi_1^2(\lambda_2)\right] + \order{\xi^3}\Big) .
\end{align}

The covariance is then obtained by integrating over the $\lambda_i$ and the distance distribution, giving (after symmetrising).
\begin{subequations}
\label{eq:cov_nonG}
\begin{align}
\frac{{\rm cov}(\mP(\delta_1),\mP(\delta_2))}{\mP(\delta_1)\mP(\delta_2)} =&\phantom{+}\overline\xi b_1(\delta_1)b_{1}(\delta_2) \\
&+\frac{\overline{\xi^2}}{2}\left[(b_{2}+q_1)(\delta_1)(b_{2}+q_1)(\delta_2)-q_1(\delta_1)q_1(\delta_2)\right]\\
\notag &+\mathcal O(\overline{\xi^3}),
\end{align}
where we defined
\begin{align}
\label{eq:b1_def}
(b_1\mathcal P)(\delta)&=\int \frac{\dd{\lambda}}{2\pi i}\phi_1(\lambda)\exp[-\lambda\delta+\phi(\lambda)] \,, \\
\label{eq:b2_def}
(b_2\mathcal P)(\delta)&=\int \frac{\dd{\lambda}}{2\pi i}\phi_1^2(\lambda)\exp[-\lambda\delta+\phi(\lambda)] \,, \\
\label{eq:q1_def}
(q_1\mathcal P)(\delta)&=\int \frac{\dd{\lambda}}{2\pi i}\phi_2(\lambda)\exp[-\lambda\delta+\phi(\lambda)].
\end{align}
\end{subequations}
For a given non-Gaussian model defined by its CGF, the functions $\phi_1,\phi_2$ can be obtained, and the resulting bias functions $b_1, b_2, q_2$ can be written as integrals as defined in equations \eqref{eq:b1_def}--\eqref{eq:q1_def}.

\subsection{The impact of relative quantities}\label{sec:relative_quantities}

The cosmic fields we work with are often normalised in some way to the mean over the entire survey or simulation. For small patches within a survey, or small simulation boxes, the mean field, e.g. $\rho$ or $\delta$ will significantly fluctuate between subvolumes, with sizeable average correlation, as illustrated in Figure \ref{fig:survey_patches_cartoon}, causing this super-sample covariance effect. In practice, the quantities which are actually used are often ``background-subtracted'' $\tilde{\delta}_i = \delta_i - \delta_s$ (in the spirit of peak-background splitting \parencite{Bardeen.etal_1986_StatisticsPeaks,MoWhite96PBS,ShethTormen99PBS} where the sample mean field $\delta_s$ is subtracted from the true value, or ``relative densities'' $\hat{\rho}_i = \rho_i/\rho_s$, such that the mean density is set $\ev{\hat\rho}=1$ within the simulation box. The average correlation function $\bar\xi$ corresponds to the variance of these sample means across the subvolumes or patches $\sigma_{\delta_s}^2$ or $\sigma_{\rho_s}^2$.

Both of these cases, the background-subtracted and the relative densities, the appropriate joint PDFs and covariances can be obtained by considering the 3-variable PDFs of the field in two cells and the sample mean field (e.g. $\mP(\rho_1, \rho_2,\rho_s)$) and subsequently marginalising over the sample mean. In the case of relative densities,
\begin{equation}
\mP(\hat{\rho}_1, \hat{\rho}_2) = \int \dd{\rho}_s \rho_s^2 \mP(\rho_1 = \hat{\rho}_1 \rho_s, \rho_2 = \hat{\rho}_2 \rho_s, \rho_s)\,,
\end{equation}
where the $\rho_s^2$ prefactor arises from the Jacobian of the change of variables. Note that $\hat\rho_i = 1+\delta_i$ has unit mean. In the case of background-subtracted fields,
\begin{equation}
\mP(\tilde{\delta}_1, \tilde{\delta}_2) = \int \dd{\delta}_s \mP(\delta_1 = \tilde{\delta}_1+\delta_s, \delta_2 = \tilde{\delta}_2 + \delta_s, \delta_s)\,,
\end{equation}
 
The covariance in these reduced variables is then constructed in the same way from the two-point PDF, by integrating via equation~\eqref{eq:covfromjointPDFmodel}.

These 3-variable PDFs can be modelled via the 3-variable CGFs and expansions of the inverse Laplace transform order by order, as presented above for the unscaled fields. To leading order and next-to-leading order, mean-subtracting or normalising the fields leads to additional terms in the bias functions.

In the background-subtracted case, the first two bias functions become, 
\begin{equation}\label{eq:bias-mean-sub}
\tilde{b}_{1,\rm NG}(\tilde{\delta}) = b_1(\tilde{\delta}) + \pdv{\log\mP(\tilde{\delta})}{\tilde{\delta}}, \quad \tilde{b}_{2,\rm NG}(\tilde{\delta}) = 2 \frac{1}{\mP(\tilde{\delta})}\pdv{b_1\mP}{\tilde{\delta}} + \frac{1}{\mP(\tilde{\delta})}\pdv[2]{\mP}{\tilde{\delta}}.
\end{equation}
The resulting covariance for the mean-subtracted field is
\begin{subequations}
\label{eq:cov_nonG_meansub}
\begin{align}
\nonumber {\rm cov}(\mP(\tilde\delta_1),\mP(\tilde\delta_2)) =&\phantom{+} \bar\xi (\tilde b_{1,\rm NG}\mP)(\tilde\delta_1)(\tilde b_{1,\rm NG}\mP)(\tilde\delta_2)\\
&+\frac{\bar\xi^2}{2}(\tilde b_{2,\rm NG}\mP)(\tilde\delta_1)(\tilde b_{2,\rm NG}\mP)(\tilde\delta_2)\\
\nonumber &+\frac{\overline{\xi^2}}{2} [(b_2\mP)(\tilde\delta_1)(q_1\mP)(\tilde\delta_2)+(q_1\mP)(\tilde\delta_1)(b_2\mP)(\tilde\delta_2)]\\
&+\frac{\overline{(\xi-\bar\xi)^2}}{2}(b_{2}\mP)(\tilde\delta_1)(b_{2}\mP)(\tilde\delta_2)\\
\notag &+ \mathcal O(\{\overline{\xi^3},\bar\xi\,\overline{\xi^2},\bar\xi^3,\bar\xi\, \overline{(\xi-\bar\xi)^2},\overline{(\xi-\bar\xi)^3}\})\,.
\end{align}
\end{subequations}
For a Gaussian field the first three lines of equation \eqref{eq:cov_nonG_meansub} do not contribute as $\phi_1(\lambda)=\lambda$ and $\phi_{n\geq 2}=0$, so we recover the Gaussian covariance
\begin{align}
\label{eq:covGauss_meansub}
{\rm cov}(\mP_{\rm G}(\tilde\delta_1),\mP_{\rm G}(\tilde\delta_2)) &=\sum_{n=2}^\infty \frac{\overline{(\xi-\bar\xi)^n}}{n!} \frac{\partial^n\mP_{\rm G}(\tilde\delta_1)}{\partial\tilde\delta_1^n} \frac{\partial^n\mP_{\rm G}(\tilde\delta_2)}{\partial{\tilde\delta_2}^n}\,,
\end{align}
which is similar in structure to the Gaussian covariance in $\delta_1$, $\delta_2$ in equation \eqref{eq:gaussian_cov_sum}. This covariance is described by a reduced variance $\sigma_{\bar{\delta}}^2 = \sigma^2 - \overline{\xi}$ and correlation $\xi_{\bar\delta}(d)=\xi(d)-\overline{\xi}$, where the reduction is set by the variance at the subvolume level, $\overline{\xi}= \sigma_{\delta_s}^2$. Additionally, the leading order ($n=1$) term vanishes due to the definition of average of powers of the correlation function \eqref{eq:meanxi}. Those averages of powers of the correlation function $\overline{\xi^n}$ and $\overline{(\xi-\overline\xi)^n}$ compared to powers of the variances $\sigma^{2n}$ and $(\sigma^2-\bar\xi)^n$ control the large-separation expansion of the covariance and follow similar a similar hierarchy to  the averages of powers of the two-point correlation function illustrated in Figure \ref{fig:average_xi_power_scaling}. This is demonstrated explicitly for the weak lensing convergence in \textcite{Uhlemann.etal_2023_ItTakes}.

In the case of rescaling densities, the bias functions also receive a modification. The leading order bias is \parencite{Bernardeau_2022_CovariancesDensity}
\begin{equation}
\hat{b}_1(\hat\rho) = b_1(\hat{\rho}) + 1 + \pdv{\log\mP(\hat\rho)}{\hat\rho}.
\end{equation}
Note that while this looks similar to the bias in the background-subtracted case $\tilde{b}_{1,\rm NG}(\tilde{\delta})$ defined in equation \eqref{eq:bias-mean-sub}, this contribution does not vanish even in the Gaussian case. Predictions for these ingredients, the one-point PDF $\mP(\rho)$ and the sphere bias $b_1(\rho)$ can be obtained from large deviation statistics \parencite{Codis2016MNRAS, Uhlemann2017MNRAS} and in the context of tree models \parencite[Appendix A of][and in Section \ref{sec:covPDF_3D}]{Bernardeau_2022_CovariancesDensity}.

\section{Super-sample covariance of the 3D matter PDF} \label{sec:covPDF_3D}


In this Section we present the theoretical modelling of the super-sample covariance of the three-dimensional matter PDF.  We have seen in Chapters \ref{chap:LDT-intro} and \ref{chap:MG-PDFs} that the matter PDF can be accurately predicted by large deviations theory, and here present a comparison between the super-sample covariance ingredients predicted from LDT and the super-sample covariance measured from separate-universe style simulations. \textcite{Uhlemann.etal_2023_ItTakes} also considers a shifted lognormal model for the matter density field (see Appendix \ref{app:sec:lognormal} for the model used) which can also be used to obtain estimates of the PDF covariance. We do not present the shifted lognormal model in detail here as it cannot be characterised by its CGF and therefore needs slightly different treatment than the other models considered in this Chapter to obtain the bias functions, but the general principle is unchanged.

\subsection{Prediction of super-sample covariance}

\subsubsection*{Measurements}
 The covariance measured from simulated boxes with fixed mean density (due to a fixed particle number) $\bar\rho$ cannot capture the super-sample effect created by having different sample densities $\rho_s$ as would arise naturally from computing the covariance of subboxes within a much larger simulation box. 
To estimate this background density driven super-sample covariance effect for the \textsc{Quijote} simulation suite, we make use of their publicly available separate-universe style `DC' runs emulating a background density contrast $\delta_b=\pm 0.035$ through changed cosmological parameters and simulation snapshot times from the separate universe approach \parencite{Sirko2005ApJ}.
The super-sample covariance between two data vector entries $D_i$ and $D_j$ can be estimated by 
\begin{equation}
\label{eq:cov_SSC_SU}
  \text{cov}^{\rm SSC}_{\rm SU}(D_i,D_j)
  = \sigma_b^2 \frac{\partial D_i}{\partial \delta_b}
               \frac{\partial D_j}{\partial \delta_b},
\end{equation}
where $\sigma_b^2$ is the variance of $\delta_b$ (here just $\delta_b^2$), and the other two
terms encode the linear response of the data vector, which can be determined from the simulations using finite differences, as was done in the Fisher analysis in Chapter \ref{chap:MG-PDFs}. The impact of super-sample covariance on the 3D matter and halo power spectra, bispectra, void size distributions and the halo mass function has been recently studied in \textcite{Bayer2022SSC}. As mentioned therein, in addition to the background density effect captured by the separate universe approach, there can be an additional super-sample effect due by tidal fields affecting higher-order statistics in a non-trivial way. For the PDF, we expect those tidal effects to be negligible due to its intrinsic symmetry and averaging.

\begin{figure}[h!t]
\centering
\includegraphics[scale=1]{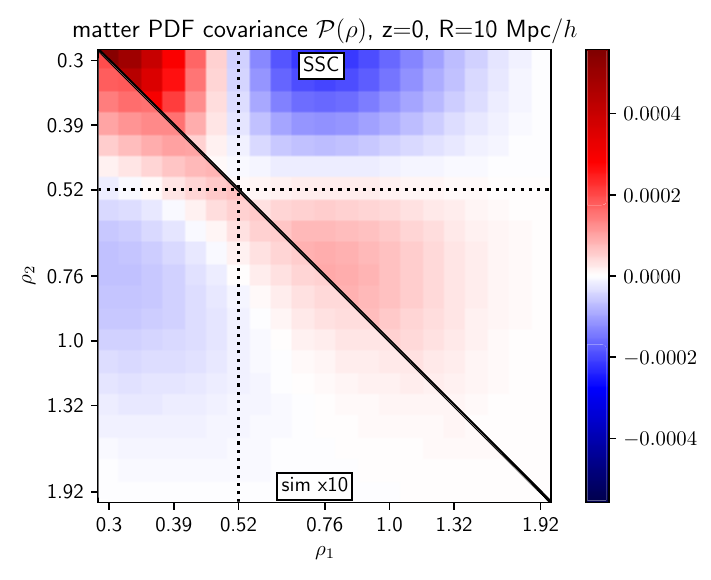}
\caption[Comparison of the super-sample covariance to measured covariance in the matte field.]{Comparison of the PDF covariance matrix contributions for $z=0$, $R=10$ Mpc$/h$ as measured from the $1 \ ($Gpc$/h)^3$\textsc{ Quijote} simulations and the super-sample covariance contribution constructed from the DC mode following equation~\eqref{eq:cov_SSC_SU}.
The upper triangle shows the super-sample covariance contribution while the lower triangle shows the measured covariance from the simulations enhanced by a factor of 10, while the diagonal from the upper panel is masked. The dotted lines indicate the PDF peak location. This Figure was published in \textcite{Uhlemann.etal_2023_ItTakes}. }
\label{fig:Quijote_SSC}
\end{figure}

In the lower triangle of the lower panel of Figure~\ref{fig:Quijote_SSC} we show the covariance matrix of the matter density PDF  showing the correlation between different density bins in the central region of the PDF. We see that, as expected, neighbouring bins are positively correlated, while intermediate underdense and overdense bins are anticorrelated with each other. The strong positive correlation in the bands along the diagonal is induced by cell overlaps as anticipated in Figure \ref{fig:rho_cov_overlap}. This behaviour is also seen in the weak lensing case considered in \textcite{Uhlemann.etal_2023_ItTakes}. 

We illustrate the separate universe super-sample  effect on the PDF covariance in the upper triangle of Figure~\ref{fig:Quijote_SSC}, showing that the super-sample covariance term completely dominates the overall covariance for a volume of $1 \, ($Gpc$/h)^3$ and will still lead to a significant effect for a volume of $\mathcal O(10) \, (\mathrm{Gpc}/h)^3$. 

\subsubsection*{Predictions}
As discussed in Section \ref{sec:relative_quantities}, when considering the joint PDF of normalised densities, the leading order effective bias $\hat{b}_1(\hat{\rho})$ receives a modification compared to the leading order bias $b_1(\rho)$
\begin{equation}\label{eq:rel_density_bias_v2}
\hat b_1(\hat\rho)=b_1(\hat\rho) + 1+\frac{\partial\log \mP(\hat\rho)}{\partial\log \hat\rho}  \,,
\end{equation} 
and this modification does not vanish even in the case of a Gaussian field. Predictions for the ingredients, the one-point PDF $\mP(\rho)$ and the sphere bias $b_1(\rho)$, can be obtained from large deviation statistics \parencite{Codis2016MNRAS,Uhlemann17Kaiser}, and hierarchical  models \parencite[see][for general hierarchical models, Section \ref{sec:minimal-tree-model} for the minimal tree model]{Bernardeau_2022_CovariancesDensity}. They can also be obtained in (shifted) lognormal models \parencite{Uhlemann.etal_2023_ItTakes, Coles1991MNRAS}.

\begin{figure}[h!t]
\centering
\includegraphics[width=0.75\columnwidth]{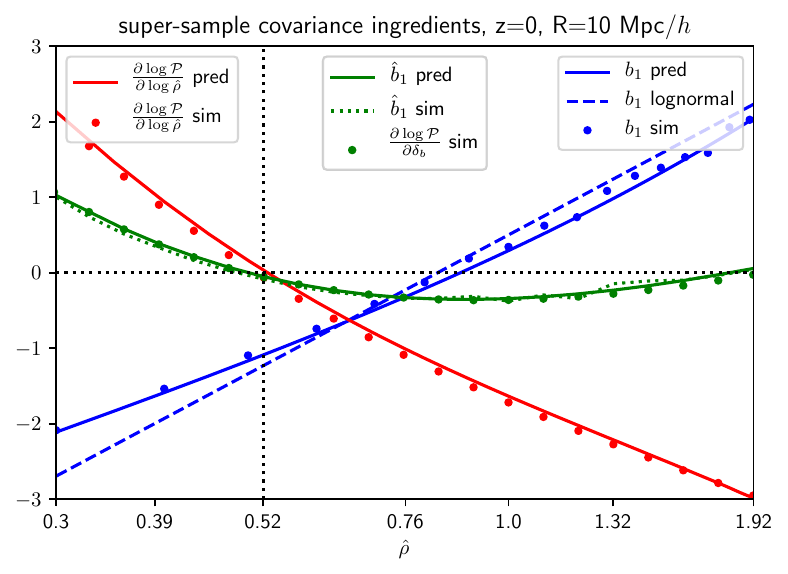}
\caption[Prediction of the ingredients for constructing the covariance.]{Comparison of the theoretically predicted sphere bias from LDT (blue solid), the lognormal approximation (blue dashed) and the measurement in one realisation of the fiducial cosmology of \textsc{Quijote} (blue data points) at radius $R=10$ Mpc$/h$, spaced by $20$ Mpc$/h$ at redshift $z=0$. 
Additionally shown is the logarithmic derivative of the PDF (red) as predicted by LDT (solid) and measured (data points) and the resulting prediction from the effective bias~\eqref{eq:rel_density_bias_v2} using the theoretical ingredients (green solid) or the simulated ones (green dotted). The predictions agrees very well with the measurements from separate universe simulations (green data points). 
The vertical dotted line indicates the peak location of the PDF. The $\rho$-range shown equals the one used in the covariance plots in Figure~\ref{fig:Quijote_SSC}. This Figure was published in \textcite{Uhlemann.etal_2023_ItTakes}.
}
\label{fig:sphere_bias_DCmode}
\end{figure}

\subsubsection*{Comparison}
 In Figure~\ref{fig:sphere_bias_DCmode} we demonstrate that the derivative of the PDF with respect to the background density (green data points) determining the super-sample covariance~\eqref{eq:cov_SSC_SU} is well predicted by the effective bias $\hat b_1$ from equation~\eqref{eq:rel_density_bias_v2} using theoretical ingredients (green solid line) or simulated ones (green dotted line). We also show the $\hat b_1$ ingredients consisting of
the first order sphere bias function $b_1$ as measured from the correlation between neighbouring spheres in one realisation (blue data points) in comparison to the large deviation theory prediction \parencite[blue solid line, following][]{Uhlemann17Kaiser} and the lognormal approximation (blue dashed line), and the logarithmic derivative of the one-point PDF as measured from finite differences (red data points) and predicted from large deviation theory (red solid line). We conclude that the super-sample covariance effect can be robustly predicted by the theory presented here and that both the LDT model and the shifted lognormal model hold promise to compute analytical covariances for the PDF of 3D clustering observables.

Notice that the zero-crossing of $\hat{b}_1$ is very close to the location of the $\hat{\rho}$-PDF peak, indicated on the covariance matrix in Figure \ref{fig:Quijote_SSC} by dotted lines. This explains the general 4-tiled structure of the super-sample covariance, as the leading order contribution to the super-sample covariance depends on the product of the sphere bias functions, as in equation \eqref{eq:cov-SSC-factorisation-biasprod}. This relates more directly to the eigendecomposition of the covariance matrix, discussed in more detail in \textcite{Uhlemann.etal_2023_ItTakes} in the context of the weak-lensing convergence, where this leading order 4-tiled structure can be removed by using background-subtracted quantities, leaving the 9-tiled structure owing to the two zero-crossings of the $b_2$ bias function. 

\subsubsection*{Applications}
The total covariance matrix is a sum of the simulated covariance, $\mathsf{C}_{\rm sim}$ and the super-sample covariance term consisting of a dyadic product, $\mathsf{C}_{\rm SSC}=v_{\rm SSC}v_{\rm SSC}^\mathsf{T}$ with $v_{\rm SSC}=\sqrt{\bar\xi}(\hat b_1\mP)(\hat\rho)$. This sum can be inverted using the Sherman-Morrison formula \parencite{ShermanMorrison1950} to obtain the precision matrix
\begin{equation}
\left(\mathsf{C}_{\rm sim} + \mathsf{C}_{\rm SSC}\right)^{-1} = \mathsf{C}_{\rm sim}^{-1} - \frac{\mathsf{C}^{-1}_{\rm sim} \mathsf{C}_{\rm SSC}\mathsf{C}_{\rm sim}^{-1}}{1 + v_{\rm SSC}^\mathsf{T}\mathsf{C}_{\rm sim}^{-1}v_{\rm SSC}}\,.
\end{equation}
The result can be used to predict the super-sample covariance impact on statistical tests like $\chi^2$ and Fisher contours, which generally leads to a degradation of parameter constraints as detailed in \textcite{Lacasa2019cov} for the case of \textit{Euclid}-like photometric galaxy clustering observables.

\section{The minimal tree model}\label{sec:minimal-tree-model}

Here we present one representative of the hierarchical tree models discussed in Chapter~\ref{chap:structure-formation}, called the ``minimal tree model'' or the ``Rayleigh-L\'{e}vy flight model''. We derive a closed form expression for the bias functions at all orders in this model, and show that its phenomenology is similar to the predictions from large deviations theory seen in the previous Section. 

Recall that a non-Gaussian field is called hierarchical if its $n$-point correlation functions take the form
\begin{equation}
\xi_n(\xx_1, \dots, \xx_n) = \sum_{\mathsf{T} \in \rm trees} Q_n(\mathsf{T}) \prod_{\mathrm{lines}\in \mathsf{T}} \xi(\xx_i, \xx_j)\,,
\end{equation}
as a sum over all possible tree diagrams on $n$ points and a product of two-point correlation functions. Such models guarantee the scaling of average $n$-point functions, such that the reduced cumulants 
\begin{equation}
S_n = \frac{\overline{\xi_n}}{\sigma^{2(n-1)}}\,,
\end{equation}
are simply numbers. The tree models are a subclass of hierarchical models where the $Q_n$ parameters can be computed locally as products over vertices
\begin{equation}
Q_n(t) = \prod_{\mathrm{vertices}\in \mathsf{T}} \nu_p\,,
\end{equation}
where $\nu_p$ is the vertex factor associated with a vertex with $p$ incoming lines (completed with $\nu_0=\nu_1=1$). In this picture, the Gaussian field we have already discussed corresponds to a hierarchical model with $\nu_{n\geq 2}=0$. The \emph{minimal tree model}  we consider here is the model in which $\nu_2$ alone is non-zero (e.g. only ``snake type diagrams'' as in Figure \ref{fig:heirarchical_4point} contribute) and the value of $\nu_2=1/2$ is fixed.\footnote{It has been shown in \textcite{Balian.Schaeffer_1989_ScaleinvariantMatter} from the behaviour of the void probability function that the only possible value for $\nu_2$ in such hierarchical models is $\nu_2=1/2$. Consideration of nearest neighbour statistics restricts the value of $Q=\nu_2>1/3$ \parencite[\S 62 of ][]{Peebles_1980_LargescaleStructure}. The value of $Q=\nu_2=1/2$ also saturates the bound placed from considering that clusters should cluster at least as strongly on galaxies from \textcite{Hamilton.Gott_1988_ClusterClusterCorrelations}. }  General hierarchical models and this minimal tree model in particular are discussed in further detail in \textcite{Bernardeau_2022_CovariancesDensity, Bernardeau2024arXiv}. The appeal of the minimal tree model is as a non-Gaussian model for $n$-point correlation functions and joint PDFs which can be characterised by its CGF (unlike lognormal models for example) for which analytic results can be obtained at all orders.

\begin{figure}[p]
\centering
\includegraphics[width=0.75\columnwidth]{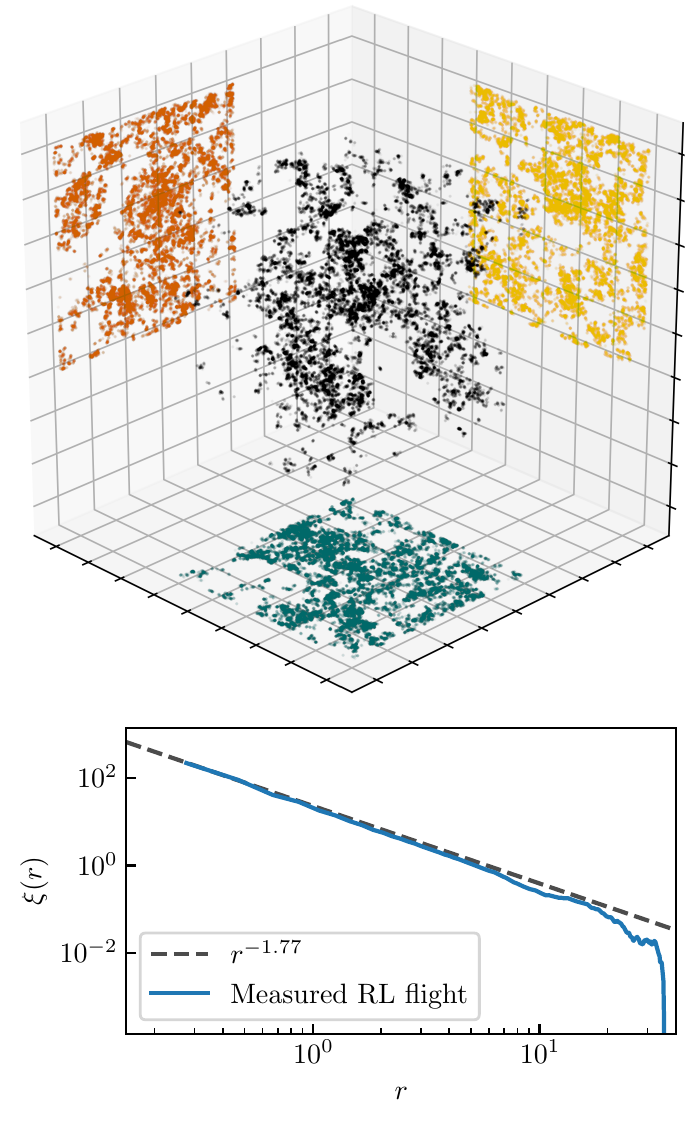}
\caption[Realisation and correlation function of a Rayleigh-L\'{e}vy flight.]{(Upper panel) A realisation of a Rayleigh-L\'{e}vy flight with $2.5\times 10^5$ steps and periodic boundary conditions in a 3D box, together with the 2D projections. The box side length is set to 100, and small-scale regularisation length $\ell_0 = 0.03$. The slope parameter is $\alpha=1.23$ such that the large scale correlation function is $\xi(r)\propto r^{-1.77}$, shown in the lower panel, which well approximates the two-point correlation of galaxy clustering \parencite{Peebles_1980_LargescaleStructure}. (Lower panel) The correlation function is calculated based on $5\times 10^5$ steps in this process. Produced using \texttt{MiSTree} \parencite{Naidoo2019}.}
\label{fig:RL-flight}
\end{figure}

An alternative perspective on this minimal tree model is as the continuous limit of a particular random walk procedure, referred to as a Rayleigh-L\'{e}vy flight. This is convenient as it provides a numerical procedure by which realisations of a field with these specific statistics can be generated. Such a process was first introduced in the cosmological context in \textcite{Peebles_1980_LargescaleStructure}, and corresponds to a random walk, placing a point at the end of each step of the walk. This random walk takes steps in a direction chosen isotropically, with step lengths drawn from a distribution with a power law cumulative distribution function 
\begin{equation}
\mathrm{CDF}(\ell) = \begin{cases} 0 & \ell < \ell_0 \\
1-\left(\frac{\ell_0}{\ell}\right)^\alpha & \ell \geq \ell_0,\end{cases}
\end{equation}
where $\ell_0$ acts as a small-scale regularisation parameter. In $D$-dimensions, this procedure results in a field with a two-point correlation function on large scales (for $k \ell_0 \ll 1$)
\begin{equation}
\xi(r=\abs{\xx_1-\xx_2}) \propto r^{\alpha-D}\ell_0^{-\alpha} / n\,,
\end{equation}
where $n$ is the number density of points in the sample.\footnote{Note that this form of the correlation function no longer holds in the case of periodic boundary conditions, in which case the two-point correlation function is no longer isotropic and receives additional corrections, see \textcite{Bernardeau_2022_CovariancesDensity} for details.} A realisation of a 3D Rayleigh-L\'{e}vy flight is shown in Figure~\ref{fig:RL-flight}, where it is clear that the resulting points are correlated in structure, with the power-law two-point correlation function shown.

Hierarchical models are entirely characterised by their two-point correlation function $\xi(\xx_1,\xx_2)$, and their vertex generating function
\begin{equation}
\zeta(\tau) = \sum_n \frac{\nu_n}{n!}\tau^n,
\end{equation}
which determines the vertices/topology factors for the higher order correlation functions. Here we use the notation $\zeta(\tau)$ for the vertex generating function, rather than $\mathcal{F}(\delta_{\rm L})$ as in Chapters \ref{chap:structure-formation}, \ref{chap:LDT-intro}, \& \ref{chap:MG-PDFs}, to be consistent with previous papers on hierarchical models e.g. \textcite{ Balian.Schaeffer_1989_ScaleinvariantMatter, Bernardeau.Schaeffer_1999_HaloCorrelations}. We also note that while many of the relationship between the CGF and the vertex generating function which come out of studying hierarchical models look similar to the results in LDT, they are formally different, with hierarchical model equations arising mostly out of pure combinatorics. For this reason as well, we use $\zeta$ and $\tau$ rather than $\mathcal{F}$ and $\delta_{\rm L}$ to represent the vertex generating function for this Chapter only. 

The minimal tree model corresponds to the vertex generating function
\begin{equation}\label{eq:minimal-tree-zeta}
\zeta_{t}(\tau) = 1 + \tau + \frac{1}{4}\tau^2 = \left(1+\frac{\tau}{2}\right)^2,
\end{equation}
as it has only $\nu_{n<3}$ non-zero.  A Gaussian field, such as that presented in Section \ref{sec:gaussian-model}, would correspond to the vertex generating function $\zeta_{\rm G}(\tau) = 1 + \tau$, being entirely determined by its two-point statistics. From this generating function, it is possible to build the $n$-point cumulant generating function based on combinatorial consistency relations, making use of the intermediate function $\tau(\xx)$ as the solution of a consistency equation \parencite{Bernardeau.Schaeffer_1999_HaloCorrelations}
\begin{equation}
\tau(\xx) = \sum_j \lambda_j \int_{V_j} \frac{\dd{\xx'}}{V_j}\xi(\xx,\xx') \zeta(\tau(\xx')),
\end{equation}
where $V_j$ represents the volume of the $j^{\rm th}$ spherical cell. Similar relations can be established for more general cell profiles, such as from Gaussian smoothing, see e.g. \textcite{Bernardeau2024arXiv}. As the defining relationship for $\tau(\xx)$ has a $\tau$ on both sides, including inside the integral, this relationship can only be solved exactly for a few simple hierarchical models. In an informal sense, we can interpret the auxiliary variable $\tau(\xx)$ as something like a density field, the integral looks something like the average value of the two-point correlation, weighted by the vertex function $\zeta$ (which in the large deviations theory picture would be the non-linear density obtained from spherical collapse). More rigorously, the value $\tau(\xx)$ can be interpreted as the sum of all tree diagrams connected to the point $\xx$ \parencite{Bernardeau.Schaeffer_1999_HaloCorrelations}.

From this intermediate function $\tau(\xx)$, the cumulant generating function can be expressed exactly as
\begin{equation}
\phi(\lambda_1, \dots, \lambda_n) = \sum_j \lambda_j \int_{V_j}\frac{\dd{\xx}}{V_j} \zeta(\tau(\xx)) - \frac{1}{2}\sum_j \lambda_j \int_{V_j} \frac{\dd{\xx}}{V_j} \tau(\xx) \zeta'(\tau(\xx)).
\end{equation}
The above result is exact for hierarchical models as a result of the combinatorics on the tree diagrams. Instead of this exact result, \textcite{Bernardeau.Schaeffer_1999_HaloCorrelations} find a mean-field approximation which assumes $\tau(\xx)$ is constant within each cell to be highly accurate. In this case the system of equations for hierarchical models simplifies
\begin{subequations}
\begin{align}
\tau_i &= \sum_j \lambda_j \xi_{ij} \zeta'(\tau_j) \label{eq:consistency-mean-field} \,,\\
\xi_{ij} &= \int_{V_i}\int_{V_j} \frac{\dd{\xx}_i}{V_i}\frac{\dd{\xx}_j}{V_j} \xi(\xx_i, \xx_j) \,,\\
\phi(\lambda_1, \dots, \lambda_n) &= \sum_j \lambda_j \left[\zeta(\tau_j) - \frac{1}{2}\tau_j \zeta'(\tau_j)\right]. 
\end{align}
\end{subequations}
Notice here that the  $\xi_{ij}$ are the average value of the two-point correlation function between points in cell $i$ and points in cell $j$. From these relationships, one can derive the joint PDF of $n$-cells. For our purposes, we need only the 1- and 2-cell CGFs to predict the covariance of the one-point PDF  \parencite[extending this method to 3 variables allows for the treatment of background-subtracted and relative quantities, but we do not present those results in detail here, refer to][]{Bernardeau_2022_CovariancesDensity}.

\subsection{The one- and two-point CGF in the minimal tree model}

Using this framework with the minimal tree model, we could obtain the $n$-cell CGF in the mean field approximation, as the implicit equation defining $\tau$ \eqref{eq:consistency-mean-field} is linear and can be solved by matrix inversion, since $\zeta(\tau)$ is quadratic for the minimal tree model. For our purposes however, we focus only on the 1-cell and 2-cell CGF, as that is all that is needed for the covariance of the matter PDF.

\begin{figure}[h!t]
\centering
\includegraphics[width=\columnwidth]{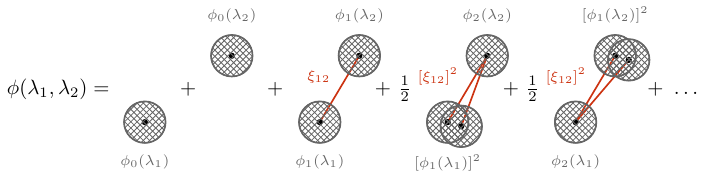}
\caption[Diagrammatic representation of the two-cell CGF expansion.]{Diagrammatic representation of the bias functions. The function $\phi_0$ is the generating function of all trees within one cell, represented with the hatched circle. The higher order functions $\phi_n$ generate all trees in one cell with $n$ external lines.  }
\label{fig:2-point-cgf-diagram}
\end{figure}

The two-point CGF can be expanded to second order in the correlation function between cells, $\xi$, in the mean field approximation in terms of the functions $\phi_0$, $\phi_1$, $\phi_2$ as in the general non-Gaussian case in equation \eqref{eq:two-point-CGF-expansion}. Using the mean field consistency equations, we can calculate $\phi_n$ in the minimal tree model, and the associated bias functions to all orders. In general for hierarchical models, including the minimal tree model, the functions $\phi_n(\lambda)$ can be interpreted as the generating function for all trees within one cell and $n$-external lines. Thus $\phi_0(\lambda)$ is the generating function for trees in one cell, simply the marginal one-point PDFs. This allows the expansion of the two-point CGF to be represented as the diagrammatic series shown in Figure \ref{fig:2-point-cgf-diagram}. This bears a resemblance to the 1-halo and 2-halo terms which appear in calculating statistics from the halo model, see e.g. \textcite{Asgari2023OJAp} for a recent review.

We begin by calculating the one-point CGF in the minimal tree model. The intermediate function $\tau_1$ in a cell is simply defined by
\begin{equation}
\tau_1 = \lambda_1 \xi_{11}  \left(1 + \frac{1}{2}\tau_1\right) \Rightarrow \tau_1(\lambda) = \frac{\lambda \sigma^2}{1-\lambda\sigma^2/2}\,,
\end{equation}
where $\xi_{11}=\sigma^2$ is the average 2-point correlation function within one cell. The one-point CGF in the minimal tree model is then (dropping the 1 subscript on quantities)
\begin{align}
\phi_{0,t}(\lambda) &= \lambda\left[ \left(1 + \frac{\tau}{2}\right)^2-\frac{1}{2}\tau\left(1 + \frac{\tau}{2}\right) \right] \nonumber \\
&= \lambda \left(1 + \frac{1}{2} \tau \right) \nonumber \\
&= \frac{\lambda}{1-\lambda \sigma^2 /2} \simeq \lambda + \frac{\sigma^2}{2}\lambda^2 + \frac{\sigma^4}{4}\lambda^3 + \order{\sigma^6}.
\end{align}
In the final line, we can expand the one-point CGF to read off the cumulants of the minimal tree model. The coefficient of the $\lambda^3$ term is $k_3/3!$ where $k_3$ is the third cumulant, and thus the reduced skewness $S_3=k_3/\sigma^4$ in the minimal tree model is $S_3^{t}=3/2$ (compared to the 34/7 value from tree order SPT as derived in Section~\ref{sec:tree-order-skewness-SPT}).

In general, one can write down the two-point CGF following this same procedure, resulting in
\begin{align}\label{eq:minimal_tree_CGF_2pt}
\phi_{t}(\lambda_1, \lambda_2) &= \frac{\lambda_1 + \lambda_2 + (\xi-\sigma^2)\lambda_1\lambda_2}{1 - (\lambda_1+\lambda_2)\sigma^2/2 + \lambda_1\lambda_2 (\xi^2-\sigma^4)/4} \nonumber \\ 
&=\frac{\phi_{0,t}(\lambda_1) + \phi_{0,t}(\lambda_2) + \xi \phi_{0,t}(\lambda_1)\phi_{0,t}(\lambda_2)}{1-\xi^2 \phi_{0,t}(\lambda_1)\phi_{0,t}(\lambda_2)/4}.
\end{align}
This two-cell CGF can be expressed entirely in terms of the one-cell CGF $\phi_{0,t}$, as written in the second line above. Expanding this two-cell CGF in powers of $\xi$, we can read off the $\phi_1$ and $\phi_2$ functions which will determine the bias functions, in precisely the same way we did in the general non-Gaussian case in Section \ref{sec:non-Gauss-NLO-cov}. 

As the minimal tree model's vertex generating function is very simple, being only quadratic in $\tau$, the only non-zero $\phi_n$ are
\begin{align}\label{eq:CGF_minimaltree}
\phi_{0,t}(\lambda) = \frac{\lambda}{1-\lambda\sigma^2/2}, \quad \phi_{1,t}(\lambda) = \phi_{0,t}(\lambda), \quad \phi_{2,t}(\lambda) = \frac{1}{2}\phi_{1,t}(\lambda) = \frac{1}{2}\phi_{0,t}(\lambda),
\end{align}
with $\phi_{n\geq 3}=0$, as in the minimal tree model there are no trees which can generate cells with 3 or more external lines, since $\nu_{n\geq 3}=0$. As all of these functions are proportional to $\phi_{0,t}$ for notational simplicity we will simply call the one-cell CGF $\phi_{0,t}(\lambda)=\phi_t(\lambda)$.

\subsection{One-point PDF of the minimal tree model}

Throughout the discussion of densities here we assume that the overall mean density is scaled to be 1. If we wished to describe the relative density in a patch of the survey with a mean $\rho_s$ instead  the one-point CGF is described by $\phi_t(\bar{\rho}\lambda;\sigma^2) = \rho_s\phi_t(\lambda;\rho_s\sigma^2)$ instead of $\phi_t(\lambda;\sigma^2)$.

\begin{figure}[h!t]
\centering
\includegraphics[scale=1]{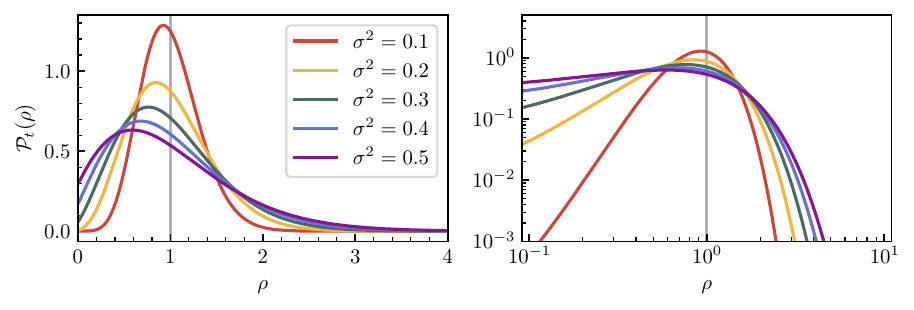}
\caption[The one-point PDF predicted by the minimal tree model.]{The one-point PDF of the matter density in the minimal tree model, as given by equation \eqref{eq:MTM-PDF}, on linear (left) and log-log (right) scales.}
\label{fig:minimal_tree_pdfs}
\end{figure}

With the full one-point CGF in analytic form, the full one-point PDF can be derived for the minimal tree model in closed form and is given by\footnote{Note that \textcite{Bernardeau_2022_CovariancesDensity} denotes the variance as $\bar{\xi}$, as the average correlation function within a cell, as opposed to our choice of $\sigma^2$. This should not be confused with our average correlation within a patch, $\bar{\xi}$ defined in equation \eqref{eq:mean-correlation}.} \parencite[][see Appendix \ref{app:aux-laplace-for-min-tree} for a derivation. Appendix \ref{app:sec:hyp-geo} defines the hypergeometric function $_0F_1$.]{Bernardeau_2022_CovariancesDensity}
\begin{equation}\label{eq:MTM-PDF}
\mP_t(\rho) = e^{-2/\sigma^2}\delta_{\rm D}(\rho) + \frac{4}{\sigma^4}\exp\left[-\frac{2}{\sigma^2}(1+\rho)\right] \ _0F_1\left(2,\frac{4\rho}{\sigma^4}\right).
\end{equation}
This PDF for different values of the variance is shown in Figure~\ref{fig:minimal_tree_pdfs}. Notice that this particular model has a non-zero void probability distribution (VPF) captured by the Dirac delta in the PDF above, highlighting that even in the continuous limit of the random walk  there will be regions of finite size which remain empty.

\subsection{Behaviour of the minimal tree bias functions}

Before deriving the bias functions to all orders, we first present the key behaviours of the leading and next-to-leading order bias functions in the minimal tree model. 

From the two-point CGF, the leading order bias functions $b_1$ and $b_2$ can be calculated according to equations \eqref{eq:b1_def}--\eqref{eq:q1_def}. Since $\phi_t=\phi_{1,t}=2\phi_{2,t}$, the only integrals which need be integrated are the exponential part of the inverse Laplace transform against powers of $\phi_t$. The lowest order bias functions for the minimal tree model are
\begin{subequations}
\label{eq:bias_tree}
\begin{align}
    \label{eq:b1_tree}
    b_{1, t}(\rho)&=\frac{_0F_1\left(1,\frac{4\rho}{\sigma^4}\right)}{_0F_1\left(2,\frac{4\rho}{\sigma^4}\right)}-\frac{2}{\sigma^2}\,,\\
\notag    b_{2, t}(\rho)&=\frac{4[\rho-1-\sigma^2 b_1(\rho)]}{\sigma^4}= \frac{4}{\sigma^2}\left[\frac{\rho-1}{\sigma^2}- b_1(\rho)\right], \\
\label{eq:b2_tree}
&= \frac{4}{\sigma^2}\left[\frac{\rho+1}{\sigma^2}-\frac{_0F_1\left(1,\frac{4\rho}{\sigma^4}\right)}{_0F_1\left(2,\frac{4\rho}{\sigma^4}\right)}\right],\\
\label{eq:q1_tree}
q_{1, t}(\rho)&=b_{1, t}(\rho)/2\,.
\end{align}
\end{subequations}
In the $\sigma^2\to 0$ limit, these expressions tend to the following, which can be further simplified for small density contrasts $\delta = \rho - 1$ and reduce to the Gaussian results
\begin{subequations}
\label{eq:bias_tree_lim}
\begin{align}
    \label{eq:b1_tree_lim}
    b_{1,t}(\rho)& \stackrel{\sigma\rightarrow 0}{\longrightarrow} \frac{2(\sqrt{\rho}-1)}{\sigma^2}\stackrel{\delta\rightarrow 0}{\longrightarrow} \frac{\delta}{\sigma^2}=b_{1,\rm G}(\delta)
    \,,\\
\label{eq:b2_tree_lim}
    b_{2, t}(\rho)& \stackrel{\sigma\rightarrow 0}{\longrightarrow} \frac{[2(\sqrt{\rho}-1)]^2-\sigma^2}{\sigma^4}
    \stackrel{\delta\rightarrow 0}{\longrightarrow} \frac{\delta^2-\sigma^2}{\sigma^4}=b_{2,\rm G}(\delta)\,.
\end{align}
\end{subequations}

\begin{figure}[h!t]
\centering
\includegraphics[scale=1]{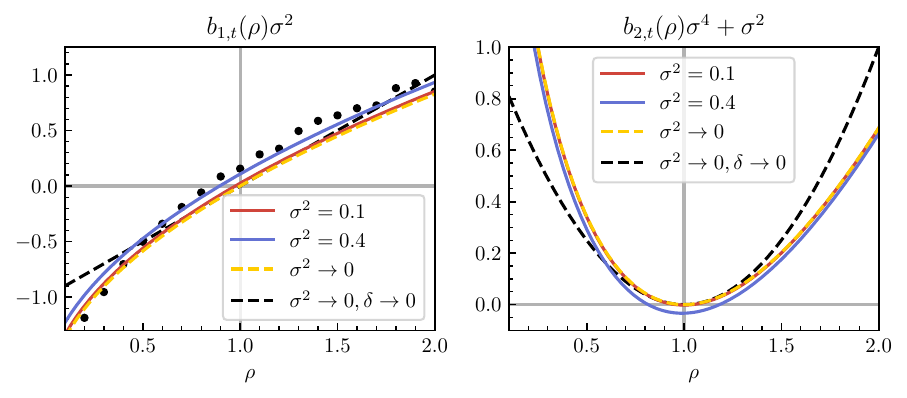}
\caption[Bias functions predicted by the minimal tree model.]{A comparison of the first two leading order bias functions from the minimal tree model~\eqref{eq:bias_tree}, $b_{1, t}$ (left panel) and $b_{2, t}$ (right panel), for different variances (red and blue lines) approaching the small-variance limit~\eqref{eq:bias_tree_lim} (yellow dashed) and the Gaussian expectation (black dashed). In the left panel we also show the measurements for $b_1$ from the \textsc{Quijote} simulations (data points).}
\label{fig:b1b2_treemodel}
\end{figure}

The two stages of this limiting behaviour are illustrated in Figure~\ref{fig:b1b2_treemodel} two values of  $\sigma^2$ (blue and red lines, corresponding to the same PDFs shown in Figure \ref{fig:minimal_tree_pdfs}) showing fast approach to the small variance limit (yellow dashed) which reproduces the Gaussian result (black dashed) for small density contrast. 
The quadratic $q_1$ term in the covariance~\eqref{eq:cov_nonG}
can be combined with the $b_1$ term, while the $q_1$ term adding to $b_{2,\rm t}$ in \eqref{eq:cov_nonG} does not change much in the case of a small variance $\sigma^2\ll 1$ and an  intermediate density regime. The behaviour of these bias functions when compared to the measured sphere bias from the \textsc{Quijote} simulations (data points) is similar to the behaviour seen in comparing the measured results to the bias predicted by large deviations theory, as shown in Figure \ref{fig:sphere_bias_DCmode}, demonstrating that this simple model has the capability of capturing key clustering features.

\subsection{Closed form bias functions for minimal tree model}

We now take advantage of the fact that the $n^{\rm th}$ order contributions to the two-point CGF $\phi_n$ in the minimal tree model have such a simple structure to compute the $n^{\rm th}$ order bias functions to arbitrary order.
Define an auxiliary version of the Laplace transform of the CGF as the integral $I_j(\rho)$:
\begin{equation}
    I_j(\rho) = \int \frac{\dd{\lambda}}{2\pi i}\exp\left[{-\lambda\rho + j\phi(\lambda)}\right].
\end{equation}
The PDF is then the value $I_1(\rho)$. In the minimal tree model, the bias functions are predicted via
\begin{equation}
    b_{n,\rm t}(\rho)\mP_{\rm t}(\rho) =  \int \frac{\dd{\lambda}}{2\pi i}\phi_{0,\rm t}^n(\lambda) \exp\left[{-\lambda\rho + \phi_{0,\rm t}(\lambda)}\right],
\end{equation}
where the one-point CGF $\varphi_{0,\rm t}(\lambda)$ is given by equation~\eqref{eq:CGF_minimaltree}.
The $n^{\rm th}$ bias function equation can also be written as 
\begin{equation}
    b_{n,\rm t}(\rho)\mP_{\rm t}(\rho) = \dv[n]{I_j(\rho)}{j}\eval_{j=1}.
\end{equation}
and for the minimal tree model the $I_j(\rho)$ can be obtained as (see Appendix \ref{app:aux-laplace-for-min-tree} for full derivation)
\begin{equation}
\label{eqn:Ij}
    I_j(\rho) = e^{-2/\sigma^2}\delta_{\rm D}(\rho) + \frac{4j}{\sigma^4}\exp\left[{-\frac{2}{\sigma^2}(j+\rho)}\right]\hypgeo{2, \frac{4j\rho}{\sigma^4}}.
\end{equation}
We can therefore obtain a closed form equation for $b_n(\rho)$ by differentiation of equation \eqref{eqn:Ij}. We will neglect the contribution from the discontinuous term given by the Dirac delta for the remainder of this Chapter. The $n^{\rm th}$ derivative of $I_j$ is given by
\begin{equation}
    \dv[n]{I_j}{j} = \sum_{k=0}^n \sum_{\ell = 0}^k \binom{n}{k}\binom{k}{\ell} f^{(n-k)}(j)g^{(k-\ell)}(j)h^{(\ell)}(j)\,,
\end{equation}
where we've split $I_j$ into three functions, $f,g,h$, which we can differentiate separately as follows
\begin{subequations}
\begin{align}
f(j) &= \frac{4j}{\sigma^4}\,,\qquad
g(j) = \exp\left[-\frac{2}{\sigma^2}(j+\rho)\right],\\
\dv[n]{f}{j} &= \frac{4}{\sigma^4}(j\delta_{n,0}+\delta_{n,1}),\\
\dv[n]{g}{j} &= \left(\frac{-2}{\sigma^2}\right)^n \exp\left[-\frac{2}{\sigma^2}(j+\rho)\right].
\end{align}
\end{subequations}
For the hypergeometric function we make use of the fact that \begin{equation}
\dv{}{z}\, \hypgeo{n, z} = \hypgeo{n+1,z}/n,
\end{equation}
where we have defined $z = 4\rho j /\sigma^4$ so that $\del_j = \frac{4\rho}{\sigma^4}\del_z$ and
\begin{subequations}
\begin{align}
    h(j) &= \hypgeo{2, \frac{4\rho j }{\sigma^4}}, \\
    \dv[n]{h}{j} &= \left(\frac{4\rho}{\sigma^4}\right)^n \frac{1}{(n+1)!} \hypgeo{2+n, \frac{4\rho j}{\sigma^4}}.
\end{align}
\end{subequations}
Putting this into the full sum we obtain (using the Kronecker $\delta$s to collapse the $k$ sum, as the only surviving terms are when $n=k$ and when $n-k=1$).
\begin{align}
    \notag\dv[n]{I_j}{j} = \frac{4}{\sigma^4} e^{-\frac{2}{\sigma^2}(j+\rho)}\sum_{k=0}^n \sum_{\ell = 0}^k &\binom{n}{k}\binom{k}{\ell}(j\delta_{n-k,0} + \delta_{n-k,1})\\
    &\times\left(\frac{-2}{\sigma^2}\right)^{k-\ell} \left(\frac{4\rho}{\sigma^4}\right)^k \frac{\hypgeo{k+2,\frac{4j\rho}{\sigma^4}}}{(k+1)!} \,.
\end{align}
Now, we separate this sum into the case where $k=n$ and when $k=n-1$,
factor out common terms only depending on $n$ and use the binomial theorem to do these sums as simply $(1-2/\sigma^2)^{n}$ and $(1-2/\sigma^2)^{n-1}$ respectively. This equation is then the form of $b_{n, t}(\rho)\mP_{ t}(\rho)$ when $j=1$. 
Dividing through by $\mP_{ t}(\rho)$ from equation~\eqref{eq:MTM-PDF} everywhere and setting $j=1$ gives us the functional form of the $n^{\rm th}$ order bias function for the minimal tree model as
\begin{align}
\label{eq:bntree}
    b_{n,\rm t}(\rho) &= \frac{1}{n!}\left[\left(\frac{4\rho}{\sigma^4}\right)\left(1-\frac{2}{\sigma^2}\right)\right]^{n-1} \left[ \frac{4\rho\left(1-\frac{2}{\sigma^2}\right)}{\sigma^4(n+1)}\frac{\hypgeo{n+2,\frac{4\rho}{\sigma^4}}}{\hypgeo{2, \frac{4\rho}{\sigma^4}}}+ \frac{\hypgeo{n+1,\frac{4\rho}{\sigma^4}}}{\hypgeo{2,\frac{4\rho}{\sigma^4}}}\right].
\end{align}
Note that, the hypergeometrics can always be re-expanded down in terms of $\hypgeo{1, z}$ and $\hypgeo{2,z}$ using the identity
\begin{equation}
    \hypgeo{n-1,z} - \hypgeo{n,z} = \frac{z}{n(n-1)}\hypgeo{n+1,z}.
\end{equation}
A recurrence relation for these bias functions were noted in \textcite{Bernardeau_2022_CovariancesDensity}. With these closed form bias functions the covariance matrix in the minimal tree model can be calculated at arbitrary order.

\section{Conclusions}

\subsection*{Summary}
In this Chapter we have shown how to compute covariances for the one-point PDF of cosmic fields on mildly non-linear scales with a particular emphasis on the matter density field. These predictions rely on the two-point joint PDF, which is integrated over all separations according to the distribution of distances in a survey area. We  present results to all orders for a Gaussian field, recovering well known features such as the Kaiser bias which describes the leading order density dependence of two-point clustering. We then present leading order and next-to-leading order results for general non-Gaussian models that can be characterised by their cumulant generating function. The super-sample covariance, which captures the field correlations due to modes larger than the survey or simulation size, is then calculated from the \textsc{Quijote} suite of separate universe style simulations. The theoretical ingredients to predict this super-sample covariance, the one-point PDF, $\mP(\rho)$, and the effective sphere bias, $\hat{b}_1$, are both shown to be well predicted by large deviations theory, allowing for the possibility to complement covariance measurements from finite-size boxes with analytic covariance predictions. Finally, the bias functions to arbitrary order are calculated in a simple hierarchical clustering model, called the minimal tree model. This model allows the covariance to be calculated to arbitrary order, and the leading order biasing functions are shown to provide similar behaviour to the large deviations theory and simulations, making it an attractive simple model which captures essential non-Gaussian behaviour.

\subsection*{Outlook}

We expect the formalism presented here can be straightforwardly adapted to describe more directly observable quantities such as spectroscopic galaxy clustering and other biased tracers \parencite{Friedrich_2021} or to weak-lensing observables such as the CMB weak-lensing convergence \parencite{Barthelemy2020postBorn} and the aperture mass function \parencite{Barthelemy2021MNRAS}. Modelling of covariances based on these observables is essential for accurate parameter estimation via the precision matrix (inverse covariance matrix), which enters in $\chi^2$ tests and Fisher forecasts \parencite{Taylor2013}. Simulations of non-standard cosmologies --- such as modified gravity as in Chapter \ref{chap:MG-PDFs} or wave dark matter as in Chapters \ref{chap:making-dm-waves} \& \ref{chap:how-classical} --- are often limited to only a few realisations or small box sizes due to computational expense, limiting the ability to use simulation informed covariances. Thus, lessons from analytic models can prove a valuable starting point for improving covariance estimates from simulations, even in regimes where the model itself becomes inaccurate \parencite{Friedrich2017precisionmatrixexp}. In particular, models where the covariance can be calculated to arbitrary order, such as the minimal tree model presented here, could provide useful starting points as they can avoid issues of creating non-invertible matrices due to using only low order approximations.

%% file: text/chapter7-wave-dynamics-in-1D.tex

\chapter{Dark matter dynamics from wave interference}\label{chap:making-dm-waves}

\minitoc

This Chapter is based on \cite{Gough.Uhlemann_2022_MakingDark} and presents a detailed analysis of a wave model of dark matter. We introduce Propagator Perturbation Theory (PPT) established in \cite{Uhlemann2019} as the framework used this Chapter and Chapter \ref{chap:how-classical}. This Chapter demonstrates that interference patterns in this wave model can be ``unwoven'' into classical trajectories which correspond to multistreaming regions. Features beyond those of a perfect fluid, such as vorticity and velocity dispersion, are also shown to be encoded in the wave interference. Regions where classical densities diverge, known as caustics, are shown to have universal scaling behaviour in the context of catastrophe theory.

\section{Introduction}

A field-level description of the large-scale structure of our Universe is one of the key challenges for upcoming galaxy surveys such as \textit{Euclid} \parencite{Euclid_mission}, Rubin Observatory LSST \parencite{LSST_mission}, and DESI \parencite{DESI_mission}, and would be a powerful tool for parametrising the cosmology-dependence of observables. Within the standard $\Lambda$CDM model, the cold dark matter (CDM) component of the Universe dominates the dynamics of structure formation. Since running fully-fledged non-linear simulations for cosmological inference is time-consuming and costly, hybrid and analytic approaches to CDM dynamics are  a crucial step on the path to field-level descriptions of the cosmic large-scale structure traced by galaxy clustering and weak lensing. 

A variety of analytic techniques exist for predicting the dynamics of cold dark matter, most based on leveraging the fact that at early times the dark matter is well described as a  perfect fluid. In standard perturbation theory (SPT, see Section \ref{sec:intro-SPT}) the density and velocity are expanded perturbatively about their background values. As it relies on the perturbative smallness of these quantities, SPT struggles to accurately describe the large densities arising from gravitational collapse. Furthermore, as cold dark matter is collisionless, fluid streams cross and create regions of (formally) infinite Eulerian density, called caustics, strongly limiting the region of applicability of this perturbative theory (see Figure \ref{fig:nbody-shellcrossing}). A useful alternative to SPT is Lagrangian perturbation theory \parencite[LPT, see Section \ref{sec:intro-LPT}, ][]{ Zeldovich1970, Buchert1989A&A, Bouchet1992ApJL,  Bouchet1995A&A, Villone2017EPJH}, which takes the displacement from fluid elements' initial positions as its perturbing quantity, avoiding the singular densities from SPT.  However, updating the gravitational potential which displaces particles requires the Eulerian density, which requires mapping between the initial (Lagrangian) and final (Eulerian) positions. A similar conversion is necessary to facilitate a comparison with galaxy survey data, which naturally lives in Eulerian space. 

In this Chapter, we consider an alternative approach to modelling cold dark matter dynamics that combines advantages of Lagrangian and Eulerian frameworks. We introduce a wavefunction $\psi$, to play the role of the dark matter field, and rely on reproducing cold dark matter dynamics in a semiclassical limit, using the propagator formalism established in \cite{Uhlemann2019, Rampf2021MNRAS}. In particular we examine how classical CDM phenomenology is encoded in the features of the wavefunction, and how it can be extracted, using a simple toy model. This toy model makes use of a ``free-particle Schr\"odinger equation'' in 1+1 dimensions \parencite{ColesSpencer2003, ShortColes2006, Uhlemann2019}, which is closely related to the classical Zel'dovich approximation (lowest order LPT). While \cite{Uhlemann2019} derives the higher order behaviour within this propagator formalism, here we focus on dissecting the wavefunction in the simplest case, which already features a rich phenomenology  and facilitates links to a well-studied classical model. Using a wavefunction in this way allows for LPT-like dynamics to be captured with direct access to Eulerian observables like the density and velocity. We consider a fully three-dimensional field produced from this formalism in Chapter \ref{chap:how-classical}.

The use of wavefunctions to model dark matter arises in a variety of other contexts in cosmology, which are physically distinct but conceptually and phenomenologically closely related to the propagator formalism used here. In studying non-relativistic self-gravitating systems, the Schr\"odinger-Poisson equation \parencite[see e.g. ][]{WidrowKaiser1993,Guth.etal_2015_DarkMatter, Marsh_2016_AxionCosmology, Hui2021, Ferreira2021A&ARv, O'Hare2024arXiv}
\begin{subequations} \label{eqn:SP-full-eqns-firsttime}
    \begin{align}
        i\hbar \partial_t \psi &= -\frac{\hbar^2}{2ma^2} \nabla^2 \psi + m V_N \psi\,, \\
        \nabla^2 V_N &= \frac{3 \Omega_m^0 H_0^2}{2}  \frac{\abs{\psi}^2 - 1}{a}.
    \end{align}
\end{subequations}
 is the principle equation of interest. These equations describe the evolution of a classical complex valued wavefunction $\psi$, which describes the dark matter field, with associated density $\abs{\psi}^2$. Here $t$ is coordinate time, spatial derivatives are with respect to comoving coordinates, $V_N$ is the Newtonian gravitational potential, and $m$ is the mass associated with the dark matter particle.

Taken as the true dynamical equations of motion, the Schr\"odinger-Poisson equation \eqref{eqn:SP-full-eqns-firsttime} describes a non-relativistic, scalar dark matter particle with mass $m$, as a limit of the Klein-Gordon equation \parencite{Hui2017} with interesting astrophysical signatures. Such dark matter candidates are often motivated from particle physics including the QCD axion \parencite{Peccei1977PhRvL}, the string axiverse \parencite[e.g.][]{Svrcek2006JHEP, Arvanitaki2010PhRvD}, and other axion-like particles \parencite{Jaeckel2022arXiv}. In the context of cosmology, these candidates often have very small mass scales ($\sim \! 10^{-22} \ \mathrm{eV}/c^2$) so that their wave phenomena are present on astrophysical scales. Though not completely interchangeable terms, such dark matter candidates are referred to as ultra-light axions (ULAs), fuzzy dark matter (FDM), Bose-Einstein condensate (BEC) dark matter or simply wave dark matter ($\psi$DM) in the literature. For a recent review of such systems see e.g. \cite{Niemeyer2020, Hui2021, Ferreira2021A&ARv}.

Alternatively, one can interpret the Schr\"odinger-Poisson equation \eqref{eqn:SP-full-eqns-firsttime} as a theoretical trick to study cold dark matter.  In the original work by Widrow and Kaiser \parencite{WidrowKaiser1993},  they propose solving the Schr\"odinger-Poisson equation as an alternative to $N$-body simulations, leveraging the fact that the phase-space distribution of $\psi$ and classical CDM obey the same equation to $\order{\hbar^2}$. In such cases, we absorb the mass $m$ into the value of $\hbar$, and take $\hbar$ to parametrise how coarsely phase-space is sampled. Within this so-called Schr\"odinger method, wave interference effects were shown to emulate the multistream phenomena of velocity dispersion and vorticity \parencite{Uhlemann2014,Kopp2017,Mocz2018} and criteria for its capability to effectively reproduce classical features were quantified  \parencite{Garny2020,Eberhardt2020}.

Equations of Schr\"odinger-Poisson type can also be derived as an alternate scheme for closing the Vlasov cumulant hierarchy which appears in CDM dynamics (see e.g. \cite{Rampf2021arXiv} for a recent review of the cosmological Vlasov-Poisson equations). If one requires that the cumulant generating function is constructed only from the density and velocity (the first two cumulants) degrees of freedom, then the resulting cumulant generating function takes the form of Schr\"odinger-Poisson \parencite{Uhlemann2018finitelygenerated}. This provides a self consistent closure scheme beyond truncation at the level of a perfect fluid. 

The propagator formalism used in this work is physically distinct from the Schr\"odinger method proposed by Widrow and Kaiser, owing to a difference in coordinates such that the natural momentum variables in the propagator formalism and the true Schr\"odinger-Poisson dynamics are related by a non-canonical transformation. This is discussed in  \cite{Uhlemann2019} and we will discuss it further in Chapter \ref{chap:how-classical} where we compare perturbative results to fuzzy dark matter results more directly. Morally however, both methods use a wavefunction to model cold dark matter and take advantage of formally having uniform resolution in position space\footnote{While in principle the wavefunction method has uniform resolution, practical computational choices might drive changes in resolution. If one only cares about the physics of halos this can be limiting, as computation must be spent resolving the phase in voids \parencite{Schive2014} or making use of a hybrid scheme as in \cite{SchwabeNiemeyer2022}. However, wave models have had success in extracting information from underdense environments such as the Lyman-$\alpha$ forest \parencite{Porqueres_2020}. For our toy model this concern is not relevant, as free evolution of wavefunctions is computationally fast and simple.}, which provides complementarity to $N$-body simulations as originally intended. While the focus of this Chapter always implicitly interprets the wave-mechanical model in the semiclassical approach, one could instead consider the wave effects as physical phenomena in the context of a true wave dark matter. While the long-term dynamics of the free Schr\"odinger equation is different from Schr\"odinger-Poisson, manifestations of wave interference effects created by the onset of multistreaming are universal features. Additionally, perturbative treatments based on the propagator formalism can be valuable for pushing the volumes of wave dark matter simulations from currently around $1$--$10$ Mpc \parencite{Schive2014,Mocz2018,Mina2020,MaySpringel2021,SchwabeNiemeyer2022} to truly cosmological scales.

This Chapter is structured as follows. In Section~\ref{sec:multi-streaming and interference} we present a 1+1D toy model for structure formation and discuss how the phenomenology of multistreaming can be understood and in the context of both classical and quantum systems. This toy model is based on a wavefunction satisfying the free Schr\"odinger equation, and corresponds to the Zel'dovich approximation. In Section~\ref{sec:unweaving_the_wavefunction} we discuss how the classical streaming behaviour can be extracted from wave interference by way of stationary phase analysis. Section~\ref{sec:phase_properties} discusses interference effects in the wave-mechanical model and how these encode information beyond perfect fluid dynamics.  The wavefunction can be separated into an ``average'' part, which describes the bulk fluid behaviour, and a ``hidden'' part, which isolates the sources of vorticity and velocity dispersion after shell crossing. Finally, Section~\ref{sec:catastrophe_theory} connects the wave properties of the wave-mechanical system to catastrophe theory, in particular to diffraction integrals which classify the properties of different singularity types. This provides both justification for studying a simple model, and quantitative results where the stationary phase analysis is insufficient. We conclude in Section~\ref{sec:conclusion}, where we also provide an outlook on the generalisations of our model to more realistic scenarios and potential applications of the splitting presented.

\section{Dark matter multistreaming, caustics, and interference}\label{sec:multi-streaming and interference}

In the following we present the phenomenology associated with multistreaming in classical cold dark matter. We introduce the propagator formalism for our wave-mechanical model, show how the classical features associated with multistreaming and shell crossing appear as wave interference, and how to extract Eulerian observables from the wavefunction. The classical toy model (sine wave collapse) is shown in Figure~\ref{fig:zeldo_phase_sheet} and compared to the wave analogue model in Figure~\ref{fig:free_schrodi_evol}.

\subsection{Coordinate system and units}

In this Chapter we present all results for an Einstein-de Sitter (EdS) universe. The entire discussion is straightforwardly extended to full $\Lambda$CDM.

We write our equations in comoving coordinates $\bm{x} = \bm{r}/a$, where $\bm{r}$ is the physical space coordinate and $a$ is the cosmic scale factor. We take our time variable to be $a$ rather than coordinate time $t$, which allows for a well defined initialisation as $a\to 0$ \parencite{Rampf2015}. We define our peculiar velocity to be $\bm{v} = \dv*{\bm{x}}{a}$, which is related to the total velocity by $\bm{U} = H\bm{r} + Ha^2 \bm{v}$. This allows for the interpretation of the Zel'dovich approximation as ballistic motion in these coordinates, as discussed in Section \ref{sec:intro-zeldovich-approximation}. The standard conjugate momentum $\bm p$ is related to our peculiar velocity by $\bm{p}/m = a^2\dv*{\bm{x}}{t} = a^{3/2}\bm{v}$. For this Chapter we also choose units such that $4\pi G\bar{\rho}_0 = 3/2$ (equivalent to choosing $\Omega_m^0 H_0^2 = 1$), which makes $\dv*{a}{t} = a^{-1/2}$ and simplifies the units in several equations. In Chapter \ref{chap:how-classical} we present these equations in a more general $\Lambda$CDM context when we use them to generate three-dimensional density fields. Unlabelled spatial derivatives are always taken to mean with respect to Eulerian coordinates e.g. $\bm{\nabla} = \bm{\nabla}_{\! \bm{x}}$, but we will occasionally explicitly label Eulerian derivatives to emphasise the space we work in or to avoid confusion.

While the focus of this Chapter is on the 1-dimensional toy model presented in the following Section, where results would hold in more dimensions they are written in vector notation, to indicate where the discussion readily generalises to more realistic systems. 

\subsection{The Zel'dovich approximation}
The Zel'dovich approximation was introduced in the context of Lagrangian Perturbation Theory in Section \ref{sec:intro-zeldovich-approximation}. We present the relevant equations in our specific units here for convenience.

Lagrangian perturbation theory (LPT) describes how particle (or fluid element) positions in the initial field, $\bm{q}$, map to positions in the final field via a displacement field, $\bm{\xi}(\bm{q})$. That is, the final position of a fluid element, $\bm{x}$, is given by the mapping
\begin{equation}\label{eqn:lagrangian_mapping}
    \bm{x}(\bm{q},a) = \bm{q} + \bm{\xi}(\bm{q},a)\,.
\end{equation}
In this view, we ``flow along with'' the fluid parcels, rather than taking a static set of coordinates to view the fluid as in the Eulerian picture.  


The mapping $\bm{x}(\bm{q},a)$ can become non-injective, taking particles from different initial positions to the same final position, a phenomenon called multistreaming. The first occurrence of multistreaming is called (first) shell crossing. Assuming our fluid describes a collisionless set of particles (as in CDM), these streams of particles flow through each other. The single and multistream regions of space are separated from each other by regions of (formally) infinite Eulerian density called caustics  \parencite[see e.g. Figure \ref{fig:nbody-shellcrossing} and Figure 3 in][]{Rampf2021arXiv}.

On large scales, the Zel'dovich approximation \parencite{Zeldovich1970} corresponding to first order LPT, performs very well at predicting the formation of structures in the cosmic web. The Zel'dovich approximation in an EdS cosmology corresponds to the displacement field
\begin{equation}\label{eqn:zeldo_displacement_field}
    \vb*{\xi}^{\rm ZA}(\bm{q},a) = -a\bm{\nabla}_{\! \bm{q}} \varphi_g^{(\text{ini})}(\bm{q})\,,
\end{equation}
where $\varphi_g^{\rm (ini)}$ is the initial gravitational potential. This form of the displacement field corresponds to constant velocity motion, provided $a$ is taken as the time variable. With this displacement field, the final position of fluid elements is simply given by ballistic motion with a constant velocity
\begin{equation}
\bm{v}^{\rm ZA}(\bm{q})=-\bm{\nabla}_{\! \bm{q}} \varphi_g^{(\text{ini})}(\bm{q})\,,
\end{equation}
set by the initial gravitational field.

\subsection{Multistreaming}
In the following we illustrate the onset of shell crossing and multistreaming in a simple 1-dimensional example in the Zel'dovich approximation.

Formally linearising the fluid variables (discussed in more detail in Section~\ref{sec:observables}) about their background values and evaluating the linearised fluid equations at arbitrarily early times $a\to 0$, it is found that analytic solutions at $a=0$ require that the boundary conditions on the initial density contrast, $\delta^{\rm (ini)}$, and the initial velocity potential, $\phi_v^{\rm (ini)}$, (which determines the velocity through $\bm{v} = \bm{\nabla} \phi_v$) must be tethered to each other \parencite{Brenier2003} 
\begin{equation}\label{eqn:ZA_boundary_conditions}
    \delta^{\rm (ini)} = 0\,, \quad \phi_v^{\rm (ini)} = - \varphi_g^{\rm (ini)}\,.
\end{equation}
These boundary conditions select for the growing-mode solutions and are vorticity free, in accordance with our requirement of a potential velocity \parencite{Rampf2015}.

\begin{figure}[h!t]
    \centering
    \includegraphics[scale=1]{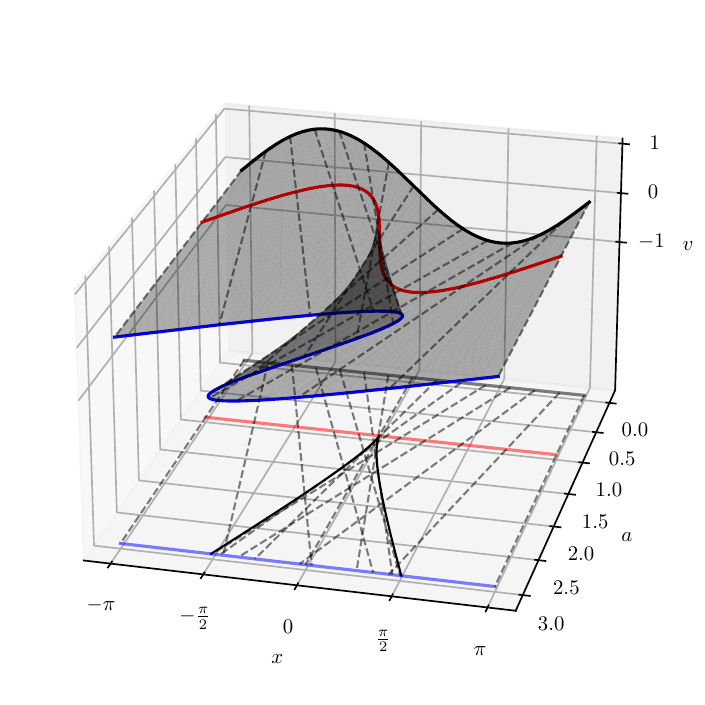}
    \caption[Phase-space evolution of a Fourier mode under the Zel'dovich approximation.]{Phase-space sheet describing the evolution of a Fourier mode under the Zel'dovich approximation described by the displacement mapping $x(q,a) = q - a \sin(q)$.\protect\footnotemark \, For times $a>1$, the phase-space sheet becomes triple valued in velocity. Projecting this phase-space sheet onto the $(x,a)$ plane results in a cusped line (the caustic line) which separates the single and triple stream regions of spacetime (given by equation~\eqref{eqn:shell_cross_region}). The dashed lines show the trajectories of a few individual particles through phase-space. The black, red, and blue curves show the phase-sheet at fixed times $a=0,1,3$. At $a=1$ (red) the phase-sheet becomes vertical, and thereafter it is multivalued.}
    \label{fig:zeldo_phase_sheet}
\end{figure}

Because of these boundary conditions, we can study how a single Fourier mode in the velocity potential evolves under the Zel'dovich approximation. Figure~\ref{fig:zeldo_phase_sheet} shows the evolution of one such mode, corresponding to an initially uniformly dense fluid with initial velocity potential $\phi_v^{\rm (ini)}(q) = \cos(q)$ so that the initial velocity is given by $v^{\rm ZA} = \nabla_{\! q} \phi_v^{\rm (ini)}(q)$. Note that the velocity variable $v$ used in the figure is a phase-space variable, related to the usual canonical momentum $p$ by $v = p/a^{3/2}$ in our units, and should not be confused for the Eulerian velocity field.

While the fluid remains single valued, the density $\rho$ can be obtained by the mapping between Lagrangian (initial) positions and Eulerian positions via conservation of mass $\rho \dd[n]{\bm{x}} = \bar{\rho} \dd[n]{\bm{q}}$. Until the first shell crossing, this can be exactly integrated, giving
\begin{equation}
    \rho = 1/\mathcal{J}\,,
\end{equation}
where $\mathcal{J} = \det (\partial\bm{x}/\partial \bm{q})$ is the Jacobian for the Lagrangian mapping. For a  1-dimensional system with initial velocity potential $\phi_v^{\rm (ini)}(q)$, this gives a density of
\begin{equation}\label{eqn:zeldo_density}
    \rho(x,a) = \frac{1}{\abs{1+a \dv[2]{}{q}\phi_v^{\rm (ini)}(q(x))}}\,.
\end{equation}
For the choice $\phi_v^{\rm (ini)}(q) = \cos(q)$ we see that this density diverges at $a=1$, corresponding to the time of shell crossing. After shell crossing there is a region of final positions, $x$, which are triple valued in velocity. Projecting the phase-space sheet onto the $(x,a)$ plane, the region separating the single-stream and the multistream regions of spacetime forms a cusp shape given by 
\begin{equation}
    (x(q),a(q)) = \left(q-\tan(q), \frac{1}{\cos(q)}\right), \quad q \in \left(-\frac{\pi}{2}, \frac{\pi}{2}\right).
\end{equation}
\footnotetext{An animated version of this Figure and its spacetime projection can be found on \href{https://commons.wikimedia.org/wiki/File:Sine_wave_collapse_in_phase_space.gif}{wikimedia commons}.}
Locations on the caustic correspond to two initial points which cross streams, and to divergences in the density field. The $q$ in this parametric form corresponds to the initial position which maps to $x(q,a)$ via equation~\eqref{eqn:lagrangian_mapping}, and is closest to the origin, which is always in the central region between $\pm \frac{\pi}{2}$. For a fixed time, the region which has shell crossed is given by
\begin{equation}
    \abs{x} < \sqrt{a^2-1}-\arccos(1/a)\,,
    \label{eqn:shell_cross_region}
\end{equation}
for $a>1$. 

We could instead study this system in full phase-space, to avoid the problem of multivaluedness. In Lagrangian coordinates the distribution function in phase-space is given by (in $n$ dimensions)
\begin{equation}\label{eqn:lagrangian_phase_space_dist} 
    f_{\rm L}(\bm{x},\bm{p}) = \int \dd[n]{\bm{q}} \delta_{\rm D}(\bm{x}-\bm{q}-\bm{\xi}(\bm{q})) \delta_{\rm D}\left[\frac{\bm{p}}{a^{3/2}}-\bm{v}^{\rm L}(\bm{q})\right],
\end{equation}
where $\bm{\xi}(\bm{q})$ is the Zel'dovich displacement field in equation~\eqref{eqn:zeldo_displacement_field} and $\bm{v}^{\rm L}(\bm{q}) = \bm{v}(\bm{x}(\bm{q},a),a)$ is the Lagrangian representation of the velocity evaluated at the Eulerian position $\bm{x}(\bm{q},a)$.

\subsection{From multiple streams to wave interference}

Rather than considering a set of classical collisionless particles or a classical fluid, we represent our dark matter field using a complex wavefunction $\psi$. 

Following in the spirit of \cite{Uhlemann2019}, we take the classical action of particles moving with constant velocity to build a propagator for our wavefunction. The classical action for a free fluid particle beginning at position $\bm{q}$ and evolving to position $\bm{x}$ at time $a$ is
\begin{equation}\label{eqn:free_action}
    S_0(\bm{q};\bm{x},a) = \frac{(\bm{q}-\bm{x})^2}{2a}\,.
\end{equation}
The associated transition amplitude for this action is the standard exponential of this action
\begin{equation}
    K_0(\bm{q};\bm{x},a) = \mathcal{N}\exp\left(\frac{i}{\hbar}S_0(\bm{q};\bm{x},a)\right),
\end{equation}
where $\mathcal{N} = (2\pi i \hbar a)^{-n/2}$ is a normalisation factor for $n$ dimensions to ensure that the propagation returns a Dirac delta distribution for time $a=0$. Here we consider $\hbar$ to be a parameter of our model, building a quantum system which should produce classical behaviour in the $\hbar \to 0$ limit (subject to subtleties discussed in Section~\ref{sec:observables}). 

In the context of quantum mechanics, the transition amplitude can be used as a propagator moving initial wavefunctions to solutions at later time
\begin{equation}\label{eqn:psi propagated}
    \psi_0(\bm{x},a) = \int \dd[n]{\bm{q}} K_0(\bm{q};\bm{x},a) \psi_0^{\rm (ini)}(\bm{q})\,,
\end{equation}
where $\psi_0^{\rm (ini)}(\bm{q}) = \psi_0(\bm{q}, a=0)$ is the initial wavefunction. Both the propagator $K_0$ and the associated wavefunction, $\psi_0$, satisfy the potential free Schrödinger equation
\begin{equation}
\label{eq:freeSchroedi}
    i\hbar \del_a \psi_0 = -\frac{\hbar^2}{2}\nabla^2\psi_0\,.
\end{equation}
In this Chapter we will only be considering this free Schrödinger evolution, using the free classical action in equation~\eqref{eqn:free_action}, and therefore will drop the 0 subscripts on quantities from here on. We note that this wavefunction model is the same as the ``free particle approximation'' introduced in \cite{ColesSpencer2003}, compared to linearised fluid in \cite{ShortColes2006} and applied to cosmic voids in \cite{Gallagher_2022_SPvoids}. In Chapter \ref{chap:how-classical} we will present how this formalism extends to a perturbative scheme called Propagator Perturbation Theory (PPT) which relies on  a time-independent Hamiltonian including an external effective potential  
\begin{equation}\label{eqn:schrodinger_with_potential}
    i\hbar \del_a \psi = \mathscr{H} \psi = -\frac{\hbar^2}{2}\nabla^2\psi + V_{\rm eff} \psi.
\end{equation}
Note that this does not take exactly the same form as the Schr\"odinger-Poisson equations arising in the context of fuzzy dark matter \eqref{eqn:SP-full-eqns-firsttime} due to the difference in time variables ($a$-time vs coordinate time). We return to this issue in Chapter \ref{chap:how-classical}.

Using the polar representation of a wavefunction, we can decompose it into $\psi(\bm{x},a) = \sqrt{\rho(\bm{x},a)}e^{i\phi_v(\bm{x},a)/\hbar}$, which is called the Madelung representation \parencite{Madelung1927}. Under this decomposition we see that the boundary conditions \eqref{eqn:ZA_boundary_conditions} require an initial wavefunction with Zel'dovich-like dynamics of the form  $\psi^{\rm (ini)}(\bm{q}) = \exp(i\phi_v^{\rm (ini)}(\bm{q})/\hbar)$. The propagator equation \eqref{eqn:psi propagated} can then be written
\begin{align} 
    \psi(\bm{x},a) &= \mathcal{N}\int \dd[n]{\bm{q}} \exp\left[{\frac{i}{\hbar}S(\bm{q};\bm{x},a)}\right]\exp\left[{\frac{i}{\hbar}\phi_v^{\rm (ini)}(\bm{q})}\right] \nonumber \\
    &= \mathcal{N} \int \dd[n]{\bm{q}} \exp\left[{\frac{i}{\hbar}\zeta(\bm{q};\bm{x},a)}\right],
    \label{eqn:psi_integral_with_zeta}
\end{align}
where $\zeta= S + \phi_v^{\rm (ini)}$ includes both the propagation of the wavefunction and its initial conditions. We note here that we take the opposite sign choice for the velocity potential to \cite{Uhlemann2019}, so that $\bm{v} = \bm{\nabla} \phi_v$ instead of $\bm{v} = - \bm{\nabla} \phi_v$.

This free wavefunction can be solved for using a variety of analytic or numerical techniques. For free evolution, solving the free Schrödinger equation with Fourier transforms at each time slice is significantly faster than computing the wavefunction at late time using the highly oscillatory integral \eqref{eqn:psi_integral_with_zeta}\footnote{For a wavefunction obeying the Schrödinger equation with a potential, the time evolution can be obtained by using a symplectic integration scheme, splitting the Hamiltonian into a kinetic and potential part (see e.g. Appendix D of \cite{Uhlemann2019} or Chapter 3 of \cite{BinneyTremaine2008}).}.  However, as we will see in Sections~\ref{sec:unweaving_the_wavefunction} and \ref{sec:catastrophe_theory}, being able to directly analyse these highly oscillatory integrals of the form~\eqref{eqn:psi_integral_with_zeta} will be of physical interest, so it is useful to develop techniques to approximate them. 

We focus on $\zeta(q;x,a)$ which depend on one-dimensional $q$ and $x$, and are analytic in $q$. In this case, we can replace the integration along the real line with \emph{any} contour in the complex plane which has the same end points without changing the value of the integral via Cauchy's theorem. In such cases, a simple way to accelerate the numerical integration is to simply step the integration contour in the direction which increases $\Im(\zeta)$ (decreasing the magnitude of the integrand). While not well suited to very complex forms of $\zeta$, and requiring some amount of tuning for the size of the displacement to take, any shift which reduces the amount of oscillation is helpful in obtaining accurate numerical evaluations of highly oscillatory integrals. A more sophisticated approach would be to consider the full steepest descent contours of $\zeta$ (see Appendix \ref{app:SPA}) which can be found using techniques from Picard-Lefschetz theory, as described in e.g. \cite{Feldbrugge2019}.

\begin{figure}[h!t]
    \centering
    \includegraphics[width=0.5\columnwidth]{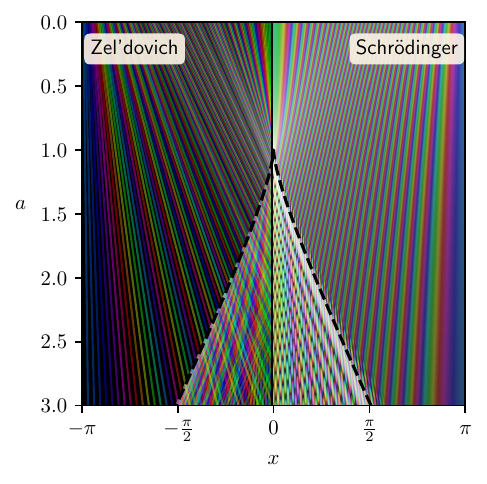}     
    \caption[Spacetime evolution in the Zel'dovich and free Schr\"odinger approximations.]{Comparison of the wavefunction evolution and classical Zel'dovich evolution with the same initial conditions. The wavefunction initial conditions are $\psi^{\rm (ini)}(q) = \exp((i/\hbar)\cos(q))$ with $\hbar = 0.01$ evolved on a grid of $(1024)^2$ (spacetime) cells. The complex value of the wavefunction is encoded through domain colouring, with brightness corresponding to amplitude, and hue corresponding to phase (the exact colouring scheme is in Appendix~\ref{app:domain_colouring}). Hence, lines of constant colour in this image correspond paths of constant phase. The Zel'dovich trajectories are coloured according to their initial phase $\cos(q)/\hbar$. For times $a>1$, interference patterns arise in the wavefunction, characterised by rapid oscillations in space and across time, corresponding to classical multistreaming. The black dashed curve separating the classically shell crossed and single-stream regions is given by equation~\eqref{eqn:shell_cross_region}. This Figure was published in \cite{Gough.Uhlemann_2022_MakingDark}. }\label{fig:free_schrodi_evol}
\end{figure}

Figure~\ref{fig:free_schrodi_evol} shows the evolution of a wavefunction corresponding the same Zel'dovich initial conditions as in Figure~\ref{fig:zeldo_phase_sheet}, compared to particle trajectories under Zel'dovich evolution. This wavefunction is defined by the initial data $\psi_0^{\rm (ini)}(q) = \exp((i/\hbar)\phi_v(q))$ with $\phi_v(q) = \cos(q)$. This wavefunction is the principal object of study in this Chapter, in particular understanding how the classical stream information is encoded in the wave phenomena. The Figure uses a domain colouring technique to show the complex value of the wavefunction at all spacetime points. The brightness corresponds to the magnitude of the $\psi$, while the hue corresponds to the phase. The Zel'dovich trajectories are coloured according to their initial phase $\cos(q)/\hbar$, which they carry along their trajectory. The wavefunction is evaluated using the contour shifting technique, and reproduces the wavefunction solved via Fourier methods in Figure 1 of \cite{Uhlemann2019}.

\begin{figure}[h!t]
    \centering
    \includegraphics[width=0.5\columnwidth]{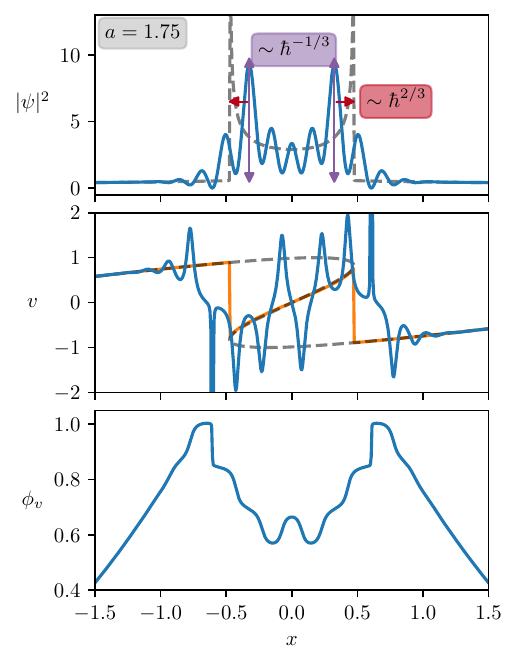}
    \caption[Interference features of the wavefunction past shell crossing.]{Features of the wavefunction $\psi$ (with $\hbar = 0.05$) with Zel'dovich initial conditions at time $a=1.75$.
    (Upper panel) The density of the wavefunction $\psi$. The dashed line corresponds to the classical Zel'dovich density at the same time. The wavefunction follows this classical density, dressing it in wave interference. The parameter $\hbar$ sets certain scaling properties, including the characteristic width and peak height of the density. The exact indices in this case will be explained in Section~\ref{sec:catastrophe_theory}. (Middle panel) The velocity associated with the wavefunction, given as the gradient of the wavefunction phase, or as the probability current divided by the density as in equation~\eqref{eqn:wavefunction_velocity_moment}. The full Zel'dovich phase-space sheet (dashed) and the classical stream-weighted average velocity (orange) are shown for comparison. Where the phase-space sheet is multivalued, the wavefunction rapidly oscillates between the classical values. (Bottom panel) The phase function of the wavefunction, which produces the velocity field in a Madelung decomposition as $v = \nabla \phi_v$. At $x\approx \pm 0.6$ the phase develops a discontinuous jump, leading to ill-defined gradient. The impact of these phase jumps will be discussed in greater detail in Section~\ref{sec:phase_properties}. This Figure was published in \cite{Gough.Uhlemann_2022_MakingDark}. 
    }
    \label{fig:fold_annotated}
\end{figure}

The features in the wavefunction system shown in Figure~\ref{fig:free_schrodi_evol} are similar to those seen in the evolution of the classical Zel'dovich mode. We see the formation of a bright (dense), cusp shaped caustic line beginning at $a=1$. The interior region of this caustic exhibits interference, characterised by spatio-temporal oscillations, which must correspond to the classical multistream regime. Of notable difference however, the caustic region no longer corresponds to infinite density, it has been regularised by the wavelength associated with the size of $\hbar$. It also now corresponds to a finite sized region, rather than being infinitely thin. These interference phenomena are shown at a constant time in Figure~\ref{fig:fold_annotated}, in comparison to the classical Zel'dovich density and stream velocities. The particular scalings of the peak density and the fringe spacing are given by certain powers of $\hbar$. These scalings are obtained by catastrophe theory, discussed in Section~\ref{sec:catastrophe_theory}. We see that this wave-mechanical model ``dresses'' the classical observables in this interference phenomena.

\subsection{Dark matter observables in phase- and real space}\label{sec:observables}

\subsubsection{Fluid equations from a wavefunction}

The Madelung representation  $\psi(\bm{x},a) = \sqrt{\rho(\bm{x},a)}e^{i\phi_v(\bm{x},a)/\hbar}$, makes connection between the wavefunction and fluid-like variables. Note that at the level of the evolution equations, this decomposition is only meaningful where the density field and phase are sufficiently smooth. If this wavefunction satisfies the free Schrödinger equation \eqref{eq:freeSchroedi}, under this Madelung decomposition the real and imaginary parts of the Schrödinger equation become a continuity and a Bernoulli equation,
\begin{subequations}\label{eqn:quantum_fluids}
\begin{align}
    &\del_a \rho + \bm{\nabla} \cdot (\rho \bm{\nabla} \phi_v) = 0\,, \\
    &\del_a \phi_v + \frac{1}{2} \abs{\bm{\nabla}\phi_v}^2 + Q = 0\,.\label{eqn:quantum_bernoulli}
\end{align}
The final term in the Bernoulli equation, 
\begin{equation}\label{eqn:quantum_pressure}
    Q = -\frac{\hbar^2}{2}\frac{\nabla^2\sqrt{\rho}}{\sqrt{\rho}}\,,
\end{equation}
\end{subequations}
is traditionally called the ``quantum pressure,'' even though strictly it does not appear as a pressure but rather a velocity dispersion term in the associated Euler equation. It would more properly be referred to as ``quantum potential'' as it appears as a potential-like term in the Bernoulli equation~\eqref{eqn:quantum_bernoulli}. In the presence of a potential in the Schr\"odinger equation, that too would appear in the Bernoulli equation. In the classical limit $Q$ vanishes and we obtain standard (potential free) fluid equations in $a$-time, assuming a potential velocity field determined by $\bm{v} = \bm{\nabla}\phi_v$. 

We note that the quantum pressure term in equation~\eqref{eqn:quantum_fluids} is important for precision cosmology, and that wave techniques which lack the quantum pressure term can differ substantially from full wave simulations. Compared to $N$-body simulations, simulations including quantum pressure cause up to $5$--$10\%$ suppression in the matter power spectrum at low redshift, particularly near the quantum Jeans scale \parencite{Veltmaat2016PhRvD, Nori_2018_AX-GADGET}. This suppression is further enhanced when the difference in initial conditions between fuzzy and cold dark matter is taken into account. While Madelung based solvers such as \texttt{AX-GADGET} \parencite{Nori_2018_AX-GADGET} can capture the role the quantum pressure plays in suppressing structure, solving the Madelung equations~\eqref{eqn:quantum_fluids} does not well capture the details of interference patterns compared to full Schr\"odinger solvers, particularly the points where $|\psi|$ is close to zero where the Madelung variables develop discontinuities while $\psi$ remains well behaved throughout its evolution \parencite{Mocz_2015_SP_SPH,Veltmaat2016PhRvD,  Hopkins_2019_numerics_FDM, SchwabeNiemeyer2022}. The interplay between these factors further motivates advances in analytic techniques for these systems, and care to be taken in running simulations. In Chapter \ref{chap:how-classical} we investigate the role of this quantum potential on statistics of the density field in full three-dimensional fields generated with PPT and LPT.

\subsubsection{Phase-space distributions for quantum systems}

To more completely characterise a wavefunction beyond the Madelung split, we consider the dynamics of a quantum state in phase-space. For a pure quantum state $\psi$ in $n$-dimensions, the Wigner  function \parencite{Wigner1932} 
\begin{equation}\label{eqn:wigner_dist}
    f_W[\psi(\xx)](\pp) = f_W(\bm{x},\bm{p})= \int \frac{\dd[n]{\tilde{\bm{x}}}}{(\pi\hbar)^n} \exp\left[\frac{2i\bm{p}\cdot \tilde{\bm{x}}}{ a^{3/2}\hbar}\right] \psi\left(\bm{x}-\tilde{\bm{x}}\right) \psi^*\left(\bm{x}+\tilde{\bm{x}}\right)\,,
\end{equation}
represents a quasi-probability distribution in phase-space (here $\psi^*$ denotes complex conjugation). The evolution of this Wigner function can be written (see Appendix \ref{app:sec:derive_wigner_eq} for a derivation)
\begin{align}\label{eq:wig-vlas-a-time}
\pdv{f_W[\psi]}{a} &= \mathscr{H} \frac{2}{\hbar}  \sin\left(\frac{\hbar}{2}(\cev{\del}_x\vec{\del}_v + \cev{\del}_v \vec{\del}_x)\right) f_W[\psi] \nonumber \\
&= \{\mathscr{H}, f_W\}_{\rm MB}^{\hbar} = \{\mathscr{H},f_W[\psi]\}_{\rm PB} + \order{\hbar^2}
\end{align}
where $\mathscr{H}$ is the classical Hamiltonian, $\{\cdot,\cdot\}_{\rm MB}^\hbar$ is the \emph{Moyal bracket} and $\{\cdot,\cdot\}_{\rm PB}$ is the classical Poisson bracket. Note that the Wigner function defined in equation \eqref{eqn:wigner_dist} is defined with $\bm{v}$ as the phase space variable, and so is conserved with respect to $a$-time, rather than the usual definition of the Wigner function with respect to coordinate time. This means the dynamical equation \eqref{eq:wig-vlas-a-time} is structurally similar to the Vlasov equation \eqref{eq:vlasov}, but formally different due to the difference in coordinates.  If we had constructed the Wigner function with the standard phase-space variables, the resulting dynamical equations are the same as the Vlasov equation to  $\order{\hbar^2}$ (the derivation of the Wigner-Vlasov equation in standard coordinate time can be found in Appendix \ref{app:sec:derive_wigner_eq}). However, the Wigner function does display some unphysical properties for a phase-space distribution function, namely that it can by negative in small localised regions (bounded to a minimum value of $-4\pi/\hbar$ for pure states). It is for this reason that the na\"ive limit $\hbar \to 0$ does not derive the classical solution in a continuous way, and the classical limit needs to be treated with care \parencite{Takahashi1989}. A theorem due to Hudson demonstrates that the only non-negative Wigner functions are those which correspond to Gaussian wavefunctions, so for generic states these negative regions necessarily exist \parencite{Hudson1974}.

To account for this, one can smooth the Wigner distribution using Gaussian filters in both position and momentum space, resulting in the coarse-grained Wigner function, or the Husimi distribution \parencite{Husimi1940}
\begin{equation}
f_H(\xx,\pp) = \bar{f}_W(\xx,\pp) = \int \frac{\dd[n]{\xx'} \dd[n]{\pp'}}{(2\pi \sigma_x\sigma_p)^n} \exp[-\frac{(\xx-\xx')^2}{2\sigma_x^2} - \frac{(\pp-\pp')}{2\sigma_p^2}] f_W(\xx',\pp').
\end{equation}
If the smoothing scales $\sigma_x, \sigma_p$ are chosen to satisfy the uncertainty principle
\begin{equation}\label{eqn:uncertainty_principle}
    \sigma_x \sigma_p \geq \frac{\hbar}{2}\,,
\end{equation} 
the resulting distribution is guaranteed to be non-negative. The evolution of this Husimi distribution differs from the evolution of the coarse-grained classical distribution by $\order{\hbar^2}$ \parencite[e.g.][]{Uhlemann2014} but corresponds to a better correspondence to the classical dynamics without the violent $\hbar$ scale oscillations.
  
 Under such a smoothing, the semiclassical $\hbar \to 0$ limit can be approached in a mathematically sound way \parencite{Lions1993,Gerard1997,Zhang2002,Athanassoulis2009,Athanassoulis2018}. Features on scales of $\order{\hbar}$ do not survive the classical limit after this smoothing. For example, wave dark matter systems can exhibit non-trivial self-similar solutions which vanish identically in the classical limit \parencite{Galazo2022}.

From this phase-space distribution, we can construct observables as momentum weighted averages (moments) of this distribution, analogously to what was done with the classical phase-space distribution in Section~\ref{sec:cosmo-fluid-equations}. For example, the first two moments of the Wigner distribution are the density and the momentum-flux
\begin{align}
    &\int \dd[n]{\bm{p}} f_W(\bm{x},\bm{p}) = \rho(\bm{x}) = \abs{\psi}^2\,, \\ 
    &\int \dd[n]{\bm{p}} \frac{\bm{p}}{a^{3/2}}f_W(\bm{x},\bm{p}) = \bm{j}(\bm{x}) = \hbar \Im(\psi^*\bm{\nabla}\psi)\,. \label{eqn:wavefunction_velocity_moment}
\end{align}
Using these kinetic moments, we can define a velocity field $\bm{v}=\bm{j}/\rho$ everywhere, even where the phase is not smooth. If one wishes to work with fluid variables even in regions where $\phi_v$ is not smooth, they should replace the quantum continuity and Bernoulli equations~\eqref{eqn:quantum_fluids} with equivalent equations using the momentum flux $\bm{j}$ \parencite{Uhlemann2014}.

The cumulants associated with these moments are extracted analogously to the classical fluid case in Section \ref{sec:cosmo-fluid-equations}. For example, the second moment of $f_W$ is
\begin{equation}
\int\dd[n]{\bm{p}} \frac{p_i p_j}{a^{3}}f_W(\bm{x},\bm{p}) = \frac{j_i(\bm{x}) j_j(\bm{x})}{\rho(\bm{x})} + \rho(\bm{x})\sigma_{ij}(\bm{x})\,,
\end{equation}
which contains a product of the first moment, and a connected part determined by the velocity dispersion $\sigma_{ij}$. These cumulants directly correspond to (Eulerian) observables: the logarithmic matter density $\ln\rho$, velocity $v_i$, and the velocity dispersion $\sigma_{ij}$. Written in terms of the density, the velocity dispersion associated with a wavefunction is
\begin{align}
    \sigma_{ij} &= \frac{\hbar^2}{4}\left(\frac{\nabla_{\! i} \rho \nabla_{\! j} \rho}{\rho^2} - \frac{\nabla_{\! i}\nabla_{\! j} \rho}{\rho}\right)= -\frac{\hbar^2}{4}\nabla_{\! i}\nabla_{\! j} \ln \rho\,, \label{eqn:sigma_ij}
\end{align}
and is directly related to the quantum ``pressure'' in equation~\eqref{eqn:quantum_bernoulli}. In this formula we can see explicitly how the wave-based model generates higher-order cumulants from derivatives of the wavefunction building blocks.

Under a Madelung split the velocity associated with a wavefunction appears potential, as $\bm{v} = \bm{\nabla} \phi_v$, however, the phase can develop discontinuities (jumps) which source vorticity. These phase jumps only occur at places where the density vanishes and the phase becomes ill-defined. The curvature in the density near the zeros of the wavefunction can then source velocity dispersion via equation~\eqref{eqn:sigma_ij}.

All the higher-order cumulants for the system can be obtained as moments of the Wigner distribution in this way, and will be non-zero after shell crossing, even when starting from initial conditions described by a perfect fluid. The procedure of extracting moments and cumulants from the wavefunction is exactly the same as extracting moments/cumulants from a Schr\"odinger-Poisson system, although the underlying evolution equation is different from the propagator formalism laid out here.

To determine the full behaviour of a system in phase-space, in principle one would have to solve the evolution equations associated with each cumulant, which becomes difficult as the evolution of the $n^{\rm th}$ cumulant depends on the $(n+1)^{\rm th}$ cumulant. The only consistent truncation to the hierarchy is the perfect fluid model, which corresponds to (in Eulerian coordinates)
\begin{equation}\label{eqn:eulerian_fluid_phase_space}
    f_{\rm fluid}(\bm{x},\bm{p}) = \frac{\rho(\bm{x})}{\bar{\rho}} \delta^{(3)}_{\rm D}\left(\frac{\bm{p}}{a^{3/2}}-\bm{\nabla}_{\! \bm{x}}\phi_v(\bm{x}) \right),
\end{equation}
where all cumulants of order 2 and higher are set to 0 (and remain 0). However, once shell crossing occurs and the system develops multiple streams, this perfect fluid description is no longer adequate to describe the dynamics of the system as a whole. At this point, the phase-space distribution cannot take the form in equation~\eqref{eqn:eulerian_fluid_phase_space}, and all higher cumulants are sourced dynamically.

The Wigner function's time evolution, induced by the time evolution of $\psi$ through the appropriate Schr\"odinger equation, naturally captures these dynamically sourced higher cumulants. This guarantees that all phase-space information is encoded in the wavefunction, even beyond shell crossing. This is in the same spirit of looking at the phase-space distribution for a fluid in Lagrangian space  (c.f. Figure~\ref{fig:zeldo_phase_sheet}). A Lagrangian fluid as defined in equation~\eqref{eqn:lagrangian_phase_space_dist} can still develop an infinite hierarchy of cumulants if $x(q)$ is a multiple-to-one mapping describing a mixture of fluid streams in Eulerian space. However, in the wavefunction model we get the same encoded information, but a more direct way to access Eulerian observables, without having to consider mapping between Lagrangian and Eulerian space.

\section{Unweaving the wavefunction}\label{sec:unweaving_the_wavefunction}

\subsection{Obtaining single-stream wavefunctions from interference}

For the free evolution of a wavefunction with Zel'dovich-like initial conditions, the multistream region is replaced with small scale wave interference. In the following we demonstrate that the shell crossed wavefunction can be decomposed into a sum of single-stream, non-interfering wavefunctions, each corresponding to a classical Zel'dovich stream.

\subsubsection{Stationary phase decomposition}

Taking a 1-dimensional wavefunction with Zel'dovich initial conditions, we write the wavefunction over spacetime as an oscillatory integral as in equation~\eqref{eqn:psi_integral_with_zeta}
\begin{equation}
    \psi(x,a) = \mathcal{N} \int \dd{q} \exp\left[{\frac{i}{\hbar}\zeta(q;x,a)}\right], \label{eqn:psi_oscillatory}
\end{equation}
where $\zeta=S + \phi_v^{\rm (ini)}$. The integrand of equation~\eqref{eqn:psi_oscillatory} is highly oscillatory and therefore the dominant contribution to the integral will come from points where $\zeta$ is stationary with respect to the integration variable $q$, in a technique known as the stationary phase approximation (Appendix~\ref{app:SPA}). It is worth noting that the \emph{phase} in the stationary phase approximation (SPA) refers to the function $\zeta$, not the phase of the wavefunction $\phi_v$. The approximation is dominated by the stationary points of $\zeta$, not $\phi_v$.

For a stationary point $q_*$ satisfying $\nabla_{\! q}\zeta(q_*;x,a) = 0$, if $\nabla^2_{\! q}\zeta(q_*;x,a) \neq 0$, the stationary phase contribution from that point reads (suppressing the $x,a$ arguments of $\zeta$)
\begin{equation}\label{eqn:psi_spa}
    \psi_{q_*}^{\rm SPA}(x,a) =  \frac{\exp(\frac{i\pi}{4}[\operatorname{sgn} (\nabla^2_{\! q}\zeta(q_*))-1])}{\sqrt{a\abs{\nabla^2_{\! q}\zeta(q_*)}}} \exp\left(\frac{i}{\hbar}\zeta(q_*)\right).
\end{equation}
The full approximation is then given by a sum over all of these stationary points,
\begin{equation} \label{eqn:psi_spa_full}
    \psi^{\rm SPA}(x,a) =\sum_{q_*} \psi_{q_*}^{\rm SPA}(x,a)\,,
\end{equation}
which in the semiclassical $\hbar \to 0$ limit obeys the asymptotic relation $\psi(x,a) \to \psi^{\rm SPA}(x,a)$.

For the choice $\phi_v^{\rm (ini)}(q) = \cos(q)$, the number of (real) stationary points is determined by the same caustic condition as equation~\eqref{eqn:shell_cross_region}, with three real roots in the shell crossed region, two real roots on the caustic line, and a single real root in the single-stream regime. Indeed this is not surprising, as for the Zel'dovich-like initial conditions we have
\begin{equation}\label{eqn:zeta}
    \zeta(q;x,a) = \frac{(x-q)^2}{2a}+\phi_v(q)\,,
\end{equation}
so the condition for $\zeta$ to be stationary with respect to $q$ is 
\begin{align}\label{eqn:stationary_condition}
    \nabla_{\! q}\zeta(q;x,a) = -\frac{x}{a} + \frac{q}{a} + \nabla_{\! q}\phi_v(q) = 0\,,
\end{align}
which is equivalent to the Zel'dovich displacement mapping $x = q + a\nabla_{\! q} \phi_v(q)$.  That is, the quantum propagation integral is dominated by the $q$ which satisfy the classical displacement mapping.

\begin{figure}[h!t]
    \centering
    \includegraphics[width=0.5\columnwidth]{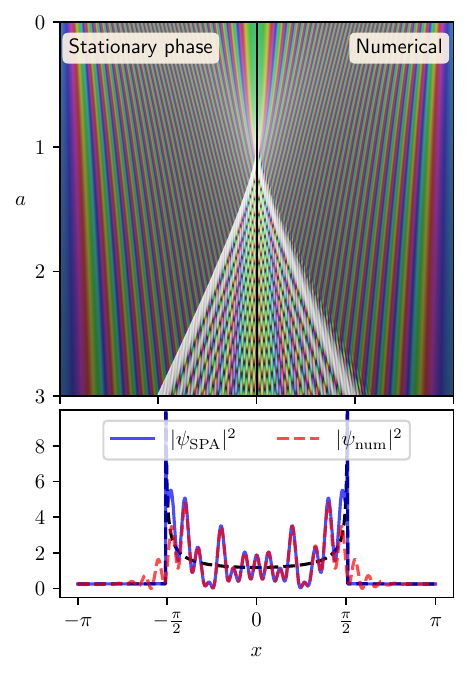}
    \caption[Comparison of the exact and stationary phase approximations of the wavefunction.]{Comparison of the stationary phase approximation (SPA) and the numerical evolution of $\psi^{\rm (ini)}(q)=\exp(i\cos(q)/\hbar)$. (Upper panel) The full spacetime evolution of this wavefunction, with $\hbar=0.01$. (Lower panel) The density profile post shell crossing of this wavefunction, with $\hbar = 0.05$. The black dashed line shows the classical Zel'dovich density, or equivalently the sum of the densities of the individual stationary phase streams. From these figures we see that the stationary phase approximation well approximates the numerical solution, except very close to the caustic boundary. This indicates that well inside the single or multistream regime that the interference can be resolved into a sum over classical trajectories. Near the caustic line, stationary phase analysis fails as the stationary points are no longer well separated in the complex plane. Behaviour close to the caustic should be studied using the normal forms of diffraction catastrophe integrals, discussed in Section~\ref{sec:catastrophe_theory}. This Figure was published in \textcite{Gough.Uhlemann_2022_MakingDark}. }
    \label{fig:compare_SPA_to_numeric_joint}
\end{figure}

Figure~\ref{fig:compare_SPA_to_numeric_joint} compares the full stationary phase analysis of the Fourier mode to the numeric integration of the wavefunction. The stationary phase approximation provides an excellent description of the wavefunction except near to the caustic lines themselves. However, well into the single or multistream regime, far away from high density caustics, this stationary phase decomposition provides a good description of the full wavefunction.

\begin{figure*}[t]
    \centering
    \includegraphics[width=\columnwidth]{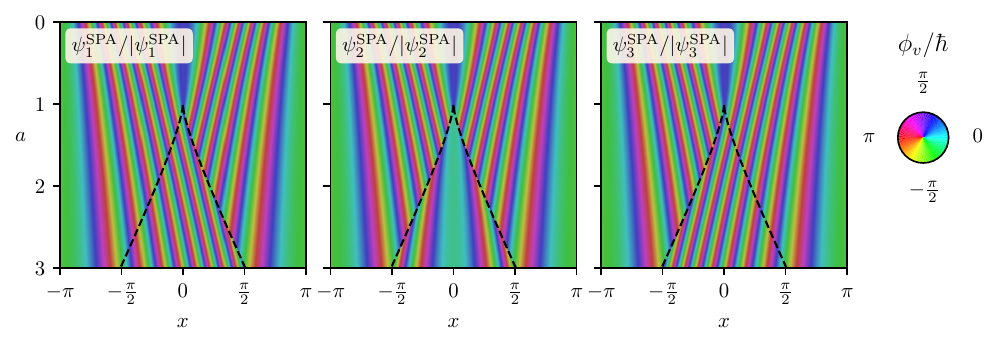}
    \caption[The stream wavefunctions corresponding to classical Zel'dovich trajectories.]{The polar angle of the normalised wavefunctions ($\hbar = 0.05$) split according to each stationary point of the function $\zeta$ in equation~\eqref{eqn:zeta}. Each of these stationary points corresponds to one of the classical Zel'dovich trajectories. Outside the cusp there is a single stationary point, corresponding to the single stream (shown in all three panels), while inside the classical cusp three wavefunctions exist, corresponding to the three classical streams illustrated in Figure~\ref{fig:zeldo_phase_sheet}. This Figure was published in \textcite{Gough.Uhlemann_2022_MakingDark}. }
    \label{fig:stream_splitting_phase_hbar0.05}
\end{figure*}

Figure~\ref{fig:stream_splitting_phase_hbar0.05} shows the individual terms in equation~\eqref{eqn:psi_spa_full} as separate wavefunctions. We see that indeed these are single-stream wavefunctions which are non-interfering and resemble individual Zel'dovich streams. Outside the classical cusp there is a single-stream wavefunction, and inside the cusp there are three stream wavefunctions, corresponding to the three sheets of the Zel'dovich phase-space sheet in Figure~\ref{fig:zeldo_phase_sheet}. This is precisely analogous to taking the geometric optics limit for an optical wave field, resolving wave phenomena as sums over classical light rays.

\begin{figure}[h!t]
    \centering
    \includegraphics[width=0.5\columnwidth]{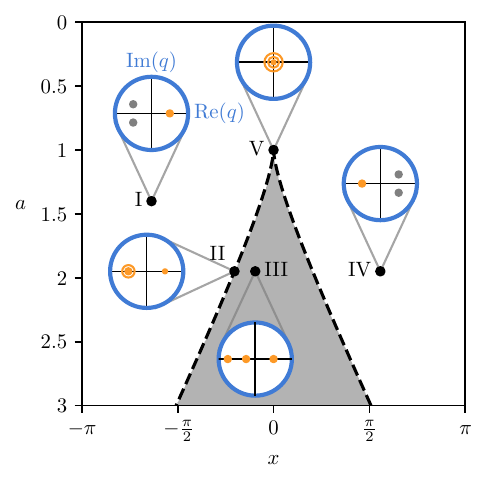}
    \caption[The structure of stationary points of $\zeta(q_*;x,a)$.]{The structure of stationary points $q_*$ which solve $\nabla_q \zeta(q_*;x,a) = 0$ (equation~\eqref{eqn:stationary_condition}) for different values of $x$ and $a$. The blue circles show the distribution of $q_*$ in the complex-$q$ plane. Beginning at point (I) in the single-stream (white) region, two solutions are complex, with only one real root. Approaching the caustic (dashed black) line, as with point (II), the two complex roots merge into a real double root. Points within the multistream (shaded) region, such as (III), have all three roots are real. Crossing the $x=0$ line causes the sign of the real part of the roots to flip, as can be seen comparing points (I) and (IV). At the cusp point (V), all three roots coalesce into a single real solution. For our stationary point analysis, we only use contributions from real stationary points (coloured in yellow in this diagram). This Figure was published in \textcite{Gough.Uhlemann_2022_MakingDark}.  }
    \label{fig:root_structure}
\end{figure}

Figure~\ref{fig:root_structure} illustrates how the stationary points are distributed in the complex-$q$ plane for different values of $x$ and $a$. By moving from the single-stream region across the caustic line, the complex stationary points coalesce into a double root, before becoming distinct real roots in the shell crossed region. This merging and splitting of roots at the caustic is why stationary phase contributions of the form \eqref{eqn:psi_spa} no longer well approximate the integral in equation~\eqref{eqn:psi_oscillatory}. On the caustic defined by $\nabla_{\! q} \zeta = \nabla_{\! q}^2 \zeta = 0$ the SPA wavefunction form~\eqref{eqn:psi_spa} does not apply, and near the caustic the stationary points are not sufficiently separated (with respect to the size of $\hbar$) in the complex plane to be treated independently. 

A more accurate asymptotic approximation can be made by considering the contributions from the complex roots along a proper steepest descent contour of the integrand, which are the same as contours of constant $\mathrm{Im}(i\zeta/\hbar)$ and as such remove the oscillatory part of the integral (see Appendix \ref{app:sec:steepest-descent} for a discussion of the general method).  The asymptotic analysis of such oscillatory integrals in the complex plane in general is difficult, as the geometry of the steepest descent contours, and the number of complex stationary points which are relevant can change suddenly in different regions of the parameter space (Stokes' phenomena). The Stokes' lines, where the steepest descent contour changes how many stationary points it intersects by one, for a cusp integral are similar to the case considered here and calculated analytically in \textcite{Wright1980}. While we restrict ourselves to simple stationary phase analysis of real roots, \textcite{Feldbrugge2019} present some techniques for automatically accounting for the changes in the nature of the stationary points and integration contours in a more rigorous manner. Figure 14 from \textcite{Feldbrugge2019} is similar to our Figure~\ref{fig:root_structure}, showing the roots, steepest descent contours, and Stokes' lines for the canonical cusp integral discussed in Section~\ref{sec:catastrophe_theory}. While the specific function these stationary points arise from is different than the function $\zeta(q;x,a)$ corresponding to our toy model, the qualitative features are the same. We focus here only on the real stationary points as they have clear physical interpretation as classical trajectories, and recover classical CDM dynamics.

The principal qualitative feature which cannot be reproduced by stationary phase decomposition is the existence of zeros in the density field \emph{outside} the shell crossed region. These zeros are of physical interest, as they result in discontinuities in the phase of the wavefunction which will be discussed in more detail in Section~\ref{sec:phase_properties}. For an optical system, zeros outside the cusp correspond to interference between a single real ray and a complex ray. Since the amplitude of the complex ray decays exponentially, it only has enough amplitude to fully destructively interfere with the real ray once outside the caustic \parencite{Wright1980}. For the free wavefunction, the interference which produces these exterior zeros occurs between the infalling part of the wavefunction (the ``real ray'', corresponding to a single stream wavefunction from the SPA decomposition), and a partial reflection of the wavefunction due to the quantum pressure term in the Bernoulli equation~\eqref{eqn:quantum_bernoulli} (the ``complex ray''). As the quantum pressure disappears in the classical limit, only density field zeros within in the multistream region remain.

\subsection{Phase-space distributions for individual streams}

To further establish the correspondence between these stationary phase contributions and classical streams, we can look at the entire phase-space distribution associated with these wavefunctions. While the individual $\psi^{\rm SPA}_{q_*}$ can be constructed without constructing a quantum phase-space distribution, it is worth looking at the correspondence in the full phase-space analysis.

\begin{figure}[h!t]
    \centering
    \includegraphics[width=0.5\columnwidth]{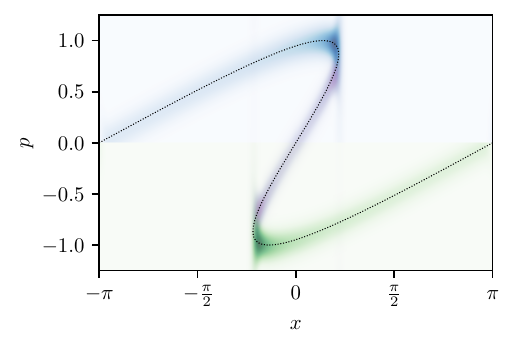}
    \caption[The Husimi phase-space distributions for the stream wavefunctions.]{The Husimi phase-space distributions for each of the individual wavefunctions associated with stationary points of $\zeta$ post shell crossing ($\hbar=0.01$). We see that this stationary phase decomposition naturally dissects the classical Zel'dovich phase-space sheet (dotted line). At the points where the phase-space sheet turns over, the Husimi functions blur in momentum space, due to the sharp localisation of the wavefunction in position space. Here we use a Gaussian filter to coarse-grain the Wigner distribution with $\sigma_x = 0.049$ and $\sigma_p= {\hbar}/({2\sigma_x})$ that satisfy the uncertainty principle \eqref{eqn:uncertainty_principle}. This Figure was published in \textcite{Gough.Uhlemann_2022_MakingDark}. }
    \label{fig:stream_split_wigner_overplot}
\end{figure}

The Husimi phase-space distributions associated with each term in the stationary phase approximation are shown in Figure~\ref{fig:stream_split_wigner_overplot}. We see that the stationary points naturally dissect the phase-space sheet into its three layers, as anticipated from looking at the phase plots in Figure~\ref{fig:stream_splitting_phase_hbar0.05}. We see also that the quantum phase-space sheet traces the characteristic S-shaped curve expected for Zel'dovich dynamics, with a characteristic width around the classical phase-space sheet owing to the uncertainty principle associated with the size of $\hbar$. The Husimi phase-space distributions are calculated using an adapted version of the \texttt{CHiMES}\footnote{\href{https://github.com/andillio/CHiMES}{https://github.com/andillio/CHiMES}} code \parencite{Eberhardt_2021_CHiMES}. We choose to plot Husimi distributions instead of Wigner distributions to emphasise the classical phase-space structure and to avoid the presence of aliasing in discrete Wigner distributions \parencite{Claasen1983, Chassande-Mottin2005}. 

The Husimi phase-space distributions of the individual stationary phase terms present an accurate tracing of the Zel'dovich phase-space sheet, except near the points where the phase-space sheet turns over. This is expected, as the points where the phase-space sheet turns over are precisely the classical caustic line, where the number of terms in the stationary phase approximation changes from 1 to 3. Near these points, the wavefunction streams develop a sharp discontinuity, localising on one or both sides of the stream, leading to a blurring in the momentum as required by uncertainty. This is expected however, as we know that near the classical caustic line the sum of the SPA terms does not accurately approximate the full wavefunction as discussed in the previous Section. This is encouraging, as it demonstrates that this splitting does not just accurately reproduce low order cumulants, but the full dynamics of the system.

\subsection{Comparing SPA streams to Zel'dovich properties}
\label{subsec:SPAstreamsZeldo}

To make the relationship between this stationary phase decomposition and the classical trajectories more explicit, we can read off the density and velocity associated with each stationary point. From equation~\eqref{eqn:psi_spa}, we see that the density of the $i^{\rm th}$ wavefunction is given by
\begin{align}
    \rho_i(x,a) &= \frac{1}{\abs{a\nabla^2_{\! q}\zeta(q_i;x,a)}} = \frac{1}{\abs{1-a\nabla_{\! q}^2 \phi_v^{\rm (ini)}(q_i(x,a))}}\,, \label{eqn:spa_density_to_zeldo}\\
    &= \frac{1}{\abs{1-a\cos(q_i(x,a))}}\,, 
\end{align}
where the final equality holds for the specific case $\phi_v^{\rm (ini)}(q)=\cos(q)$ considered for our Zel'dovich mode. This is precisely the same density profile one predicts from the Zel'dovich approximation (c.f. equation~\eqref{eqn:zeldo_density}).

\begin{figure}[h!t]
    \centering
    \includegraphics[width=0.5\columnwidth]{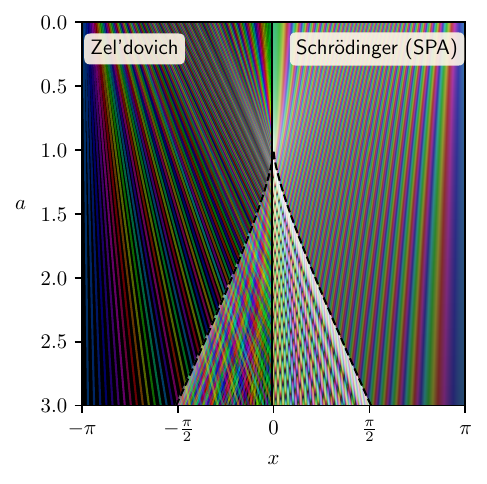}
    \caption[Comparison of classical Zel'dovich trajectories to the stationary phase approximation for the  wavefunction.]{Comparison of classical Zel'dovich trajectories to the stationary phase approximation for a  wavefunction (with $\hbar=0.01$) evolving under the free Schr\"odinger equation. The stationary phase wavefunction is the proper analogue to this classical system, having the exact same density and velocity profiles, and forming the same shell crossed region. This Figure was published in \textcite{Gough.Uhlemann_2022_MakingDark}. }
    \label{fig:zeldo_vs_SPA}
\end{figure}

The velocity field associated with each stream can be extracted by taking gradients (with respect to $x$) of the phase in equation~\eqref{eqn:psi_spa}. This recovers the velocity of the $i^{\rm th}$ stream at position $x$ and time $a$,
\begin{equation}
    v_i(x, a) = \nabla_{\! x} \zeta(q_i,x,a)\eval_{q_i = \rm const} = \frac{x-q_i}{a}\,,
\end{equation}
which is the constant velocity motion expected from the Zel'dovich approximation.\footnote{We note that the velocity can also be obtained as $\bm{j}/\rho$ using the expression for $\psi$ given in equation~\eqref{eqn:psi_spa_full}, as is done in Appendix C of \textcite{Uhlemann2019}.} This is completely generic, not requiring specific initial conditions. We thus see that the density and velocity of the streams extracted from stationary phase \emph{exactly} match those of a Zel'dovich system with the same initial conditions. Because of this exact correspondence, the stationary phase decomposition is the correct way to upgrade individual Zel'dovich streams into wavefunctions. Figure~\ref{fig:zeldo_vs_SPA} shows these two analogue systems side by side. Both of these systems have precisely the same extent of the shell crossed region, with the SPA wavefunction not presenting the same finite width to the caustic line that the full numerically evolved wavefunction does.

As a note of caution however, we point out that the phase of the individual stationary phase streams in equation~\eqref{eqn:psi_spa} is \emph{not} simply the initial phase transported to the final position via the displacement mapping. The phase of $\psi^{\rm SPA}_{q_i}$ also contains time propagation terms responsible for converting the Lagrangian derivative determining velocity into a Eulerian one. Starting from the link between Eulerian and Lagrangian velocity potentials,
\begin{subequations}
\label{eq:velpot_Eulerian}
\begin{equation}
    \nabla_{\! x}\phi_v^{\rm E}(x) = \pdv{q}{x}\nabla_{\! q}\phi^{\rm (ini)}_v(q) = \frac{1}{\mathcal{J}}\nabla_{\! q}\phi_v^{\rm (ini)}(q)\,,
\end{equation}
with the Jacobian $\mathcal{J}=\partial x/\partial q$, we can integrate to get the Eulerian velocity potential $\phi_v^{\rm E}$,
\begin{align}
    \phi_v^{\rm E}(x(q,a)) &= \int \dd{q} \nabla_{\! q}\phi_v^{\rm (ini)}(q) \mathcal{J} \\
    &= \cos(q) - \frac{a}{2} \cos^2(q) + C\,,
\end{align}
\end{subequations}
where the second line holds for $\phi_v^{\rm (ini)}(q) = \cos(q)$ and $C$ is an integration constant.

``Zel'dovich wavefunctions,'' constructed from the Zel'dovich density and the Eulerian phase, 
$
\psi^{\rm (Z)}_i(x,a) = \sqrt{\rho_i(x,a)} \exp(\frac{i}{\hbar}\phi_v^{\rm E}(x(q_i,a)) )
$
reproduce the wavefunctions of the individual SPA streams. So while the colouring scheme used for the particle trajectories in Figures~\ref{fig:free_schrodi_evol} and \ref{fig:zeldo_vs_SPA} provides intuition for the similarities between the dynamics, the appropriate phase for one of the component wavefunctions at spacetime position $(x,a)$ is not simply the phase $\phi^{\rm (ini)}(q_i(x,a))$ as if carried by particles, but the Eulerian phase $\phi^{\rm E}_v(x(q_i,a))$. The explicit difference between these phases can be seen in the left panel of Figure~\ref{fig:zeldo_loop_avg_phase}, with the Lagrangian velocity potential evaluated at a Eulerian position shown in blue and the proper Eulerian velocity potential (or SPA wave phase) in orange. 

\subsection{Comparison to other wave techniques}\label{sec:comparision_other_wave_techs}
Here we briefly summarise how the stationary phase decomposition presented here relates to similar looking models used in the literature. 

\textcite{Veltmaat_2018_FDM_halos} and \textcite{SchwabeNiemeyer2022} interface $N$-body simulations and full Schr\"odinger-Poisson solvers to study individual halos. Both of these methods rely on translating $N$-body particles into wavepackets to build a wavefunction which sets the initial and boundary conditions for the Schr\"odinger-Poisson solver on small scales. They use classical wavefunctions \parencite{Wyatt_2005_quantum_trajectories} for individual particle wavepackets to construct an overall wavefunction. In \textcite{Veltmaat_2018_FDM_halos} wavepackets in position space are used to obtain a phase from the combined wavefunction, while the amplitude is determined by the classical density in order to erase the interference of overlapping wavepackets which is present even in the single-stream regime. This problem could be avoided by using wavepacket `beams' localised in phase-space, for which \textcite{SchwabeNiemeyer2022} present an approximation for obtaining a phenomenological wavefunction including interference from collapsed halos. Our SPA split for the wavefunction corresponds to assigning classical wavefunctions to the classical fluid streams instead of individual particles.

\textcite{Lague_2021_FDM_LPT} approach modelling mixed dark matter containing FDM and CDM by adapting LPT. Their approach is to use the CDM LPT displacement but adjust for the difference in the linear growth using a scale-dependent transfer function for fuzzy dark matter. This incorporates the quantum pressure at the linear level while neglecting non-linear effects in the displacement described in Appendix B of \textcite{Uhlemann2014}. Our propagator approach transports a dark matter wave function thus incorporating the wave nature in the evolution, which is lost when treating FDM as particles moved by a modified displacement field.

\section{Wave interference effects}\label{sec:phase_properties}

We now turn our attention to another piece of interesting phenomenology associated with multistreaming: the dynamical production of vorticity and higher order cumulants like velocity dispersion. In the wave-mechanical model, these features arise from interference and are concentrated in regions where the wavefunction vanishes and topological defects form. In Section~\ref{sec:hidden_features} we show that the  interference features can be isolated in a ``hidden'' part of the wavefunction, which decorates the ``average'' part of the wavefunction describing the Zel'dovich fluid behaviour.

\subsection{Phase jumps and vorticity}

\begin{figure}[h!t]
    \centering
    \includegraphics[width=0.5\columnwidth]{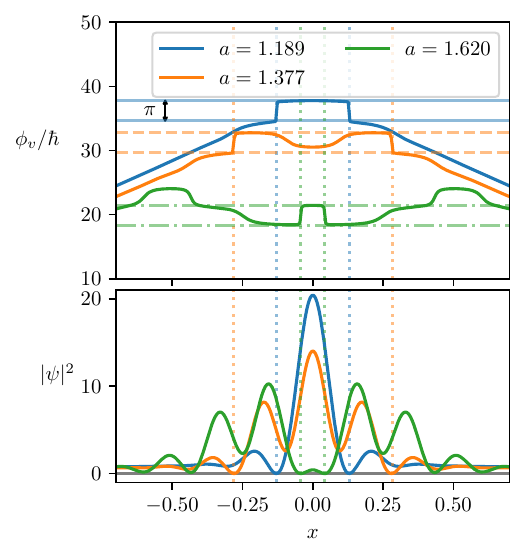}
    \caption[The discontinuous wavefunction phase and associated density zeros.]{The unwrapped phase (upper panel) and density (lower panel) of $\psi$ at times past shell crossing when defects occur (with $\hbar=0.05$). The spatial phase discontinuity is always $\pm \pi$, as indicated by the horizontal lines, which are spaced apart by $\pi$. These phase discontinuities source the non-potential part of the velocity, and occur at points where the density vanishes (as seen in the lower panel). The narrow regions around the zeros in density source velocity dispersion through equation~\eqref{eqn:sigma_ij}. The times were chosen to show the first two phase jumps occurring outside the cusp (as seen in Figure~\ref{fig:bp_located}),  and the first interior phase jump. This Figure was published in \textcite{Gough.Uhlemann_2022_MakingDark}. }
    \label{fig:phase_jumps_1d}
\end{figure}

As discussed in Section~\ref{sec:observables}, post shell crossing, all cumulants are sourced dynamically. In the wave-mechanical model, these higher order cumulants are  hidden in the oscillations of the wavefunctions and can be extracted from moments of the Wigner distribution function \eqref{eqn:wigner_dist}. 
In addition to these higher order cumulants, such as velocity dispersion, shell crossing sources vorticity, even for initially irrotational velocity fields. In the classical system this vorticity is due to multistream averaging, which results in a non-potential velocity field. In the wave-mechanical model, the non-potential parts of the velocity field occur at isolated points where the density vanishes. At these so-called branch points, the  wavefunction can develop phase jumps that cause infinite spatial gradients. As shown in Figure~\ref{fig:phase_jumps_1d}, the spatial dislocation of the phase is a jump of $\pm \pi$ as the complex value of the wavefunction passes through zero \parencite{Hui2021}. Since the wavefunction is single valued, the winding of $\phi_v/\hbar$ around a closed loop $\gamma$ must be an integer multiple of $2\pi$ leading to a quantisation of $\phi_v$ \parencite{Feynman1958, Gross1961}. This can be expressed as the integral quantisation condition
\begin{equation}\label{eqn:quantised_vorticity}
    \Gamma =  \oint_{\gamma} \bm{\nabla} \phi_v \cdot \dd{\bm{x}} = 2\pi\hbar n\,, \quad n\in \mathbb{Z}\,,
\end{equation}
related to vorticity via Stokes' theorem $\Gamma = \int_\Sigma (\bm{\nabla} \times \bm{v}) \cdot \hat{\bm{n}} \dd{A}$, where $\Sigma$ is the surface bounded by the curve $\gamma$, which has normal vector $\hat{\bm{n}}$ and area element $\dd{A}$. In 2-dimensional systems these vortices are points and in 3-dimensional systems closed vortex loops trace the defects in the phase of the wavefunction. The association of these branch points and vanishing of the wavefunction with the sourcing of vorticity and higher order cumulants is precisely why the Madelung formalism breaks down post shell crossing.

These vortex cores are of phenomenological interest for wave dark matter  signatures in collapsed structures, for example the way that such vortex lines interact with the presence of solitonic cores in fuzzy dark matter halos \parencite{hallock_vortex_2011, Rindler-Daller2012MNRAS, Hui2021JCAP, Hui2021,   Schobesberger2021MNRAS},   or in their role in cosmic filament spins \parencite{Alexander2021arXiv}. Near the vortex cores, the velocity scales as $v\sim 1/r_\perp$, owing to equation~\eqref{eqn:quantised_vorticity}, while the density profile near a branch point scales as $\abs{\psi}^2 \sim r_{\perp}^2$, where $r_\perp$ is the perpendicular distance from the vortex core/line \parencite{Hui2021JCAP}.

For the 1+1D toy model, we do not develop true vorticity, as in one spatial dimension the condition~\eqref{eqn:quantised_vorticity} is trivially satisfied. Nevertheless, the localised phase jumps create a non-potential velocity field and the associated vanishing of the density induces velocity dispersion, both of which are classically produced by multistream averaging. The derivatives of the density about these zeros produces velocity dispersion via equation~\eqref{eqn:sigma_ij}.

\subsubsection{Locating the phase jumps}

\begin{figure}[h!t]
    \centering
    \includegraphics[width=0.5\columnwidth]{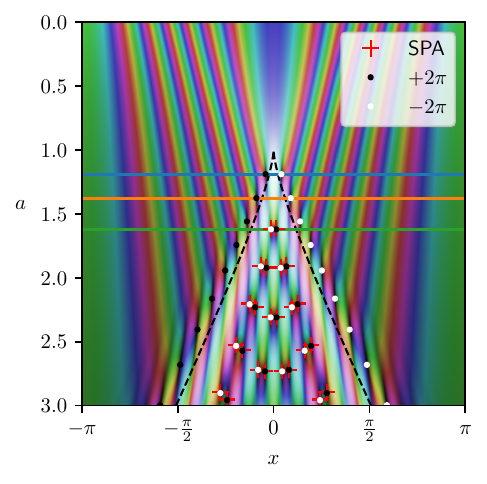}
    \caption[Spacetime location of branch points in the wavefunction.]{Location of branch points in the evolution of the free wavefunction ($\hbar=0.05$). The branch points predicted by the stationary phase approximation (SPA) are shown as red crosses, and accurately capture the position of the interior branch points from the full wavefunction. Points are coloured according to their (spacetime) circulation value in the diagram via equation~\eqref{eqn:quantised_vorticity}. The horizontal lines indicate the times chosen for the spatial phase profiles shown in Figure~\ref{fig:phase_jumps_1d}. This Figure was published in \textcite{Gough.Uhlemann_2022_MakingDark}. }
    \label{fig:bp_located}
\end{figure}

Figure~\ref{fig:bp_located} shows the evolution of the wavefunction, with the spacetime positions of the phase jumps located, and identified with their sign. While the Figure shows the time-evolution of a 1-dimensional system, we make use of the 2D spacetime diagram to find these branch points. Numerically, we located these branch point by masking the phase to only include regions of low density (since branch points only occur where $\abs{\psi} = 0$). Then the circulation (in spacetime) $\Gamma$ can be calculated by discrete line integral, and the branch points are taken to be the centroids of the connected regions where the circulation is $\pm 2 \pi$. The same procedure could instead be applied to constant time slices of a 2-dimensional wavefunction, locating vortex cores.

Figure~\ref{fig:bp_located} demonstrates that the phase jumps predicted from the SPA decomposition accurately locate the branch points inside the classical caustic. The first branch points to form are a pair outside the classical caustic line, owing to interference from the partial reflection of the wavefunction due to the quantum pressure term in equation~\eqref{eqn:quantum_bernoulli}. We will first focus on the interior branch points and their relation to multistreaming and return to the exterior branch points in Section~\ref{sec:catastrophe_theory}.

\subsubsection{Conserved quantities associated with vorticity}
In a classical fluid system, the Kelvin-Helmholtz theorem \parencite{Helmholtz1858,Kelvin1869} requires that the circulation of the velocity field,
\begin{equation}
    \Gamma = \oint_{\gamma(a)}\bm{v}\cdot \dd{\bm{x}} = \int_{\Sigma(a)} (\bm{\nabla} \times \bm{v}) \cdot \hat{\bm{n}}\dd{A},
\end{equation}
is conserved. Here $\gamma(a)$ is a loop, bounding a surface $\Sigma(a)$ at time $a$ with associated normal vector $\hat{\bm{n}}$ and area element $\dd{A}$. The integration surface is taken to be transported by the fluid flow from an initial surface $\Sigma^{\rm (ini)}$. In particular, this means that an initially irrotational flow will conserve its initial value,
\begin{equation}
    \Gamma = \int_{\Sigma^{\rm (ini)}}(\bm{\nabla} \times \bm{v}^{\rm (ini)}) \cdot \hat{\bm{n}}\dd{A} = 0\,.
\end{equation}
This requires that any vorticity produced in the multistream regime arises only from stream averaging. 

Under certain circumstances the conservation of circulation also applies to quantum and semiclassical systems \parencite{Damski2003}. For sufficiently smooth initial conditions one can apply the same definition of $\Gamma$, using the velocity defined by $\bm{j}/\rho$ and ensuring that the integral contour only goes through regions where the velocity is well defined, as we did in equation~\eqref{eqn:quantised_vorticity}. However, for a loop $\gamma(a)$, the conservation of circulation only holds if such a loop initially only goes through points where the velocity is well defined, and will evolve only through such points \parencite{Damski2003}, thus excluding the branch point locations, where a relaxed quantisation condition~\eqref{eqn:quantised_vorticity} applies.
For an initially irrotational system with $\Gamma=0$ these dislocations must occur in pairs  \parencite[called rotons][]{Savchenko1999} such that
\begin{equation}
    \int (\bm{\nabla} \times \bm{v}) \cdot \hat{\bm{n}} \dd{A} = 2\pi (n_+ - n_-) \hbar = 0\,, \quad n_\pm \in \mathbb{N}\,,
\end{equation}
if the surface is taken to be the entire space to ensure global conservation. The 1+1-dimensional system shown in Figure~\ref{fig:bp_located} resembles this pair creation property, with the spacetime circulation being globally conserved throughout the evolution.

For the 1+1D toy model, the circulation is trivially satisfied on an interval and the appropriate conserved quantity is the Poincar\'e-Cartan invariant from Hamiltonian dynamics (e.g. Chapter 9 of \textcite{Arnold1978_classicalmechanics} or Chapter 3 of \textcite{heller_semiclassical_2018}). As our semiclassical model has a natural Hamiltonian structure (see e.g. equation~\eqref{eqn:schrodinger_with_potential}), this conserved quantity remains valid for higher dimensions. The Poincar\'e-Cartan invariant is formulated by extending the phase-space, relying on the time $a$ as an extra variable conjugate to (minus) the Hamiltonian $\mathscr{H}$. The quantity
\begin{equation}\label{eqn:PC_invariant}
    \tilde{\Gamma}(\gamma) = \oint_\gamma \bm{v} \cdot \dd{\bm{x}} - \mathscr{H}\dd{a},
\end{equation}
is then conserved along the Hamiltonian flow and generalises the circulation theorem from fluid dynamics.

\begin{figure}[h!t]
    \centering
    \includegraphics[width=0.5\columnwidth]{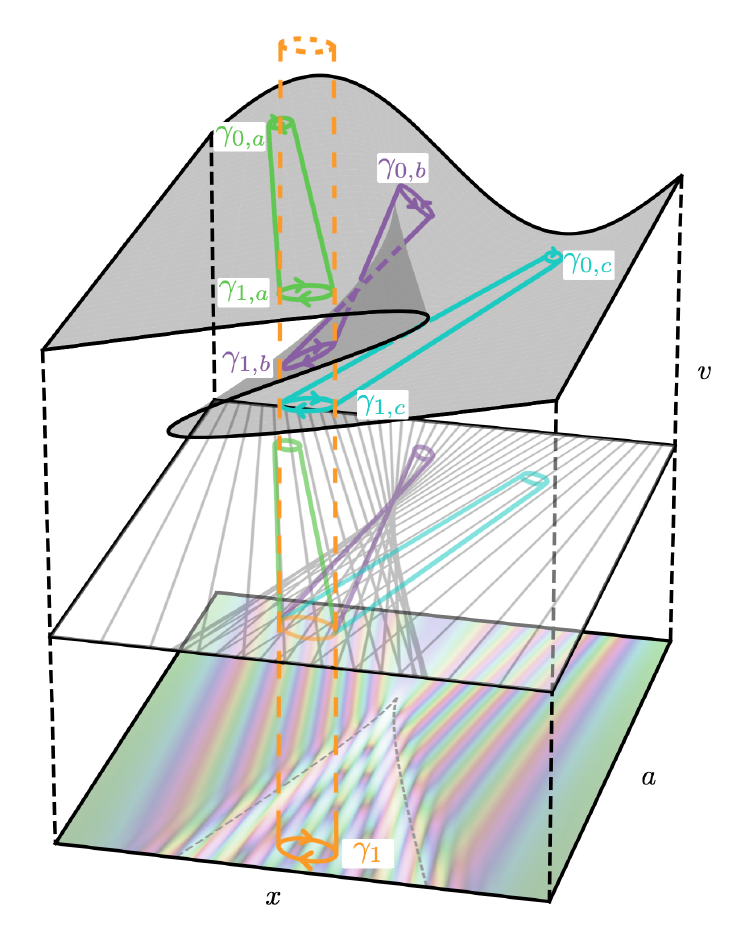}
    \caption[Illustration of inadmissible spacetime loop for the Poincar\'e-Cartan invariant.]{Illustration of an inadmissible spacetime loop for the Poincar\'e-Cartan invariant. Shown are the phase-space sheet for the system (top, as in Figure~\ref{fig:zeldo_phase_sheet}), together with projections onto spacetime for both classical trajectories (middle), and the wave mechanical model (bottom). The orange contour $\gamma_1$ drawn in spacetime is not appropriate to calculate the invariant, as it corresponds to three different loops in phase-space, labelled $\gamma_{1,i}, i\in\{a,b,c\}$. As such, the orange spacetime loop cannot be evolved uniquely by Hamiltonian flow to relate it to another loop via the invariant. Since the $\gamma_{1,i}$ loops are local in phase-space, each has a unique time evolution (shown on both the phase-space sheet and in projection), and can be evolved backwards by Hamiltonian flow to loops $\gamma_{0,i}$. The invariant is conserved along each of these loops individually, e.g. $\tilde{\Gamma}(\gamma_{0,i})=\tilde{\Gamma}(\gamma_{1,i})$.  Admissible loops drawn in spacetime must stay in the singe-stream region (classically), and must not include branch points (in the quantum case), which guarantees unique evolution when projected to phase-space. This Figure was published in \textcite{Gough.Uhlemann_2022_MakingDark}. }
    \label{fig:PC_invariant}
\end{figure}
Figure~\ref{fig:PC_invariant} shows a schematic diagram of loops in phase-space, demonstrating that not all loops in spacetime can be used to calculate the Poincar\'e-Cartan invariant. The orange contour in spacetime corresponds to three different loops in phase-space, and cannot be evolved through Hamiltonian flow. This essentially amounts to asking ``which $\bm{v}$'' to chose in equation~\eqref{eqn:PC_invariant}. Loops in spacetime which are entirely outside the classically shell crossed region (and stay away from the outside branch points) are uniquely mapped to loops on the phase-space sheet, and therefore can be related to their initial values. This can be seen by considering a loop in the bottom projection of Figure~\ref{fig:PC_invariant} in spacetime fully outside the caustic and projecting it upwards. Such a loop intersects the phase-sheet exactly once, which then allows unique time evolution via the Hamiltonian flow. Such loops conserve the value of the Poincar\'e-Cartan invariant $\tilde{\Gamma}$ until the Hamiltonian flow makes them intersect the classical caustic.

\subsection{Hidden features beyond a perfect fluid}\label{sec:hidden_features}

We now show that it is possible to decouple the presence of these phase jumps and the associated oscillating density from an ``average'' wavefunction capturing the stream-averaged fluid density and velocity. 

\subsubsection{The average wavefunction from SPA decomposition}

The stationary phase decomposition provides a way to construct the ``average wavefunction'' which captures the bulk fluid dynamics. We can construct the density weighted mean velocity in the multistream region, and require the phase of the ``average wavefunction'' produces this mean velocity.
As seen in Section~\ref{subsec:SPAstreamsZeldo}, the densities and velocities of the individual stationary phase wavefunctions are exactly those expected classically. Thus, we can straightforwardly construct the density-weighted mean velocity from the stationary phase streams 
\begin{subequations}
\label{eq:average_phase}
\begin{equation}
    \bar{v}(x,a) = \frac{\sum\limits_i \rho_i(x,a) v_i(x,a)}{\sum\limits_j \rho_j(x,a)}\,,
\end{equation}
where these sums run over the appropriate number of streams at position $(x,a)$. We define an ``average velocity potential'' for this mean velocity by $\bar{v} = \nabla_{\! x}\phi_{\rm avg}$ and calculate this in Fourier space,
\begin{equation}
    \phi_{\rm avg} = \mathcal{F}^{-1}\left[\frac{1}{ik^2}k\cdot \mathcal{F}[\bar{v}]\right],
\end{equation}
\end{subequations}
where $\mathcal{F}[\cdot]$ is a fast Fourier transform. From this smooth potential we construct an ``average wavefunction'' 
\begin{equation}
\label{eq:psi_avg}
    \psi_{\rm avg} = \sqrt{\sum_i \rho_i} \exp(\frac{i}{\hbar}\phi_{\rm avg}),
\end{equation}
representing a hypothetical perfect fluid with the classical density and mean velocity. 

\begin{figure*}[h!t]
    \centering
    \includegraphics[width=\columnwidth]{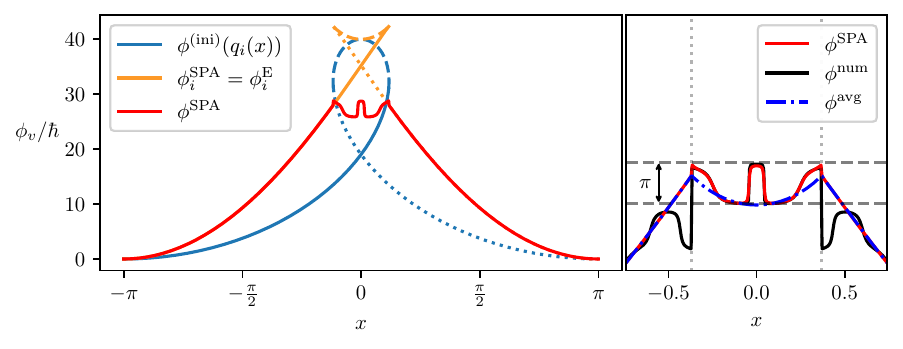}
    \caption[Phases of individual stream and stream-average wavefunctions.]{(Left) The phases of individual streams of the wavefunction model (with $\hbar=0.05$) past shell crossing, at the time of the first interior phase jump. We show the phase of the individual SPA stream wavefunctions, $\phi^{\rm SPA}_i$ (which are equal to the Eulerian velocity potentials from equation~\eqref{eq:velpot_Eulerian}) in comparison to the Lagrangian phase $\phi^{\rm (ini)}(q_i)$ plotted against the Eulerian position $x(q_i)$. In the single-stream region, the Eulerian and SPA phases are equal (the red line completely overlaps the orange). The linestyles indicate the portions corresponding to different initial positions/stationary points $q_i(x,a)$. The phase of the full SPA wavefunction is also shown. (Right) A zoomed in view of the phase in the shell crossed region. The phase of the numerical wavefunction, $\phi^{\rm num}$, is shown in comparison to the stationary phase approximation. Note that the discontinuity in $\phi^{\rm num}$ at the caustic boundary is not physical, the phase has simply been shifted up by $2\pi$ to align the profile with $\phi^{\rm SPA}$. The phase associated with the mean fluid velocity, $\phi_{\rm avg}$, well describes the overall profile, allowing the phase jumps to be isolated in $\phi_{\rm hid} = \phi_v-\phi_{\rm avg}$. This Figure was published in \textcite{Gough.Uhlemann_2022_MakingDark}. }
    \label{fig:zeldo_loop_avg_phase} 
\end{figure*}

Figure~\ref{fig:zeldo_loop_avg_phase} shows the individual phases of the SPA wavefunctions, together with the phase of the full SPA at the time of an interior phase jump. Additionally, we show the Lagrangian velocity potential evaluated at a Eulerian position $\phi^{\rm (ini)}_i = \cos(q_i(x,a))/\hbar$ in blue and the proper Eulerian velocity potential, which agrees with the phases of the individual SPA streams $\phi_i^{\rm (SPA)}$ in orange. The right panel zooms into the shell crossed region, comparing the full numerical phase of the wavefunction (black) and the stationary phase approximation to the phase (red), which match extremely well. The phase associated with the mean velocity (blue dot-dashed) well describes the overall profile of the phase.

We will show that the hidden features of the SPA wavefunction describing the wave-mechanical system decorate the Zel'dovich density and velocity, encoded in an average wavefunction from equation~\eqref{eq:psi_avg}, with wave interference effects. Those interference effects can be captured in a ``hidden wavefunction''
\begin{equation}
    \label{eq:psi_hidden}
    \psi_{\rm hid}=\frac{\psi_{\rm SPA}}{\psi_{\rm avg}}=\sqrt{\rho^{\rm hid}
}\exp\left(\frac{i}{\hbar}\phi_{\rm hid}\right),
\end{equation}
which produces multistream phenomena of a non-potential velocity from the phase jumps and a velocity dispersion from the oscillating density.

\subsubsection{Hidden phase associated with non-potential velocity}

We seek a decomposition of the phase of the wavefunction into a smooth and a ``hidden''  part, where the smooth part of the phase produces the potential part of the velocity field. The remaining ``hidden'' phase then contains the discontinuities which, in 2- or 3-dimensional systems would source vorticity. Such a phase decomposition has already been put forward for 2-dimensional optical systems in \textcite{Fried1998}, as reviewed in Appendix~\ref{app:hid_phase_fried}. That form of the hidden phase is not appropriate for our 1-dimensional toy model, but could be applicable to 2-dimensional extensions.

We take the smooth part of our phase to be the phase of the average wavefunction, defined in equation~\eqref{eq:average_phase}. We then define our hidden phase as the difference between the full phase of the wavefunction (either predicted by stationary phase or calculated numerically) and this smooth phase to isolate the phase jumps,
\begin{equation}
    \phi_{\rm hid} = \phi_v - \phi_{\rm avg}\,.
\end{equation}

\begin{figure*}[h!t]
    \centering
    \includegraphics[width=\columnwidth]{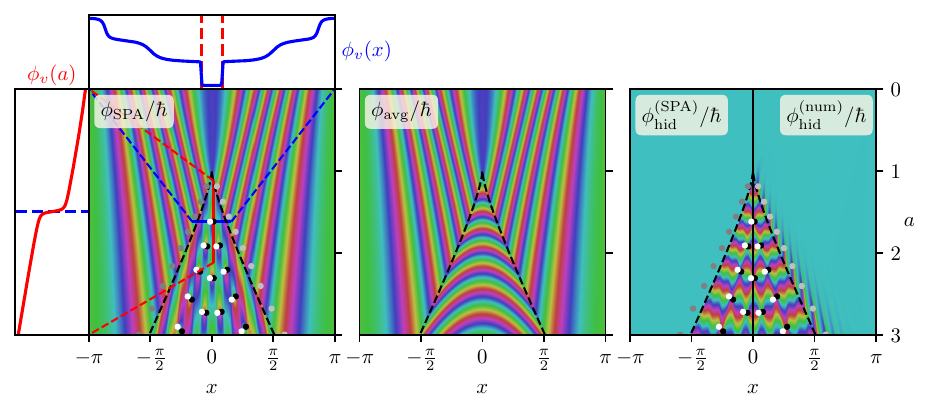}
    \caption[Spacetime evolution of the average and hidden wavefunctions.]{The spacetime evolution of the polar angle of the full stationary phase wavefunction (with $\hbar=0.05$, (left)), the ``average phase'' associated with the mean velocity (middle), and the discontinuous ``hidden phase'' (right). The left panel also shows the spatial and temporal phase profiles through a branch point (both of which jump by $\pm\hbar\pi$ upon crossing the branch point). The hidden phase in the right panel is calculated as using the stationary phase approximation, $\phi_{\rm hid}^{\rm (SPA)}=\phi_{\rm SPA}-\phi_{\rm avg}$ (left side), and numerically, $\phi_{\rm hid}^{\rm (num)}=\phi_{\rm num}-\phi_{\rm avg}$ (right side). The black and white points shown in the outer panels are the numerically located branch points of the full wavefunction, coloured according to their spacetime circulation as in Figure~\ref{fig:bp_located}. The hidden phase in the numerical case does not identically vanish outside the cusp, due to the presence of branch points outside the cusp which stationary phase analysis cannot produce (which are coloured light and dark grey to emphasise this point). This Figure was published in \textcite{Gough.Uhlemann_2022_MakingDark}. }
    \label{fig:phase_profiles_and_hid}
\end{figure*}

Figure~\ref{fig:phase_profiles_and_hid} shows both the smooth average phase $\phi_{\rm avg}$ and the jumpy hidden phase $\phi_{\rm hid}$ over the entire spacetime region. We see that if the hidden phase is constructed as $\phi_{\rm hid}^{\rm (SPA)}=\phi_{\rm SPA} - \phi_{\rm avg}$, then it vanishes identically in the single-stream region. This is expected as the average velocity before shell crossing is identical to the single-stream velocity. If we instead construct the hidden phase using the phase of the numerically solved wavefunction, $\phi_{\rm hid}^{\rm (num)}=\phi_{\rm num}-\phi_{\rm avg}$, then presence of branch points outside the classical caustic causes a nonzero hidden phase  in the single-stream region, which the stationary phase analysis cannot capture (as seen in the final panel of Figure~\ref{fig:phase_profiles_and_hid}).

\subsubsection{Hidden density associated with velocity dispersion}

Analogous to the hidden phase, we can write a ``hidden density'' as the part of the density which dresses the Zel'dovich dynamics in wave-phenomena. The hidden density $\rho^{\rm hid } = \abs{\psi^{\rm SPA}}^2/\sum_i \rho_i$, describes the oscillations which trace the Zel'dovich density.

\begin{figure}[h!t]
    \centering
    \includegraphics[width=0.5\columnwidth]{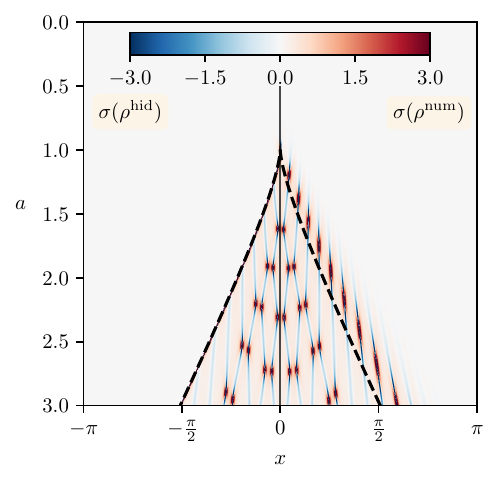}
    \caption[The spacetime evolution of the scalar velocity dispersion associated with the wavefunction.]{The scalar velocity dispersion $\sigma$ associated with the wavefunction $\psi$ (for $\hbar = 0.05$) calculated by equation~\eqref{eqn:sigma_ij} using the hidden density $\rho^{\rm hid} = \abs{\psi^{\rm SPA}}^2/\sum_i \rho_i$ (left) and the full (numerically solved) density $\rho^{\rm num} = \abs{\psi^{\rm num}}^2$ (right). The curvature near the zeros in the density field (which correspond to the branch points) sources velocity dispersion. This Figure was published in \textcite{Gough.Uhlemann_2022_MakingDark}. }
    \label{fig:velocity_dispersion}
\end{figure}

The velocity dispersion associated with a wavefunction is defined in equation~\eqref{eqn:sigma_ij}, and is generally a tensor. However, in one spatial dimension, the velocity dispersion is simply a scalar quantity, sourced by the curvature in the (log) density,
\begin{equation}\label{eqn:scalar_vel_disp}
    \sigma = \frac{\hbar^2}{4}\nabla_{\! x}^2 \ln{\rho}\,.
\end{equation}
Figure~\ref{fig:velocity_dispersion} shows this scalar velocity dispersion using both the full numerically evolved density $\rho^{\rm num}$, and the hidden density $\rho^{\rm hid}$. We see that the velocity dispersion is well captured by the hidden density in the multistream region, and that a positive velocity dispersion is sourced by the positive curvature in the density at the locations of the branch points. 

Since the velocity dispersion is proportional to the curvature of the density, the minima of $\sigma$ seen in Figure~\ref{fig:velocity_dispersion} correspond to the maxima of the density field, while the maxima of $\sigma$, are sourced by the minima (the zero points) of the oscillating density. When coarse grained over, these small regions of negative velocity dispersion will be compensated by the positive velocity dispersion at the branch point, and will reproduce the velocity dispersion seen in the Zel'dovich approximation, for example as seen in Figure 1 of \textcite{BuehlmannHahn2019}.

The fact that the velocity dispersion can become negative in the wave-mechanical case owes to the fact that is properly thought of as a modification to the Newtonian potential, or an stress-energy term, in the fluid equations~\eqref{eqn:quantum_fluids}. In particular, the Euler equation associated with equation~\eqref{eqn:quantum_bernoulli} reads
\begin{align}
    \partial_a v_i + (\bm{v}\cdot\bm{\nabla})v_i &= - \nabla_{\! i}\left(-\frac{\hbar^2}{2}\frac{\nabla^2\sqrt{\rho}}{\sqrt{\rho}}\right)  \nonumber \\
    &= -\frac{1}{\rho}\nabla_{\! j}\left(\rho\sigma_{ij}\right) \,.
\end{align}
Interpreted in this way, as an additional source of stress-energy, $\sigma_{ij}$ can be negative in the wave-mechanical model.

To make this connection to the internal energy of the system more explicit, consider the expectation value of the quantum pressure $Q$ defined in equation~\eqref{eqn:quantum_pressure}. This can be written as the integral (in $n$ spatial dimensions)
\begin{subequations}
\begin{equation}
    \langle Q \rangle = \int \dd[n]{\bm{x}} \rho(\bm{x}) Q(\bm{x})
    =\int \dd[n]{\bm{x}} \rho \cdot \frac{\hbar^2}{2}\frac{(\bm{\nabla}\sqrt{\rho})^2}{\rho}\,,
\end{equation}
making use of $\rho=|\psi|^2$ and integration by parts  in the second step. We recognise this new part of the integrand as the internal energy of a wavefunction \parencite{Yahalom2018MolPh}
\begin{equation}
    \varepsilon_q = \frac{\hbar^2}{2}\frac{(\bm{\nabla}\sqrt{\rho})^2}{\rho} = \frac{\hbar^2}{8}(\bm{\nabla}\ln\rho)^2\,,
\end{equation}
such that $\langle Q \rangle = \langle \varepsilon_q \rangle$. In addition to this global relation between the quantum pressure and the internal energy, we can locally relate them to the trace of the velocity dispersion
\begin{equation}
    \sigma_{ii} = Q + \varepsilon_q\,.
\end{equation}
\end{subequations}

For our 1+1D model, the scalar velocity dispersion $\sigma$ is determined by the hidden wavefunction (as seen in Figure~\ref{fig:velocity_dispersion}), which gives rise to the internal energy via the 1-dimensional versions of these relations between $\sigma_{ij}$, $Q$, and  $\varepsilon_q$.

\bigskip

In this Section we directly examined the non-potential velocity and velocity dispersion, to illustrate the principal effects of the phase jumps and the oscillatory density. Higher order cumulants are determined similarly, through higher derivatives of the wavefunction, which will be largely driven by the ``hidden wavefunction'' introduced here. We have seen that the hidden wavefunction can accurately locate the interior branch points and reproduce the velocity dispersion associated with the vanishing density near those points. These interior branch points are the only ones relevant in the classical limit as they correspond to the classically shell crossed region with multistreaming phenomenology. In the next section we examine the exterior branch points, and other features which the stationary phase approximation alone cannot capture.

\section{Catastrophe theory}\label{sec:catastrophe_theory}

The caustic region formed by our wavefunction is very analogous to caustics formed by optical wave fields undergoing focusing. Leveraging this mathematical similarity, we turn to the theory of diffraction catastrophes, which classifies the types of stable caustics and their properties, to understand the regions near to the caustic, precisely where the stationary phase decomposition fails.

\subsection{Classical catastrophe theory}

Classical catastrophe theory describes the singularities of differentiable mappings. For example, for a Zel'dovich Fourier mode, the phase-space to position space mapping $(v,x,a) \mapsto (x,a)$ develops singularities where the sheet folds over itself, as well as at the singular point where the sheet starts to twist. Importantly however, catastrophe theory is concerned with \emph{stable} singularities, where perturbing the mapping only slightly does not remove the character of the singular point. It does this by determining a set of standard \emph{generating polynomials}, $\zeta(\bm{s};\bm{C})$, which generate differentiable mappings such that all stable singularities look locally like one of these standard forms. The parameters $\bm{C} = (C_1, \dots, C_m)$ are called the \emph{control parameters}, and $\bm{s} = (s_1, \dots, s_n)$ are called the \emph{state parameters}. The control parameters are quantities on which the classical trajectories/rays depend, such as spatial position or describing the surface initial conditions are provided on. The state parameters are internal, and define the gradient mapping which produces the singularity. Roughly, it is the stationary points of $\zeta$ with respect to $\bm{s}$ which determine the number of classical trajectories/rays which meet at the singularity.

The elementary catastrophes were originally classified and named in \textcite{Thom1994}, but a more systematic and complete classification is due to Arnol'd \parencite{Arnold1973, Arnold1975}, which provided each generating polynomial a symbol related to the Coxeter reflection groups. In 1+1 dimensions, the only stable caustics are (using Thom's naming system) the fold and the cusp corresponding to the generating polynomials
\begin{align}
    \zeta_{\rm fold}(s; C_1) &= \frac{s^3}{3}+ C_1 s\,, \label{eqn:zeta_fold}\\
    \zeta_{\rm cusp}(s; C_1, C_2) &= \frac{s^4}{4} + C_2 \frac{s^2}{2} + C_1 s\,. \label{eqn:zeta_cusp}
\end{align}

Both of these singularities are present in the Zel'dovich phase-space sheet in Figure~\ref{fig:zeldo_phase_sheet}. The fold point corresponds to the point at shell crossing, where the sheet begins to twist, corresponding to three trajectories intersecting. This fold point exists for a single moment in time, before splitting into a pair of fold lines which move apart from each other (called the ``unfolding'' of the catastrophe). Each point on the fold is characterised by the intersection of two trajectories. In this way, we have the heuristic identification of the spacetime parameters $(x,a)$ relating to the control parameters $(C_1, C_2)$, and the velocity phase-space variable playing the role of the state parameter $s$. Equally, one could consider the singularity in the $(q,x,a)\mapsto (x,a)$ mapping, which has the same structure as the phase-space sheet in Figure~\ref{fig:zeldo_phase_sheet}, treating the initial position $q$ as the state parameter. This second view will prove to be more directly useful in relating the integral form of our wavefunction to the standard catastrophe forms. We will refine this correspondence in the following Sections.

This classical catastrophe theory has been applied to the cosmic web skeleton of caustics under the Zel'dovich approximation \parencite{ArnoldShandarinZeldovich1982, Hidding2014, Feldbrugge2014, Feldbrugge2018}. In these, the classical Zel'dovich divergences and their statistics are classified and identified by examining the eigenvalues and eigenvectors of the deformation tensor, the gradient of the Lagrangian displacement field $\bm{\xi}$, with respect to Lagrangian coordinates. In the Zel'dovich approximation, this deformation tensor directly encodes the tidal information of the initial gravitational potential.

\subsection{Diffraction catastrophe theory}

Instead of simply defining smooth surfaces and differentiable mappings, the generating polynomials $\zeta$ can instead be used to generate oscillatory integrals. This allows the classification of classical caustics to be extended into the classification of caustics associated with wave phenomena. We refer to such integrals as diffraction catastrophe integrals. These integrals retain the underlying skeleton from the classical catastrophes they are built from, but now dressed with wave interference. Such systems were studied series of optical experiments in the 1970s \parencite{Berry1977_focusing, Berry1977_finestructure,BerryNyeWright1979, BerryUpstill1980}.

The optical field, $u$, for a monochromatic source with frequency $\nu$ will act as an analogue for the wavefunction $\psi$ with ``quantumness'' $\hbar$ (such that $\hbar$ acts like an inverse frequency). The eikonal ($\nu \to \infty$) limit plays the role of the semiclassical ($\hbar\to 0$) limit. The intensity of the optical field is given by $\abs{u}^2$, and interference in the optical field is produced by mixing of the complex phase. A standard diffraction catastrophe integral is of the form\footnote{The prefactor of $\nu^{n/2}$ in this integral is necessary compared to the dimensionless forms listed in e.g. \textcite{BerryUpstill1980} so that far from the catastrophe, the stationary phase approximation removes the $\nu$ dependence of the amplitude.} (notice the resemblance to equation~\eqref{eqn:psi_oscillatory})
\begin{equation}
    u(\bm{C};\nu) = \left(\frac{\nu}{2\pi}\right)^{n/2}\int \dd[n]{\bm{s}} \exp\left[i\nu\zeta(\bm{s};\bm{C})\right].
    \label{eqn:diffraction_integral}
\end{equation}

As we have already seen, these oscillatory integrals will be mainly dominated by their stationary points, except very close to the singular point. Since we are always interested in the stationary points relative to the state parameters, by $\zeta'(\bm{s};\bm{C})$ we will always mean $\bm{\nabla}_{\! \bm{s}}\zeta(\bm{s};\bm{C})$. This is precisely how the universality of catastrophe theory arises, as for any function $\zeta$ which has a singularity, the integral will be dominated by the form of $\zeta$ near the singular point in the $\nu \to \infty$ limit. Therefore, $\zeta$ will be dominated by the leading order terms in its Taylor expansion about that point. For any system which includes a given catastrophe, a (smooth) coordinate transform bringing the expansion into one of the standard generating polynomials is guaranteed to exist \parencite{Arnold2012_catastrophebook}.

\subsubsection{Properties of diffraction integrals}

Diffraction catastrophe theory also classifies certain topological properties of these integrals, which are invariant under diffeomorphism. This means that while the initial conditions in our wavefunction do not exactly give rise to any of the standard generating polynomials $\zeta(\bm{s};\bm{C})$ with their propagation and initial conditions, certain properties of the wavefunction are preserved in the coordinate change to bring them into standard form.

The relevant topological quantities for our interests are the indices which describe the maximum intensity of the caustic (which corresponds to the amount of regularisation the Zel'dovich divergences receive), and the fringe spacing in different directions. These scalings can be recovered by changing coordinates within the generating polynomial $\zeta(\bm{s};\bm{C})$ to remove the frequency dependence from the integrand. Doing so results in a relation of the following form
\begin{equation}
    u(\bm{C};\nu) = \nu^{\beta}u(\nu^{\sigma_1}C_1,\dots,\nu^{\sigma_m}C_m;\nu=1)\,,
\end{equation}
where now the quantities $\tilde{C}_i = \nu^{\sigma_i}C_i$ appearing on the right hand side are control parameters with dimensions of $\nu$ restored (note that the associated power of $\nu$ for each control parameter will be different).

The index $\beta$, called the \emph{singularity index} \parencite[introduced in][]{Arnold1975}, describes the maximum intensity at the catastrophe, since the intensity of the light scales as
\begin{equation}
    \abs{u}^2 \propto \nu^{2\beta}\,.
\end{equation}
The indices $\sigma_i$ are called the \emph{fringe exponents},  \parencite[introduced in][]{Berry1977_focusing}, and control the spacing of the fringes in the $i^{\rm th}$ control parameter direction. Note that since the dimensional coordinates scale as $\tilde{C}_i\sim \nu^{\sigma_i}C_i$, the fringe spacing in the $i^{\rm th}$ direction is $\order{\nu^{-\sigma_i}}$ in physical coordinates as $\nu\to\infty$. The sum of the fringe exponents, $\gamma = \sum_i \sigma_i$, is called the \emph{fringe index} and describes the (hyper-)volume scaling of the size of the singularity in control space. All the indices $\beta$, $\sigma_i$, and $\gamma$ are invariant under diffeomorphism.

\subsubsection{The fold catastrophe}

The fold catastrophe is the simplest non-trivial catastrophe, with just a single control parameter. The diffraction integral of the fold is given by using the fold generating polynomial~\eqref{eqn:zeta_fold} in the diffraction catastrophe integral~\eqref{eqn:diffraction_integral}
\begin{equation}
    u_{\rm fold}(C_1;\nu) = \sqrt{\frac{\nu}{2\pi}} \int_{-\infty}^{\infty} \dd{s} \exp\left[i\nu\left(\frac{s^3}{3}+C_1 s\right)\right].
\end{equation}
As an illustrative example, we can obtain  the singularity and fringe indices by changing the integration variable to $s=\nu^{-1/3}t$ to remove the $\nu$ dependence from the highest order term in the integrand. This leaves
\begin{subequations}
\begin{align}
    u_{\rm fold}(C_1;\nu) &= \frac{\nu^{1/6}}{\sqrt{2\pi}}\int_{-\infty}^\infty \dd{t}\exp\left[i\left(\frac{t^3}{3}+\nu^{2/3}C_1 t\right)\right] \nonumber \\
    &= \nu^{1/6} u_{\rm fold}(\nu^{2/3}C_1;\nu=1) = \sqrt{2\pi} \nu^{1/6} \operatorname{Ai}\left[\nu^{2/3}C_1\right],
\end{align}
\end{subequations}
which recovers a singularity index $\beta=1/6$ and the fringe exponent (and index) $\sigma_1 = \gamma = 2/3$. We also note that the fold catastrophe is simply the well known Airy function (up to appropriate normalisation).

\subsubsection{The cusp catastrophe}

The cusp catastrophe has two control parameters, and is given by the using the cusp generating polynomial~\eqref{eqn:zeta_cusp} in the diffraction catastrophe integral~\eqref{eqn:diffraction_integral}
\begin{subequations}
\label{eq:int_cusp}
\begin{align}
    u_{\rm cusp}(C_1, C_2;\nu) &= \sqrt{\frac{\nu}{2\pi}}\int_{-\infty}^{\infty} \dd{s} e^{i\nu\zeta_{\rm cusp}(s;C_1,C_2)}\,,  \\
    \zeta_{\rm cusp}(s;C_1,C_2) &= \frac{s^4}{4} + C_2\frac{s^2}{2} + C_1 s\,.
\end{align}
\end{subequations}

By changing the integration variable $s=\nu^{-1/4}t$, we can remove the $\nu$ dependence in the same way as with the fold and recover the catastrophe indices. This produces
\begin{subequations}
\begin{align}
    u_{\rm cusp}(\bm{C}; \nu) &= \nu^{1/4}u_{\rm cusp}(\nu^{3/4}C_1, \nu^{1/2}C_2;\nu=1)\,, \\
    &= \frac{(4\nu)^{1/4}}{\sqrt{2\pi}}\operatorname{Pe}(4^{1/4}\nu^{3/4}C_1, \nu^{1/2}C_2)\,,
\end{align}
\end{subequations}
where $\operatorname{Pe}(A, B)$ is the Pearcey integral \parencite{Pearcey1946},
\begin{equation}
    \operatorname{Pe}(A,B) = \int_{-\infty}^\infty \dd{y} \exp(i(y^4 + Ay^2 + By))\,.
\end{equation}
This analysis gives us a singularity index $\beta = 1/4$, and the fringe exponents $\sigma_1 = 3/4, \sigma_2=1/2$, resulting in a fringe index $\gamma = 5/4$ for the cusp catastrophe. 

\begin{figure}[h!t]
    \centering
    \includegraphics[width=0.5\columnwidth]{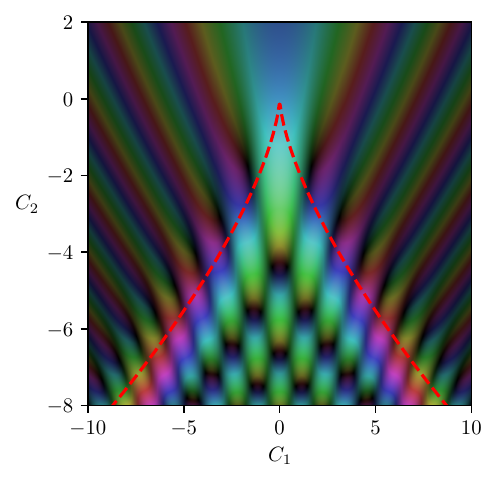}
    \caption[The cusp catastrophe integral.]{The cusp catastrophe $u_{\rm cusp}(C_1, C_2;\nu=1)$ from equation~\eqref{eq:int_cusp} in control parameter space, coloured according to the same domain colouring as the wavefunction. This is numerically calculated using the contour shifting technique described in Section~\ref{sec:multi-streaming and interference}. This Figure was published in \textcite{Gough.Uhlemann_2022_MakingDark}. }
    \label{fig:cusp_integral_numeric}
\end{figure}

This cusp catastrophe integral, shown in Figure~\ref{fig:cusp_integral_numeric}, appears visually very similar to the spacetime evolution of the wavefunction with Zel'dovich initial conditions, with a cusped line separating a three ray region from a single ray region.

The equation of the caustic line is determined through the joint conditions
\begin{equation}
    \zeta'(s;\bm{C}) = 0\,, \quad \zeta''(s;\bm{C}) = 0\,.
\end{equation}
where the first condition says that you lie on the phase-space sheet (a cubic equation in this case) and the second condition is where the projection mapping becomes singular. In control parameters (eliminating $s$), the classical caustic line is given by
\begin{equation}\label{eqn:diff_cusp_caustic_condition}
    4C_2^3 + 27 C_1^2 = 0\,,
\end{equation}
a semi-cubic parabola.

\subsection{Mapping the wavefunction to standard catastrophes}\label{sec:fringe_properties_of_wavefunction}

We now wish to map the full wavefunction $\psi(x,a)$ of our single Fourier mode to these standard caustic forms. The role of the wavelength (inverse frequency) from the diffraction catastrophes is now played by the size of $\hbar$.

\begin{figure}[h!]
    \centering
    \includegraphics[width=\columnwidth]{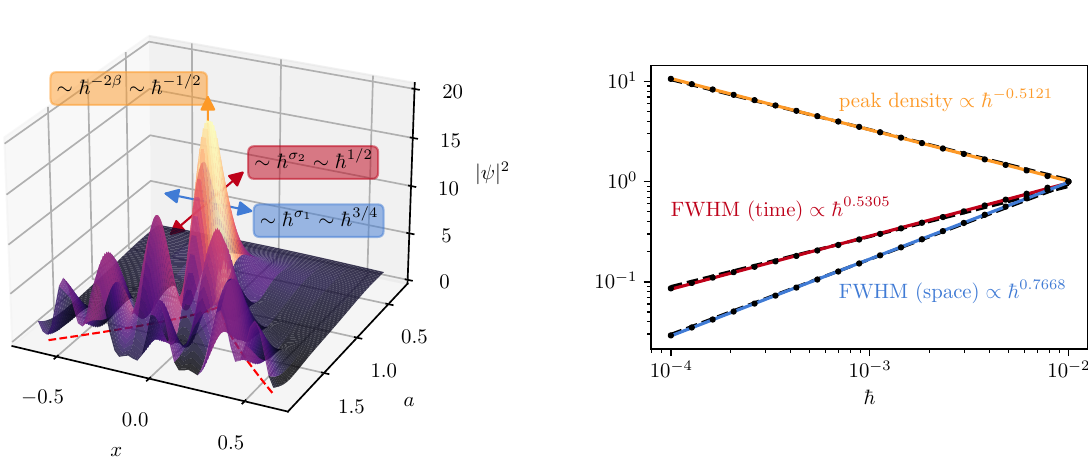}
    \caption[The catastrophe theory indices and scalings for the wavefunction. ]{Density of the wavefunction with Zel'dovich initial conditions for $\hbar = 0.05$. This demonstrates how the peak density, as well as the delay of shell crossing and the spatial width of the wavefunction at the cusp point scale with $\hbar$, related to the singularity and fringe indices for the cusp catastrophe. (Lower panel) The peak height and characteristic widths around the cusp point for different values of $\hbar$, relative to their values at $\hbar=0.01$. To characterise the temporal and spatial widths we take the full width half maximum on constant space or time surfaces through the point of maximum density. The fit $\hbar$ scalings for each of these properties are shown as the solid coloured lines, while lines with the slopes anticipated from the catastrophe indices (made to go through a central data point) are shown as black dashed lines, which are mostly covered by the coloured lines. The measured $\hbar$ exponents agree well with the catastrophe theory prediction over two orders of magnitude in $\hbar$.  This Figure was published in \textcite{Gough.Uhlemann_2022_MakingDark}. }
    \label{fig:3d_cusp_annotated}
\end{figure}

We write the full wavefunction using the free propagator as before
\begin{equation}\label{eqn:psi_for_catastrophe_expansion}
    \psi(x,a) = \mathcal{N}\int \dd{q} \exp\left(\frac{i}{\hbar}\left[\frac{(q-x)^2}{2a} + \cos(q)\right]\right).
\end{equation}
By Taylor expanding the cosine from the initial conditions to fourth order, we can recover a term quartic in $q$, which resembles the quartic generating function $\zeta_{\rm cusp}(s,\bm{C})$ (higher order terms in the Taylor expansion are suppressed by powers of $\hbar$, providing the universality promised by catastrophe theory). Scaling our coordinates so this leading order term is the same as the standard form, we can read off the mapping between the (dimensional) control parameters and our spacetime parameters
\begin{subequations}\label{eqn:control_of_spacetime}
\begin{align}
    \tilde{C}_1(x,a) &= -6^{1/4}\hbar^{-3/4}\frac{x}{a}\,, \\
    \tilde{C}_2(x,a) &= 6^{1/2}\hbar^{-1/2}\frac{1-a}{a}\,,
\end{align}
\end{subequations}
or their inversion
\begin{subequations}
\begin{align}
    a(\tilde{C}_1,\tilde{C}_2) &= \frac{1}{1+\sqrt{\hbar/6}\tilde{C}_2}\sim 1-\sqrt{\frac{\hbar}{6}}\tilde{C}_2\,, \label{eqn:c2_of_spacetime} \\
    x(\tilde{C}_1, \tilde{C}_2) &= -\frac{\tilde{C}_1}{\tilde{C}_2}\frac{\hbar^{3/4}}{6^{1/4}}\frac{1}{1+\sqrt{\hbar/6}}\sim -\frac{\tilde{C}_1}{\tilde{C}_2}\frac{\hbar^{3/4}}{6^{1/4}}\,.
\end{align}
\end{subequations}
The coordinate relations~\eqref{eqn:control_of_spacetime} recover the cusp fringe exponents, which we can see by expanding about the cusp point $(x,a)=(0,1)$. Writing $x = \delta x$ and $a = 1+\delta a$, and the coordinate relation becomes $\tilde{C}_1 \sim \hbar^{-3/4}\delta x$ and $\tilde{C}_2 \sim \hbar^{-1/2}\delta a$, leading to $\sigma_1 = 3/4, \sigma_2 = 1/2$ as expected from the cusp. The required coordinate transforms in equation \eqref{eqn:control_of_spacetime} are not simple rescalings, both because the cusp catastrophe happens at $a=1$ rather than 0, and a natural mixing due to the propagator containing $x/a$ terms. Applying this coordinate transformation, we obtain the following integral relationship between our wavefunction and the standard cusp integral,
\begin{align}
    \psi(x,a) \approx \exp&\left(i\left[\frac{x^2+2a}{2\hbar a}-\frac{\pi}{4}\right]\right)\frac{6^{1/4}}{a^{1/2} \hbar^{1/4}} \times \nonumber \\ 
    &u_{\rm cusp}(\tilde{C}_1(x,a),\tilde{C}_2(x,a);\nu=1)\,,
\end{align}
which correctly recovers the singularity index of $\beta = 1/4$. A schematic diagram showing the density from this wavefunction is shown in Figure~\ref{fig:3d_cusp_annotated}, together with the catastrophe scalings. We compare the numerical peak height, temporal width, and spatial width of the cusp peak over two orders of magnitude in $\hbar$. We use the full width half maximum (FWHM) on constant time/space surfaces through the maximum peak as a proxy for the characteristic widths. The best fit scaling exponents are in good agreement with the scalings anticipated by catastrophe theory, demonstrating that this analysis does extract the correct scaling behaviour.

In principle one could perform a similar set of coordinate transformations along each time slice after shell crossing, bringing the form of $\psi(x,a=\text{const}>1)$ into an integral relationship with $u_{\rm fold}$. Providing the details of the coordinate transforms required is tedious and provides little physical insight into the system beyond doing this for the cusp integral. Appendix~\ref{app:fold_coord_change} demonstrates how to perform these coordinate changes on the standard cusp integral to recover the fold contribution. While one could explicitly recover the relevant coordinate transformations for the wavefunction by similar methods, we can instead rely on the existence of such coordinate changes and simply write down the relevant scalings. The fold indices $\beta_{\rm fold} = 1/6$ and $\sigma_{\rm fold}=2/3$ now finally explain the scalings presented in Figure~\ref{fig:fold_annotated} back in Section~\ref{sec:multi-streaming and interference}.

The coordinate relation equation~\eqref{eqn:c2_of_spacetime} provides the delay in the onset of shell crossing from $a=1$, which scales as $\hbar^{1/2}$. They also provide the fact that the peak density reached for the wavefunction scales as $\hbar^{-1/2}$ (owing to $\beta_{\rm cusp} = 1/4$). From looking at constant time slices of the wavefunction post shell crossing, we see the peaks of the wavefunction are offset from the Zel'dovich caustic positions by $\hbar^{2/3}$ (the ``finite thickness'' of the caustic), and the peak intensity of the wavefunction at fixed time ($a>1$) scales as $\hbar^{-1/3}$ (as shown in Figure~\ref{fig:fold_annotated}).  Importantly, the fringe indices provide the characteristic widths which can provide zeros in the density outside the classically shell crossed region. This demonstrates that the exterior branch points occur on length scales $\order{\hbar^{\sigma_i}}$ from the classical caustic, for the appropriate fringe exponent. As $\sigma_i > 0$ for all the elementary catastrophes (see e.g. \textcite{BerryUpstill1980, Feldbrugge2019} for the $\sigma_i$ of catastrophes in higher dimensions), these exterior branch points do not survive the classical limit. This is an important recognition, as the position of these exterior dislocations cannot be predicted by a stationary phase analysis of the real roots alone, although they could be estimated by steepest descents including contributions of one complex root as in \textcite{BerryNyeWright1979}.

The (spatial) fringe spacings  in cylindrical collapse simulations performed by \textcite{Lague_2021_FDM_LPT} have been found to be well approximated by half the de Broglie wavelength, taking the velocity as the root-mean-square Zel'dovich velocity at shell crossing. Viewed as a scaling in $\hbar$, this result corresponds to $\sigma = 1$ in the language of catastrophe theory. This is larger than any of the fringe indices of the elementary catastrophes, which all have $\sigma < 1$. However, cylindrical collapse is a highly symmetric scenario, which can lead to very different scaling behaviour catastrophe theory cannot predict  \parencite[in the same way a spherical lens can achieve much stronger focusing which breaks when the lens is perturbed][]{BerryUpstill1980}. We suspect that perturbing the collapse from cylindrical would result in the focus splitting into separate caustics, each of which has scalings predicted by catastrophe theory.  

With these wave properties near the caustics, and the stationary phase decomposition described in Section~\ref{sec:unweaving_the_wavefunction}, we have a powerful dictionary characterising our wavefunction over all regions of spacetime.

\section{Conclusions}\label{sec:conclusion}

\subsection*{Summary}
 We have provided a detailed analysis of a single Fourier mode evolving under the Zel'dovich approximation making use of a wavefunction obeying the free Schr\"odinger equation. We demonstrate that the interfering region of the wavefunction, corresponding to classical multistreaming, can be resolved into a sum over three stream wavefunctions, each corresponding to a classical Zel'dovich stream as shown in Figure~\ref{fig:stream_splitting_phase_hbar0.05}. These stream wavefunctions are recovered as stationary points of the function $\zeta$ from equation~\eqref{eqn:zeta}, which encodes the dynamics of the propagation of the wavefunction, and its initial conditions. By visualising phase-space in Figure~\ref{fig:stream_split_wigner_overplot}, we show that this wavefunction decomposition naturally dissects the phase-space distribution into the classical streams, without having to explicitly construct a quantum phase-space. Such a decomposition validates the ``golden rule of semiclassics'' described in \textcite{heller_semiclassical_2018}: ``quantum amplitudes are to be approximated as the sum of square roots of classical probabilities, with phases given by classical actions.''

We analyse the interference properties of this Fourier mode, and provide a decomposition of the phase into a smooth average potential, which encodes the stream-averaged velocity of the fluid, and a discontinuous ``hidden phase'' which would source vorticity in a 2- or 3-dimensional system. These phase jumps accompany the zeros in the density field, which source velocity dispersion in the interference region and can be captured in a ``hidden density''.

In the regions near the high density caustics, where this separation into Zel'dovich streams fails, we have made explicit connections between our semiclassical wave model and standard ``diffraction catastrophes'', which  classify the wave phenomena arising around caustics. In particular, this predicts the peak density and the fringe spacing about the Zel'dovich divergences illustrated in Figure~\ref{fig:fold_annotated} and \ref{fig:3d_cusp_annotated}.  While the exact fringe spacings and density profiles will not follow these catastrophe scalings in the fully non-linear dynamics, we expect that their qualitative features should persist on mildly-nonlinear scales. The full classification of diffraction integrals provides a universality to these results, at least in 1+1 dimensions.

\subsection*{Outlook} This Chapter has focused on a simple toy model in 1+1 dimensions, under the Zel'dovich approximation. While very simple, the results here naturally extend to more realistic cases, and display certain universal features. The unweaving of a free wavefunction into the Zel'dovich streams based on stationary points of $\zeta(\bm{q};\bm{x},a)$ is not restricted to 1+1 dimensions, and readily generalises to any number of spatial dimensions. Near the caustics, in higher dimensions, one has to consider a larger set of diffraction catastrophes than considered here, but these are fully classified and their normal forms can be found in e.g. \textcite{BerryUpstill1980, Feldbrugge2019}. In this way, the cosmic caustic web in two or three dimensions, as examined in \textcite{ArnoldShandarinZeldovich1982, Hidding2014, Feldbrugge2014, Feldbrugge2018}, could be dressed in wave phenomena via diffraction catastrophe integrals in equation~\eqref{eqn:diffraction_integral}. Using the universal profiles of these caustics has been suggested for probing the mass of a fuzzy dark matter particle with tidal streams on galactic scales \parencite{Dalal2021JCAP}.

There is potential to extend this beyond free wavefunctions using Propagator Perturbation Theory (PPT) from \textcite{Uhlemann2019}. By replacing the free propagator $K_0 = e^{i S_0/\hbar}$ with a higher-order propagator, one can build a higher-order function $\zeta_n$, whose stationary points determine the classical behaviour. This perturbatively encodes the effect of tidal effects relevant in a higher dimensional system. For example, using the next to leading order propagator, the stationary phase approximation would return the results of 2LPT.

The smooth phase recovered from the stationary phase approximation as detailed in Section~\ref{sec:phase_properties} potentially holds use in extending PPT beyond shell crossing. The perturbing element of PPT is an effective potential $V_{\rm eff}$, which in the single-stream region is the (using our sign convention) sum of the gravitational and velocity potentials. At lowest order, this vanishes (hence why we study free Schr\"odinger), and at next to leading order it is time independent. However, past shell crossing, the velocity potential develops  discontinuities, making it an unattractive perturbation variable. The smooth profile here could be used in the interference region, allowing for the effective potential to continue to have meaning past shell crossing in the spirit of a post-collapse perturbation theory that has been formulated for CDM \parencite{Colombi2015,TaruyaColombi2017,Pietroni2018,BuehlmannHahn2019, Saga_2022_LPT_Vlasov}. It could also allow to bridge from the perturbative cosmic web regime \parencite[at early times and on large scales considered here and in ][]{Uhlemann2019} to the asymptotic dynamics in the solitonic core regime \parencite[at late times and on small scales as described by][]{Zimmermann2021PhRvD,Zagorac2022PhRvD, Taruya_2022_corehalo}.

These wave-mechanical models also present an attractive application complementing the power of $N$-body simulations for understanding CDM dynamics. The Husimi phase-space distributions shown in Figure~\ref{fig:stream_split_wigner_overplot} demonstrate that the wave-mechanical system samples phase-space in a complementary way to $N$-body simulations. While $N$-body simulations sample discrete points along a cold phase-space sheet and trace their evolution, the wave-mechanical model follows a uniform density sheet with finite thickness (determined by $\hbar$). This is particularly useful in underdense regions, where large-scale $N$-body simulations effectively lose resolution (see e.g. \textcite{AnguloHahn2022} for a recent review). The propagator formalism adopted here has been applied to field-level inference in the Lyman-$\alpha$ forest \parencite{Porqueres_2020} and extended to two components \parencite{Rampf2021MNRAS} for setting initial conditions for Eulerian simulations including dark matter and baryons \parencite{Hahn_2021_ICs_2fluid}. We use these techniques in the following Chapter to investigate the impact of initial conditions vs dynamics on density statistics in wavelike dark matter systems using PPT and LPT.

%% file: text/chapter8-how-classical-is-FDM.tex

\chapter{Density statistics in wave models of dark matter}\label{chap:how-classical}

\minitoc

In this Chapter we examine the impact of wave dynamics and initial conditions on the statistics of 3D forward modelled density fields run with both Lagrangian Perturbation Theory and Propagator Perturbation Theory. In this way, this Chapter synthesises the research themes of my PhD research by looking at statistics of the density field in the context of wave mechanical models of dark matter.

\section{Introduction}

As introduced in Chapter \ref{chap:making-dm-waves}, wave-based models of dark matter are of interest both as potential fundamental descriptions of wavelike/ultralight of dark matter and an approximation of cold dark matter dynamics.

Dark matter which is fundamentally wavelike on astrophysical or cosmological scales has gained a large amount of interest as a dark matter candidate motivated by both particle physics \parencite{Peccei1977PhRvL, Svrcek2006JHEP, Arvanitaki2010PhRvD, Hui2017, Jaeckel2022arXiv} and proposing solutions to small scale challenges in $\Lambda$CDM \parencite{Douspis:2019, DiValentino:2021, Perivolaropoulos:2021}. Current constraints on various particle properties of axion models specifically, combining both particle and astrophysical constraints can be found in \textcite{AxionLimits, O'Hare2024arXiv}. 

While a multitude of models exist which can encode self interactions or further degrees of freedom, here we concern ourselves with the simplest fundamentally wavelike dark matter candidate, a non-relativistic, ultralight scalar field, which we will refer to throughout as Fuzzy Dark Matter (FDM). Fully non-linear descriptions of FDM are currently limited to box sizes of $1$--$10$ Mpc at $z\simeq 3$ for boson masses of $m\sim 10^{-23} \ \mathrm{eV}/c^2$ \parencite{MaySpringel2021, May.Springel_2023_HaloMass}. As such, investigations of the impact of FDM on larger cosmological scales require approximation techniques which neglect the full dynamics. Many of these techniques rely on modelling only the suppression of power on small scales which is present in the initial conditions of FDM, and neglect the interference effects and role of the quantum potential in the dynamics of the dark matter, an approached that has been dubbed ``classical FDM'' \parencite{Dome.etal_2023_CosmicWebElongation, Dome.etal_2023_CosmicWebDissection}. As such, those interference effects are not fully represented in the final density distribution of the dark matter.

\begin{figure}[p]
    \centering
     \captionsetup{width=0.9\textwidth}
    \includegraphics[width=\textwidth]{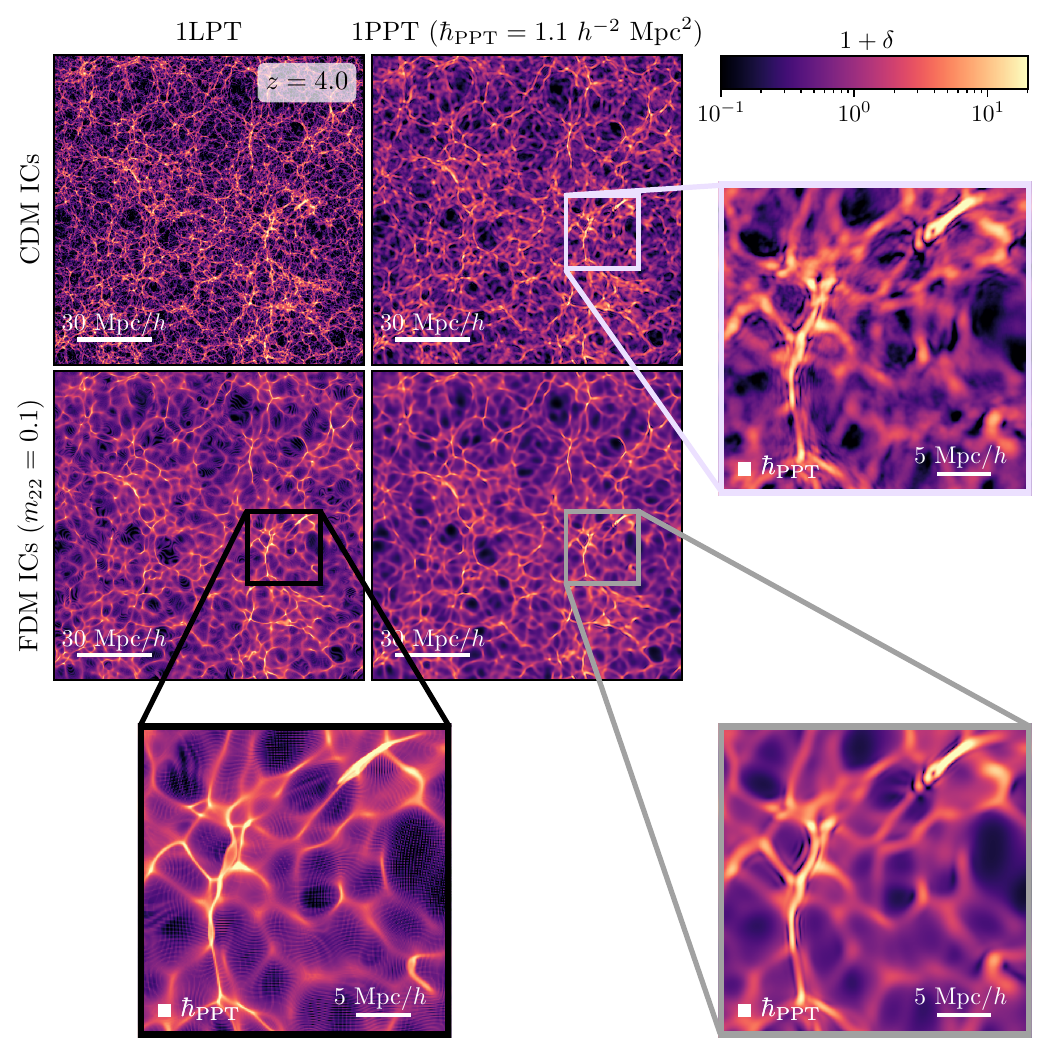}
    \caption[The density field for CDM and FDM initial conditions evolved with LPT and PPT.]{Projected $0.125 \ \mathrm{Mpc}/h$ thick slice of the dark matter density field at $z=4.0$. All plots are shown on the same logarithmic colour scale. These simulations have a box size of $128 \ \mathrm{Mpc}/h$ and a grid resolution of $0.125 \ \mathrm{Mpc}/h$. The LPT runs traced the positions of $(1024)^3$ particles. The zoomed inset regions are $30 \ \mathrm{Mpc}/h$ on a side. The size of the semiclassical parameter $\hbar_{\rm PPT}$ is shown as a white square in the zoomed plots. Columns in the main 2-by-2 correspond to which perturbation theory method is used: classical LPT fluid dynamics or wave PPT dynamics. The rows correspond to the initial conditions, either standard CDM initial conditions, or fuzzy dark matter initial conditions which suppress structure on small scales. The mass for the FDM initial conditions is chosen to be $10^{-23} \ \mathrm{eV}/c^2$. Notice that while classical evolution on FDM initial conditions produces a similar large scale cosmic web to the PPT+FDM ICs case, but by construction it cannot produce interference ripples seen in the wave evolution cases.}  
     \label{fig:slices-of-log-delta}
\end{figure}

Figure~\ref{fig:slices-of-log-delta} shows a slice of the density field produced from the perturbative forward model used in this work. We can directly see that while classical FDM (the black outlined inset) does produce the overall smoothing seen in the full wave physics case (grey inset), by construction it cannot produce the interference fringes which occur due to the quantum potential. This work aims to investigate how justified such approaches are on large cosmological scales, separating the impact of dynamics and initial conditions on larger scales than can be probed with full numerical simulations. 

\subsection{Numerical challenges for FDM}

The principle challenge in numerically solving the Schr\"odinger-Poisson equations \eqref{eqn:SP-full-eqns-firsttime} for the wavefunction while resolving wave interference scales are strict space and time resolution requirements which are dependent on the mass of the FDM particle. For pseudospectral methods such as those used in \parencite{MaySpringel2021, May.Springel_2023_HaloMass} the time step requirement is
\begin{equation}
	\Delta t < \operatorname{min}\left(\frac{4}{3\pi} \frac{m}{\hbar} a^2 (\Delta x)^2, 2\pi \frac{\hbar}{m} \frac{1}{\abs{V_{\rm N, max}}}\right),
\end{equation}
where $\Delta x$ is the spatial resolution and $V_{\rm N, max}$ is the maximum of the gravitational potential. This is imposed to avoid aliasing in the wavefunction where the phase would increase by more than $2\pi$ in adjacent cells. Different numerical schemes will have different exact prescriptions for this form, but all have the scaling that $\Delta t \sim \Delta x^2$ owing to the diffusive nature of the Schr\"odinger equation. The FDM particle mass itself imposes a restriction on the spatial resolution through its de Broglie wavelength. If we wish to resolve velocities of size $v_{\rm max}$ in the simulation the spatial grid needs to resolve the de Broglie wavelength associated with that velocity
\begin{equation}
\Delta x < \frac{\hbar}{m}\frac{\pi}{v_{\rm max}}\,.
\end{equation}

These joint constraints on $\Delta x$ and $\Delta t$ are what make direct, large box, simulations of fuzzy dark matter so computationally expensive, as going to cosmological boxes $\order{100} \ h^{-1} \ \rm Mpc$ cannot be achieved by simply decreasing the spatial/temporal resolution.

\subsection{Summary of numerical approaches}

A variety of different numerical approaches and codes for FDM exist. The most accurate of these are full wavefunction solvers  which solve the Schr\"odinger-Poisson equations for the complex wavefunction e.g. \parencite{Schive2014, MaySpringel2021, May.Springel_2023_HaloMass}.  \textcite{MaySpringel2021} examines the impact of changing the dynamics between FDM and $N$-body CDM simulations (on fixed CDM initial conditions) in 10 Mpc/$h$ box at $z=3$ with $m=0.35$--$0.7 \times 10^{-22} \ \mathrm{eV}/c^2$.  \textcite{May.Springel_2023_HaloMass} extends this analysis to also consider the impact of changing the initial conditions. They find that the initial conditions mostly impact the early evolution of the power spectrum and by $z=5$--$3$, non-linear growth has mostly made up for the initial suppression, with the non-linear FDM power spectrum gaining a positive bump on small scales dues to wave interference effects.

\texttt{PyUltraLight} \parencite{Edwards.etal_2018_PyUltraLightPseudospectral} and \texttt{UltraDark.jl} \parencite{Glennon.etal_2023_SimulationsMultifield} are also pseudospectral wave solvers. \textcite{Luu2024MNRAS} and \textcite{Glennon.etal_2023_SimulationsMultifield} also describe modelling multiple axion fields, motivated by the fact that the axion-like particles arising from the string axiverse \parencite{Svrcek2006JHEP, Arvanitaki2010PhRvD} generically span a spectrum of masses. 

Instead of solving the Schr\"odinger-Poisson equations for the complex wavefunction, a variety of approaches leverage the fact that the Schr\"odinger-Poisson equations can be rewritten in the form of hydrodynamical equations and make use of Smoothed Particle Hydrodynamic (SPH) techniques to solve for the fluid variables. Using SPH in this way is advocated for by e.g. \parencite{Mocz_2015_SP_SPH, Marsh_2015_NonlinearHydrodynamics}. One such code is \textsc{Ax-Gadget}  \parencite{Nori_2018_AX-GADGET}, a modification of \textsc{Gadget-2} \parencite{Springel_2005_CosmologicalSimulation} to accommodate for quantum pressure. SPH techniques can reach box lengths of $\sim \! 50 \ h^{-1}$ Mpc \parencite{Zhang.etal_2018_ImportanceQuantum}, but are known to fail getting the details of the  interference patterns correct, as the fluid variables become ill behaved in regions of low density as we saw in Section \ref{sec:phase_properties} \parencite[see also ][]{Veltmaat2016PhRvD, Zhang.etal_2019_CosmologicalSimulation}. Different SPH approaches also disagree on the role of the quantum potential   with \textcite{Nori_2018_AX-GADGET} finding that the quantum potential suppresses the matter power spectrum while \textcite{Veltmaat2016PhRvD} finds enhancement on small scales.

Other techniques use a hybrid approach, such as \textsc{Axionyx} \parencite{Schwabe.etal_2020_SimulatingMixed} that extends the \textsc{Nyx} code \parencite{Almgren.etal_2013_NyxMassively} to solve systems with mixed fuzzy and cold dark matter, solving the FDM component via pseudospectral methods and running $N$-body for the CDM component. This was recently used in \textcite{Lague.etal_2023_CosmologicalSimulations} for example, which simulates mixed CDM and FDM in 1--30 Mpc/$h$ boxes, with FDM masses of $10^{-25}$--$10^{-21} \ \mathrm{eV}/c^2$. \textcite{SchwabeNiemeyer2022} also takes this approach, solving $N$-body dynamics on large scales and full Schr\"odinger-Poisson on smaller scales to study individual halos. Both of these methods rely on translating the $N$-body particles into wavepackets at some point, as discussed in Section \ref{sec:comparision_other_wave_techs}. The work in \textcite{Dome.etal_2023_CosmicWebDissection,Dome.etal_2023_CosmicWebElongation} studies ``classical'' fuzzy dark matter in a 40 $h^{-1}$ Mpc box between $z\simeq 3.5$--$5.5$, where the impact of the FDM is only captured by the suppressed initial conditions before running standard $N$-body evolution. As a result the final density fields cannot have interference patterns present in them on any scale as seen in Figure \ref{fig:slices-of-log-delta}.

\subsection{Summary of theoretical approaches}

The computational expense of solving the Schr\"odinger-Poisson equations on cosmological scales has motivated a number of methods to make analytic or perturbative progress on such wavelike systems.

\textcite{Li.etal_2019_NumericalPerturbative} attempts a perturbative expansion directly in the wavefunction $\delta\psi=\psi-\psi_{\rm bkgd}$, however they demonstrate that the requirement that the smallness of $\delta\psi$ generally breaks down before the smallness of fluid variables $\delta$, $\bm{u}$, limiting the applicability of this direct approach.

There are also approaches which are based around modifying  modelling tools used in the usual CDM context. \textcite{Lague_2021_FDM_LPT} adjusts the particle displacements from Lagrangian Perturbation Theory (LPT) to account for the fact that linear growth in FDM is scale dependent. This is then applied to models of mixed dark matter. While this incorporates the quantum pressure at the linear level, it neglects non-linear effects in the displacement, as described in Appendix B of \textcite{Uhlemann2014}, and fails to produce interference fringes in the final density field.

Approaches based on the Effective Field Theory of Large Scale Structure (EFTofLSS) have also been tried, such as in \textcite{ManouchehriKousha.etal_2024_EffectiveField}. These approaches bundle the effects of the wave interference into the counterterms introduced in EFT (such as effective sound speed). This approach however relies on expanding the quantum potential as a series in powers of $\delta$, which breaks down where $\delta=-1$, see the quantum potential term defined in  equation \eqref{eqn:SP-fluid-eqns:Bernoulli}. Their simulations also follow the ``classical FDM'' approach of running $N$-body simulations on FDM suppressed initial conditions, failing to produce interference effects.

The approach used in this Chapter is Propagator Perturbation Theory (PPT), which perturbatively solves for the propagator which describes how final wavefunctions are constructed from initial wavefunctions in terms of an effective potential. To lowest order, this approach describes free wave propagation in $a$-time, and at next-to-leading order the effective potential is time-independent, making these equations tractable to solve. We saw leading order PPT in Chapter \ref{chap:making-dm-waves} and present PPT in more detail in Section \ref{sec:PPT-theory}.

\section{Wave dark matter}

In this Section we sketch the relevant cosmological wave equations in this investigation, the true Schr\"odinger-Poisson equations for an ultra-light non-relativistic scalar field and the Schr\"odinger equations which come out of the Propagator Perturbation formalism seen in Chapter \ref{chap:making-dm-waves}.

The de Broglie wavelength associated to a particle of mass $m$ is given by
\begin{equation}
\lambda_{\rm dB} = \frac{2\pi\hbar}{m v} = 1.21 \ \mathrm{kpc} \left[\frac{10^{-22}\ \mathrm{eV}/c^2}{m}\right] \left[\frac{1 \rm \ km \ s^{-1}}{v}\right],
\end{equation}
which motivates the definition $m_{22} = m/(10^{-22} \mathrm{eV}/c^2)$ as this sets the de Broglie wavelength at astrophysically relevant scales. Throughout this Chapter we will quote masses as values of $m_{22}$.

\subsection{Cosmological Schr\"odinger-Poisson equations}

We consider the simplest case of wave dark matter, a single scalar field without self-interactions in the non-relatavistic limit. The cosmological Schr\"odinger-Poisson equations are \parencite[see e.g. ][]{WidrowKaiser1993,Guth.etal_2015_DarkMatter, Marsh_2016_AxionCosmology, Hui2021, Ferreira2021A&ARv, O'Hare2024arXiv}
\begin{subequations} \label{eqn:SP-full-eqns}
    \begin{align}
        i\hbar \partial_t \psi &= -\frac{\hbar^2}{2ma^2} \nabla^2 \psi + m V_N \psi\,, \\
        \nabla^2 V_N &= \frac{3 \Omega_m^0 H_0^2}{2}  \frac{\abs{\psi}^2 - 1}{a}.
    \end{align}
\end{subequations}
where $V_N$ is the Newtonian gravitational potential, $\Omega_m^0$ is the current matter energy density and $H_0$ is the current Hubble parameter. The wavefunction described here is co-moving, meaning that the physical matter density would scale as $a^{-3/2}$ times this wavefunction. 

The Schr\"odinger-Poisson equations can be recast into fluid-like variables via the Madelung representation \parencite{Madelung1927}. By writing $\psi = \sqrt{1+\delta}\exp(i m \phi /\hbar)$, the Schr\"odinger-Poisson equations become the continuity equation and a modified Bernoulli equation:
\begin{subequations}\label{eqn:SP-fluid-eqns}
    \begin{align}
        &\del_t \delta + \frac{1}{a^2} \div{\left[(1+\delta) \grad \phi \right]} = 0 \,, \label{eqn:SP-fluid-eqns:continuity} \\
        &\del_t \phi + \frac{1}{2a^2} \abs{\grad \phi}^2 = - V_N  + \frac{\hbar^2}{2a^2}\frac{\nabla^2 \sqrt{1+\delta}}{\sqrt{1+\delta}}\,, \label{eqn:SP-fluid-eqns:Bernoulli} \\
        & \nabla^2 V_N = \frac{3\Omega_m^0 H_0^2}{2a}\delta\,, \label{eqn:SP-fluid-equns:Poisson}
    \end{align}
\end{subequations}
where $\delta$ is the density contrast and $\phi$ is the velocity potential which generates the peculiar velocity\footnote{Note that this is not the typical peculiar velocity $\bm{U}=\dv*{\xx}{\tau} = a\dv*{\xx}{t}$ as was used in the introductory chapter.} $\bm{u} = \dv{\xx}{t} = \grad\phi$. The final term appearing in the Bernoulli equation \eqref{eqn:SP-fluid-eqns:Bernoulli} is a source of velocity dispersion which doesn't appear in the equations for a standard fluid, referred to as the ``quantum pressure'' or ``quantum potential'' as discussed in Chapter \ref{chap:making-dm-waves}.

Linearising these equations in Fourier space leads to a second order equation for $\delta$ \parencite{Ferreira2021A&ARv}
\begin{equation}\label{eq:FDM-linear-growth}
\ddot{\delta}_k + 2 H(t) \dot{\delta}_k + \left[ \left(\frac{\hbar}{m}\frac{k^2}{2a^2}\right)^2 - \frac{3}{2}\Omega_{\rm FDM}(t)H^2(t)\right]\delta_k =  0,
\end{equation}
which sets the Jean's scale as \parencite{Marsh_2016_AxionCosmology}
\begin{equation}
k_J = 66.5 a^{1/4} \left(\frac{\Omega_{\rm FDM}^0h^2}{0.12}\right)^{1/4} m_{22}^{1/2} \ \rm Mpc^{-1}\,,
\end{equation}
where perturbations with $k<k_J$ grow in the linear regime while those with $k>k_J$ oscillate due to the quantum potential balancing against gravity. Solutions to equation \eqref{eq:FDM-linear-growth} determine the linear growth factors $D_\pm(k,a)$ for FDM which now no longer factorise into a spatial part and a time dependent part as they did with CDM. The exact growing mode solution to \eqref{eq:FDM-linear-growth} is \parencite{Marsh_2016_AxionCosmology}
\begin{equation}
D_+^{\rm FDM}(k,a) = \frac{3\sqrt{a}}{\tilde{k}^2} \sin(\frac{\tilde{k}^2}{\sqrt{a}}) + \left[\frac{3a}{\tilde{k}^4} - 1\right]\cos(\frac{\tilde{k}^2}{\sqrt{a}} ) \,,
\end{equation}
where $\tilde{k}=k/\sqrt{m H_0}\propto k/k_J$. In the limit $\tilde{k}\ll 1$ this recovers $D_+^{\rm FDM}\sim a$ which is the usual CDM growing mode.

\subsection{Initial conditions}

Adding an additional scalar field to the content of the universe changes the initial conditions which set the linear power spectrum compared to a standard $\Lambda$CDM cosmology. The equations which couple the dynamics of this scalar field to the standard fluid content of the universe (CDM, baryons, neutrinos, photons etc) can be found in e.g. \textcite{Hlozek.etal_2015_SearchUltralight}. For the purposes of structure formation, the principle effect of replacing the cold dark matter with an ultralight field is in the linear matter power spectrum, which is suppressed on small scales. This suppression is usually quantified relative to the standard CDM power spectrum such that
\begin{equation}
    P_{\rm FDM}(k,z) = T^2_{\rm FDM}(k,z) P_{\rm CDM}(k,z).
\end{equation}
The transfer function $T_{\rm FDM}(k,z)$ can be approximated by a redshift independent expression \parencite{Hu.etal_2000_ColdFuzzy}
\begin{equation}
    T_{\rm FDM}(k)  \approx \frac{\cos (x_J^3(k))}{1+x_J^8(k)}, \quad x_J(k) = 1.61 m_{22}^{1/18} k / k_{\rm J, eq}\,,
\end{equation}
where $k_{\rm J, eq} = 9 m_{22}^{1/2} \rm \ Mpc^{-1}$ is the Jeans scale at matter-radiation equality. Alternatively, the power spectrum of the axion field can be calculated by directly solving the Boltzmann hierarchy with the axion fluid equations included, as in the code \texttt{axionCAMB}\footnote{\url{https://github.com/dgrin1/axionCAMB}}\parencite{Hlozek.etal_2015_SearchUltralight}.

\subsection{Propagator Perturbation Theory}\label{sec:PPT-theory}

In this Section we present the equations of Propagator Perturbation Theory (PPT), extending the free Schr\"odinger equation used in  Chapter \ref{chap:making-dm-waves}. In addition to discussing the NLO perturbative potential, we provide some commentary on the unit systems used. Because in this Chapter we discuss connections to true wave dark matter more directly, we will denote the semiclassical parameter in PPT as $\hbar_{\rm PPT}$ rather than simply $\hbar$ as was done in Chapter \ref{chap:making-dm-waves}.

Propagator Perturbation Theory introduces a semiclassical wavefunction $\psi$ to describe the dark matter field, using a semiclassical parameter $\hbar_{\rm PPT}$ which controls the amount of wave behaviour in the system, akin to the combination $\hbar/m$ in the FDM scenario. In the semiclassical limit $\hbar_{\rm PPT}\to 0$ the dynamics of 1PPT reproduce Zel'dovich approximation (1LPT) dynamics, as seen in the previous Chapter, and can straightforwardly extract quantities in Eulerian space. At higher order, 2PPT produces similar dynamics to 2LPT in the semiclassical limit, but can also protect certain quantities such as the generation of spurious vorticity to higher perturbative order \parencite{Uhlemann2019}. PPT was established in \textcite{Uhlemann2019}, extended to two cold fluids in \textcite{Rampf2021MNRAS, Hahn_2021_ICs_2fluid} and applied as an in the context of generating initial conditions for cosmological simulations in \textcite{Michaux.etal_2021_AccurateInitial}. While this formalism is fundamentally based on different underlying physical assumptions than ``true'' Shr\"odinger-Poisson/FDM systems as discussed above, it does retain non-linear wavelike phenomena in its evolution, rather than relying on classical fluid evolution or simply changing linear growth. 

The PPT equations are usually written in units such that $4\pi G\bar{\rho}_0 = \frac{3}{2}$ (equivalent to setting $\Omega_m^0 H_0^2 = 1$). Here we keep the physical constants more explicit to aid in mapping between PPT results and more standard PT results. 

The dynamical equations of PPT are
\begin{subequations}\label{eqn:PPT-schrodinger-poisson}
\begin{align}
    i\partial_a \hbar_{\rm PPT}\psi &= -\frac{\hbar_{\rm PPT}^2 }{2}\nabla^2 \psi + V_{\rm eff} \psi\,, \label{eqn:PPT-schrodinger} \\
    V_{\rm eff} &= \frac{3 \Omega_m^0 H_0^2}{2a}(\phi_v + \varphi_g)\,, \\
    \nabla^2 \varphi_g &= \frac{1}{a}\left(\abs{\psi}^2-1\right)\,, \label{eqn:PPT_poisson}
\end{align}
\end{subequations}
where the effective potential encodes both the gravitational effects and expansion effects through the Hubble drag in the velocity potential $\phi_v$.\footnote{Note that we take the sign convention of \textcite{Gough.Uhlemann_2022_MakingDark} where $\bm{v} = \grad\phi_v$ rather than the convention of \textcite{Uhlemann2019} where $\bm{v} = -\grad\phi_v$.}  It is this effective potential $V_{\rm eff}$ which is solved for perturbatively in PPT. Notice that PPT takes the scale factor $a$ to be its natural time variable, rather than cosmic-time $t$.\footnote{See \textcite{Rampf2021MNRAS} for generalisation to $D$-time where $D$ is the linear growth factor.}

Under the Madelung transformation $\psi = \sqrt{1+\delta}\exp(i\phi_v/\hbar_{\rm PPT})$, these equations become the following fluid equations
\begin{subequations} \label{eqn:PPT-fluid-equations}
    \begin{align}
        &\partial_a \delta + \grad \cdot [(1+\delta)\grad\phi_v] = 0\,,\label{eqn:PPT-continuity}\\
        &\partial_a \phi_v + \frac{1}{2} \abs{\grad\phi_v}^2 = -V_{\rm eff} + \frac{\hbar_{\rm PPT}^2}{2} \frac{\nabla^2 \sqrt{1+\delta}}{\sqrt{1+\delta}}\,,  \label{eqn:PPT-Bernoulli}\\
        &\nabla^2 \varphi_g = \frac{\delta}{a}\,. \label{eqn:PPT-poisson-fluid}
    \end{align}
\end{subequations}

The initial conditions of PPT (and LPT) were discussed in Chapter \ref{chap:making-dm-waves} where as $a\to 0$ the density contrast is required to vanish and the initial velocity (with respect to $a$-time) is proportional to the initial gravitational potential
\begin{equation}\label{eqn:ZA_boundary_conditions-howclassicalchap}
    \delta^{\rm (ini)} = 0\,, \quad \phi_v^{\rm (ini)} = - \varphi_g^{\rm (ini)}\,.
\end{equation}

In these coordinates, the effective potential becomes very simple. To lowest perturbative order, the effective potential vanishes, making the equations of 1PPT simply a free Schr\"odinger equation (in $a$-time). At NLO, the effective potential is non-zero, but is ($a$)-time independent, making these dynamical equations straightforward to solve.

The free Schr\"odinger/1PPT dynamics where this effective potential vanishes describe the same dynamics as 1st order Lagrangian Perturbation Theory (LPT) or the Zel'dovich approximation \parencite{Zeldovich1970}, where particles are given velocities (relative to $a$-time) set by the initial gravitational potential, then free stream until shell crossing. This has also been studied as the ``free-particle Schr\"odinger equation'' \parencite{ColesSpencer2003, ShortColes2006, Gallagher_2022_SPvoids}. The correspondence of the free wavefunction of 1PPT to  Zel'dovich approximation dynamics is explicitly demonstrated in \textcite{Gough.Uhlemann_2022_MakingDark} (and Chapter \ref{chap:making-dm-waves}).

\subsubsection{The role of time variables}
Here we comment on the difference in coordinates between the PPT wave dynamics and the true FDM style equations (which are the same as the Widrow-Kaiser style application of Schr\"odinger-Poisson to modelling CDM). This is also discussed in \textcite{Uhlemann2019}.

The PPT equations  take the scale factor $a$ as the time variable, rather than the cosmic time $t$ as in equations \eqref{eqn:SP-full-eqns}. This impacts the natural velocity/momentum variable in the two pictures: in PPT it is $\bm{v} = \dv*{\xx}{a} = \grad \phi_v$ which differs from the  peculiar velocity $\bm{u} = \dv*{\xx}{t} = \dot{a}a\bm{v} = Ha^2\bm{v}$ which is generated by the wavefunction phase in the FDM scenario \eqref{eqn:SP-fluid-eqns:Bernoulli}. This leads to the identification $\phi = \dot{a} a^2 \phi_v$ as the mapping between the phase in the FDM picture and the PPT picture. These variable changes amount to a non-canonical transformation in phase space, and are what lead to the differing powers of $a$ in the quantum potentials in the two formulations. Specialising to an Einstein-de Sitter universe, where $\dot{a} \overset{\rm EdS}{=} H_0 a^{-1/2}$ and $\Omega_m^0 = 1$, the relationship between these wavefunction phases is  $\phi  \overset{\rm EdS}{=}H_0 a^{3/2}\phi_v$. 

Rewriting the fluid equation \eqref{eqn:SP-fluid-eqns} for FDM in $a$-time, we can make more direct comparison between these systems. The continuity equation becomes the same as that in PPT \eqref{eqn:PPT-continuity}, under the mapping between phase variables. Defining the rescaled the gravitational potential by $\varphi_g \overset{\rm EdS}{=} 2V_N/(3H_0^2)$, the FDM Poisson equation \eqref{eqn:SP-fluid-equns:Poisson} becomes the PPT Poisson equation \eqref{eqn:PPT-poisson-fluid}. The Bernoulli equation \eqref{eqn:SP-fluid-eqns:Bernoulli} becomes 
\begin{align}
\del_a \phi_v + \frac{1}{2}\abs{\nabla\phi_v}^2 
&\overset{\mathrm{EdS}}{=} -\frac{3}{2a}(\varphi_g + \phi_v) + \frac{\hbar^2}{2a^3H_0^2} \frac{\nabla^2\sqrt{1+\delta}}{\sqrt{1+\delta}} .
\end{align}
Comparing this to the PPT Bernoulli equation \eqref{eqn:PPT-continuity} we see that the quantum potential term carries an additional factor of $a^3$ in EdS spacetimes (more generally a factor of $(\dot{a}a^2)^{2}$), leaving the $\hbar_{\rm PPT}$ to be interpreted as a time dependent coarse graining scale, or because the quantum potential in the true FDM case is controlled by the boson mass, $\hbar_{\rm PPT}$ acts like a time dependent mass.

\subsubsection{Mapping $\hbar_{\rm PPT}$ to FDM mass}

As the PPT velocity potential $\phi_v$ generates the velocity $\bm{v}=\dv*{\xx}{a}$ which has dimensions of length, $\hbar_{\rm PPT}$ must then have dimensions of $[\mathrm{length}]^2$. The mapping between between the semiclassical parameter $\hbar_{\rm PPT}$ and the physical FDM particle mass $m_{22}$ isn't straightforwardly 1-to-1 due to the different time evolutions of the quantum potential terms (effectively the $\hbar_{\rm PPT}$ time dependence). Additionally, numerical applications of these systems require different initial conditions, with true FDM requiring initial conditions set by the physical dynamics at some fixed early time, while perturbation theories formally evolve from an initial time of $a\to 0$ on tethered initial conditions like those discussed in Chapter \ref{chap:making-dm-waves} in equation \eqref{eqn:ZA_boundary_conditions}.

For the purposes of this investigation we take a code-driven approach to choosing $\hbar_{\rm PPT}$ associated with a mass $m_{22}$, which we discuss in Section~\ref{sec:howw-fuzzy-numerics}. Here we outline ideas for an analytic matching below which could be implemented in future work.

The na\"ive approach to matching the FDM particle mass and the $\hbar_{\rm PPT}$ value is to simply equate the strength of the quantum potential terms in equations \eqref{eqn:SP-fluid-eqns:Bernoulli} and \eqref{eqn:PPT-Bernoulli} at some matching redshift, $z_m$. This results in
\begin{equation}\label{eq:mass_hbar_mapping}
    m_{22} \frac{\hbar_{\rm PPT} }{(\mathrm{Mpc}/h)^2} = 2.41\times 10^{-4} (1+z_m)^{3/2} \left(\frac{h}{0.7}\right)\left(\frac{\Omega_m^0}{0.31}\right)^{-1/2}.
\end{equation}

If we fix $m_{22}=0.1$ as the smallest value of FDM particle mass of interest at  $z=4$, the resulting $\hbar_{\rm PPT}\sim 0.04 \ h^{-2} \rm \ Mpc^2$, which is too small to resolve on a reasonable grid scale without aliasing in the wavefunction. Conversely, fixing $\hbar_{\rm PPT} = 1.1 \ h^{-2} \rm \ Mpc^2$ results in FDM particle masses of $m_{22}\sim 10^{-3}$ which are already ruled out by CMB+galaxy clustering \parencite[which constrain $m\gtrsim 10^{-24} \ \mathrm{eV}/c^2$][]{Hlozek.etal_2015_SearchUltralight, Hlozek2018MNRAS} and by Lyman-$\alpha$ data \parencite[which limit $m\gtrsim 10^{-21} \ \mathrm{eV}/c^2$][]{Irsic2017PhRvL,Armengaud2017MNRAS, Kobayashi2017PhRvD}. These mass constraints can be evaded somewhat if FDM makes up only a fraction of the dark matter content or in the presence of self-interaction terms.

However, the matching prescription \eqref{eq:mass_hbar_mapping} is crude as the physical effect of interest is the integrated impact of the quantum potential term on the density field. Neglecting the spatial variation due to density, the integrated quantum potential scales as $a^{-1/2}$, which cannot be extended all the way to $a\to 0$ as the perturbation theory would require. However, as the quantum potential should only be relevant once enough curvature in the density field is sourced---which will later cause interference in regions of classical shell crossing---this integration should only be carried back to the onset of classical shell crossing. Assuming the final redshift of interest lies between $z=0$--$4$ and the integration is carried back to redshift $z=6$--$10$, a factor of 20--200 can easily be generated, pushing the masses associated with the $\hbar_{\rm PPT}$ scale into the physically relevant region. A more precise analysis of what redshift shell crossing becomes sufficiently large to affect this could for example be investigated in a simple one-dimensional comparison between PPT and a full Schr\"odinger-Poisson solver.

\section{Numerical methods for perturbative forward modelling}\label{sec:howw-fuzzy-numerics}

To examine the effects of initial conditions and evolution dynamics we produce a set of perturbative forward models using a mildly modified version of the initial conditions code \texttt{monofonIC}\footnote{\url{https://bitbucket.org/ohahn/monofonic/src/master/}} \parencite{Hahn_2021_ICs_2fluid, Michaux.etal_2021_AccurateInitial}. This is intended to generate initial conditions for cosmological simulations using either $n$LPT to generate particle displacements or $n$PPT to generate the initial density and velocity fields. For our analysis we run \texttt{monofonIC} in 1-fluid mode (dark matter only) with the cosmological parameters shown in Table \ref{tab:cosmo_params_for_monofonic} in a box of $(128 \ h^{-1} \ \rm Mpc)^3$ on a $(1024)^3$ grid. For the LPT runs we run with one particle per grid point.  Note that while baryon, radiation, and neutrino fractions are included in Table \ref{tab:cosmo_params_for_monofonic}, they are used only in setting the appropriate linear power spectrum, the final output field is dark matter only.

\begin{table}[h!t]
    \centering
    \begin{tabular}{ccccc}
        \hline
         $\Omega_m$  &   $\Omega_b$  & $\Omega_c$ & $\Omega_r$ & $h$\\
         0.3158 & 0.0494 & 0.264979 & $7.99185\times 10^{-5}$ & 0.67321 \\
        \hline
        \hline
         $\sum m_\nu$ [$\mathrm{eV}/c^2$] & $\Omega_\nu$ &  $n_s$ & $A_s$ & $\sigma_8$\\
         0.06 & 0.001423 & 0.9661 & $2.094\times 10^{-9}$ & 0.8102\\
        \hline
    \end{tabular}
    \caption[Cosmological parameters used to produce density fields from \texttt{monofonIC}.]{Cosmological parameters used to produce our density fields with \texttt{monofonIC}. We set the value of  $A_s$ rather than fixing $\sigma_8$, the value of $\sigma_8 = 0.8102$ is the extrapolated value at $z=0$.}
    \label{tab:cosmo_params_for_monofonic}
\end{table}

The output density fields from \texttt{monofonIC} are generated in the following steps:
\begin{enumerate}
	\item Provide a transfer function for the total matter at the target output redshift $T_m(k,z_{\rm t})$ to \texttt{monofonIC}. 
	\item Generate a Gaussian random field with a power spectrum matching the provided target transfer function, call this $\delta_{\rm code}(\xx,a_{\rm t})$. As we're in the perturbative limit, this factorises into $\delta(\xx,a) = D_+(a)C_+(\xx)$. The spatial part of this product is solved by
	\begin{equation}
	C_+(\xx) = \frac{\delta_{\rm code}(\xx,a_{\rm t}) }{D_+(a_{\rm t})}\,,
	\end{equation}
	\item The initial gravitational potential is found by solving the Poisson equation for $\varphi_g$ then backscaling by the CDM linear growth factor taking the $a\to 0$ limit
	\begin{equation}
	\varphi_g^{\rm (ini)}(\xx) = \frac{\nabla^{-2}\delta_{\rm code}}{D_+(a_{\rm t})} \lim_{a\to 0}\frac{D_+(a)}{a}.
	\end{equation}
	\item This initial gravitational potential is then used in the relevant PT scheme (either LPT or PPT) to determine the displacement field or the initial wavefunction and effective potential. The perturbative evolutions are then evolved from these initial conditions to the ``start redshift'' (which will be the final output of this process). In the PPT case the density is constructed as simply $\abs{\psi}^2$ on this grid. 
	\item For the classical dynamics (LPT) runs, the density fields are constructed from particle displacements via \texttt{Pylians3}\footnote{\url{https://pylians3.readthedocs.io/en/master/index.html}} \parencite{Pylians} using the cloud-in-cell mass assignment scheme. 
	
\end{enumerate}

Slices of the final density fields at $z=4$ are shown in Figure~\ref{fig:slices-of-log-delta}. All our output fields have the same fixed target redshift of $z=4$ and varying ``start redshifts''\footnote{The term ``start redshift'' is the name given by \texttt{monofonIC}, as its standard use is in setting initial conditions for $N$-body simulations, which would start at $z_{\rm start}$. When used as a forward model, this ``start redshift'' is the output redshift of interest, as all the perturbative schemes ``start'' at $a=0$.} (we will refer to these as output redshifts) to investigate the time evolution of certain quantities. All runs which we compare directly are run from the same random seed, providing the same white noise field which is scaled by the target transfer function.

The CDM and FDM linear power spectra which are supplied to \texttt{monofonIC} were calculated using  \texttt{AxionCAMB}\footnote{\url{https://github.com/dgrin1/axionCAMB}} \parencite{Hlozek.etal_2015_SearchUltralight}, an extension to the standard Boltzmann solver \textsc{CAMB} \parencite{Lewis.Bridle_2002_CosmologicalParameters}. As \texttt{axionCAMB} requires both axions and standard cold dark matter, the fully CDM/FDM transfer functions were calculated by using an axion fraction of $10^{-7}$ and $(1-10^{-7})$ respectively. The FDM power spectra were specified with a boson mass of $m_{22}=0.1$ to accentuate the wave effects on the final field. These will be referred to as the CDM ICs and FDM ICs throughout.

By default, if \texttt{monofonIC} calculates PPT fields, it will choose the minimum value of $\hbar_{\rm PPT}$ that is allowed by the initial gravitational potential to avoid aliasing, in order to match as closely to CDM results as possible, we have modified the \texttt{monofonIC} code to allow specification of the size of $\hbar$ which can be larger than this. As \texttt{monofonIC} specifies the box size in $h^{-1} \ \rm Mpc$ units, the value of $\hbar_{\rm PPT}$ is always specified in $(h^{-1} \ {\rm Mpc})^2$. We choose to run on a fixed size of $\hbar_{\rm PPT}$ which is about 10\% larger than the minimum value in the CDM ICs, which ensures that the same $\hbar_{\rm PPT}$ on different random seeds or different initial conditions would also avoid aliasing.\footnote{The minimum value of $\hbar_{\rm PPT}$ on the random seed used for   this Chapter was 1.04619 for CDM ICs and 0.987302 on FDM ICs.}

We choose $\hbar_{\rm PPT} = 1.1 \ (\mathrm{Mpc}/h)^2$, which has a similar effect on the power spectrum as a boson mass of $m_{22}=0.1$ on perturbative scales, as shown in Figure~\ref{fig:Pk-z4}. As generally $\hbar_{\rm PPT}$ is inversely related to the FDM particle mass, this is the closest $\hbar_{\rm PPT}$ which was feasible to give effects similar to reasonably physical FDM masses.

\section{Statistics of the density field}

In this Section we present statistics measured on the density field produced from \texttt{monofonIC} as described above. In principle one could also calculate statistics on the velocity fields predicted with these forward models (directly in LPT and through the wavefunction in PPT), which would be  a natural extension to this work.

As much as possible we stick to the same colouring scheme for the different statistics investigated in this Section. Colour (or marker shape) is used to differentiate perturbation scheme, representing whether wave dynamics are present or not. The line style is used to differentiate initial conditions, with solid lines for CDM ICs and dashed for FDM ICs. Transparency is used in a few plots for further categorisation, e.g. perturbation theory order or number vs mass weighting.

\subsection{Power spectrum}

\begin{figure}[h!t]
    \centering
\includegraphics[scale=1]{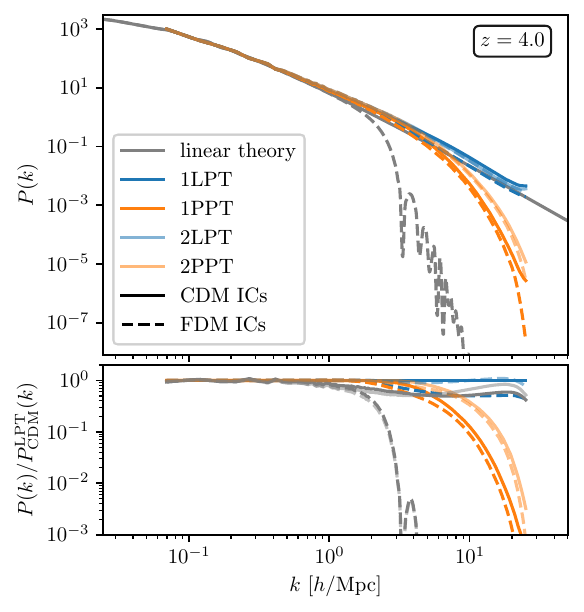}
    \caption[Four way comparison of the matter power spectrum.]{Four way comparison of the power spectrum from LPT (CDM) and PPT (wave) perturbative simulations with either FDM or CDM initial conditions in a $L=128 \ \mathrm{Mpc}/h$ box. The target linear power spectrum for CDM/FDM is shown as the grey lines. The lower plot shows the ratio of the power spectrum to the classically evolved CDM initial conditions power spectrum. This plot uses the same colouring as Figure~6 in \textcite{May.Springel_2023_HaloMass} for easy comparison. The paler results show the 2PT results, bold lines 1PT results.}
    \label{fig:Pk-z4}
\end{figure}

The dependence of the power spectrum on initial conditions and dynamics in a full Schr\"odinger-Poisson solver has been presented in e.g. Figure 6 of \textcite{May.Springel_2023_HaloMass}. In Figure~\ref{fig:Pk-z4} we plot the measured power spectrum at $z=4$ in a four way comparison between the 1LPT/1PPT dynamics and the CDM/FDM initial conditions, as well as the linear theory power spectra produced by \texttt{axionCAMB}. This is done to validate that the chosen value of $\hbar_{\rm PPT}=1.1 \ h^{-2} \, \mathrm{Mpc}^2$ has a similar effect to turning on the full wave dynamics seen in \textcite{May.Springel_2023_HaloMass}, at least on perturbative scales. \textcite{May.Springel_2023_HaloMass} use a $m_{22}=0.7$ compared to our choice of $m_{22}=0.1$ so we see suppression in linear power on slightly larger scales than their analysis. At this redshift the effect of changing the initial conditions is very small. Switching from classical to wave evolution does suppress the power at high $k$ as expected and seen in the fully non-linear simulations. On scales which are present in both our investigation and the fully non-linear simulations in \textcite{May.Springel_2023_HaloMass}, we find that the amount of suppression is similar for the masses and redshift chosen, with  $P^{\rm 1PPT}(k)/P^{\rm 1LPT}_{\rm CDM}(k)$ of order $10^{-1}$ between $k\sim 1$--$10$ $h$/Mpc.

\textcite{May.Springel_2023_HaloMass} find the ratio of power in wave simulations to $N$-body of 0.3--1 at $k\sim 10 \ h /\rm Mpc$ for redshifts $z=3$--$5$. Our power spectra fall into that range for both 1PPT and 2PPT, indicating that the impact of $\hbar_{\rm PPT}$ is reasonably similar to non-linear wave evolution of a FDM wavefunction with $m_{22}=0.1$. This validates our choice for $\hbar_{\rm PPT}$, and we use this value for the rest of this analysis.

\subsection{Matter PDFs and skewness}

The non-linear evolution of gravitational collapse causes the final density field to become non-Gaussian in structure, no longer being fully characterised by its power spectrum, as we have seen in Chapter \ref{chap:structure-formation}. A particularly simple class of non-Gaussian statistics which recover this information is the distribution of matter density in spheres of radius $R$. We use spherical top-hat filters in real space to smooth the density field, corresponding to the Fourier space window function
\begin{align}
    \delta_R(k) &= \tilde{W}(kR) \delta(k)\,, \\
    \tilde{W}(kR) &= \frac{3}{(kR)^3}\left(\sin(kR)-kR\cos(kR)\right). \nonumber
\end{align}
as the matter PDF in with top-hat smoothing is well modelled in the quasilinear regime by large deviations theory (as seen in Chapter \ref{chap:LDT-intro}). 

\begin{figure}[h!t]
    \centering
    \includegraphics[width=\textwidth]{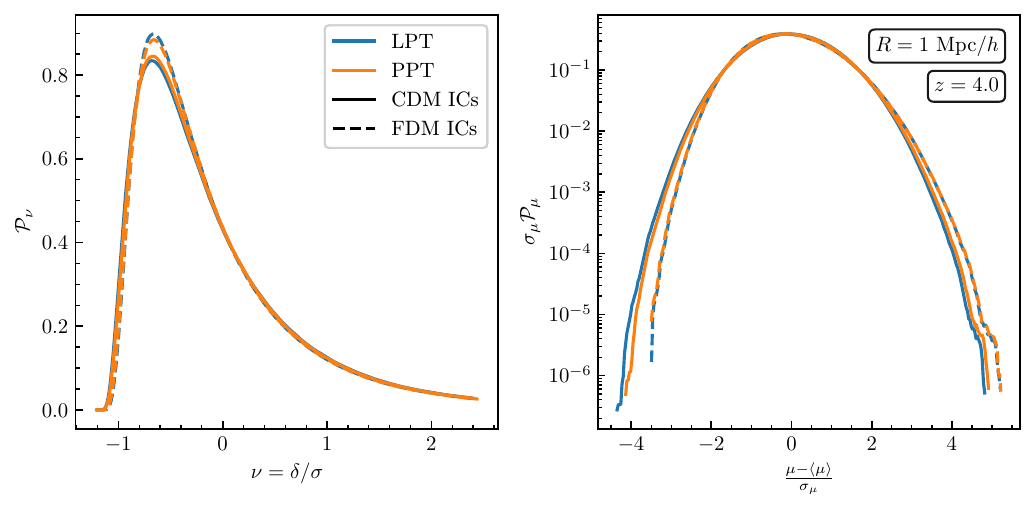}
    \caption[Four way comparison for the matter PDF in spheres of $R= 1 \ h^{-1} \ \rm Mpc$ at $z=4$.]{Four way comparison for the matter PDF in spheres of $R= 1 \ h^{-1} \ \rm Mpc$ at $z=4$.  (Left panel) The PDF of the rarity $\nu=\delta_R/\sigma(R)$ where $\sigma^2(R)$ is the variance of the density field on scale $R$. (Right panel) the PDF of the rarity relative to the log density $\mu = \log(1+\delta)$. Note that suppressed initial conditions (dashed lines) introduce additional skewness to the PDF. }
    \label{fig:matter_pdf_mu_nu_r1_z4}
\end{figure}

Figure \ref{fig:matter_pdf_mu_nu_r1_z4} shows the PDF of the rareness of the density $\nu=\delta/\sigma$ and the log density $\mu=\log(1+\delta)$ while varying the evolution dynamics and the initial conditions in spheres of radius $R= 1 \ h^{-1} \ \rm Mpc$ (8 times the grid scale). As changing the initial conditions changes the final variance on scale $R$, by plotting the PDF of the rareness the PDFs are normalised to have the same width. After normalising the width of the PDF it becomes evident that the FDM IC simulations (dashed lines) have an additional skewness compared to the simulations run on CDM initial conditions at $R= 1 \ h^{-1} \ \rm Mpc$, similar to the matter PDFs in extended cosmologies in Figure~\ref{fig:mg_pdf}. Changing the dynamics on this scale doesn't have a large impact, with the main impact being produced by changing the initial conditions.  

The skewness seen in these PDFs can be well predicted by perturbation theory, as we have seen in Chapter \ref{chap:structure-formation} for SPT and in the context of the minimal tree model in Chapter \ref{chap:covariance}. The reduced cumulants
\begin{equation}
S_n = \frac{\ev{\delta^n}_c}{\ev{\delta^2}_c^{n-1}}\,,
\end{equation}
can be predicted from perturbation theory. In the case of an unsmoothed density field and an EdS universe these are constant to tree order in standard perturbation theory. When smoothed with a top-hat filter, the reduced cumulants acquire a correction which depends only on the linear variance. SPT predicts the reduced skewness $S_3(R)$ \parencite[][see also Appendix \ref{app:sec:S3-smoothing-SPT}]{Bernardeau_1994_EffectsSmoothinga}
\begin{equation}
S_3^{\rm tree, SPT}(R) = \frac{34}{7} + \dv{\log\sigma_{\rm L}^2(R)}{\log R}\,,
\end{equation}
However, as the density fields considered in our analysis are not full non-linear simulations the recovered reduced skewness will differ from this value, owing to the difference in dynamics. It is known that $n$LPT only accurately reproduces cumulants up to $S_{n+1}$ \parencite{Munshi.etal_1994_NonlinearApproximations}. The calculation for reduced skewness can be adapted to the Zel'dovich approximations/1LPT by replacing the 2nd order perturbation kernel $F_2^{\rm SPT} \to F_2^{\rm ZA}$ in the calculation of $\ev{\delta^3_R}$ where \parencite{Scoccimarro.Frieman_1996_LoopCorrections} 
\begin{align}
F_2^{\rm SPT}(\kk_1, \kk_2) &= \frac{5}{7} + \frac{1}{2}\frac{\kk_1\cdot\kk_2}{k_1 k_2}\left(\frac{k_1}{k_2} + \frac{k_2}{k_1}\right) + \frac{2}{7}\frac{(\kk_1\cdot \kk_2)^2}{k_1^2 k_2^2}, \\
F_2^{\rm ZA}(\kk_1, \kk_2) &= \frac{1}{2}\frac{(\kk_1+\kk_2)\cdot(\kk_1+\kk_2)}{k_1^2 k_2^2}.
\end{align}
This replacement does not affect the smoothing term, only the bare value of the skewness, leading to
\begin{equation}
S_3^{\rm tree, ZA}(R) = 4 + \dv{\log\sigma_{\rm L}^2(R)}{\log R}.
\end{equation}
This result can also be derived via vertex generating functions  \parencite{Bernardeau1995ApJ}.
As the linear variance is an integral over the linear power spectrum
\begin{equation}
\sigma_{\rm L}^2(R) = \int \frac{\dd{k}}{2\pi^2} W(kR)^2 P_{\rm L}(k)\,,
\end{equation}
the variance for the field run on FDM initial conditions at a fixed physical scale $R$ will be smaller than the variance in the field run on CDM due to the suppression in linear power.

\begin{figure}[h!t]
    \centering\includegraphics[width=0.5\textwidth]{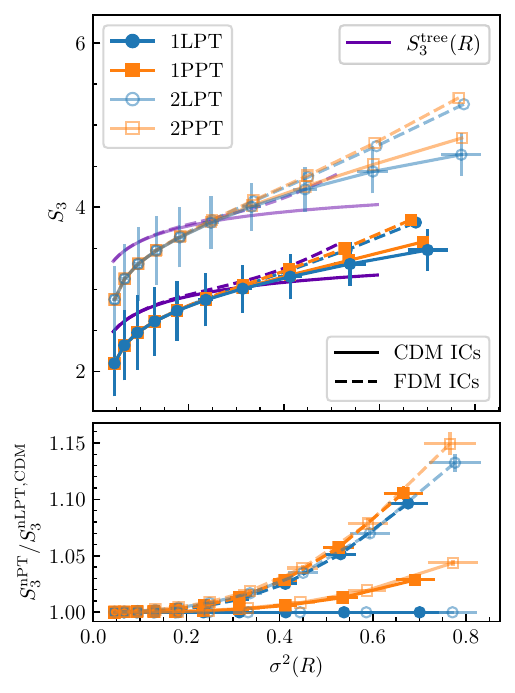}
    \caption[The reduced skewness measured with fluid and wave evolution on CDM and FDM initial conditions.]{(Top panel) Reduced skewness as a function of  variance at $z=4$, measured on top-hat smoothing scales $1$--$10 \ h^{-1} \ \rm Mpc$, log-spaced in $R$. Paler lines/open symbols show the results from 2nd order PT. The error bars are the standard deviation of the measured quantities across 8 subboxes. The errors are only shown on one line on the top plot to reduce clutter, but errors on other lines are similar. While the errors are relatively large (comparable with the scatter from measuring this across different realisations), the relative splitting (shown in the lower panels) have smaller scatter across realisations, demonstrating that wave effects generally enhance $S_3$ compared to LPT and CDM, in both smoothing ICs and dynamics. The purple lines (with no markers) are the tree order predictions for $S_3(R)$, which depend only on the linear variance. }
    \label{fig:S3_1column}
\end{figure}

Figure~\ref{fig:S3_1column} shows the measured reduced skewness against the measured variance $\sigma^2(R)$ on physical scales between 1--10 Mpc/$h$. The paler lines/open symbols show that the 2nd order PT simulations recover the vertical shift corresponding to $S_3^{\rm tree, SPT}=34/7$ instead of the $S_3^{\rm tree, ZA}=4$. We see also that on small scales (higher variance) the FDM initial conditions produce a higher reduced skewness, as anticipated from the PDFs in Figure~\ref{fig:matter_pdf_mu_nu_r1_z4}. While the standard deviation across subboxes of the simulation is (as indicated by the errorbars) is comparable with this enhancement, the ratio to the LPT+CDM ICs simulation is more robust, indicating that while the entire $S_3$ line has scatter comparable to the errorbars shown across realisations, the relative enhancement seen is not simply numerical noise. However we can see that the enhancement in $S_3$ due to the initial conditions appears to be in line with the amount of enhancement anticipated from the smoothing term $\dv*{\log \sigma_{\rm L}^2}{\log R}$. This means that the change in the skewness is largely captured by the change in the suppression from the change in the linear power spectrum. There is a small enhancement to the skewness by introducing wave dynamics but this is subdominant to the change in the overall variance from the initial conditions. This agrees with the findings of \textcite{Dome.etal_2023_CosmicWebDissection} in their classical FDM simulations, where they find simulations with higher suppression in their initial conditions have higher values of $S_3$, even on scales significantly smaller than considered here.

The enhancement of non-Gaussian features such as the (reduced) skewness for fields with less small scale power is expected by considering the dynamics of the smoothed density field in Lagrangian space \parencite{Bernardeau_1994_SkewnessKurtosis, Bernardeau1995ApJ}. At a fixed smoothing scale $R$, regions which are overdense in the final field evolved from larger regions via collapse, while underdense grow from smaller regions. Thus in fields with less small scale power, this asymmetry between over- and underdense regions is stronger than fields with power on small scales, enhancing the skewness as seen. This effect is similar to changing the effective tilt in the primordial power spectrum, as for a linear power spectrum $P_{\rm L}(k)\propto k^n$, the tree order skewness becomes
\begin{equation}
S_3^{\rm tree}(R) = S_3^{\rm tree}(0) + n(R).
\end{equation}
The removal of small scale power by smoothing on order Mpc scales is known to improve the accuracy of the Zel'dovich approximation by reducing the amount of shell crossing, known as the \emph{truncated Zel'dovich approximation} \parencite{Melott1994_TruncatedZA}.

From the matter PDF and the reduced skewness it appears that the averaging of the density in cells efficiently erases most of the dynamical difference between LPT and PPT, at least in this perturbative regime. However, exactly how strongly separated the dynamical and initial condition effects on $S_3$ are in PPT depends on understanding the $\hbar_{\rm PPT}$ to mass mapping better, as a larger $\hbar_{\rm PPT}$ can cause enhancement similar to the FDM initial conditions even from cold initial conditions, generating the skewness purely through dynamics (see Appendix \ref{app:sec:largerhbar}).

\subsection{Critical points of the density field}

Elements of the cosmic web can be defined and identified in a variety of ways.  These elements are usually referred to by their visual nature, knots/peaks for the compact and densest points, line-like filaments which extend in one dimension, and  sheets or walls which bound the underdense voids. These elements of the cosmic web are traced by luminous sources such as galaxies, as seen in Figure \ref{fig:2mass-galaxies}. The NEXUS/NEXUS+ algorithm \parencite{Cautun.etal_2013_NEXUSTracing} assigns a cosmic web element to every point in a volume a classification based on multiscale analysis of the Hessian of cosmic fields. This was applied in the context of classical FDM in \textcite{Dome.etal_2023_CosmicWebDissection}. Classifications based on different fields, such as the tidal or velocity shear fields give rise to the cosmic T-web \parencite{Forero-Romero.etal_2009_DynamicalClassification, Aycoberry.etal_2023_TheoreticalView} and V-web \parencite{Hoffman.etal_2012_KinematicClassification, Cui.etal_2018_LargescaleEnvironment}. Other classifications include identifying persistent topological structures via Morse methods \parencite{Colombi.etal_2000_TreeStructure, Sousbie_2011_PersistentCosmic, Sousbie.etal_2011_PersistentCosmic}, segmenting the density field  \parencite[the SpineWeb formalism, ][]{Aragon-Calvo.etal_2010_SpineCosmic}, and counting phase space folds \parencite[the ORIGAMI method, ][]{Falck.etal_2012_ORIGAMIDelineatinga}. The statistics of critical points in the density field have also been investigated recently in the context of  the minimal tree model discussed in Chapter \ref{chap:covariance} \parencite{Bernardeau2024arXiv}.

For our purposes we concern ourselves with critical points of the smoothed density field, $\delta_R$.\footnote{In this Section we continue to smooth the density field with spherical top-hat filters, as in the previous section. This is in contrast to many other critical point studies which use a Gaussian smoothing filter.} The Hessian matrix of the smoothed density field
\begin{equation}
    H_{ij}(\xx) = \nabla_i\nabla_j \delta_R(\xx)\,,
\end{equation}
encodes the curvature of the density field in 3 dimensions. Critical points of the density field (defined by $\grad\delta_R(\xx_c) = 0$), can be classified by the number of positive eigenvalues of $\mathsf{H}$ at that point, leading to the classification (schematically represented in Figure \ref{fig:critical_points}):
\begin{figure}[h!t]
\centering
\includegraphics[scale=1]{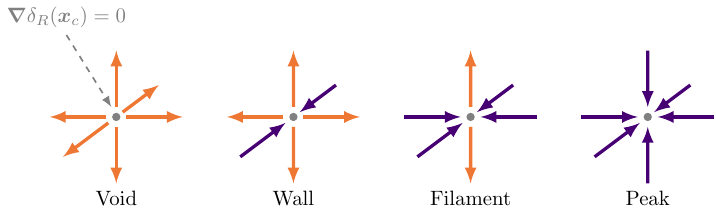}
\caption[Schematic diagram for critical point classification. ]{Classification of cosmic web elements via critical points of the smoothed density field. The directions/colours of the arrows indicate the sign of the eigenvalues $\lambda$ of the Hessian matrix $H_{ij}$ evaluated at the critical point. Outwards/orange arrows correspond to $\lambda>0$ while inwards/purple arrows correspond to $\lambda<0$.}
\label{fig:critical_points}
\end{figure}
\begin{itemize}
    \item 0 positive eigenvalues, Peak/Node/Knot, $\mathcal{P}$
    \vspace{-8pt}
    \item 1 positive eigenvalues, Filament, $\mathcal{F}$
    \vspace{-8pt}
    \item 2 positive eigenvalues, Wall, $\mathcal{W}$
    \vspace{-8pt}
    \item 3 positive eigenvalues, Void, $\mathcal{V}$
\end{itemize}
Note that in other critical point analyses  \parencite[e.g. ][]{Forero-Romero.etal_2009_DynamicalClassification, Hoffman.etal_2012_KinematicClassification, Carlesi.etal_2014_HydrodynamicalSimulations, Cui.etal_2018_LargescaleEnvironment, Libeskind.etal_2018_TracingCosmic, Suarez-Perez.etal_2021_FourCosmic, Aycoberry.etal_2023_TheoreticalView} a small non-zero threshold for the eigenvalues of the relevant matrix is used for better visual agreement between the field and the located cosmic web elements. 

We run the extrema finding code \texttt{py\_extrema}\footnote{\url{https://github.com/cphyc/py_extrema}} (\cite{Shim.etal_2021_ClusteringCritical}, see also Appendix G of \cite{Gay.etal_2012_NonGaussianStatistics}) on the density fields produced by \texttt{monofonIC} after smoothing. This calculates the derivatives and Hessian of the smoothed density field in Fourier space and then discards multiple critical points of the same kind which are found to lie in the same pixel. As the derivatives are calculated in Fourier space and require periodic boundary conditions we have to compute them on the full field. The smoothed density field is downsampled from a $(1024)^3$ grid to $(256)^3$ grid before searching for critical points for memory reasons. The density fields used in this Section produced with 1LPT/1PPT with CDM and FDM initial conditions.

\subsubsection{Total number of critical points}

\begin{figure}[h!t]
    \centering\includegraphics[width=\columnwidth]{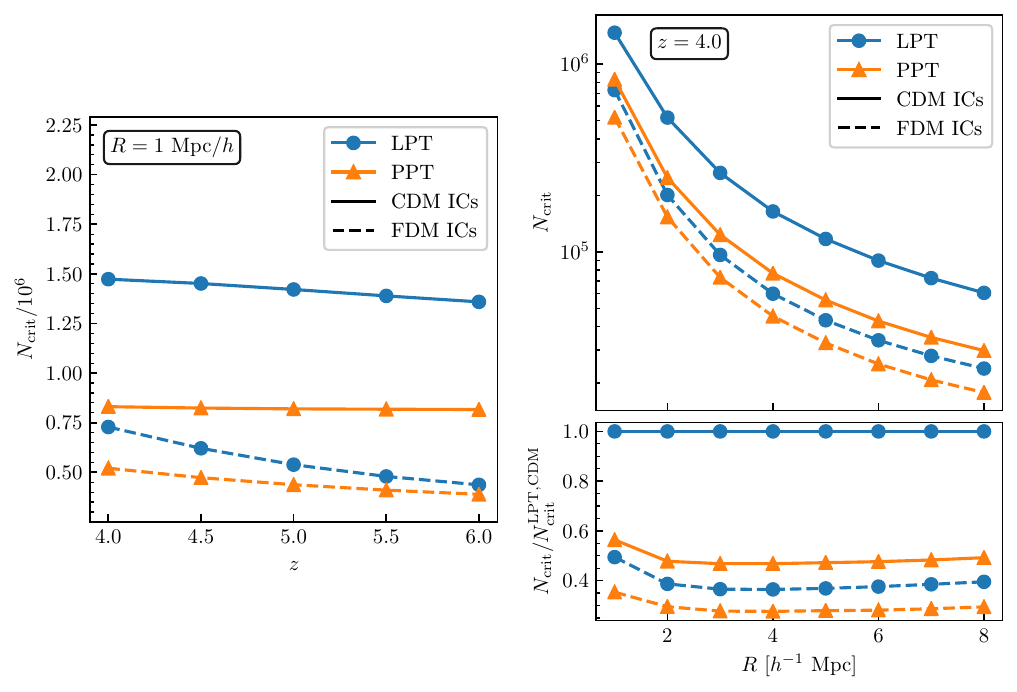}
    \caption[Total number of critical points in the density field.]{(Left panel) Total number of critical points in a density field smoothed in $R=1 \ h^{-1} \ \rm Mpc$ spheres. (Right panels) Total number of critical points at $z=4$ as a function of top-hat smoothing scale $R$. The lower panel shows the ratio to the number of critical points in the LPT + CDM ICs simulation, which is roughly constant on scales larger than $1 \ h^{-1} \ \rm Mpc$. The error on the mean based on 8 subboxes is smaller than the marker size in all panels.}
    \label{fig:crit_points_Ntot_evo}
\end{figure}

Figure~\ref{fig:crit_points_Ntot_evo} shows the evolution of the total number of critical points identified in the simulation box over time as a function of both time (left) and scale (right). We see that at all redshifts, any amount of smoothing in the density field, whether from initial conditions or introducing wave dynamics decreases the total number of critical points. This is mirrored as the smoothing scale is increased, which also decreases the total number of critical points. However for radii larger than about $1 \ h^{-1} \ \rm Mpc$, the ratio of number of critical points in the different simulation runs remains roughly constant. We additionally see a stronger redshift evolution of the number of critical points in classical dynamics compared to wave dynamics. Higher amounts of smoothing decrease the total number of critical points, however the relative fraction of different critical points do not all decrease at the same rate, as we see reflected in the ratios and number fractions between different types of critical points in Figures \ref{fig:crit_points_ratio_evolution} \& \ref{fig:crit_points_fraction_evolution}. 

We note as well that the effect of the dynamics (LPT vs PPT, blue vs orange) in this case is stronger than seen in the one-point statistics on the same scale. While the change to the initial conditions provides the bulk of the suppression in the number of critical points, there is a larger separation between the LPT+FDM ICs (blue dashed) and PPT+FDM ICs (orange dashed) cases than seen in the reduced skewness in Figure \ref{fig:S3_1column}, indicating that critical points are more sensitive to wave interference effects, as expected from a probe of derivatives of the density field.

\subsubsection{Fraction of cosmic web elements}

In a Gaussian random field the ratio of critical points of different kinds is exactly solvable. The number of peaks and voids (or filaments and walls) are equal, while the ratio of filaments to peaks (or walls to voids) is \parencite{Bardeen.etal_1986_StatisticsPeaks}
\begin{equation}
    \frac{N^{\rm G}_{\mathcal{P}}}{N^{\rm G}_{\mathcal{V}}}=\frac{N^{\rm G}_{\mathcal{F}}}{N^{\rm G}_{\mathcal{W}}}=1, \quad \frac{N^{\rm G}_{\mathcal{F}}}{N^{\rm G}_{\mathcal{P}}}=\frac{N^{\rm G}_{\mathcal{W}}}{N^{\rm G}_{\mathcal{V}}}= \frac{29\sqrt{15}+18\sqrt{10}}{29\sqrt{15}-18\sqrt{10}}\,,
\end{equation}
and therefore, the number fractions for a Gaussian random field $f_i^{\rm G}=N_{i}^{\rm G}/N_{\rm crit}^{\rm G}$ are
\begin{align}
    f_\mathcal{P,V}^{\rm G} =  \frac{29- 6\sqrt{6}}{116} \approx 12.3\%, \quad f_\mathcal{F,W}^{\rm G} = \frac{29+ 6\sqrt{6}}{116} \approx 37.7\%\,.
\end{align}

\begin{figure}[h!t]
    \centering
\includegraphics[width=\columnwidth]{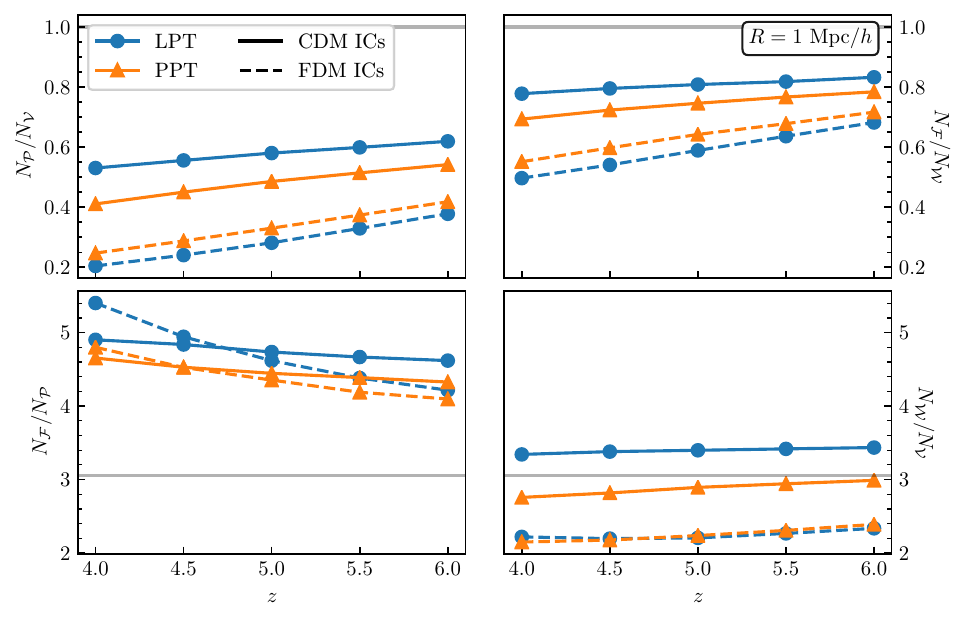}
    \caption[The ratios of critical points of different types.]{Ratio of number of critical points of different types. The grey lines show the ratios for a Gaussian random field.}
    \label{fig:crit_points_ratio_evolution}
\end{figure}

Figure~\ref{fig:crit_points_ratio_evolution} shows these ratios as measured on $R=1 \ h^{-1} \ \rm Mpc$ together with the Gaussian random field values. We see that generally the fully classical (LPT + CDM ICs) systems are the closer to the Gaussian ratios, with either FDM ICs or wave dynamics introducing higher deviation from the Gaussian ratios. This is consistent with the skewness results, which indicate that FDM initial conditions and wave dynamics both act to produce a more non-Gaussian field, though the dynamical effects have a stronger impact on the non-Gaussianity as measured from the critical points than the skewness.

\begin{figure}[h!t]
    \centering
\includegraphics[width=\columnwidth]{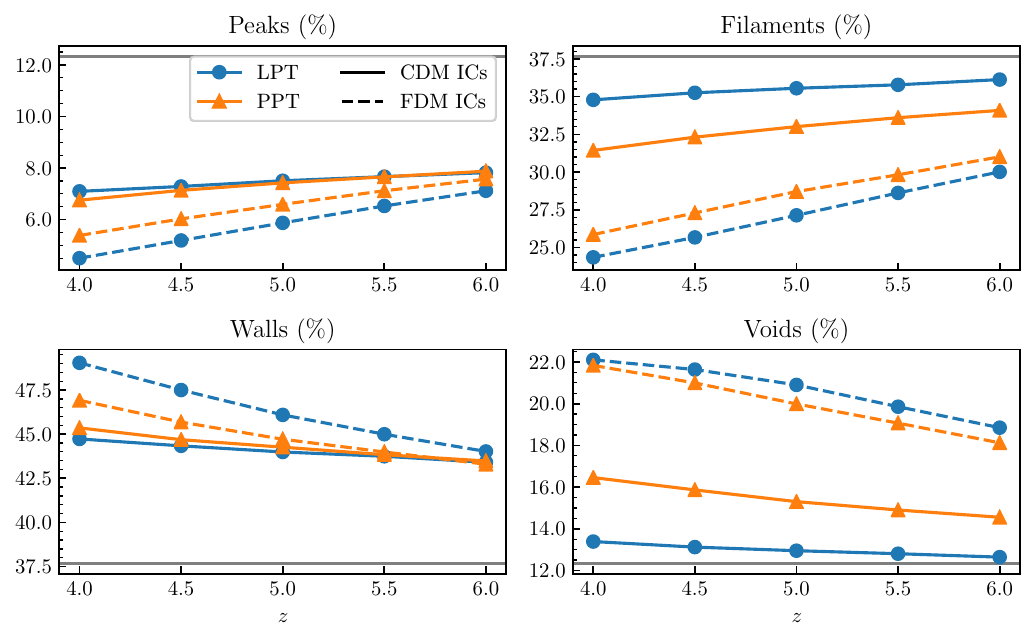}
    \caption[Evolution of the fraction of critical points of a given type.]{Evolution of the fraction of critical points on $R=1 \ h^{-1} \ \rm Mpc$ of a given type. The horizontal grey lines show the fraction for a Gaussian random field. Both wave dynamics (orange) and FDM initial conditions (dashed lines) move these fractions further from their Gaussian values.}
    \label{fig:crit_points_fraction_evolution}
\end{figure}

Figure~\ref{fig:crit_points_fraction_evolution} shows the evolution of the fraction of critical points in 1 $h^{-1}$ Mpc spheres of different types across the different cosmologies. 
%
We see that at all redshifts and for all of the cosmologies, the underdense type critical points (voids and walls) are both enhanced from the fractions which would be expected in a Gaussian field ($\sim 12\%$ for voids and $\sim 37\%$ for walls) with smoothed initial conditions particularly enhancing this enhancement. While we only show the results for 1LPT/1PPT in Figure \ref{fig:crit_points_fraction_evolution}, the same behaviour is seen in the $2^{\rm nd}$ order perturbative results as well.

From these individual number fractions, we see that while the FDM ICs or PPT dynamics cases produce fewer critical points than the fully classical case, as seen in Figure \ref{fig:crit_points_Ntot_evo}, the decrease is not uniform across different types of critical points. We see from Figure \ref{fig:crit_points_fraction_evolution} that underdense type critical points (voids and walls) are pushed further from their Gaussian predicted values than filaments and peaks. This can be explained by two effects. The removal of power on small scales in the FDM ICs cases reduces the gravitational potential, making it more difficult for structures to collapse in the first place, preferentially making void and wall type critical points more abundant over filaments and peaks. Secondly, the quantum pressure term in the wave dynamics case (equation~\ref{eqn:PPT-continuity}) acts to reduce shell crossing compared to the LPT dynamics case, which reduces the overdense type critical points more.

We note that \textcite{Dome.etal_2023_CosmicWebDissection} examined the mass and volume fractions of different cosmic web elements in the context of classical FDM (analagous to our LPT + FDM ICs case) and found that their volume fraction of voids decreased as the initial conditions were suppressed, contrary to our findings here. However, their cosmic web classification uses the NEXUS+ algorithm to classify all points into a cosmic web element, based on the signs of the Hessian over a range of scales, rather than critical points specifically, so direct comparison of these results is difficult. Much of their analysis also takes place on significantly smaller scales ($\sim 39 \ h^{-1}\  \rm kpc$, their grid scale) than we consider, where more non-linear dynamics is relevant.

\subsubsection{Environment spilt PDFs}

\begin{figure}[h!t]
\centering
\includegraphics[scale=1]{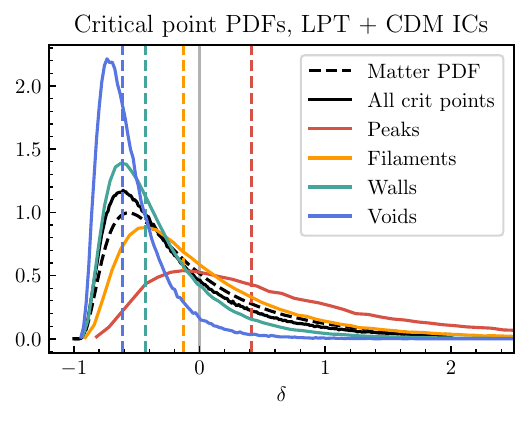}
\caption[The overdensity PDFs in different critical point environments.]{The PDFs of the overdensity of different critical points of the fully classical (LPT + CDM initial conditions) density field at $z=4$ when smoothed on $R=1$ Mpc/$h$. Each individual PDF is normalised. The PDF of the entire matter field (black dashed), and of all critical points (black solid) are shown for comparison. Coloured vertical lines show the median overdensity for the critical points of that type. The grey vertical line divides over- and underdense regions.}
\label{fig:crit_point_env_split_pdfs}
\end{figure}

\begin{figure}[h!t]
    \centering
\includegraphics[width=\columnwidth]{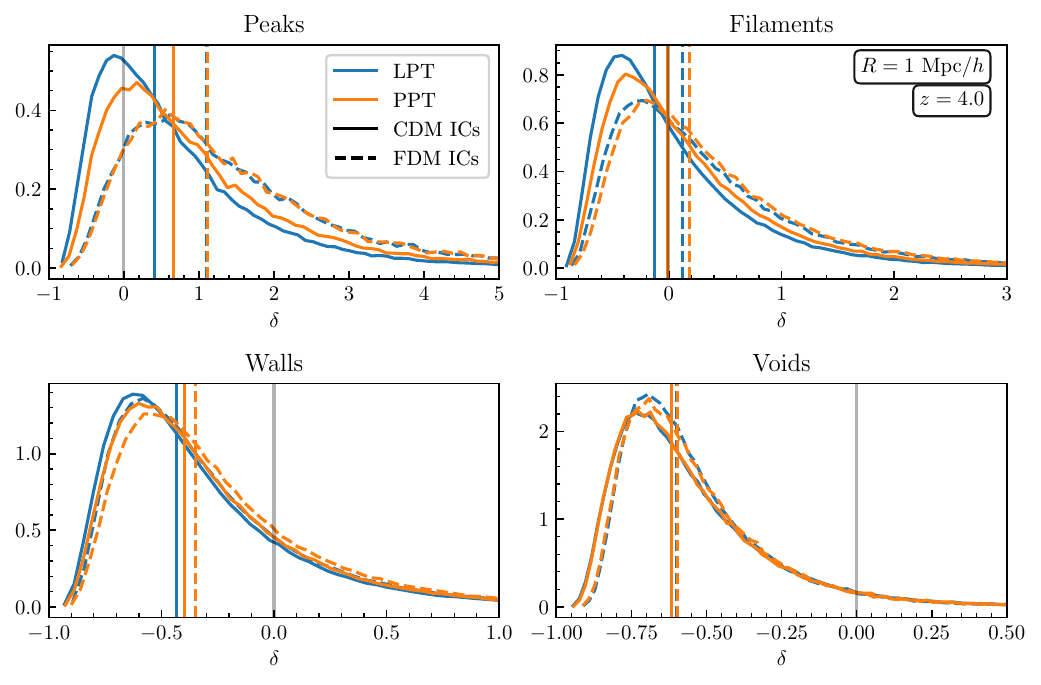}
    \caption[The overdensity PDFs in different environments for different initial conditions and dynamics.]{The PDFs of the density found in different critical point environments, linearly interpolated between bins. The vertical lines show the median value of $\delta$ for that cosmology, the grey vertical line divides under- and over-dense environments. Environments with more shell crossing, such as peaks, have a larger shift when wave dynamics are turned on, due to the quantum potential having a larger effect. }
    \label{fig:crit_point_PDFs_2x2}
\end{figure}

In addition to the number of critical points of certain types we can also examine the mass distribution of different critical points. Figure \ref{fig:crit_point_env_split_pdfs} shows the  overdensity PDFs for different critical point environments $z=4$ on $R=1 \ h^{-1} \ \rm Mpc$ scales in the fully classical (LPT + CDM initial conditions) density field. 
We see that as expected, peaks are the densest critical points, while voids are the most underdense. We see the variation of these PDFs across cosmologies in Figure~\ref{fig:crit_point_PDFs_2x2}. We see that wave dynamics and FDM ICs both increase the density of peaks, shifting the PDFs to the right. Interestingly, while the number of voids was heavily impacted by introducing FDM ICs or wave dynamics, as seen in Figure~\ref{fig:crit_points_fraction_evolution}, the void PDF is relatively stable compared to the other environment changes. These effects can be understood by the effect of the quantum potential,
\begin{equation}
Q = - \frac{\hbar_{\rm PPT}^2}{2}\frac{\nabla^2 \sqrt{1+\delta}}{\sqrt{1+\delta}},
\end{equation}
as it is dependent on the curvature of the (square root) of the density, and therefore is largest in regions of high curvature. We see that for cold initial conditions (solid lines) the critical points involving more shell crossing have larger shifts to their PDFs due to this quantum pressure. However, on FDM initial conditions, the PDFs of critical point environments seem less sensitive to the role of this quantum potential, even in the peaks where it was previously the strongest. We expect this is due to the suppressed initial conditions removing the critical points on the smallest scales, which also correspond to the highest curvature environments where quantum pressure would be the strongest.

The behaviour of the PDFs we see on these scales is quite different to those found in \textcite{Dome.etal_2023_CosmicWebDissection} in the context of classical FDM, where they find peaks to have the most stable PDF across changing cosmologies, and their voids have the most sensitivity to changing cosmology. However this is not immediately surprising as their PDFs are extracted on much smaller scales ($0.04$ vs $1$ Mpc/$h$), and their environments are quantified differently (for example, their void PDFs quantify all the matter in the void, while ours consider the PDF of densities at the void-type critical points, roughly the density at the centre of voids). While the behaviour between different environments is quite different, the overall effect of shifting all PDFs towards more overdense values is seen in both cases.

We note as well that in the fully non-linear case, the PDF of peak type critical points is likely to look different on small scales, as in \textcite{Dome.etal_2023_CosmicWebDissection}, as a consequence of virialisation creating stable halo profiles which feed into the peak PDF. Since our work considers 1PPT, equivalent to the Zel'dovich approximation in the $\hbar_{\rm PPT}\to 0$ limit, our model will not produce virialised objects due to overshooting, and thus we do not see the stability of the peak PDF discussed in \textcite{Dome.etal_2023_CosmicWebDissection}.

\section{Conclusions}

\subsection*{Summary}
In this Chapter we present an analysis of density fields produced by  two perturbative forward models on scales larger than where running full Schr\"odinger-Poisson solvers is possible. The models considered encode either classical fluid dynamics (LPT) or wave dynamics (PPT) and were run from both CDM and FDM initial conditions to disentangle effects of initial conditions from dynamical interference effects. We validate that the wave-perturbation scheme used provides similar effects on the power spectrum to fully non-linear Schr\"odinger-Poisson simulations.

For statistics of density environments such as the matter PDF in spheres on mildly non-linear scales or the reduced skewness $S_3$, we show that the principle difference between fully classical (LPT + CDM ICs) and fully wave (PPT + FDM ICs) is driven mostly by the suppression of the initial conditions. This validates applicability of the ``classical fuzzy dark matter'' approach on large perturbative scales at least in the context of these sorts of averaged cell statistics.

Analysis of the critical points of the density field show that these density field extrema are more sensitive to interference and quantum pressure effects than averaged one-point statistics. This is expected as both the quantum potential and these extrema are sensitive to derivatives of the underlying density field. Both suppression of the initial power and wave interference affect the total number and relative fraction of critical points, decreasing the amount of gravitational collapse and shell crossing, preferentially erasing critical points which require more shell crossing (peaks and filaments) compared to underdense type critical points (voids and walls). The one-point statistics of these different critical point environments also clearly demonstrates the role of the quantum pressure, with more collapsed regions producing a larger shift in the PDF compared to the fully classical case.

\subsection*{Outlook}

The skewness in wave simulations is mostly captured by the change in linear variance due to the initial conditions suggests that modelling of the matter PDF by Large Deviation Theory (Chapter \ref{chap:LDT-intro}) could be adapted to FDM universes, similarly to how it was adapted for modified gravity \parencite{Cataneo.etal_2022_MatterDensity}. Because of the success of LDT as a probe of extended cosmologies as presented in Chapter \ref{chap:MG-PDFs}, a similar forecast or inference based on the matter PDF and related quantities could be a successful complement to existing statistics in detecting wavelike dark matter effects. Successful modelling of the dark matter PDF in FDM cosmologies could then be straightforwardly extended to observables such as biased tracers or weak lensing statistics, as has been done in $\Lambda$CDM cosmologies.

A variety of other statistics could be extracted from this forward model beyond those presented in this work, further pushing the question of whether the quantum potential imprints signatures on any large scale statistics which  are not captured by simply suppressing the initial conditions. By using the phase information of the PPT wavefunction statistics related to the velocity and velocity dispersion could be extracted. Spatial correlations on the critical points of the density field could provide additional information to the simple number counts and one point statistics presented here. Additionally, analysing how sensitive our results are to the particular cosmic web identification method would be a natural extension to this work and could provide insights into the structure of the cosmic web in fuzzy cosmologies.

These wave-mechanical models present an appealing complementary approach to large scale studies of fuzzy dark matter. With further theoretical work to motivate the mapping between $\hbar_{\rm PPT}$ and the FDM particle mass $m$, propagator perturbation theory could be used as a less numerically intensive method of generating simulations of wave dark matter on large scales. Rather than hybrid approaches which patch large scale $N$-body with small scale Schr\"odinger-Poisson solvers, it could be possible to patched this wave based perturbation theory on large scales with Schr\"odinger-Poisson on small scales, in a similar spirit to \texttt{COLA} \parencite{Tassev.etal_2013_SolvingLarge} and \texttt{Hi-COLA} \parencite{Wright.etal_2023_HiCOLAFast}. Large scale simulations of wave dark matter which maintain interference effects on all scales would be an important tool in determining how best to test and constrain if the fundamental nature of dark matter is particle-like or wave-like from astrophysical data.

%% file: text/chapter9-conclusions.tex

\chapter{Conclusions}\label{chap:conclusions}

For convenience,  we summarise the content and conclusions of the research chapters in this thesis, before presenting some general conclusions about this research as a whole.

\section{Research chapter summaries}

\subsection*{The matter PDF in extended cosmologies (Chapter \ref{chap:MG-PDFs})}

This Chapter presents an analysis of the three-dimensional matter PDF as a probe of extensions to standard cosmology, through modified gravity and dynamical dark energy. We begin by presenting the basic dynamics of the specific modified gravity and dark energy models considered: $f(R)$ gravity, DGP gravity, and parametrised dark energy. Next we present how each of these theories impacts the ingredients in the large deviations theory model of the PDF: the linear and non-linear variance of the (log)-density field, and the spherical collapse dynamics. The main result is that all the matter PDF in all cases is well described by the large deviations theory (LDT) model introduced in Chapter \ref{chap:LDT-intro}, only needing to modify the linear theory terms which enter into the model. We then quantify the constraining power of the matter PDF for a \textit{Euclid}-like survey volume in comparison to the matter power spectrum via a Fisher forecast. We see excellent complementarity between these probes, where the additional information from the matter PDF is able to help parameter degeneracies in  the matter power spectrum. The  addition of the PDF  provides an improvement of up to six times the power spectrum alone in detecting deviations from standard $\Lambda$CDM cosmology, allowing for $>\,5\sigma$ detections in both models of modified gravity. This demonstrates that not only does the matter PDF contain additional non-Gaussian information not found in the matter power spectrum, but that the PDF is very efficient at capturing essential non-Gaussian information.

\subsection*{Modelling covariance matrices for matter PDFs (Chapter \ref{chap:covariance})}

This Chapter presents a model for calculating the covariance matrices of one-point statistics such as the matter PDF used in Chapter \ref{chap:MG-PDFs}. Such covariance matrices are important for accurate modelling of cosmological data, statistical tests, and parameter inference. In particular we present a method to predict covariance matrices for the PDF of densities in spherical cells from knowledge of the joint PDF of the densities in two cells. The general structure of these covariance matrices is shown to depend on the amount of overlap between the cells, representing the trade off between error on the value of PDF itself and correlation between bins of the PDF. For non-overlapping cells, the joint two-cell PDF can be represented in terms of a set of  ``bias'' functions, which describe the density dependence of two-point clustering. These two-cell PDFs can then be expanded as a series in the average two-point correlation function between cells. Leading and next-to-leading order results are derived for a Gaussian and general non-Gaussian model, recovering the well known Kaiser bias for a Gaussian field. This analytic model for the covariance is shown to be able to predict the ``super-sample covariance'' using the LDT model of the PDF, which simulation informed covariance matrices cannot typically account for. We also derive the bias functions at all orders for a hierarchical clustering model known as the ``minimal tree model'' which demonstrates essential clustering phenomenology.

\subsection*{Dark matter dynamics from wave interference (Chapter \ref{chap:making-dm-waves})}

This Chapter presents a detailed analysis into the connection between a wave model of dark matter and multistreaming in cold dark matter. It focuses on the  evolution of a one-dimensional system undergoing gravitational collapse, described in the Zel'dovich approximation and first order Propagator Perturbation Theory, which models the dark matter field as a single wavefunction. We demonstrate explicitly that the interference seen in the wave model directly corresponds to regions of classical multistreaming, and that the wave interference can be unwoven to recover the classical fluid streams. The interference patterns are also shown to encode other phase-space information, characterising behaviour beyond that of a perfect fluid, which is stored in the discontinuities in the wavefunction's phase. The wavefunction can then be decomposed into a smooth part which describes the bulk fluid flow, and a discontinuous ``hidden part'' which encodes information about higher cumulants of the phase-space distribution including velocity dispersion. The interference patterns near the classical caustics are connected to diffraction catastrophe theory, and shown to have certain universal scalings. Despite the specific model analysed in this Chapter being a toy model, these scalings are expected to hold into the non-linear regime and for models which describe fundamentally wavelike dark matter.

\subsection*{Density statistics in wave models of dark matter (Chapter \ref{chap:how-classical})}

This Chapter examines the impact of wave dynamics and the suppressed initial power expected from ultralight (fuzzy) dark matter candidates on statistics in the density field in the late universe. This is done in the perturbative regime, comparing the results of Lagrangian Perturbation Theory to those from the wave based Propagator Perturbation Theory also used in Chapter \ref{chap:making-dm-waves}. This provides a complementary analysis to existing studies of wavelike dark matter which must either restrict themselves to smaller simulation volumes or make different theoretical approximations. The forward model used in this Chapter allows larger box sizes than can be simulated with full Schr\"odinger-Poisson solvers, while maintaining wave interference features, unlike many other semi-analytic approaches. One-point statistics of the density field, such as the matter PDF and the reduced skewness are found to be well approximated by changing the initial conditions only, an approximation referred to as ``classical fuzzy dark matter''. However, the statistics of the critical points in the density field, taken as proxies for elements of the cosmic web, appear to be more sensitive to difference in dynamics, owing to the presence of the ``quantum potential'' hinting that neglecting interference phenomena could have an impact on statistics of the cosmic web in wavelike dark matter cosmologies.

\section{General conclusions and outlook}

Cosmology is an exciting field to be doing research in at the moment. The past few decades have established the $\Lambda$CDM model, which has been very successful at accurately predicting a wide variety of cosmological data, allowing us to map out the history of the Universe over the last 13 billion years. Despite this, we still do not understand the fundamental nature of two of the key ingredients of this model, dark matter and dark energy. There are also are several tantalising  discrepancies between observations and data which potentially point to new physics. The large-scale structure of the universe in particular is an exciting probe for the future of cosmology, if we are able to accurately understand its history and distribution we can unlock even more about the fundamental nature of our universe. Even just during the course of my PhD several cosmological surveys have begun, including \textit{Euclid} which launched in 2023, and DESI which released its first set of cosmological parameter results while I was writing this thesis \parencite{DESICollaboration2024arXiv_cosmoconstraints}. Even more will come online soon, for example the Legacy Survey of Space and Time (LSST) at the Vera Rubin Observatory which is scheduled to have its first light next year in 2025. 

A common theme throughout this thesis is the scale and density dependence of the clustering of matter. This is important to the statistical side of the project through the spherical collapse mapping which enters the LDT model of the matter PDF, and the bias functions which enter the joint two-cell PDF used to construct covariance matrices. Since new fundamental physics change the dynamical equations for matter in an explicitly density dependent way, understanding and modelling these environmental effects is important. In modified theories of gravity, this density dependence enters into the fifth force and the screening mechanism, while in theories of ultralight dark matter, they enter in the form of the quantum potential term. We have seen that the techniques presented in this thesis can model and measure this density dependence in the statistics of matter, and that these statistics allow us to constrain the parameters of the new fundamental physics. There is still significant scope for further investigation along these lines, by testing the accuracy, sensitivity, and predictive power of these statistics in more realistic scenarios and in the context of other changes to fundamental physics.

The work presented in this thesis has focused on the statistics and dynamics of the matter field, which is not directly observable from real data, but is an important stepping stone in preparing for this era of precision cosmology. The LDT model for the PDF has already been adapted to model weak lensing statistics and biased tracers. Improvements and extensions to these models remains an active area of research in the PDF team of collaborators I have worked with over the course of my PhD. There is still a lot of work which can be done to fully prepare these techniques for new data, in particular the modelling of  various systematic effects such as baryonic feedback or galaxy intrinsic alignments. Another key takeaway from this research has been the utility of different complementary approaches for learning about the non-linear regime. While $N$-body and hydrodynamical simulations have been an extremely important tool in learning about non-linear regime, they have not made analytic and semi-analytic methods redundant in this field. This is particularly important in cases where one wishes to test the effects of new physics, where running enough large-scale simulations to obtain enough sensitivity to cosmological parameters can be prohibitively expensive. Here (semi)-analytic techniques can act either to provide approximate answers directly, or as a tool which lowers the computational cost of running more sophisticated numerical methods. Even in standard cosmologies, the assumptions entering into different numerical and theoretical models can change where these models are useful. We should take insights from many different approaches to get the full picture how these complex systems work.

All said, the future of cosmology is packed with potential discoveries about some of the oldest questions of humanity, and I am honoured to have spent the last four years contributing to this search for understanding.

%% file: text/appendix-master.tex
\chapter{Mathematical conventions}\label{app:mathematical_notation}

\section{Fourier transforms}\label{app:sec:fourier}
 We use the following Fourier transform conventions
\begin{align}
\mathcal{F}[f](\bm{k}) &= \tilde{f}(\bm{k}) = \int \dnx e^{-i\bm{k}\cdot\bm{x}} f(\bm{x})\,, \\
\mathcal{F}^{-1}[\tilde{f}](\bm{x}) &= f(\bm{x}) = \int \dnkk e^{i \bm{k}\cdot\bm{x}} \tilde{f}(\bm{k})\,.
\end{align}
With these conventions we have the following useful properties:
\begin{itemize}
\item Transform of derivatives
\begin{align}
\FFT{\Del_{\xx} f}(\kk)  &= i\kk \cdot \FFT{f}(\kk) = i\kk \cdot \tilde{f}(\kk)\,, \\
\IFFT{\Del_{\kk} \tilde{f}}(\xx) &= -i\xx \cdot \IFFT{\tilde{f}(k)}(x) = -i\xx\cdot f(\xx).
\end{align}
\item Dirac delta representations
\begin{align}
\delta_{\rm D}(\xx-\bm{a}) &= \int \dnkk e^{i\kk\cdot (\xx-\bm{a})} = \IFFT{1}(\xx)\,, \\
(2\pi)^n \delta_{\rm D}(\kk-\bm{a})&= \int \dnx e^{-i(\kk-\bm{a})\cdot\xx} = \FFT{1}(\kk)\,.
\end{align}
These can be obtained by double expanding the sifting property of the Dirac delta
\begin{align}
f(\bm{a}) &= \int \dkk e^{i\kk\cdot \bm{a}} \tilde{f}(\kk) \nonumber\\
&= \int \dnkk e^{i\kk\cdot \bm{a}}  \int \dnx e^{-i\kk\cdot \xx}f(\xx) \nonumber \\
&= \int \dnkk \dnx e^{i \kk \cdot (\xx-\bm{a})} f(\xx) \nonumber\\
&\seteq \int \dnx \delta_{\rm D}(\xx-\bm{a}) f(\xx).
\end{align}

\item Convolution theorem.
Define the convolution between two functions as
\begin{equation}
(f*g)(x) \defeq \int \dd{x'} f(x') g(x-x') = (g*f)(x).
\end{equation}
To make it clear that convolution is symmetric, it is sometimes advantageous to write
\begin{equation}
(f*g)(x) = \int \dd{x_1}\dd{x_2} \delta_{\rm D}(x-x_{12})f(x_1)g(x_2),
\end{equation}
where $x_{12} = x_1 + x_2$. Consider the Fourier transforms of products and of convolutions
\begin{equation}
(2\pi)^n \FFT{fg}(\kk) = (\FFT{f}*\FFT{g})(\kk),
\end{equation}
\begin{equation}
\FFT{f*g}(\kk) = \FFT{f}(\kk) \FFT{g}(\kk),
\end{equation}
\begin{equation}
\IFFT{\tilde{f}\tilde{g}}(\xx) = f*g(\xx),
\end{equation}
\begin{equation}
\IFFT{\tilde{f}*\tilde{g}}(\xx) = (2\pi)^n \FFT{f}(\kk)\FFT{g}(\kk).
\end{equation}
Or slightly more compactly:
\begin{align*}
(2\pi)^n \FFT{fg} &= \tilde{f}*\tilde{g}, \\
\FFT{f*g} &= \tilde{f}\tilde{g}, \\
\IFFT{\tilde{f}\tilde{g}} &= f*g ,\\
\IFFT{\tilde{f}*\tilde{g}} &= (2\pi)^n fg.
\end{align*}
\end{itemize}

\subsection{Power spectra}\label{app:sec:power-spectra}
We use the definition
\begin{equation}
\ev{\tilde{\delta}(\kk)\tilde{\delta}(\kk')} = (2\pi)^n\delta_{\rm D}(\kk + \kk')P(k)\,,
\end{equation}
for the power spectrum, to keep the factors of $2\pi$ with the momentum space Dirac delta in keeping with the Fourier conventions above. However, many places define simply
\begin{equation}
\ev{\tilde{\delta}(\kk)\tilde{\delta}(\kk')} = \delta_{\rm D}(\kk + \kk')P(k).
\end{equation}
such that the power spectrum measures the root mean square Fourier amplitude of the field for real fields.

The conventions for both the power spectrum and Fourier transforms in \textcite{Bernardeau:2002} swapping the role of the $2\pi$s. This affects how the variance is defined for example. The (unsmoothed) linear variance is by our conventions 
\begin{align}
\sigma_{\rm L}^2 = \ev{\delta_L^2(\xx)} &= \int \frac{\dd[3]{\kk_1}\dd[3]{\kk_2}}{(2\pi)^3(2\pi)^3} e^{i(\kk_1+\kk_2)\cdot \xx} \underbrace{\ev{\delta(\kk_1)\delta(\kk_2)}}_{(2\pi)^3\delta_{\rm D}(\kk_1+\kk_2)P_{\rm L}(k_1)} \nonumber \\
&= \int \frac{\dd[3]{\kk_1}}{(2\pi)^3} e^{i \kk_1 \cdot \xx} P_L(k_1) \nonumber \\
&= \int \frac{\dd{k}}{2\pi^2} k^2 P_{\rm L}(k)
\end{align}
where the last line comes out of doing the angular integral to get $4\pi k^2$ and now we integrate just on the magnitude of $k$. With the role of the $2\pi$s swapped in both places, the linear variance is instead
\begin{equation}
\sigma_{\rm L}^2 = \int \dd{k} 4\pi k^2 P_{\rm L}(k).
\end{equation}

Whether $(2\pi)^n$ in the definition of $\ev{\delta(k)\delta(k')}$ should match the power of the $(2\pi)^n$ in the Fourier conventions determines whether the correlation function in real space is the Fourier transform of the power spectrum or not. For example, if we keep my Fourier conventions but define the power spectrum $\ev{\delta(k)\delta(k')} = P(k)\delta_{\rm D}(k+k')$ then we have
\begin{equation}
\xi(r) = \ev{\delta(\xx)\delta(\xx+\bm{r})} = \int \frac{\dd[3]{\kk_1}}{(2\pi)^3} \frac{\dd[3]{\kk_2}}{(2\pi)^3} e^{i\kk_1\cdot \xx} e^{i \kk_2\cdot(\xx+\bm{r})} \underbrace{\ev{\tilde{\delta}(\kk_1)\tilde{\delta}(\kk_2)}}_{P(k_2)\delta_{\rm D}(\kk_1+\kk_2)}\,,
\end{equation}
which, when integrating out $\kk_1$ gives
\begin{equation}
\xi(r) = \ev{\delta(\xx)\delta(\xx+\bm{r})} =\frac{1}{(2\pi)^3} \int \frac{\dd[3]{\kk_2}}{(2\pi)^3}  e^{i\kk_2\cdot \bm{r}} P(k_2) = \frac{1}{(2\pi)^3} \mathcal{F}[P(\kk_2)](\bm{r})\,,
\end{equation}
where, if we'd used consistent Fourier conventions and power spectrum definitions, this would have just been the Fourier transform, which is nicer.

\subsection{Window functions}\label{app:window-functions}

For obtaining smoothed fields, different window functions are commonly used. In this thesis smoothing is exclusively done with real space top-hat filters unless otherwise specified, as these filters are well suited to LDT and certain analytic properties for SPT.

Smoothing in real space is done with a convolution
\begin{equation}
\delta_R(\xx) = \int \dnx' W_R(\xx-\xx')\delta(\xx') = (W_R * \delta)(\xx)\,,
\end{equation}
which in Fourier space is simply a product.
\begin{equation}
\delta_R(\kk) = \tilde{W}_R(\kk)\delta(\kk).
\end{equation}

\textbf{Gaussian filters} are Gaussian in both real and Fourier space
\begin{align}
W_R(\xx) &= \frac{1}{(2\pi R^2)^{n/2}}\exp(-\frac{x^2}{2R^2}) \nonumber \\
\tilde{W}_R(\kk) &= \exp(-\frac{1}{2}R^2 k^2).
\end{align}
These filters are convenient analytically as they are nicely differentiable in both real and Fourier space, and some analytic properties can be derived for Gaussianly smoothed fields (such as the statistics of critical points as discussed in Chapter~\ref{chap:how-classical}). In numerical applications however their infinite extent in both real and Fourier space can work against them, introducing the need for truncation and difficulty navigating masked data.

\textbf{Spherical top-hat} filters in real space (3D)
\begin{align*}
W_R(\xx) &= \frac{3}{4\pi R^3} \Theta\left(\abs{\xx}-R\right) \\
\tilde{W}_R(\kk) &= \frac{3}{4\pi R^3} \int_{\R^3} \dd[3]{\xx} e^{-i\kk\cdot\xx} \Theta(\abs{x}-R) \\
&= \frac{3}{4\pi R^3}  \int_{\phi=0}^{2\pi}\int_{\cos\theta=-1}^1\int_{r=0}^R  r^2  e^{-i k r \cos\theta} \dd{r}\mathrm{d}\cos\theta \dd{\phi} \\
&=  \frac{3}{2 R^3} \int_0^R \dd{r} r^2 \frac{2\sin(kr)}{kr} \\
&= \frac{3}{(kR)^3} \left(\sin(kR)-kR\cos(kR) \right) = 3\sqrt{\frac{\pi}{2}} \frac{J_{3/2}(kR)}{(kR)^{3/2}}
\end{align*}
while in 2D (relevant for gravitational lensing analysis for example) we have
\begin{align*}
\tilde{W}_R(\kk) &= \frac{1}{\pi R^2} \int_0^{2\pi} \int_0^R r e^{-ikr\cos\phi} \dd{r}\dd{\phi} \\
&= \frac{1}{\pi R^2} \int_0^R r (2\pi J_0(kr)) \dd{r} \\
&= \frac{2J_1(kR)}{kR}
\end{align*}
where $J_n(z)$ is the $n^{\rm th}$ order Bessel function of the first kind. Appendix C of \textcite{Bernardeau:2002} has various mathematical properties of $N$-dimensional spherical top-hat filters.

It is sometimes useful to match features between fields smoothed with different filters. The appropriate prescription will vary depending on the case of interest, but we note here that
\begin{itemize}
\item Matching the FWHM of the a Gaussian and real-space top-hat filter in Fourier space requires
\begin{equation}
R_{\rm TH} \simeq \frac{2.49826}{\sqrt{2\ln2}}R_{\rm G} \approx 2.121 R_{\rm G}.
\end{equation}
\item Matching the mass enclosed by the filters (in 3D space) produces a ratio of
\begin{equation}
R_{\rm TH} \simeq 1.56 R_{\rm G}.
\end{equation}
\end{itemize}

\section{Moments and cumulants of probability distributions} \label{app:prob-distributions}
Here we review certain standard statistical quantities. For a continuous random variable $X$ with an associated probability density function $\mP_X$, the moments of $X$ are expectation values of $X$ relative to this probability density
\begin{equation}
M_n = \ev{X^n} = \int \mP_X(X=x) x^n \dd{x}.
\end{equation}
If the $\mP$ is for example a phase-space distribution $f$, then the moments with respect to the momentum $\pp$ are physical quantities such as the density and momentum flux.\footnote{Throughout this work we assume the ergodic hypothesis where we can replace ensemble averages with spatial averages over one realisation of a random field. This requires that spatial correlations decay sufficiently rapidly with separation that many statistically independent volumes are contained in one realisation.}  The moment generating function is a formal power series in some auxiliary variable $\lambda$ where the coefficients are these moments, 
\begin{equation}
\mathcal{M}_X(\lambda) = \sum_{n=0}^\infty \frac{\ev{X^n}}{n!}\lambda^n = 1 + \lambda\ev{X} + \frac{\lambda^2}{2}\ev{X^2} + \dots
\end{equation}
which provides a useful calculation tool as from the moment generating function the $n^{\rm th}$ moment can be obtained by taking the $n^{\rm th}$ derivative and evaluating at $\lambda=0$. The moment generating function is related to the PDF by Laplace transform
\begin{equation}
\mathcal{M}_X(\lambda) = \ev{e^{\lambda X}} = \int_{-\infty}^{\infty} e^{\lambda X} \mP_X(x) \dd{x}.
\end{equation}

Similar to the moments of a distribution are the cumulants of a distribution which are the ``connected part'' of the moments. This is understood in the following sense, part of the information content in a moment of a distribution simply comes products of the lower moments. The ``connected part'' is the independent part which is obtained by subtracting off a sum over partitions, where each partition then involves products of those corresponding moments. That is
\begin{align}
\langle X_1 X_2 \dots X_n \rangle_{c} = &\phantom{+}\langle X_1 X_2 \dots X_n \rangle \nonumber \\
&- \sum_{\mathcal{S}\in\mathcal{P}_n} \prod_{\sigma\in\mathcal{S}} \ev{X_{\sigma(1)}\dots X_{\sigma(\abs{\mathcal{S}})}}
\end{align}
where $\mathcal{S}$ is any proper partition of $\{1,\dots, n\}$, and $\sigma$ is a subset of that partition. I find this is best understood through examples. The second cumulant is simply the variance (the same as the second central moment)
\begin{align}
k_2 &= \langle X^2\rangle_c = \langle X^2\rangle - \langle X\rangle \langle X\rangle.
\end{align}
The third cumulant is
\begin{align}
k_3 = \langle X^3\rangle_c &= \langle X^3\rangle - 3 \langle X^2\rangle \langle X \rangle. 
\end{align}
where the factor of three arises from the 3 different ways that $XXX$ can be partitioned into a set of 2 $X$s and 1 $X$. This partitioning can be represented diagramatically, as we see in the discussion of correlation functions in Chapter~\ref{chap:structure-formation}.

The cumulant generating function\footnote{One could instead simply define the cumulants as the coefficients which are generated by $\log\mathcal{M}(\lambda)$ and then find the connected part interpretation.} (CGF), $\phi_X(\lambda)$ is again a power series with the cumulants as coefficients, and is related to the moment generating function by exponentiation\footnote{For readers familiar with QFT and Feynman diagrams, the moment generating function is similar to the path integral functions partition functions $Z_j$ which generate vacuum correlation values $\langle 0 \vert \phi(\xx_1)\dots \phi(\xx_n)\vert 0 \rangle$ by successive differentiation, while the cumulant generating function is usually written $W_j = \log Z_j$ and generates only the \emph{connected} Feynman diagrams. There is usually a slight difference as the partition function is usually defined as the Fourier transform rather than the Laplace transform, requiring different powers of $i$, and changing technical convergence effects, but the spirit is the same.}

\begin{align}
\phi_X(\lambda) &= \ln\mathcal{M}_X(\lambda) = \sum_{n=1}^\infty \frac{\ev{X^n}_c }{n!}\lambda^n  = k_1 \lambda + \frac{1}{2}k_\lambda^2 + \dots \\
k_n &= \dv[n]{\phi_X}{\lambda}\Big\vert_{\lambda = 0}
\end{align}

The cumulants are simpler objects to work characterise a random variable. For example, if a random variable has a Gaussian PDF, its moments are all non-zero, but moments of degree 3 or higher are simply combinations of the mean and variance of the distribution (this is the basis for Wick's theorem for Gaussian fields). In contrast, the only non-zero cumulants of a Gaussian are $k_1,k_2$, with all higher cumulants vanishing.

Note that when the moment/cumulant generating function exists, the associated PDF is uniquely defined by it. However this does not mean that the moments/cumulants uniquely define the PDF, as it is possible for all the moment to exist but the limit which defines the generating function to not converge, as is the case for the log-normal distribution.

For joint distributions the joint cumulant generating function reads
\begin{subequations}
\begin{align}
\phi_{X,Y}(\lambda_1, \lambda_2) &= \log(\mathcal{M}_{X,Y}(\lambda_1,\lambda_2)) = \log \ev{e^{\lambda_1 X + \lambda_2 Y}} \\
&= \sum_{p,q=0}^\infty \frac{\ev{X^pY^q}}{p!q!}\lambda_1^p\lambda_2^q - 1.
\end{align}
\end{subequations}
This then allows the CGF to be defined on any linear combination of random variables. The PDF can then be obtained from the CGF via inverse Laplace transform \parencite[see][for a detailed derivation of this]{Balian.Schaeffer_1989_ScaleinvariantMatter, Bernardeau_2013_EvolutionLargescale}
\begin{equation}
P_X(x) = \int_{\alpha - i \infty}^{\alpha + i \infty} \frac{\dd{\lambda}}{2\pi i} \exp(-\lambda x  + \phi_X(\lambda)).
\end{equation}

Cumulants and moments can be expressed in terms of each other via the partial Bell polynomials $B_{n,k}$
\begin{subequations}
\begin{align*}
k_{n} &= \sum_{k=1}^n (-1)^{k-1} (k-1)! B_{n,k}(M_1,\dots,M_{n-k+1}) \,, \\
M_n &= \sum_{k=1}^n B_{n,k}(k_1, \dots , k_{n-k+1}).
\end{align*}
\end{subequations}

\subsection{Reduced cumulants and scaled CGF}

Taking the cosmic density field as the random field of interest, the quantities
\begin{equation}
S_n = \frac{\ev{\delta^n}_c}{\ev{\delta^2}_c^{n-1}} = \frac{k_n}{k_2^{n-1}}
\end{equation}
are called the reduced cumulants, where $k_2 =\sigma^2$ is the variance of the density. These are of importance in this work as they are independent of cosmology down to mildly non-linear scales \parencite{Peebles_1980_LargescaleStructure, Baugh.etal_1995_ComparisonEvolution}, and are at the root of Large Deviations Theory, which we apply to the matter PDF. The scaled cumulant generating function (SCGF) is defined with these reduced cumulants as the coefficients, in the limit of vanishing variance,
\begin{equation}
\varphi_X(\lambda) = \lim_{k_2\to 0} \sum_{n=0}^\infty S_n \frac{\lambda^n}{n!} = \lim_{k_2\to 0} k_2 \,  \phi_X\left(\frac{\lambda}{k_2}\right).
\end{equation}

\section{Hypergeometric functions}\label{app:sec:hyp-geo}

For general information on special functions and properties of these functions, see \cite{NIST:DLMF}.

The Gauss hypergeometric function is defined by a power series
\begin{equation}
{}_{2}F_{1}(a,b;c;z) = \sum_{k=0}^\infty \frac{(a)_k (b)_k z^k}{(c)_k k!} = 1 + \frac{ab}{c}\frac{z}{1!} + \frac{a(a+1)b(b+1)}{c(c+1)}\frac{z^2}{2!} + \dots,
\end{equation}
where $(n)_k$ is the rising Pochhammer symbol (like a factorial but the terms raise by 1 instead of decreasing by 1)
\begin{equation}
(n)_k = \begin{cases} 
	1 & k=0 \\
	n(n+1)\dots(n+k-1) & k>0.
\end{cases}
\end{equation}
These Pochhammer symbols can also be represented via the Gamma function as
\begin{equation}
(n)_k = \frac{\Gamma(n+k)}{\Gamma(n)}.
\end{equation}

This can be generalised the generalised hypergeometric series ${}_{p}F_{q}$ which takes $p$ arguments in the numerator and $q$ arguments in the denominator of the series
\begin{equation}
{}_{p}F_q(a_1,\dots,a_p;b_1,\dots,b_q;z) = \sum_{k=0}^\infty \frac{(a_1)_k\dots(a_p)_k}{(b_1)_k \dots (b_q)_k} \frac{z^k}{k!}.
\end{equation}

Such functions come up as solutions to differential equations of the form
\begin{equation}
z(z-1)\dv[2]{w}{z} + [c- (a+b+1)z] \dv{w}{z} - ab w = 0,
\end{equation}
which has three regular singular points at $z=0,1,\infty$. Any second order linear differential equation with 3 singular points can be converted to this form by change of variables. Solutions to this differential equation are built out of this function ${}_{2}F_{1}$ and polynomials in $z$ times it.

Another specific hypergeometric function of use to is the  hypergeometric function
\begin{equation}
{}_0F_1(a;z) = \sum_k \frac{z^n}{(a)_k k!} = \sum_k \frac{\Gamma(a)}{\Gamma(k+a)\Gamma(k+1)}z^n
\end{equation}
which arises in the minimal tree model for the matter covariance in Chapter \ref{chap:covariance}. This solves the differential equation
\begin{equation}
\left[z\dv[2]{}{z} + a \dv{}{z} - 1\right] F = 0.
\end{equation}
This hypergeometric function is closely related to the better known Bessel functions
\begin{equation}
J_n(x) = \frac{(\frac{1}{2}x)^2}{n!} {}_0F_1\left(n+1;-\frac{1}{4}x^2\right) ,
\end{equation}
where $J_n(x)$ is the Bessel function of the first kind.  The regularised confluent hypergeometric function ${}_0\tilde{F}_1(a;z) = {}_0F_1(a;z)/\Gamma(a)$ is related to the modified Bessel functions $I_n(x)$ by
\begin{equation}
{}_0\tilde{F}_1(a;z) = z^{\frac{1-a}{2}} I_{a-1}(2\sqrt{z}).
\end{equation}

\section{Asymptotic evaluation of integrals}\label{app:SPA}

If there is one non-linear problem which physicists are well equipped to handle, it is the simple harmonic oscillator and relatedly, the evaluation of Gaussian integrals. Here we leverage this power to evaluate integrals of the form
\begin{equation}
I(\nu) = \int \dd{t} f(t) \exp(\nu g(t)),
\end{equation}
where $\nu$ is some large parameter. Integrals of this form appear frequently in physics. In this work the relevant examples are
\begin{enumerate}
\item The theorems of LDT rely on evaluation of integrals with rate function $\psi$ and a large driving parameter $n$, which is given by the $1/\sigma^2$ in the context of cosmological PDFs.
\item The path integral formulation of quantum mechanics relies on integrals over $e^{i S/\hbar}$ where $S$ is a classical action. The semiclassical limit is then obtained in the $\hbar\to 0$ limit.
\item The wave field in optics can be expressed as the oscillatory Fresnel-Kirchhoff integral. The geometric optics limit is obtained as the wavenumber $\nu\to\infty$. 
\end{enumerate}

Such integrals can be evaluated in the $\nu\to \infty$ limit as asymptotic series in powers of $\nu$ through a variety of techniques. Here we only discuss the leading order term, which is known as the Saddle Point Approximation. Further details on these methods and their extensions can be found in standard texts on asymptotic analysis e.g. \textcite{BenderandOrszag}.

\subsection{Laplace's method}

Laplace's method applies to real integrals
\begin{equation}
I(\nu) = \int_a^b \dd{t} f(t) \exp(\nu g(t))\,,
\end{equation}
where $f,g$ are both real functions and the integration range is along the real line. The idea behind this method is that the main contribution to this integral as $\nu$ becomes large is from points where $g(t)$ is maximised, as the integrand exponentially drops off away from such points.  For concreteness we take the location of this maximum $t=t_*$ to not occur at the endpoint $a < t_* < b$, though similar arguments can be made in those cases. We then approximate the integral $I(\nu)$ by restricting the range of integration to a small region around the stationary point
\begin{equation}
    I(\nu) \sim \int_{t_*-\epsilon}^{t_*+\epsilon} \dd{t} f(t) e^{\nu g(t)}.
\end{equation}
We then Taylor expand both $f$ and $g$ about this stationary point, then further Taylor expand the exponential to bring down any terms which are more than quadratic in $t$:
\begin{subequations}
\begin{align}
    I(\nu) &\sim \int_{t_*-\epsilon}^{t_*+\epsilon} \dd{t} f(t) e^{\nu g(t)} \\ 
    &\sim \int_{t_*-\epsilon}^{t_*-\epsilon} \dd{t} (f(t_*) + \dots)  e^{\nu\left(g(t_*) +\frac{1}{2}g''(t_*)(t-t_*)^2 + \dots\right) }.
\end{align}
\end{subequations}
The final trick comes from realising that we can now push the limits of the integral back out to  $\pm \infty$ and at the cost of introducing only exponentially small errors, 
\begin{equation}
I(\nu) \sim \int_{-\infty}^{\infty} \dd{t} (f(t_*) + \dots)  e^{\nu\left(g(t_*) +\frac{1}{2}g''(t_*)(t-t_*)^2 + \dots\right) }\,,
\end{equation}
which is now a Gaussian integral that can be evaluated exactly
\begin{equation}
    I(\nu) \sim f(t_*) e^{\nu g(t_*)} \sqrt{\frac{2\pi}{-\nu g''(t_*)}} + \dots.
\end{equation}
Higher order terms (with respect to $\nu$)  in this series could be obtained by keeping further terms in the Taylor expansions of $f(t)$ and $\exp(\nu g(t))$, then performing the resulting moments of Gaussian integrals exactly.

If there are multiple stationary points, and they are sufficiently separated compared to $\nu^{-1}$, then the total integral can be approximated by simply summing over these stationary points.

\subsection{Method of stationary phase}

The method of stationary phase is similar in spirit to Laplace's method, and deals with integrals with oscillatory integrands
\begin{equation}
    I(\nu) = \int_a^b \dd{t} f(t) e^{i \nu g(t)}.
\end{equation}
The strategy for such integrals follows the same spirit as Laplace's method. In this case, the integrand is not exponentially suppressed as we move away from stationary points of $g$, as the magnitude of the integrand is controlled by the size of $f$. However, since the integrand is so oscillatory, it will largely cancel itself out. Near points $t_*$ where $g'(t_*)=0$, the frequency of oscillations decreases, leading to such points dominating the integral. The method to obtain this leading contribution then follows the same steps as in Laplace's method:
\begin{enumerate}
    \item Restricting the integration range to be isolated around each stationary point (provided the stationary points are sufficiently separated compared to $\nu^{-1}$).
    \item Taylor expanding $f$ and $g$ about $t=t_*$, then further Taylor expanding the exponential term so that only quadratic or lower terms are left in the exponent.
    \item Replacing the limits of the integral by $\pm\infty$, and performing the resulting Gaussian integrals.
\end{enumerate}
The contribution from a single stationary point in this case is
\begin{align}\label{eqn:SPA_1d}
    I_{t_*}(\nu) =& \sqrt{\frac{2\pi}{\nu\abs{g''(t_*)}}}f(t_*)  \exp\left( i\nu g(t_*) + \frac{i\pi}{4} \operatorname{sgn}(g''(t_*))\right),
\end{align}
which has the same structure as the result from Laplace's method, but with an additional phase factor $\exp(\pm i \pi)$ depending on the sign of the second derivative, called the Morse index.

We note for completeness two useful direct extensions to equation~\eqref{eqn:SPA_1d}. Firstly, if the first $(p-1)$ derivatives of $g$ are all 0 at a stationary point, but $g^{(p)}(t_*)\neq 0$, the contribution receives a modification \parencite{BenderandOrszag}
\begin{align}
    I_{t_*}^{\rm SPA}(\nu) = f(t_*) \left[\frac{p!}{\nu \abs{g^{(p)}(t_*)}}\right]^{\frac{1}{p}} \frac{\Gamma(\frac{1}{p})}{p} \, \exp\left(i\nu g(t_*) + \frac{i\pi}{4}\operatorname{sgn}(g^{(p)}(t_*))\right).
\end{align}

Secondly, we can extend this formalism to integrals over $n$-dimensional space. In this case
\begin{align}
    I_{t_*}^{\rm SPA}(\nu) = \frac{f(\bm{t}_*)}{\abs{\det\mathsf{H}(\bm{t}_*)}^{1/2}}\exp\left(i\nu g(\bm{t}_*) + \frac{i\pi}{4}\operatorname{sgn}(\mathsf{H}(\bm{t}_*))\right),
\end{align}
where $\mathsf{H}_{ij} = \pdv*{g}{t_i}{t_j}$ is the Hessian matrix (provided $\mathsf{H}(\bm{t}_*)\neq 0$). The signature of the Hessian is the difference between the number of negative and positive eigenvalues. Similar results hold for Laplace's method.

As a note of caution, the stationary phase approximation is not well suited to obtaining terms higher than this leading contribution in $\nu$. It performs worse than its equivalent for non-oscillatory integrals as in Laplace's method. This is because in Laplace's method, the errors introduced by steps such as pushing the integration range to $\infty$ introduce exponentially small corrections, while in stationary phase, the corrections are generally only algebraically small \parencite[see Chapter 6 of][]{BenderandOrszag}. Thus higher order terms depend both on further terms in the Taylor expansion about the stationary points, \emph{and} on contributions from non-stationary points. For this reason, obtaining next order (in $\nu$) corrections should be done either by asymptotic matching, or by considering the full steepest descent contour.

\subsection{Method of steepest descent}\label{app:sec:steepest-descent}

The most general form of these integrals is
\begin{equation}
I(\nu) \sim \int_\gamma \dd{t} f(t) e^{\nu g(t)}, 
\end{equation}
where now $f$ and $g$ are allowed to be complex and integration is performed over a curve in the complex $t$-plane. Evaluation of such integrals takes advantage the freedom to deform the integration contour without changing the integral by Cauchy's theorem (provided the functions are analytic). If we write $g(t)=\phi(t) + i \psi(t)$ where $\phi,\psi$ are both real functions, the integral becomes 
\begin{equation}
I(\nu) \sim \int_\gamma \dd{t} f(t) e^{\nu (\phi(t) + i \psi(t) )}.
\end{equation}
We now wish to choose a new integration contour $\gamma'$ such that along that contour $\psi(t)$ is constant, removing the oscillations in the integral
\begin{equation}
I(\nu) \sim e^{i \nu \psi_*}\int_{\gamma'} \dd{t} f(t) e^{\nu \phi(t) }.
\end{equation}
where $\psi_*$ is the value of $\psi$ along $\gamma'$. Then the remaining integral can be approximated via Laplace's method. 

Writing the complex integration variable as $t=x+iy$, the Cauchy-Riemann conditions for $g(t)$ to be analytic read
\begin{equation}
\pdv{\phi}{x} = \pdv{\psi}{y}, \quad \pdv{\phi}{y} = - \pdv{\psi}{x}\,,
\end{equation}
and the functions $\psi$ and $\psi$ are both harmonic, with $\nabla^2 \phi = \nabla^2 \psi = 0$. This means that the only critical points of $\phi$ which can occur are saddle points (as maxima or minima would require $\del_{x}^2\phi \del_y^2 \phi > 0$). 

The contours which have constant $\psi = \mathrm{Im}(g)$ must have $\nabla \psi$ perpendicular to that contour, and by the Cauchy-Riemann conditions, $\nabla \phi$ must then be along said contour. Thus the contours of constant $\psi$ are also contours of steepest descent (or ascent) of $\phi$. Thus the method outlined above is usually referred to as the method of steepest descent.

We note briefly also that the ``thimbles'' of Picard-Lefschetz theory \parencite[e.g.][]{Feldbrugge2019} are these steepest descent/ascent contours, together with a prescription for how to flow from one contour to another, to find such contours.

%
%
%

\section{Multi-index notation}\label{sec:multi-index_notation}
Multi-indices make certain results in multivariable calculus much more notationally compact and transparent. A multi-index $\alpha = (\alpha_1, \dots, \alpha_N) \in \Z_{\geq 0}^N$ is a list of $N$ non-negative integers. We define \emph{size} of a multi-index $\alpha$ by
\begin{equation}
\abs{\alpha} = \sum_{i=1}^N \alpha_i = \alpha_1 + \dots + \alpha_N.
\end{equation}
We define the \emph{factorial} of a multi-index, $\alpha !$ as the product of the factorials of each index
\begin{equation}
\alpha! = \prod_{i=1}^N \alpha_i ! = \alpha_1 ! \dots, \alpha_N!.
\end{equation} 
For a vector $\xx \in \R^N$ we define the exponentiation by a multi-index as
\begin{equation}
\xx^\alpha = \prod_{i=0}^N x_i^{\alpha_i} = x_1^{\alpha_1}\dots x_N^{\alpha_N}.
\end{equation}
We define the $\alpha$\th ~partial derivative in an analogous way, writing
\begin{equation}
\del_{\xx}^{\alpha} = \del_{x_1}^{\alpha_1}\dots\del_{x_N}^{\alpha_N}.
\end{equation}

With this notation we can write down generalisations to the multinomial and Taylor's theorems in a very neat and compact way. The multinomial theorem in $\R^N$ becomes
\begin{equation}
(x_1 + \dots + x_N)^m = \sum_{\abs{\alpha}=m} \frac{m!}{\alpha!}(x_1+\dots+x_N)^\alpha.
\end{equation}
Note that writing this more explicitly
\begin{equation}
(x_1 + \dots + x_N)^m = \sum_{\alpha_1+\dots+\alpha_N=m}\frac{m!}{\alpha_1!\dots\alpha_N!}x_1^{\alpha_1}\dots x_N^{\alpha_N},
\end{equation}
Taylor's theorem becomes
\begin{equation}
f(\xx) = \sum_{\abs{\alpha}=0}^\infty \frac{(\xx-\xx_0)^\alpha}{\alpha!}\del^\alpha f\eval_{\xx=\bm{x_0}}.
\end{equation}

\chapter{Further perturbation theory calculations}

\section{Linear growth factor }\label{app:growth_hypgeometric}
The linear growth factor can be written in terms of hypergeometric functions (defined in Appendix~\ref{app:sec:hyp-geo}) in the following way.

Linearised density perturbations obey
\begin{equation}
\ddot{ \delta }+ 2 H(t) \dot{\delta} = \frac{3\Omega_m(t) H^2(t)}{2}\delta,
\end{equation}
which defines the growth factor. Letting $\lambda(a) = \Omega_\Lambda / \Omega_m(a)$, $\lambda_0 = \lambda(a=0)$, the growing mode solution is \parencite{ Demianski.etal_2005_EvolutionDensity}
\begin{equation}
D_+(a) = a \sqrt{1+\lambda_0 a^3} {}_{2}F_{1}\left(\frac32, \frac56, \frac{11}{6}, -\lambda_0 a^3\right),
\end{equation}
where $ {}_{2}F_{1}(a,b,c,z)$ is the Gauss hypergeometric function. The decaying mode decays as
\begin{equation}
D_-(a) = \sqrt{1+\lambda_0 a^3} a^{-3/2}.
\end{equation}
For small scale factors, the growing mode can be expanded as
\begin{equation}
D_+(a) = a - \frac{2}{11}\lambda_0 a^3 + \order{a^7}.
\end{equation}
This relation can be inverted to write
\begin{equation}
a(D_+) = D_+ + \frac{2}{11}\lambda_0 D_+^3 + \order{D_+^7}.
\end{equation}

\section{Fluid equations in Fourier space}\label{app:fluid-equations-in-fourier}

In this Section calculations are done assuming an Einstein-de Sitter background, such that $D_+=a$. This largely follows \textcite{Bernardeau:2002, Jain.Bertschinger_1994_SecondOrderPower}. I will also follow their time variable choices, using often using conformal time rather than coordinate time as is used mostly in the body of this thesis. 

The fluid equations in real space are
\begin{subequations}
\begin{align}
\delta' + \frac{1}{a}\bm{\nabla}\cdot [(1+\delta)\uu] &= 0, \\
\uu' + \mathcal{H}\uu + (\uu \cdot\bm{\nabla})\uu &= -\grad\Phi_N, \\
\nabla^2 \Phi_N = \frac{4\pi G \bar{\rho}}{a^2}\delta &= \frac{3}{2} \Omega_m(\tau)\mathcal{H}^2(\tau) \delta,
\end{align}
so the associated equation for the velocity divergence is
\begin{equation}
\theta' + \mathcal{H}\theta + \frac{3}{2}\Omega_m(\tau)\mathcal{H}^2(\tau) \delta = - \bm{\nabla}\cdot [(\uu \cdot \bm{\nabla})\uu]
\end{equation}
\end{subequations}

We want to rewrite these in Fourier space. Fourier transforming the continuity equation
\begin{align}
\delta'(\kk) +\theta(\kk) = -i\kk \cdot \mathcal{F}[\delta \uu ](\kk) = -i\kk \cdot (\mathcal{F}[\delta]*\mathcal{F}[\uu])(\kk)
\end{align}
where we've applied the convolution theorem in the last equality. Writing this out as an integral, an noting that since $\theta = \bm{\nabla}\cdot \uu$, in Fourier space we have $\uu(\kk) =-i\kk \theta(\kk) / k^2$ the continuity equation reads
\begin{align}
\delta'(\kk) +\theta(\kk)  &= -i \kk \cdot \int \dd[3]{\kk_1}\delta(\kk-\kk_1) \theta(\kk_1) \frac{-i\kk_1}{k_1^2} \nonumber \\
&= \int \dd[3]{\kk_1} \delta_{\rm D}(\kk - \kk_{12}) \delta(\kk_2) \theta(\kk_1) \underbrace{\left(\frac{k_{12} \cdot \kk_1}{\kk_1^2}\right)}_{\alpha(\kk_1,\kk_2)}
\end{align}
where the function $\alpha(\kk_1,\kk_2)$ is called a mode coupling function as it couples together different Fourier modes of $\delta$ and $\theta$. 

To obtain the Fourier Bernoulli equation, consider Fourier transforming the term
\begin{align}
\mathcal{F}[\bm{\nabla}\cdot( (\uu \cdot \bm{\nabla})\uu)] &= \mathcal{F}\left[\bm{\nabla}\cdot \left(\frac12 \grad\vert\uu\vert^2 - \uu \times \grad \times \uu  \right)\right]  \nonumber \\
&=\mathcal{F}\left[ \frac12 \nabla^2 u^2 \right] = \frac{1}{2}\mathcal{F}[\nabla^2 (\uu \cdot \uu)] \nonumber \\
&= -\frac{1}{2}k^2 \mathcal{F}[\uu \cdot \uu](\kk) \nonumber \\
&= -\frac12 k^2 (\mathcal{F}[\uu] * \mathcal{F}[\uu])(\kk) \nonumber \\
&= -\frac{1}{2}k^2 \int \dd[3]{\kk_1}\dd[3]{\kk_2} \delta_{\rm D}(\kk-\kk_{12}) \frac{k_{12}^2(\kk_1\cdot\kk_2))}{2k_1^2 k_2^2} \theta(\kk_1) \theta(\kk_2)
\end{align}
thus the full Bernoulli equation in Fourier space is
\begin{equation}
\theta'(\kk) + \mathcal{H}(\tau)\theta(\kk) + \frac{3}{2}\Omega_m(\tau)\mathcal{H}^2(\tau)\delta(\kk) = \int \dd[3]{\kk_1}\dd[3]{\kk_2} \delta_{\rm D}(\kk-\kk_{12})  \underbrace{\frac{k_{12}^2 (\kk_1\cdot\kk_2)}{2k_1^2 k_2^2} }_{\beta(\kk_1, \kk_2)} \theta(\kk_1) \theta(\kk_2)
\end{equation}

The mode coupling functions then define the perturbation kernels $F_n, G_n$. A recursive relation between them can be found by inserting the perturbative ansatz into the fluid equations and matching perturbative orders, as in \textcite{Jain.Bertschinger_1994_SecondOrderPower}.

\section{$F_2$ and $\nu_2$ in SPT}\label{app:sec:f2-nu2}
In SPT, the symmetrised perturbation kernel $F_2$ is
\begin{align}
F_2^{(s)}(\kk_1, \kk_2) &= \frac{5}{7}\alpha^{(s)}(\kk_1,\kk_2) + \frac{2}{7}\beta^{(s)}(\kk_1, \kk_2) \nonumber \\
&= \frac{5}{7} + \frac{1}{2}\frac{(\kk_1 \cdot \kk_2)}{k_1 k_2}\left(\frac{k_1}{k_2} + \frac{k_2}{k_1}\right) + \frac{2}{7}\frac{(\kk_1 \cdot \kk_2)^2}{k_1^2 k_2^2}.
\end{align}
The angular average of the perturbation kernel (which determines the tree-order skewness) can then be obtained by integrating over the angle between $\kk_1, \kk_2$
\begin{equation}
\nu_2 = \frac{2!}{ (4\pi)^2} \times \int \dd{\Omega_1}\dd{\Omega_2} F_2(\kk_1,\kk_2)\,,
\end{equation}
where $\dd{\Omega_i} = \sin\theta_i \dd{\theta_i}\dd{\phi}_i$ is the solid angle element. Since $F_2$ depends only on the angle between the vectors, we change to $\dd{\Omega}\dd{\Omega_{12}}$ to obtain
\begin{equation}
\nu_2 = \frac{2!}{ (4\pi)^2} \times \int \dd{\Omega} \times \int \dd{\Omega_{12}} F_2(\kk_1,\kk_2).
\end{equation}
Since the angular dependence of $F_2$ comes from $\kk_1\cdot\kk_2 = k_1 k_2\cos\theta_{12}$, we make use of
\begin{equation}
\int \dd{\theta} \sin\theta \cos\theta = 0,  \quad \int \dd{\theta} \sin\theta \cos^2\theta = \frac{2}{3}\,,
\end{equation}
which means that the angular average of $F_2$ is
\begin{equation}
\nu_2 = 2 \left[ \frac{5}{7} + 0 + \frac{2}{7}\times \frac{1}{3}\right] = \frac{34}{17}.
\end{equation}

\section{Matching spherical collapse index to skewness} \label{app:spherical_collapse_to_skew}

Consider the fitting formula for spherical collapse
\begin{equation}\label{eqn:appendix-spherical-collapse-parametrise}
    \mathcal{F}(\delta_{\rm L}) = 1 + \delta_{\rm NL}^{\rm SC}(\delta_{\rm L}) = \left(1-\frac{\delta_{\rm L}}{\nu_{\rm SC}}\right)^{-\nu_{\rm SC}}\,,
\end{equation}
which is parametrised by some parameter $\nu_{\rm SC}$. The expansion of this density leads to the following for the skewness
\begin{equation}
    \delta_{\rm NL} = \delta_{\rm L} + \frac{1}{2!}\frac{\nu_{\rm SC} + 1}{\nu_{\rm SC}}\delta_{\rm L}^2 +  \frac{1}{3!} \frac{(\nu_{\rm SC}+1)(\nu_{\rm SC}+2)}{\nu_{\rm SC}^2} \delta_{\rm L}^3 + \dots\,,
\end{equation}
that is, the $n^{\rm th}$ vertex factor for this fitting formula gives:
\begin{equation}
    \nu_n = \frac{\nu_{\rm SC}(\nu_{\rm SC}+1)\dots(\nu_{\rm SC}+(n-1))}{\nu_{\rm SC}^{n-1}}\,,
\end{equation}
so, the predicted tree-order skewness for this expansion is
\begin{equation}
    S_3^{\rm tree, fitting} = 3\nu_2 = 3 \frac{\nu_{\rm SC} + 1}{\nu_{\rm SC}} = 3\left(1 + \frac{1}{\nu_{\rm SC}}\right).
\end{equation}

This means to match the tree-order skewness predicted from SPT by using this fitting formula, we should choose a spherical collapse index $\nu_{\rm SC}$ related to the PT vertex $\nu_2$ by
\begin{equation}
    \nu_{\rm SC} = \frac{1}{\nu_2-1} 
\end{equation}
which in 3D is 
\begin{equation}
    \nu_{\rm SC} = \frac{1}{34/21-1} = \frac{21}{13} \approx 1.615.
\end{equation}

\section{Smoothing contribution to skewness}\label{app:sec:S3-smoothing-SPT}

In this Section we present the calculation of the result that the reduced skewness at tree order in SPT, when smoothed with spherical top-hat filter in 3D is \parencite{Bernardeau_1994_SkewnessKurtosis}
\begin{equation}
    S_3^{\rm tree, SPT}(R) = \frac{34}{7} + \dv{\log\sigma_R^2}{\log R}.
\end{equation}
Anticipating this result we write down a couple of useful things. The variance on scale $R$ is given as
\begin{equation}
    \sigma_R^2 = \int \frac{\dd{k}}{2\pi^2} k^2 P(k) W(kR)^2.
\end{equation}
The log-derivative smoothing term is 
\begin{align}
    \dv{\log \sigma_R^2}{\log R} &= \frac{R}{\sigma^2}\dv{\sigma^2}{R} \\
    &= \frac{1}{\sigma^2}\int \frac{\dd{k}}{2\pi^2} k^2 P(k) R \dv{}{R}[W(kR)^2 ]\\
    &= \frac{1}{\sigma^2}\int \frac{\dd{k}}{2\pi^2}  k^2 P(k)\cdot (2kR)\cdot W(kR)W'(kR)\,,
\end{align}
where prime here denotes differentiation with respect to the argument.

The ability to write the smoothing term in closed form relies on properties of Bessel functions (which are the Fourier transform of real space spherical smoothing kernels). The identities we will need to use involve the functions
\begin{align}
\mathscr{P}(\kk_1, \kk_2) &= \left[1 + \frac{\bm{k}_1\cdot \bm{k}_2}{k_1^2}\right] \,, \\
\mathscr{Q}(\kk_1, \kk_2) &= \left[1 - \frac{(\bm{k}_1\cdot \bm{k}_2)^2}{k_1^2k_2^2}\right]\,,
\end{align}
which arise in the integrals
\begin{align}\label{eq:top-hat-identities}
\int \frac{\dd{\Omega_{12}}}{4\pi}W(\abs{\bm{k}_1+\bm{k}_2}) \mathscr{P}(\kk_1, \kk_2) &= W(k_1)W(k_2) + \frac13 k_2 W'(k_2)W(k_1) \,,\\
\int \frac{\dd{\Omega_{12}}}{4\pi}W(\abs{\bm{k}_1+\bm{k}_2}) \mathscr{Q}(\kk_1, \kk_2) &= \frac23 W(k_1)W(k_2).
\end{align}

The (symmetrised) 2nd order SPT density kernel function can be written in terms of these functions
\begin{align}
F_{2}^{(s), \rm SPT}(\kk_1, \kk_2) &= \frac{5}{7} + \frac{1}{2}\frac{\kk_1\cdot\kk_2}{k_1^2k_2^2}\left(\frac{k_1}{k_2} + \frac{k_2}{k_1}\right) + \frac{2}{7}\frac{(\kk_1\cdot\kk_2)^2}{k_1^2 k_2^2}\\
&= \frac{1}{2}\left[\mathscr{P}(\kk_1, \kk_2) + \mathscr{P}(\kk_2, \kk_1)\right] - \frac{2}{7}\mathscr{Q}(\kk_1, \kk_2)\,,
\end{align}

The tree order skewness corresponds to $\ev{\delta_R^3(\xx)} = 3 \ev{\delta^{(1)}_R(\xx)\delta^{(1)}_R(\xx)\delta^{(2)}_R(\xx)} $ as we've seen. The second order smoothed density can be written (taking $\xx=0$ by homogeneity allows us to drop the exponential in the Fourier transform)
\begin{align}
\delta^{(2)}_R(\xx) &= (\delta^{(2)}* W_R)(\xx) = \mathcal{F}^{-1}[\delta^{(2)}(\kk)W(kR)] \nonumber \\
&= \int \frac{\dd[3]{\kk}\dd[3]{\kk_1}\dd[3]{\kk_2}}{(2\pi)^3} (2\pi)^3 \delta_{\rm D}(\kk-\kk_{12}) W(kR)\delta_{\rm L}(\kk_1)\delta_{\rm L}(\kk_2) F_2(\kk_1,\kk_2) \nonumber \\
&= \int \frac{\dd[3]{\kk_1}\dd[3]{\kk_2}}{(2\pi)^6} \, W(\abs{\kk_1+\kk_2}R) \, \delta_{\rm L}(\kk_1) \, \delta_{\rm L}(\kk_2) \, F_2(\kk_1,\kk_2).
\end{align}
Inserting this into the tree order skewness we obtain
\begin{align}
\langle \delta_R^3\rangle_c^{\rm tree} = 3 \int \frac{\dd[3]{\kk_{1,2,3,4}}}{(2\pi)^{12}}\, W(\abs{\kk_1+\kk_2}R) \, W(k_3 R) \,  &W(k_4 R) \, F_2(\kk_1, \kk_2) \times \nonumber \\
&\ev{\delta_{\rm L}(\kk_1)\delta_{\rm L}(\kk_2)\delta_{\rm L}(\kk_3)\delta_{\rm L}(\kk_4)}_c.
\end{align}
Because of the structure of $F_2(\kk_1,\kk_2)$, if we pair $\kk_1$ and $\kk_2$ together via Wick's theorem, the contribution to the integral will vanish. The other terms contribute equally if we use the symmetrised $F_2$ kernel function, allowing us to write the four point correlation
\begin{align}
\ev{\delta_{\rm L}(\kk_1)\delta_{\rm L}(\kk_2)\delta_{\rm L}(\kk_3)\delta_{\rm L}(\kk_4)}_c &= 2 \ev{\delta_{\rm L}(\kk_1)\delta_{\rm L}(\kk_3)} \ev{\delta_{\rm L}(\kk_2)\delta_{\rm L}(\kk_4)} \nonumber \\
&= 2 \cdot  (2\pi)^6 \delta_{\rm D}(\kk_1+\kk_3)   \delta_{\rm D}(\kk_2+\kk_4) P(k_1) P(k_2).
\end{align}
Putting this in to the skewness integral and integrating out the Dirac deltas leaves
\begin{align}
\langle \delta_R^2\rangle_c^{\rm tree} = 6 \int \frac{\dd[3]{\kk_{1,2}}}{(2\pi)^{6}}  \, W(k_1 R) \, P_{\rm L}(k_1) \, W(k_2 R)  \, P_{\rm L}(k_2) \,  W(\abs{\kk_1+\kk_2}R) \, F_2(\kk_1,\kk_2) 
\end{align}

We now change to polar coordinates, $\dd[3]{\kk}_i = k_i^2 \dd{k}_i\dd{\Omega}_i$. Note as well that since $F_{2}(\kk_1, \kk_2)$ and $W(\abs{\kk_1+\kk_2}R)$ depend only on the relative angle between $\kk_1, \kk_2$ we can do one of the angular integrals immediately to obtain $4\pi$, and then only integrate only over the angle between $\kk_1, \kk_2$
\begin{align}
\langle \delta_R^2\rangle_c^{\rm tree} = 6 \int \frac{\dd{k_1}\dd{k_2} }{(2\pi)^6}  \, &k_1^2 P_{\rm L}(k_1)W(k_1 R)  \, k_2^2 P_{\rm L}(k_2)W(k_2 R) \times \nonumber \\
&4\pi \int \dd{\Omega_{12}} W(\abs{\kk_1+\kk_2}R) \, F_2(\kk_1,\kk_2).
\end{align}
We now make use of the properties of the window function to solve the angular part of this integral. Defining the angular part
\begin{align}
I = (4\pi)^2 \int \dd{\Omega_{12}} W(\abs{\kk_1+\kk_2}R)  \left[\mathscr{P}^{(s)}(\kk_1, \kk_2) - \frac{2}{7}\mathscr{Q}(\kk_1,\kk_2) \right] \,,
\end{align}
and applying the identities~\eqref{eq:top-hat-identities}
\begin{align*}
I &= (4\pi)^2 \times \left[ \frac{17}{21} W(k_1R) W(k_2 R) + \frac{1}{6}k_1R W'(k_1 R) W(k_2 R) + \frac{1}{6}k_2R W'(k_2 R) W(k_1 R)  \right].
\end{align*}
The full skewness integral then becomes
\begin{align}
\langle \delta_R^3\rangle_c^{\rm tree}  = &\phantom{+} \frac{34}{17}\int \frac{\dd{k_1}}{2\pi^2} k_1^2 P_{\rm L}(k_1) W^2(k_1 R) \int \frac{\dd{k_2}}{2\pi^2} k_2^2 P_{\rm L}(k_2) W^2(k_2 R) \nonumber \\
&+ \int \frac{\dd{k_1}}{2\pi^2}  \, k_1^2 P_{\rm L}(k_1) W(k_1 R) (k_1 R)W'(k_1 R) \int \frac{\dd{k_2}}{2\pi^2} k_2^2 P_{\rm L}(k_2) W^2(k_2 R)  + (k_1 \leftrightarrow k_2)\,, \nonumber
\end{align}
which we can simplify by recognising the linear variance and the derivative of the linear variance (noting we can simply double the second line to account for the symmetrisation)
\begin{align}
\langle \delta_R^3\rangle_c^{\rm tree}  &= \frac{34}{17}(\sigma_R^2)^2 + \sigma_R^2 R \dv{ \sigma_R^2}{ R} \nonumber \\
&= (\sigma_R^2)^2 \left[ \frac{34}{17} + \frac{R}{\sigma^2_R}\dv{ \sigma_R^2}{ R}\right]\,,
\end{align}
which reproduces the tree order reduced skewness
\begin{equation}
S_3^{\rm tree, \, SPT} = \frac{34}{17}  + \dv{\log \sigma_R^2}{\log R}.
\end{equation}

\chapter{Appendix for Chapter \ref{chap:covariance}}

\section{Deriving the auxillary Laplace transform in the minimal tree model} \label{app:aux-laplace-for-min-tree}

Here we calculate the auxiliary version of the Laplace transform of the CGF as the integral $I_j(\rho)$ used in Chapter~\ref{chap:covariance} to obtain the PDF and bias functions for the minimal tree model. This calculation was presented in \cite{Uhlemann.etal_2023_ItTakes}, but here contains a correction to correctly derive the Dirac delta term of the PDF. The integral of interest is:
\begin{equation}
    I_j(\rho) = \int_{-i\infty}^{i \infty} \frac{\dd{\lambda}}{2\pi i}\exp\left[{-\lambda\rho + j\phi(\lambda)}\right].
\end{equation}
The PDF is then the value $I_1(\rho)$, and the bias functions for the minimal tree model are obtained by successive derivatives with respect to $j$,
\begin{equation}
    b_{n,\rm t}(\rho)\mP_{\rm t}(\rho) = \dv[n]{I_j(\rho)}{j}\eval_{j=1}.
\end{equation}
The minimal tree one-point CGF is
\begin{equation}
\phi(\lambda) = \frac{\lambda}{1-\lambda\sigma^2/2}\,,
\end{equation}
which we can insert into the integral. Recall that if a complex function $f(z)$ has a pole of order $n$ at $z=c$, the residue associated with that pole can be obtained by
\begin{equation}
\mathrm{Res}_f(c) = \frac{1}{(n-1)!}\lim_{z\to c}\dv[n-1]{}{z}\left[ (z-c)^n f(z)\right].
\end{equation}
Motivated by this we rewrite the one-point CGF as
\begin{equation}
\phi(\lambda) = \frac{-2\lambda/\sigma^2}{\lambda - 2/\sigma^2}\,,
\end{equation}
and we then can expand the exponential as a series in 
\begin{equation}
    e^{j\phi(\lambda)} = \sum_{n=0}^\infty \frac{1}{n!}(j\phi)^n = \sum_{n=0}^\infty \frac{1}{n!} \left(-\frac{2j\lambda}{\sigma^2}\right)^n \frac{1}{(\lambda - 2/\sigma^2 )^n}.
\end{equation}
The $n=0$ term does not have a simple pole, and takes the form
\begin{equation}
\int_{-i\infty}^{i\infty} \frac{\dd{\lambda}}{2\pi i}e^{-\lambda \rho}\,,
\end{equation}
which is simply a Dirac delta term $\delta_{\rm D}(\rho)$(consider $\lambda \to -i\lambda$ and compare to the Fourier representations of the Dirac delta in Appendix \ref{app:sec:fourier}). The total integral is thus (after exchanging the sum and integral)
\begin{equation}
I_j(\rho) = \delta_{\rm D}(\rho) + \frac{1}{2\pi i}\sum_{n=1}^\infty \int \dd{\lambda} e^{-\lambda\rho}\frac{1}{n!} \left(\frac{2j\lambda}{\sigma^2}\right)^n \frac{1}{(\lambda-2/\sigma^2)^n}.
\end{equation}

Choose the integration contour to be up the imaginary $\lambda$ axis and closing towards the positive real axis since the pole is at $(2/\sigma^2)^n$. Along this contour we can apply Cauchy's residue theorem to each term in this sum (calling the integrand $f_n$ for brevity)
\begin{equation}
    I_j(\rho) = \delta_{\rm D}(\rho) + \frac{1}{2\pi i } \sum_{n=1}^\infty \int_{-i\infty}^{i\infty} \dd{\lambda} f_n(\lambda) =\delta_{\rm D}(\rho)  -\sum_{n=1}^\infty \text{Res}_{f_n}(2/\sigma^2).
\end{equation}

To calculate the $n^{\rm th}$ residue we need the $(n-1)^{\rm th}$ derivative of the product appearing in this sum. We make use of 
\begin{subequations}
\begin{align}
    \dv[k]{}{\lambda} e^{-\lambda \rho} &= (-\rho)^k e^{-\lambda \rho} \,,\\
	\dv[k]{}{\lambda}\left[\left(-\frac{2j}{\sigma^2}\right)^n\lambda^n\right] = \left(-\frac{2j}{\sigma^2}\right)^n \frac{n!}{(n-k)!}\lambda^{n-k} &= \frac{n!}{(n-k)!}\left(-\frac{2j}{\sigma^2}\right)^k\left(-\frac{2j\lambda}{\sigma^2}\right)^{n-k}\,,
\end{align}
\end{subequations}
then apply the generalised product rule,
\begin{equation}
\dv[n-1]{}{\lambda}\left[ g(\lambda)h(\lambda) \right] = \sum_{k=0}^{n-1} \binom{n-1}{k} \dv[k]{g}{\lambda} \dv[n-k]{h}{\lambda}\,,
\end{equation}
to obtain the $n^{\rm th}$ residue at $\lambda = 2/\sigma^2$, we obtain
\begin{equation}
    \text{Res}_{f_n}(2/\sigma^2) = e^{-2\rho/\sigma^2} \sum_{k=0}^{n-1}\frac{1}{k!(k+1)!(n-k-1)!}(-\rho)^k \left(-\frac{2j}{\sigma^2}\right)^{n-1-k}\left(-\frac{4j}{\sigma^4}\right)^{k+1}.
\end{equation}
Putting this into the sum of residuals and shifting the sum of residuals to start at $n=0$ the resulting double sum can be factored as a product via Cauchy sums
\begin{equation}
    \sum_{n=0}^\infty \text{Res}_{f_{n}} (2/\sigma^2) = -\frac{4j}{\sigma^4}e^{-2\rho/\sigma^2} \left[\sum_{\ell = 0}^\infty \frac{1}{\ell (\ell+1)!}\left(\frac{4j\rho}{\sigma^4}\right)^\ell\right] \left[\sum_{r=0}^\infty \frac{1}{r!}\left(-\frac{2j}{\sigma^2}\right)^r\right].
\end{equation}
We can resum these sums in square brackets using hypergeometric functions, so we finally arrive at
\begin{equation}
    I_j(\rho) = \delta_{\rm D}(\rho) + \frac{4j}{\sigma^4}e^{-\frac{2}{\sigma^2}(j+\rho)} \hypgeo{2, \frac{4j\rho}{\sigma^4}}.
\end{equation}
To obtain the PDF, we need to evaluate $I_j$ at $j=1$ and ensure the PDF is appropriately normalised. We do this by introducing a coefficient in front of the Dirac delta term such that the integral over the entire PDF over the range $\rho\in[0,\infty)$ is equal to 1. The continuous part of the PDF obtained by taking the continuous part of $I_{j=1}(\rho)$,
\begin{equation}
 \mP_{\rm continuous}(\rho) = \frac{4}{\sigma^4}e^{-\frac{2}{\sigma^2}(1+\rho)} \hypgeo{2, \frac{4\rho}{\sigma^4}}.
\end{equation}
The integral over the continuous part of the PDF is given by
\begin{align}
\int_0^\infty \mP_{\rm continuous}(\rho)\dd{\rho} &= \frac{4}{\sigma^4}e^{-2/\sigma^2} \int_0^\infty e^{-2/\sigma^2} \hypgeo{2, \frac{4\rho}{\sigma^4}} \dd{\rho} \nonumber \\
&= \frac{4}{\sigma^4}e^{-2/\sigma^2}  \sum_{n=0}^\infty \frac{1}{\Gamma(n+2)\Gamma(n+1)}\int_0^\infty e^{-2\rho/\sigma^2} \left(\frac{4\rho}{\sigma^4}\right)^n \dd{\rho} \nonumber \\
&= \frac{4}{\sigma^4} e^{-2/\sigma^2}  \sum_{n=0}^\infty \frac{(2/\sigma^2)^n}{\Gamma(n+2)\Gamma(n+1)}\int_0^\infty e^{-2\rho/\sigma^2} \left(\frac{2\rho}{\sigma^2}\right)^n \dd\rho \nonumber \\
&= \frac{2}{\sigma^2} e^{-2/\sigma^2}  \sum_{n=0}^\infty \frac{(2/\sigma^2)^n}{(n+1)! n!}\int_0^\infty e^{-x} x^n \dd{x} \nonumber \\
&= e^{-2/\sigma^2}  \sum_{n=0}^\infty \frac{(2/\sigma^2)^{n+1}}{(n+1)!}
\end{align}
where we have integrated the series expansion of the hypergeometric function term by term. We then recognise that we can rewrite the sum as
\begin{equation}
\sum_{n=0}^\infty \frac{(2/\sigma^2)^{n+1}}{(n+1)!} = \sum_{n=0}^\infty\frac{(2/\sigma^2)^{n}}{(n)!} - 1 =  e^{2/\sigma^2}-1
\end{equation}
such that the integral over the continuous part of the PDF is simply $1-e^{-2/\sigma^2}$. To make the integral over the total PDF equal to 1, we introduce a factor $e^{-2/\sigma^2}$, such that the final PDF in the minimal tree model is given by
\begin{equation}
 \mP(\rho) = e^{-2/\sigma^2}\delta_{\rm D}(\rho) + \frac{4}{\sigma^4}e^{-\frac{2}{\sigma^2}(1+\rho)} \hypgeo{2, \frac{4\rho}{\sigma^4}}.
\end{equation}
The bias functions are obtained by derivatives as described in Chapter \ref{chap:covariance}.

We note here that a pair of identities presented in Appendix C of \cite{Bernardeau2024arXiv} which can be used to derive this PDF more quickly but were unknown at the time of writing \cite{Uhlemann.etal_2023_ItTakes}. They are
\begin{equation}
\int \frac{\dd{\lambda}}{2\pi i} \exp(-\lambda \rho + \frac{A}{\lambda_c - \lambda}) = \delta_{\rm D}(\rho) + A e^{-\lambda_c \rho} {}_0\tilde{F}_1(2; A\rho),
\end{equation}
and
\begin{equation}
\int \frac{\dd{\lambda}}{2\pi i} \left(\frac{1}{\lambda_c - \lambda}\right)^n\exp(-\lambda \rho + \frac{A}{\lambda_c - \lambda}) = e^{-\lambda_c\rho} \left(\dv{}{A}\right)^n [A \, {}_0\tilde{F}_1(2; A\rho)],
\end{equation}
for $n\geq 1$. Here ${}_0\tilde{F}_1(a;z)$ is the regularised confluent hypergeometric function defined as ${}_0\tilde{F}_1(a;z)={}_0F_1(a;z)/\Gamma(a)$. Since in our case we are interested in $\tilde{F}_1(2; A\rho)$ and $\Gamma(2)=1$, we can simply replace the regularised hypergeometric function with the ordinary one. This identity immediately leads to the form of the PDF, and differentiation with respect to $A$ obtains the bias functions. One still has to do the integration over the continuous part to obtain the prefactor for the Dirac delta term.

\section{Shifted lognormal model} \label{app:sec:lognormal}
Here we detail the specific shifted lognormal model used in \cite{Uhlemann.etal_2023_ItTakes} in the context of the weak lensing convergence $\kappa$ and the matter density contrast $\delta$. 

A bivariate shifted lognormal joint PDF for the density contrast $\delta$ is parametrised by the variance $\sigma_\delta^2$ and the two-point correlation function $\xi_\delta = \xi_{12,\delta}(d)$ between the cells, and a shift parameter $s$ related to the skewness, which we will keep fixed. We define a Gaussian-distributed variable $g(\delta)$ with zero-mean
\begin{equation}
g(\delta) = \ln(\frac{\delta}{s}+1)-\mu_{\rm G} = \ln(\frac{\delta}{s}+1) + \sigma_{\rm G}^2\,,
\end{equation}
where $\mu_{\rm G}=-\sigma_{\rm G}^2/2$ is fixed to ensure that $\ev{\delta}=0$ and the Gaussian variance $\sigma_{\rm G}^2=\ln(1+\sigma_\delta^2/2)$. This can then be used to construct the shifted lognormal univariate PDF
\begin{equation}
\mP_{\rm LN}(\delta; \sigma_{\rm G}^2, s) = \frac{\Theta(\delta + s)}{\sqrt{2\pi}\sigma_{\rm G}(\delta + s)} \exp[-\frac{g(\delta)^2}{2\sigma_{\rm G}^2}]\,,
\end{equation}
where $\Theta$ is the Heavyside step function such that the PDF is set to zero for $\delta \leq -s$. The shift parameter $s$ can be set to produce the desired skewness of the PDF\parencite[see the formula for $\lambda$ in ][]{Xavier2016}
\begin{align}\label{eq:xavier_lambda}
    s &= \frac{\sigma_\delta}{\tilde{\mu}_3}\left(1+y(\tilde{\mu}_3)^{-1}+y(\tilde{\mu}_3)\right)-\langle \delta\rangle \approx 3\frac{\sigma_\delta^4}{\langle\delta^3\rangle}, \\
    y(\tilde{\mu}_3) &= \left[\frac{2+\tilde{\mu}_3^2+\tilde{\mu}_3\sqrt{4+\tilde{\mu}_3^2}}{2}\right]^{1/3}\approx 1\,, \notag
\end{align}
where $\langle \delta\rangle$ is the desired mean, $\sigma$ the target variance and $\tilde \mu_3=\langle\delta^3\rangle/\sigma_\delta^3$ the skewness. The Gaussian limit is reproduced by sending $s/\sigma_\delta^2\to \infty$. The standard lognormal model \parencite[e.g.][]{Coles1991MNRAS} is recovered for $s=1$. 

The bivariate shifted lognormal model is then expressed in terms of the correlation $\xi_{\rm G}=\ln(1+\xi_\delta/s^2)$ and the determinant $D = \sigma_{\rm G}^4 - \xi_{\rm G}^2$ as
\begin{equation}
\mP_{\rm LN}(\delta_1, \delta_2; \sigma_{\rm G}^2, \xi_{\rm G}, s) = \frac{\Theta(\delta_1 + s)\Theta(\delta_2 + s)}{2\pi\sqrt{D}(\delta_1+s)(\delta_2+s)} \exp[-\frac{\sigma_{\rm G}^2}{2}(g(\delta_1)^2 + g(\delta_2)^2) + \frac{\xi_{\rm G}}{D}g(\delta_1)g(\delta_2)].
\end{equation}

The leading order and NLO bias functions for the shifted lognormal model cannot be obtained in the same manner as presented in Chapter \ref{chap:covariance}, as it is not well described in terms of a CGF. However, since the two-point PDF is expressible in closed form we can instead simply do a brute force expansion in powers of the correlation function to obtain bias functions from this PDF. This requires an extra step compared to simply expanding the PDF in powers of $\xi_{\rm G}$ as the Gaussian field correlation terms $\xi_{\rm G}$ must also be expanded in terms of the underlying $\delta$ correlation $\xi_{\delta}$. To next-to-leading order, this expansion produces (in the $\xi_{\rm G}(\xi_\delta)/\sigma_{\rm G}^2\ll 1$ limit)
 \begin{subequations}
    \begin{align}
    \label{eq:joint_LN_biasexp}
    \frac{\mP_{\rm LN}(\delta_1,\delta_2;\xi_\delta)}{\mP_{\rm LN}(\delta_1)\mP_{\rm LN}(\delta_2)}
    =&\phantom{+} 1+\left(\frac{\xi_\delta}{s^2}-\frac{\xi_\delta^2}{2s^4}\right)\frac{g(\delta_1)}{\sigma_{\rm G}^2}\frac{g(\delta_2)}{\sigma_{\rm G}^2}\\
    \notag &+\frac{\xi_\delta^2}{2s^4}\frac{\left(g(\delta_1)^2 -\sigma_{\rm G}^2\right)}{\sigma_{\rm G}^4}\frac{\left(g(\delta_2)^2-\sigma_{\rm G}^2\right)}{\sigma_{\rm G}^4}\\
    \notag &+\mathcal O(\xi_\delta^3).
    \end{align}
    \end{subequations}
    In principle, on can obtain the bias results to any order for this model using results for the Gaussian bias functions at arbitrary order.

\chapter{Appendix for Chapter \ref{chap:making-dm-waves}}

\section{Domain colouring}\label{app:domain_colouring}
We represent complex numbers, $z = \abs{z}e^{i\theta}$, with the argument, $\theta\in [0,2\pi)$, as points in hue, lightness, saturation (hls) colour space via the mapping
\begin{equation}
    \begin{pmatrix}
    h \\ l \\ s
    \end{pmatrix} = \begin{pmatrix}
    \theta/(2\pi) \\ 1-0.5^\abs{z} \\ 0.5
    \end{pmatrix}.
\end{equation}
Each of $\{h,l,s\}$ lies in the range $0$ to $1$. These hls values can then be mapped to standard RGB values (or any other colour space) in the usual way (c.f. \cite{Wegert2012}).

\section{Derivation of the hidden phase (following Fried 1998)} \label{app:hid_phase_fried}

\cite{Fried1998} derives an explicit expression for the hidden phase based on the location of the branch points in an optical field (equation~\eqref{eqn:hidden_phase}). Such a phase profile could be applicable to studying constant time snapshots for a 2-dimensional system, where now the branch points act as vortex cores \parencite[as shown in][]{Uhlemann2019}. Below we present the derivation of this hidden phase from this work.

The wavefunction $\psi$ plays the role of the optical field, while the velocity field $\bm{v}$ velocity field plays the role of the ``gradient field''  of the principal valued phase of the optical field, $\bm{g}$.

Consider a 2-dimensional velocity field, $\bm{v}$, with a single branch point located at  $\bm{r}_{\rm BP}$ with a circulation of $\pm 2\pi$ as defined by equation~\eqref{eqn:quantised_vorticity}. We guarantee that the circulation is always $\pm 2 \pi$ and not some other integer multiple by constructing it from the ``principal valued'' phase function. By introducing an auxiliary third dimension, $z$, we can write the circulation integral as the condition,
\begin{equation}
    \hat{\bm{e}}_z \cdot \bm{\nabla} \times \bm{v}(\bm{r}) = \pm 2\pi \delta_{\rm D}(\bm{r}-\bm{r}_{\rm BP})\,.
    \label{eqn:curl_v_BP}
\end{equation}
Performing a Helmholtz decomposition on the velocity field, we obtain the incompressible and irrotational parts of the velocity
\begin{equation}
    \bm{v} = \bm{\nabla} \phi_s + \bm{\nabla} \times \bm{H}\,.
\end{equation}
where now $\phi_s$ and $\bm{H}$ are smooth functions, in contrast to the raw phase which describes this (optical) field, which contains jumps from the branch points.

The scalar potential, $\phi_s$, determines the divergence of the velocity field,
\begin{equation}
    \bm{\nabla} \cdot \bm{v} = \nabla^2 \phi_s\,.
\end{equation}
For our 2-dimensional system, we can write the vector potential as a scalar potential in the auxiliary dimension,
\begin{equation}
    \bm{H}(\bm{r}) = h(\bm{r}) \hat{\bm{e}}_z\,. 
\end{equation}
Since the divergence of $\bm{H}$ vanishes (as it is solenoidal), we can rewrite the condition in equation~\eqref{eqn:curl_v_BP} as
\begin{equation}
    \nabla^2 h(\bm{r}) = \mp 2\pi \delta_{\rm D}(\bm{r}-\bm{r}_{\rm BP})\,,
\end{equation}
which is solved by 
\begin{equation}
    h(\bm{r}) = \mp \log(\abs{\bm{r}-\bm{r}_{\rm BP}})\,.
    \label{eqn:hertz_function}
\end{equation}

To then write velocity field as the gradient of some (principal valued) velocity potential $\bm{v} = \bm{\nabla} \phi$ with some split into a ``smooth'' and a ``hidden'' part, $\phi = \phi_{\rm sm} + \phi_{\rm hid}$, we make the choice that $\phi_{\rm sm}$ should be equal to the scalar potential $\phi_{s}$ from the Helmholtz decomposition. Then, the equation for the hidden phase
\begin{equation}
    \bm{\nabla} \phi_{\rm hid} = \bm{\nabla} \times \bm{H}\,,
\end{equation}
becomes the pair of equations
\begin{subequations}
\begin{align}
    \del_x \phi_{\rm hid}(x,y) &= -\del_y [-h(x,y)]\,, \\
    \del_y \phi_{\rm hid}(x,y) &= \del_x [-h(x,y)]\,,
\end{align}
\end{subequations}
which can be interpreted as the Cauchy-Riemann conditions for a single complex analytic function
\begin{equation}
    f(z = x+iy) = -h(x,y) + i \phi_{\rm hid}(x,y)\,,
\end{equation}
with real part $-h$ and imaginary part $\phi_{\rm hid}$. Using the expression for $h(\bm{r})$ in equation~\eqref{eqn:hertz_function}, the hidden phase for a single branch point is
\begin{equation}
    \phi_{\rm hid}(\bm{r}) = \Im\{\pm\log\left[(x-x_{\rm BP} + i (y-y_{\rm BP})\right]\}.
\end{equation}

If instead the system contains multiple branch points, each will contribute a term of this form to the overall hidden phase. Since we are concerned with irrotational systems, these branch points must be created in pairs such that the number of positive and negative branch points is equal. Consider $N$ pairs of branch points, with the $i^{\rm th}$ point associated with circulation $\pm 2 \pi$ located at $\bm{r}_i^\pm = (x_i^\pm, y_i^\pm)$. The total hidden phase from these branch points is then
\begin{equation}
    \phi_{\rm hid}(\bm{r}) = \Im\left\{\log\left[\frac{\prod\limits_{n=1}^{N}(x-x_n^+) + i (y-y_n^+)}{\prod\limits_{m=1}^{N}(x-x_m^-) + i (y-y_m^-)} \right]\right\}.
    \label{eqn:hidden_phase}
\end{equation}

\section{Contribution of the fold to the cusp diffraction integral} \label{app:fold_coord_change}

Away from the singular cusp point, the cusp catastrophe unfolds into a pair of fold lines. On lines of constant $C_2<0$, near to the classical caustic line, the integral is dominated by the contribution from the folded part of the phase-space sheet. This section demonstrates how to obtain the contribution from the fold to the cusp integral on these constant $C_2$ lines.
Start with the cusp diffraction integral
\begin{subequations}
\begin{align}
        u_{\rm cusp}(C_1,C_2) &= \sqrt{\frac{\nu}{2\pi}} \int \dd{s} \exp(i\nu\zeta_{\rm cusp}(s;C_1,C_2))\,, \\
        \zeta_{\rm cusp}(s;C_1,C_2) &= \frac{s^4}{4} + C_2\frac{s^2}{2} + C_1 s\,.
\end{align}
\end{subequations}
Consider a horizontal slice through this integral at constant $C_2<0$. We want to evaluate the contribution along this line near the caustic line described by equation~\eqref{eqn:diff_cusp_caustic_condition}. On the actual caustic line, there are 2 stationary points of $\zeta$, one single and one double root. We are interested in the double root, which produces the fold. The stationary points along the caustic line satisfy:
\begin{equation}
    s_*^3 + C_2 s_* + \left(\frac{4}{27}\abs{C_2}^{3}\right)^{1/2} = 0\,,
\end{equation}
where we have written $C_1 = C_1(C_2)$ using equation~\eqref{eqn:diff_cusp_caustic_condition}. This has two solutions in $s$, 
\begin{equation}\label{eqn:s1_s2_appendix}
    s_* = \sqrt{\frac{-C_2}{3}}, -2\sqrt{\frac{-C_2}{3}}\,.
\end{equation}
The positive stationary point is the double root if $C_1>0$, and the other is if $C_1<0$. Generically, call the double root $s_2$. Change coordinates in the integral $t=s-s_2$, so that the fold occurs at $t=0$. Under this transformation, the cusp integral has the form
\begin{equation}
    u_{\rm cusp} = \sqrt{\frac{\nu}{2\pi}}\int\dd{t} \exp(i\nu\left[\frac{t^4}{4}+At^3 + Bt^2 + C t + D\right]),
\end{equation}
(we will not list explicit forms of all intermediate variables, as this section is simply to show that such a set of transformations can be done). Now we want to remove the $\nu$ dependence in a way which lets us analyse the fold contribution to this integral coming from $t=0$. Under the coordinate change $t=y/(3A\nu)^{1/3}$ the cubic term will become $y^3/3$ resembling the fold integral. The resulting integral is
\begin{align}\label{eqn:cusp_transformed_1}
    u_{\rm cusp} = \sqrt{\frac{\nu}{2\pi}}\frac{1}{(3A\nu)^{1/3}} &\int\dd{y} \exp(i\frac{\nu^{-1/3}}{4(3A)^{4/3}}y^4)\times \nonumber \\ 
    &\exp(i\left[\frac{y^3}{3} + \tilde{B}y^2 + \tilde{C} y + \tilde{D}\right]).
\end{align}

We then notice that the quartic term is suppressed in the $\nu\to \infty$ limit, and we can expand the exponential 
\begin{align}
    \exp(i\frac{\nu^{-1/3}}{4(3A)^{4/3}}y^4) \sim 1 + \order{\nu^{-1/3}}.
\end{align}
So to leading order in $\nu$ we can neglect the quartic term's contribution to the integral in equation~\eqref{eqn:cusp_transformed_1} at $y=0$. Note that this is the same trick used in standard stationary phase analysis, but here we retain the cubic term in the exponent, where in stationary phase we would expand all exponential terms of cubic or higher order and then perform the resulting Gaussian integrals. 

We can turn the resulting equation into the standard fold integral by taking $y = x - \tilde{B}$, which will transform this cubic into a depressed cubic (one without a quadratic term). This results in the contribution
\begin{align}
    \hat{u} = \frac{\nu^{1/6}}{\sqrt{2\pi}(3A)^{1/3}} \int\dd{x} \exp\left(i\left[\frac{x^3}{3} + Ex + F\right]\right),
\end{align}
to the cusp integral. We have written $\hat{u}$ here rather than $u_{\rm cusp}$ to remind us that this is an approximation to the contribution only around $s=s_2$, rather than a full analysis of the cusp integral. This fold contribution $\hat{u}$ is now in the form of the standard fold diffraction integral
\begin{align}
    \hat{u} &= \frac{e^{iF}\nu^{1/6}}{(3A)^{1/3}}u_{\rm fold }(E;\nu=1)\,, \\
    &= \sqrt{2\pi}\frac{e^{iF}}{(3A)^{1/3}}\nu^{1/6}\operatorname{Ai}(E)\,.
\end{align}
The explicit forms of $E$ and $F$ are 
\begin{align}
    E &= \nu^{2/3}\frac{3^{2/3} \left(12 C_{1} s_{2} - C_{2}^{2} + 3 s_{2}^{2} \left(2 C_{2} + s_{2}^{2}\right)\right)}{36 s_{2}^{4/3}}\,, \\
    F &= \nu \frac{ \left(- 18 C_{1} C_{2} s_{2} + C_{2}^{3} + 9 s_{2}^{2} \left(6 C_{1} s_{2} - C_{2}^{2} + C_{2} s_{2}^{2}\right)\right)}{108 s_{2}^{2}}\,.
\end{align}
Notice that $E\sim \nu^{2/3}$, correctly recovering the fringe index for the fold catastrophe.

One could perform a similar analysis on the full wavefunction with cosine initial conditions~\eqref{eqn:psi_for_catastrophe_expansion} on constant time slices post shell crossing. Doing this would result in a set of coordinate relations between $(x,a)$ and the single control parameter of the standard fold integral. Doing this explicitly is significantly messier than the case here, requiring an expansion of cosine around the caustic line described in equation~\eqref{eqn:shell_cross_region}. Additionally, the wavefunction contribution cannot be written in closed form since the solutions  $s_2$ cannot be written down analytically (which was only possible here because $\zeta_{\rm cusp}'=0$ is a cubic equation).

\section{A brief note on $\star$-products }

The treatment of quantum mechanics in phase-space is significantly less well known than the  usual Schr\"odinger or Heisenberg pictures of quantum mechanics.  \textcite{Case_2008_WignerFunctions} is a good introduction to the Wigner function and associated Weyl transform. 

For readers who wish to go deeper into the underlying mathematical structure of quantum mechanics in phase-space, the introduction of Wigner-Weyl quantisation in \textcite{Marino_2021_AdvancedTopics} provides motivation for why the Wigner function has the specific form that it does. It begins by motivating the Weyl transform of a function $W(f)$ as a kind of non-commutative Fourier transform, then finding a binary operation $\star$, called the $\star$-product (pronounced ``star product'')
\begin{equation}
f\star g = f \exp(\frac{i\hbar}{2}\overset{\leftrightarrow}{\Lambda}) g, \quad \overset{\leftrightarrow}{\Lambda} = \overset{\leftarrow}{\del}_x\overset{\rightarrow}{\del}_x - \overset{\leftarrow}{\del}_p \overset{\rightarrow}{\del}_x\,,
\end{equation}
which obeys a kind of ``convolution theorem'' relation with Weyl transforms analogously to ordinary function convolution with Fourier transforms,
\begin{equation}
W(f)W(g) = W(f\star g).
\end{equation}

The Moyal bracket is then defined as the commutator relative to this $\star$-product, rather than ordinary products
\begin{equation}
\{f,g\}_{\rm MB}^{\hbar} = [f,g]_{\star} = f\star g - g \star f.
\end{equation}
The Wigner function in general is defined as the Weyl transform of the density matrix $\hat\rho$, which can encode states with statistical uncertainty as well as quantum uncertainty. For a pure state, this density matrix is $\hat\rho_\psi = \vert \psi \rangle \langle \psi \vert$. The evolution equation from the Heisenberg picture of quantum mechanics
\begin{equation}
i\hbar \pdv{\hat\rho_\psi}{t} = [\mathscr{H},\hat\rho_\psi]\,,
\end{equation}
with a Hamiltonian $\mathscr{H}$ can then be lifted to phase-space by Wigner transforming both sides, leaving
\begin{equation}
i \hbar \pdv{f_W}{t} = [\mathscr{H}, f_W]_{\star} = \{\mathscr{H}, f_W\}_{\rm MB}^\hbar.
\end{equation} 

\textcite{ Curtright.etal._2013_ConciseTreatise} and \textcite{Curtright.Zachos_2011_QuantumMechanics} provide more details and a guide to the literature on this subject.

\section{Deriving the Wigner equation}\label{app:sec:derive_wigner_eq}
This Section makes explicit the derivation of the time evolution of the Wigner distribution and compares it to the time evolution of a phase-space distribution obeying the usual Vlasov equation. This can be done more elegantly using operator notation or the $\star$-algebra formulation, but here we present a more elementary and explicit derivation here.

Begin with the definition of the Wigner distribution function (here using the standard momentum variable which is conjugate to position with respect to coordinate time)
\begin{equation}
f_W(\xx,\pp,t) \defeq \paren{\frac{1}{2\pi\hbar}}^3\int\dd[3]{\bm{y}} e^{i\pp\cdot\bm{y}/\hbar} \psi^*(\xx+\bm{y}/2) \psi(\xx-\bm{y}/2).
\end{equation}
The wavefunctions appearing here obey the Schr\"{o}dinger equation in expanding coordinates
\begin{subequations}
\begin{align}
\pdv{\psi}{t} &= \frac{i\hbar}{2ma^2}\Lap \psi - i \frac{m}{\hbar} V\psi \,, \\
\pdv{\psi^*}{t} &= -\frac{i\hbar}{2ma^2}\Lap \psi^* + i \frac{m}{\hbar} V\psi^*,
\end{align}
\end{subequations}
with some potential $V$.

For the time being, let us rescale $\hbar$ to absorb the factor of $m$. We can then directly calculate the time derivative of the Wigner distribution function (we will also omit arguments of functions where they are not needed for brevity)
\begin{equation}
\del_t f_W = \paren{\frac{1}{2\pi\hbar}}^3\int\dd[3]{\bm{y}} e^{i\bm{p}\cdot\bm{y}/\hbar}  \brac{\psi^*(\bm{x}+\bm{y}/2)\del_t \psi(\bm{x}-\bm{y}/2) + \psi(\bm{x}-\bm{y}/2)\del_t \psi^*(\bm{x}+\bm{y}/2)}\,,
\end{equation}
now apply the Schr\"{o}dinger equation, grouping kinetic terms and potential terms together
\begin{equation}
\del_t f_W = I_T + I_V\,,
\end{equation}
where $I_T$ is the kinetic part of the integral, and $I_V$ is the potential part
\begin{subequations}
\begin{align}
I_T &= \paren{\frac{1}{2\pi\hbar}}^3\frac{i\hbar}{2a^2}\int\dd[3]{\bm{y}} e^{i\bm{p}\cdot\bm{y}/\hbar} \brac{\psi^*(\bm{x}+\bm{y}/2)\del_x^2 \psi(\bm{x}-\bm{y}/2) + \psi(\bm{x}-\bm{y}/2)\del_x^2 \psi^*(\bm{x}+\bm{y}/2)} \,, \\
I_V &= \paren{\frac{1}{2\pi\hbar}}^3\frac{i}{\hbar}\int\dd[3]{\bm{y}} e^{i\bm{p}\cdot\bm{y}/\hbar}  \brac{V(\bm{x}+\bm{y}/2)-V(\bm{x}-\bm{y}/2)}\psi(\bm{x}-\bm{y}/2)\psi^*(\bm{x}+\bm{y}/2)\,,
\end{align}
\end{subequations}

It will prove slightly notationally easier to work with a different form of the Wigner function, where we redefine our integration variable $\bm{y} \mapsto 2\bm{y}$. In these variables we then have that the Wigner function is
\begin{equation}
f_W(\bm{x},\bm{p}) = \paren{\frac{1}{\pi\hbar}}^3\int \dd[3]{\bm{y}} e^{2i\bm{p}\cdot\bm{y}/\hbar}\psi(\bm{x}-\bm{y})\psi^*(\bm{x}+\bm{y}).
\end{equation}

\subsection*{Kinetic term}
For brevity, let $\tilde{\bm{x}}_\pm = \bm{x}\pm\bm{y}$ and functions $F(\tilde{\bm{x}}_\pm) = F_\pm$.

For the kinetic term, we want to massage this into something which looks like the Vlasov term $\bm{p}\cdot \del_{\bm{x}} f_W$. To do this, we not that we want to have just single $x$ derivatives acting on the wave functions, rather than double derivatives. This suggest doing some sort of integration by parts. To do this, we note the useful relation which allows us to swap $x$ and $y$ derivatives
\begin{equation}
\del_x F(x\pm y) = \pm \del_y F(x\pm y).
\end{equation}

Using this identity to replace one of the $\del_x$ in the kinetic term we can then preform integration by parts on $y$.
\begin{align}
I_T &= \frac{i\hbar}{2a^2} \paren{\frac{1}{\pi\hbar}}^3 \int\dd[3]{\bm{y}} e^{2i\bm{p}\cdot\bm{y}/\hbar} \brac{\psi^*_+\del_x^2\psi_- - \psi_-\del_x^2\psi_+}  \\
&= \frac{i\hbar}{2a^2} \paren{\frac{1}{\pi\hbar}}^3 \int\dd[3]{\bm{y}} e^{2i\bm{p}\cdot\bm{y}/\hbar} \brac{-\psi^*_+\del_y\del_x\psi_- - \psi_- \del_y\del_x\psi_+}\,,
\end{align}

We can now integrate by parts with respect to $\bm{y}$. The boundary terms will be of the form $e^{2i \bm{p}\cdot\bm{y}/\hbar}\psi^{(*)}(\tilde{\bm{x}}_\pm)\del_x\psi^{(*)}(\tilde{\bm{x}}_\mp)$ which will vanish at infinity due to the vanishing of both the wavefunction and its derivative. Swapping the derivative term we then obtain by parts (notice the sign change from parts)
\begin{equation}
I_T = \frac{i\hbar}{2a^2} \paren{\frac{1}{\pi\hbar}}^3 \int\dd[3]{\bm{y}} \brac{\del_x\psi^*_+\del_y\paren{e^{2i \bm{p}\cdot\bm{y}/\hbar} \psi_-}+ \del_x\psi_-\del_y \paren{e^{2i \bm{p}\cdot\bm{y}/\hbar} \psi^*_+}}\,,
\end{equation}
\begin{align}
I_T  = \frac{i\hbar}{2a^2} \paren{\frac{1}{\pi\hbar}}^3 \int\dd[3]{\bm{y}} e^{2i \bm{p}\cdot\bm{y}/\hbar} \Bigg\{ &\frac{2i\bm{p}}{\hbar}\paren{\psi_-\del_x\psi^*_+ + \psi^*_+ \del_x\psi_-}  \nonumber \\
&+ \del_x\psi^*_+\del_y\psi_- + \del_x\psi_- \del_y\psi^*_+\Bigg\}. \label{eqn:wigner_kinetic_partial_work}
\end{align}
Notice that the term on the second line of \refeqn{eqn:wigner_kinetic_partial_work} vanishes since $\del_y\psi_- = -\del_x\psi_-$ and $\del_y\psi^*_+ = \del_x\psi^*_+$. We note that the term in the first line of \refeqn{eqn:wigner_kinetic_partial_work} is simply the product  rule on $\psi_-\psi^*_+$. This means we can write the kinetic term as
\begin{equation}
I_T = -\frac{\bm{p}}{a^2} \del_x \brac{\paren{\frac{1}{\pi\hbar}}^3 \int\dd[3]{\bm{y}}e^{2i\bm{p}\cdot\bm{y}/\hbar}\psi_- \psi^*_+}\,,
\end{equation}
which leads us to the final form (restoring factors of $m$)
\begin{equation}
I_T =  -\frac{\bm{p}}{ma^2} \del_x f_W\,, \label{eqn:time_evolve_wigner_kinetic}
\end{equation}
which is precisely of the form of the kinetic term in the Vlasov equation.

\subsection*{Potential term}
For the potential term we will make use of the same form of the Wigner function as above. 
\begin{equation}
I_V = \frac{1}{(\pi\hbar)^3}\frac{i}{\hbar} \int \dd[3]\bm{y} e^{2i\bm{p}\cdot\bm{x}/\hbar}\brac{V(\bm{x}+\bm{y}) - V(\bm{x}-\bm{y})} \psi(\bm{x}-\bm{y}) \psi^*(\bm{x}+\bm{y})
\end{equation}
We're hoping that this turns into something like $V(x)\del_p f_W$, so we need to Taylor expand the difference in potentials to get it entirely in terms of $\bm{x}$. To do this we use the multi-index version of Taylor's theorem (see \refsec{sec:multi-index_notation}). Focusing on the difference in potentials we obtain
\begin{equation}
V(\bm{x}+\bm{y})-V(\bm{x}-\bm{y}) = \sum_{\abs{\alpha}=0}^\infty \frac{1}{\alpha!} \del_x^\alpha V\eval_{\bm{x}}\brac{\bm{y}^\alpha - (-\bm{y})^\alpha}.
\end{equation}
where $\alpha \in \Z_{\geq 0}^3$. Note that when the multi-index has an even magnitude the terms in the Taylor expansion vanish, so we are left summing only odd terms. The potential term is then
\begin{equation}
I_V = \frac{2i}{(\pi\hbar)^3\hbar}\sum_{\abs{k}=0}^\infty \int\dd[3]{\bm{y}} e^{2i\bm{p}\cdot\bm{y}} \frac{\bm{y}^{2k+1}}{(2k+1)!}\del_x^{2k+1}V(\bm{x}) \psi(\bm{x}-\bm{y})\psi^*(\bm{x}+\bm{y}).
\end{equation}

Now we notice that we can write the $\bm{y}^{n}$ terms as successive derivatives of the exponential part with respect to $\bm{p}$. Using the identity
\begin{equation}
\bm{y}^n e^{2i\bm{p}\cdot\bm{y}/\hbar} =\paren{\frac{\hbar}{2i}}^n \del_p^n  e^{2i\bm{p}\cdot\bm{y}/\hbar} \,,
\end{equation}
and noting that $\paren{\frac{\hbar}{2i}}^{2k+1} = \frac{(-1)^k}{i}\paren{\frac{\hbar}{2}}^{2k+1}$, we can write the potential term as
\begin{align}
I_V &= \frac{2}{\hbar} \sum_{\abs{k}=0}^\infty (-1)^k\frac{\del_x^{2k+1}V(\bm{x})}{(2k+1)!}\paren{\frac{\hbar}{2}}^{2k+1}\underbrace{\frac{1}{(\pi\hbar)^3}\int \dd[3]{\bm{y}} \del_p^{2k+1} e^{2i\bm{p}\cdot\bm{y}} \psi(\bm{x}-\bm{y})\psi^*(\bm{x}+\bm{y})}_{\del_p^{2k+1} f_W}  \nonumber \\
&= \frac{2}{\hbar} \sum_{\abs{k}=0}^\infty (-1)^k\frac{\del_x^{2k+1}V(\bm{x})}{(2k+1)!}\paren{\frac{\hbar}{2}}^{2k+1} \del_p^{2k+1} f_W(\bm{x},\bm{p})\,,
\end{align}
Now notice that we are summing an (operator) series of the form $(-1)^k Z^{2k+1}/(2k+1)!$ which is the series for $\sin(Z)$. We can make this clearer by ``peeling off'' the functions $V(\bm{x})$ and $f_W(\bm{x},\bm{p})$. We use $\vec{\del}$ to denote derivatives acting to their right, and $\cev{\del}$ to denote derivatives acting on all functions appearing on their left. Therefore we have
\begin{align}
I_V &= \frac{2}{\hbar} V(\bm{x}) \paren{\sum_{\abs{k}=0}^\infty \frac{(-1)^k}{(2k+1)!} \paren{\frac{\hbar}{2}}^{2k+1} \cev{\del}_x^{2k+1}\vec{\del}_p^{2k+1}}f_W(\bm{x},\bm{p}) \nonumber \\
&= \frac{2}{\hbar} V(\bm{x})\sin\paren{\frac{\hbar}{2}\cev{\del}_x\vec{\del}_p}f_W(\bm{x},\bm{p}) \,, \label{eqn:time_evolve_wigner_potential}
\end{align}

\subsection*{Putting it all together}
The full time evolution of the Wigner distribution function is then the sum of this kinetic and potential term. Using the results from \refeqn{eqn:time_evolve_wigner_kinetic} and \ref{eqn:time_evolve_wigner_potential} we then have 
\begin{align}
\del_t f_W(\bm{x},\bm{p}) &= I_T + I_V \nonumber \\
&= -\frac{\bm{p}}{ma^2}\vec{\del}_x f_W(\bm{x},\bm{p}) + mV(\bm{x}) \cdot \frac{2}{\hbar}\sin\paren{\frac{\hbar}{2}\cev{\del}_x\vec{\del}_p}f_W(\bm{x},\bm{p}).
\end{align}
Note that if one formally expands the sine function, we recover the Vlasov equation to $\order{\hbar^2}$. As a last manipulation, it would be nice to recast this evolution equation in something like a Hamiltonian form with a bracket on phase-space. To do this we need to write something like the Hamiltonian $\frac{p^2}{ma^2}+mV$, and we should expect that the sine of the operators should appear in the first term. We can do this by writing the linear $\bm{p}$ term as
\begin{equation}
\bm{p} = \frac{p^2}{2}\sin(\cev{\del}_p)\,,
\end{equation}
since the lowest order term in the expansion of the $\sin$ gives us the linear term in $\bm{p}$ and all higher derivatives vanish. We therefore have
\begin{equation}
\del_t f_W(\bm{x},\bm{p}) = \frac{p^2}{2ma^2} \sin\paren{-\cev{\del}_p}\vec{\del}_x f_W(\bm{x},\bm{p})+ mV(\bm{x}) \cdot \frac{2}{\hbar}\sin\paren{\frac{\hbar}{2}\cev{\del}_x\vec{\del}_p}f_W(\bm{x},\bm{p}).
\end{equation}
As each $\cev{\del}_p$ acting on $p^2$ gives us 0 beyond leading order, we are free to include the $\vec{\del}_x$ inside the sine function as any higher term will be 0 anyway (NB we can also insert a factor of $\hbar/2$ inside by multiplying by $2/\hbar$ on the outside since only the linear order term is non-zero)
\begin{equation}
\del_t f_W(\bm{x},\bm{p}) = \frac{p^2}{2ma^2} \frac{2}{\hbar}\sin\paren{-\frac{\hbar}{2}\cev{\del}_p\vec{\del}_x} f_W(\bm{x},\bm{p})+ mV(\bm{x}) \cdot \frac{2}{\hbar}\sin\paren{\frac{\hbar}{2}\cev{\del}_x\vec{\del}_p}f_W(\bm{x},\bm{p}).
\end{equation}
Even though sine is not a linear function, we can combine these two terms, because the terms $(\cev{\del}_x\vec{\del}_p - \cev{\del}_p\vec{\del}_x)^n$ for $n\geq 2$ will make all the minus terms which have greater than two $\cev{\del}_p$ terms in it disappear, which leads us to the final form:
\begin{equation}
\del_t f_W(\bm{x},\bm{p}) = \brac{\frac{p^2}{2ma^2}+V(\bm{x})}\frac{2}{\hbar}\sin\paren{\frac{\hbar}{2}(\cev{\del}_x\vec{\del}_p - \cev{\del}_p\vec{\del}_x)} f_W(\bm{x},\bm{p}).
\end{equation}
Defining the \emph{Moyal} bracket for two functions on phase-space 
\begin{equation}
\curly{f(\bm{x},\bm{p}),g(\bm{x},\bm{p})}_{\rm MB}^{\hbar} = \frac{2}{\hbar} f \sin\paren{\frac{\hbar}{2}(\cev{\del}_x\vec{\del}_p-\cev{\del}_p\vec{\del}_x)}g. \label{eqn:Moyal_bracket}
\end{equation}
we can recast this evolution equation as
\begin{equation}
\del_t f_W = \curly{\mathscr{H},f_W}_{\rm MB}^\hbar = \curly{\mathscr{H},f_W}_{\rm PB} + \order{\hbar^2}
\end{equation}
where $\curly{\cdot, \cdot}_{\rm PB}$ is the standard Poisson bracket.

There is a whole branch of mathematics which deals with the Moyal bracket and how unique such structures are on Poissonian manifolds called \emph{deformation quantisation}. For the purposes of this thesis however, the Moyal bracket in \refeqn{eqn:Moyal_bracket} is the only one which we will consider.

\chapter{Appendix for Chapter \ref{chap:how-classical}}
\section{Effect of larger $\hbar_{\rm PPT}$}\label{app:sec:largerhbar}

\begin{figure}[h!t]
    \centering
    \includegraphics[width=\textwidth]{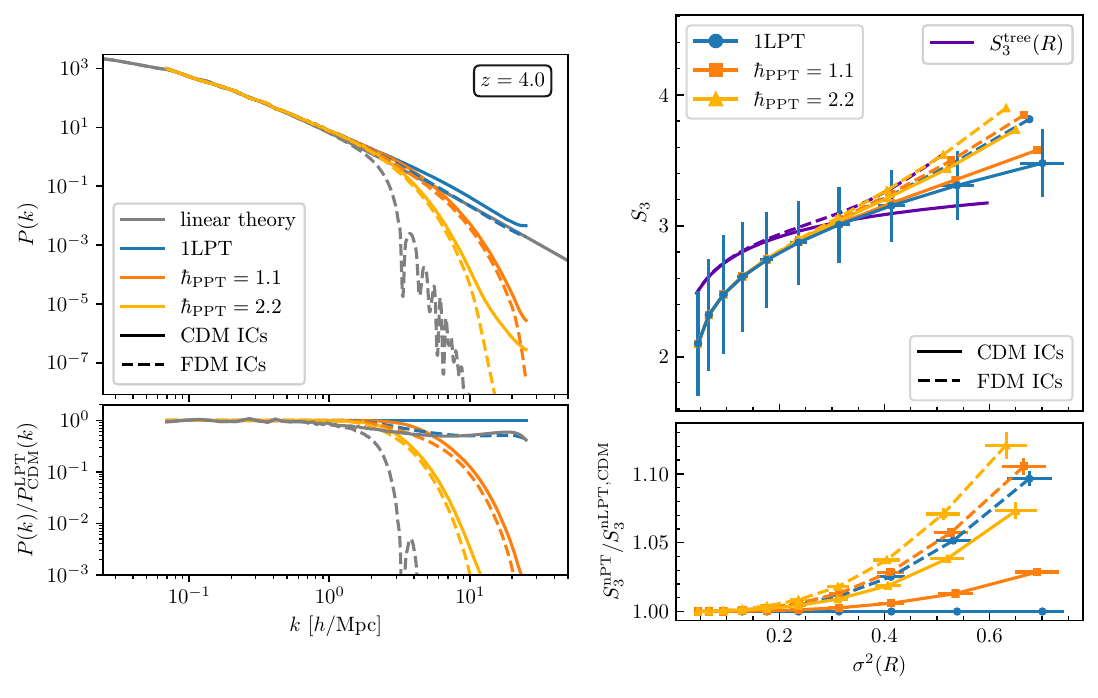}
    \caption[Matter power spectrum and reduced skewness with two values of $\hbar_{\rm PPT}$.]{(Left panel) Same as Figure~\ref{fig:Pk-z4} but with addtional $\hbar_{\rm PPT}$ which is twice the size of the one used in the main analysis. (Right panel) Same as Figure~\ref{fig:S3_1column} with the additional $\hbar_{\rm PPT}$. The larger value of $\hbar_{\rm PPT}=2.2 \ h^{-2} \ \rm Mpc^{2}$ is able to produce the skewness of the LPT + FDM ICs case even on cold initial conditions.}
    \label{fig:S3_1column_2hbars}
\end{figure}

Figure \ref{fig:S3_1column_2hbars} shows the power spectrum and reduced skewness as measured from density fields produced using 1LPT/1PPT via \texttt{monofonIC} as described in Chapter \ref{chap:how-classical}. Using a value of $\hbar_{\rm PPT} = 2.2 \, h^{-2} \ \rm Mpc^2$ (twice the value used in the main analysis of Chapter \ref{chap:how-classical}), it is possible to produce skewness enhancements similar to those due to changing the initial conditions through wave dynamics alone (compare the yellow/triangle solid line to the blue/circle or orange/square dashed lines). This indicates more need to understand the appropriate mapping between $\hbar_{\rm PPT}$ and FDM effective mass for setting initial conditions. However, even in this large $\hbar_{\rm PPT}$ case, we see that suppressing the initial conditions does continue to enhance the skewness similarly to in the cases considered in the main Chapter.

